\documentclass[a4paper,11pt]{article}
\usepackage{fullpage}
\usepackage{bm}
\usepackage[utf8]{inputenc}
\usepackage{amsmath,mathtools}
\usepackage{amsfonts}
\usepackage{booktabs}
\usepackage{graphics}
\usepackage{xcolor}
\usepackage{algpseudocode}
\usepackage{algorithm}
\usepackage[colorlinks]{hyperref}
\usepackage{multirow}
\usepackage{caption}
\usepackage{subcaption}
\usepackage{adjustbox}
\usepackage{natbib}
\hypersetup{colorlinks,linkcolor={black},citecolor={black}}
\usepackage{verbatim}

\usepackage{enumitem}
\usepackage{setspace}
\usepackage{graphicx} % Required for inserting images
\usepackage{appendix}
\usepackage{pdflscape}  

\newlist{steps}{enumerate}{1}
\setlist[steps, 1]{label = Step \arabic*:}
\def\blue #1{{\color{blue}{#1}}}

\date{\today}

\newcommand{\bX}{{\bm X}}
\newcommand{\bZ}{{\bm Z}}
\newcommand{\mcO}{\mathcal{O}}
\newcommand{\bbeta}{\boldsymbol{\beta}}
\newcommand{\btheta}{\boldsymbol{\theta}}

\newcommand{\bgamma}{\boldsymbol{\gamma}}

\begin{document}
%\maketitle

{ \centering{\bf \Large
 Single-index  Semiparametric Transformation  Cure Models with Interval-censored Data}\\
}

\vskip 7mm
{\centering {\bf Xiaoru Huang$^1$, Tonghui Yu$^2$, Xiaoyu Liu$^{1*}$}
\\}
{\centering {$^1$ School of Economics, Jinan University, Guangzhou, China }\\
	$^2$ International Center for Interdisciplinary Statistics, School of Mathematics, Harbin Institute of Technology, Harbin, China\\}

\let\thefootnote\relax\footnotetext{* Corresponding author. %Tel.: .... ;  Fax: +86 .....\\
\\
\indent {\it E-mail address: xyliu0075@jnu.edu.cn} .}

\begin{abstract}
 %In this paper, we consider a class of flexible single-index semiparametric transformation cure models for the analysis of interval-censored data. 
Interval-censored data commonly arise in medical studies when the event time of interest is %observed within specific time intervals.
only known to lie within an interval.
In the presence of a cure subgroup, 
%In the context of cure models with interval-censored data, 
conventional mixture cure
models typically assume a logistic model for the uncure probability  
and a proportional hazards model %traditionalsurvival function 
for the susceptible subjects. However, in practice,
%In practical applications, 
the assumptions of parametric form for the uncure probability
and the proportional hazards model for the susceptible may not always be satisfied.
In this paper, we propose a class of flexible single-index semiparametric transformation cure
models for interval-censored data, where a single-index model and a semiparametric transformation model are utilized for the uncured and conditional survival probability, respectively,
%a single-indexmodel for the probability of being uncured, alongside a class of semiparametric transformation models that assess the survival distribution of susceptible subjects. 
%This methodology circumvents the possible curse of dimensionality associated with incidence and offers a class of more flexible models compared to traditional logistic-based cure models, 
encompassing both the proportional hazards cure and
proportional odds cure models as specific cases. We approximate the single-index function
and cumulative baseline hazard functions via the kernel technique and splines, respectively,
and develop a computationally feasible expectation-maximisation (EM) algorithm, facilitated by a four-layer gamma-frailty {Poisson} %passion 
data augmentation. %The large-scale 
Simulation studies demonstrate the satisfactory performance of our proposed method, compared to the {spline-based approach and the
classical logistic-based mixture cure models.} The application of the proposed methodology
is illustrated using the Alzheimer's dataset.
 
%The numerical studies and The application of real data illustrates the satisfactory performance of the proposed approach.   
\end{abstract}


\section{Introduction}
% Outline of introduction
% \begin{itemize}
%     \item brief introduces the data structure of interest. It may be better to start with a motivating example(interval-censored data+ cure fraction)
%     \item existing research on cure models with interval-censored data, the drawback of the existing research.
%     \item existing research regarding the flexible mixture cure models(under right and interval-censored), discuss the latency and incidence
%     \item The contribution of the this work(what have you done)
%     \item organization of this paper.
% \end{itemize}

%The major drawback of existing cure rate models %is that they are limited to parametric and semiparametric models.More specifically,

%In many clinical trials, a proportion of subjects may be cured (insusceptible) to the event of interest or disease relapse. %such as oncology research.

The conventional survival analysis assumes that all study subjects are susceptible to the event of interest. However, in many clinical trials, a proportion of subjects may be cured %and thus become risk-free 
of a specific event. For instance, in the progression of Alzheimer’s disease (AD), a fraction of the patients diagnosed with mild cognitive impairment (MCI), which is recognized as a transitional stage, may never convert to AD. 
%Alzheimer’s disease (AD), an irreversible and progressive neurodegenerative disease, is the most common cause of dementia. Mild cognitive impairment (MCI) serves as a transitional stage in the progression of AD, and individuals diagnosed with MCI encounter an elevated risk of developing the disease\citep{langa2014diagnosis}. Identifying the risk factors associated with the duration time from MCI to the onset of AD is essential for early prognosis and proper clinical intervention. However, a proportion of MCI patients may never convert to AD;
%Some MCI patients may remain stable or even improve cognitive function over time. 
Identifying the risk factors associated with the duration time from MCI to the onset of AD is essential for early prognosis and proper clinical intervention\citep{oulhaj2009predicting,scolas2016variable}. Moreover, %in the Alzheimer’s Disease Neuroimaging Initiative (ADNI) research,
the exact AD conversion time is typically known to have occurred between two consecutive visits since the 
%an additional feature of these data is that 
patients underwent regular follow-up assessments to monitor the onset of AD \citep{scolas2016variable}. %, and the exact conversion time was recorded between two consecutive visits without being observed exactly \citep{scolas2016variable}.%,weiner2017alzheimer}.%,weiner2017recent}. To handle such data, it is
Disregarding these features of survival data may lead to biases, and a more accurate approach that incorporates the cure fraction with interval censoring is required.

 Literature on cure survival data mainly consists of two classes of models: the mixture cure model \citep{boag1949maximum} and the proportional time cure model \citep{tsodikov1996stochastic}. %(or non-mixture cure model). 
%Under the promotion time cure model, much research has been conducted on methods for analyzing interval-censored data, for example, \cite{liu2009semiparametric, hu2013efficient, lam2013semiparametric}, among others. Under the mixture cure model, there is extensive literature on this approach for interval-censored data \citep{ma2009cure,wang2012two,zhou2018computationally,chen2019semiparametric,liu2021generalized}. 
%Among them, %the mixture cure model assumes that the survival function of population is a mixture of two components:
The mixture cure model, which posits that a population's survival function comprises two components: the survival function of susceptible subjects (latency) and the probability of being uncured (incidence), 
%It allows for the direct interpretation of risk factors both on the cured and uncured individuals and 
has been investigated intensively for right censored data \citep{kuk1992mixture,sy2000estimation,peng2000nonparametric,lu2004semiparametric,fang2005maximum,mao2010semiparametric}.
%Although the proportional time cure maintains the proportional hazards structure, the mixture cure model allows for the direct interpretation of risk factors both on the incidence and latency components, thus has been investigated intensively. 
In the context of interval censoring, most of the literature concentrates on assessing the impact of covariates on the conditional survival function of susceptible subjects using conventional survival models. These include the %\citep{kim2008cure,ma2009cure,ma2010mixed,zhou2018computationally,chen2019semiparametric, liu2021generalizend}.For example, the Cox
 proportional hazards (PH) model \citep{kim2008cure,ma2009cure,ma2010mixed}, additive hazard model \citep{wang2021additive}, accelerated failure time models \citep{lam2005semiparametric}, {generalized odds rate models \citep{zhou2018computationally}, semiparametric transformation models \citep{chen2019semiparametric} and generalized accelerated failure time models \citep{liu2021generalized}.}
 % and numerous survival models  \citep{zhou2018computationally,chen2019semiparametric,liu2021generalized}. % semiparametric transformation models \citep{chen2019semiparametric}, and generalized accelerated failure time models \citep{liu2021generalized} have been studied extensively for in the literature. 
While various survival models have been employed to analyze the latency component, relatively few studies have explored the incidence.
%The mixture cure model comprises two sub-models: one representing the probability of notbeing cured (incidence), and the other representing the conditional survival function for susceptible individuals (latency). 

%As discussed by \cite{banerjee2004parametric}, one disadvantage of the proportional time cure model in comparison with the mixture cure model is that the proportional time cure cannot distinguish between the long-term (related to cure probability) and short-term (related to conditional survival function) effects of covariates. Thus the mixture cure model is the most commonly used method.
%The association of the risk factors with the probability of being uncured is usually modelled via the logistic regression or log-log link function. For example, {\color{red} add citation}

For the incidence, the association of the risk factors with the probability of being uncured 
is usually modelled via the logistic regression or log-log link function\citep{kuk1992mixture,
sy2000estimation,lam2005estimating}. 
While the logistic model is easy to implement and allows for straightforward interpretation, its rigorous parametric assumptions may be inappropriate for modelling uncured probabilities in some practical contexts. To relax the parametric assumption and enhance flexibility, nonparametric and semiparametric models have been explored. \cite{wang2012two} considered the nonparametric smoothing spline analysis of variance (ANOVA) model, which was extended by \cite{xu2014nonparametric} and \cite{lopez2017nonparametric}. \cite{xie2021mixture} and \cite{aselisewine2023integration} investigated machine learning techniques for incidence modelling, including neural network estimators and decision tree-based classifiers. Most existing literature has focused primarily on right-censored data, with limited research available on interval-censored data.
%\cite{wang2012two} proposes a nonparametric smoothing splines approach for both the cure probability and the hazard rate function. %\cite{aselisewine2023integration} model the cure probability via a decision tree-based classifier, and develop an expectation-maximization algorithm to estimate the model parameters.ome nonparametric methods have been proposed for cure survival data, %such as the nonparametric smoothing spline analysis of variance (SS ANOVA) models \cite{wang2012two} proposed a nonparametric smoothing spline analysis of variance (ANOVA) model, which has been extended to incorporate a continuous covariate and both discrete and continuous covariates by \cite{xu2014nonparametric} and \cite{lopez2017nonparametric}, respectively. More recently,  machine learning approaches have been employed for the incidence, including the neural network estimator \citep{xie2021mixture} and decision tree-based classifier \citep{aselisewine2023integration}.

The fully nonparametric models mentioned above lack interpretability and may encounter the “curse of dimensionality” issue when the dimension of the covariates is large. Therefore, a more flexible semiparametric single-index model is required. 
%A possible alternative for the incidence is the single-index model, 
The model is computationally tractable with numerous covariates and allows for assessing the relative importance of covariates by examining the absolute values of their coefficients. The single-index approach has been investigated in the context of survival data \citep{lu2006class,sun2008polynomial,gorst2013independent}.
%{\color{green}\cite{lu2006class} defined a class of “partially linear single-index" survival models that are more flexible in dealing with covariates than classical proportional hazards regression models. \cite{sun2008polynomial} proposed a partially linear single-index proportional risk regression model that simulates both linear and nonlinear covariates of logarithmic risks in proportional risk models. \cite{gorst2013independent} proposed a computationally efficient independent screening method for survival data, investigating its sure screening property within single-index hazard rate models with ultrahigh dimensions, and evaluating it through simulations and real gene expression data.}
%{\color{red}add reference, especially for survival analysis. \color{green}done}
%To avoid the curse of dimensionality issue appearing in the fully nonparametric model and provide interpretability, 
Recently, \cite{amico2019single} introduced the single-index model for incidence, assuming a proportional hazards model on the latency component with right-censored data. Extending this work, %in the context of right censoring. 
\cite{lee2024partly} proposed two single-index modelling frameworks for both the incidence and latency components %for right censored data,
retaining the proportional hazards assumption for the latter. % constraint applied to the latency component. 
However, these methodologies are limited to right-censored data. % and cannot handle interval-censored observations. 
%Compared with right-censored data,
{Unlike right-censoring, interval-censored data, where the exact event time is only known to lie within an interval, 
 inherently provides less information and introduces both theoretical and computational challenges \citep{sun2006statistical}. 
For instance, %with right-censored data
in right-censored settings,
the partial likelihood function (which eliminates the baseline hazard) and the martingale process theory are standard tools for regression parameter estimation and inference. Such approaches, however, are generally inapplicable to interval-censored data.}
Furthermore, the imposition of a proportional hazards constraint on the latency may limit the model's flexibility and practical applicability.

In this paper, we developed a flexible single-index semiparametric transformation mixture cure model for interval-censored data (SMCI) to address potential non-monotonic and non-logistic patterns in the incidence while providing a more general survival function for the latency. The kernel smoothing approach is employed to estimate the unknown link function of the incidence, and the spline technique is utilized to approximate the unspecified, increasing function within the transformation model in the latency. Conditional on the frailty variable induced by a transformation model, we propose a four-layer data augmentation procedure and develop an expectation-maximization (EM) algorithm for parameter estimation. For comparison, we also consider the splines approach for the incidence. Additionally, we provide results about the 
model identifiability.

%In this paper, we developed a single-index mixture cure model for interval-censored data(SMCI) with a single-index model for incidence probability and a more flexible semiparametric transformation model for the progression of the event time of the uncured subjects. We derive the MLE for all parameters and baseline unknown function.In the derivation process, we use a Nadaraya-Watson kernel estimator to estimate the link function of incidence and employ the monotone splines technique to approximate the unspecified baseline cumulative hazard function in the transformation model. A complete-data likelihood function is constructed by treating the uncured indicator, the frailty variable induced by a transformation model, and a sequence of Poisson random variables under the four-layer data augmentation method as missing data. Based on this approach, we develop an EM-type algorithm. The single-index structure relaxes the parametric logistic assumption and avoids curse-of-dimensionality problems. Also, the proposed method provides a flexible framework for modeling the failure times of the susceptible subjects, including PH cure, PO cure, and GORMC cure models as its special cases.}

The remainder of the paper is structured as follows. Section \ref{sec2} presents a single-index mixture cure model for interval-censored data and discusses the model identifiability. Section \ref{sec3} introduces the four-layer data augmentation procedure and the detailed EM algorithm. 
%estimation procedure, and provides a detailed introduction to the proposed maximum likelihood estimation process and its corresponding implementation algorithm.
The finite sample performance and real data analysis are illustrated in Section \ref{sec4} and \ref{sec5}, respectively. %introduce simulation studies and analysis of ADNI data. Finally, 
Section \ref{sec6} provides a summary and discussion of the entire text. All technical proofs are provided in the \hyperref[Appendix]{Appendix}.

\section{Data, Model and Likelihood}\label{sec2}
\subsection{Model and Likelihood }
%{\color{red} Describe the data structure and notation first}

%{\color{blue}For individual $i$, where $i = 1,\dots, n$, let $T_i$ be a nonnegative random variable representing the event time of interest, and $(V_{i1}, \ldots, V_{i,K_i})$ the random observation times. Write $\Tilde{V}=(V_{i0},V_{i1}, \ldots, V_{i,K_i}, V_{i,K_i+1})$, where $V_{i0}=0$ and $V_{i,K_i+1}=\infty$.Under mixed case interval censoring, only the time interval $(L_i, R_i)$ are available, where $L_i=\max\{V_{ik}:V_{ik}<T_i\}$ and $R_i=\min\{V_{ik}: V_{ik}\ge T_i\}$. Note that $L_i=0$ if the subject is left-censored,  and $R_i=\infty$ if the subject is right-censored. Define interval censoring indicators as $\delta_{L_i}=\mathbf{1}(L_i=0)$, $\delta_{I_i}=\mathbf{1}(0<L_i< R_i<\infty)$ and  $\delta_{R_i}=\mathbf{1}(R_i=\infty)$, where $\mathbf{1}(\cdot)$ is the indicator function, and it is stipulated that $\delta_{R_i}+\delta_{I_i}+\delta_{L_i}=1$. To account for the presence of the cure subgroup, we introduce the latent cure indicator $B_i=\mathbf{1}(T_i<\infty)$, with $B_i= 1$ if the individual is susceptible, and 0 otherwise. }
Let $T$ be a nonnegative random variable representing the event time of interest. To account for the presence of the cure subgroup in the population, we introduce the latent cure indicator $B$, with $B= 1$ if the individual is susceptible, and 0 otherwise. {%G cure indicators, 
The survival time $T$ can then be expressed as %of the population is given by 
$T= (1-B)\infty +B\times T^*$, where $T^*$ is the survival function of susceptible subjects.
{Under mixed case interval censoring, denote $(V_1, \ldots, V_K)$ the random sequence of observation times, with $K$ being the random number of observation times.
Write $\Tilde{V}=(V_{0},V_{1}, \ldots, V_{K}, V_{K+1})$, where $V_{0}=0$ and $V_{K+1}=\infty$.}
%For interval censored data, only the time interval $(L, R)$ are available, 
Assume $(L, R)$ be the smallest interval that contains $T$, i.e., $L=\max\{V_{k}:V_{k}<T\}$ and $R=\min\{V_{k}: V_{k}\ge T\}$. Note that $L=0$ if the subject is left-censored,  and $R=\infty$ if the subject is right-censored. Define interval censoring indicators as $\delta_{L}=\mathbf{1}(L=0)$, $\delta_{I}=\mathbf{1}(0<L< R<\infty)$ and  $\delta_{R}=\mathbf{1}(R=\infty)$, where $\mathbf{1}(\cdot)$ is the indicator function. % with the value of 1 representing that $T_i$  is left-, interval- and right-censored, respectively, and 0 otherwise. 
%To account for the presence of the cure subgroup, we introduce the latent cure indicator $B$, with $B= 1$ if the individual is susceptible, and 0 otherwise. 
Let $\bX$ and $\bZ$ be $d_1$-dimensional and $d_2$- dimensional vectors of covariates associated with the cure fraction and the event occurrence time of a susceptible group, respectively. Note that $\bX$ and $\bZ$  may overlap or differ. 
%{\color{red}In a failure time study where there may be a cured or non-susceptible subgroup.} Let $T$ denote the failure time of interest and $B$ be the latent binary variable indicating the cure status, where $B =1$ represents the subject is susceptible and 0 otherwise. In the context of the mixture cure model framework, the failure time can be decomposed as $T= BT^*+ (1-B)\infty$, where $T^*< \infty$ denotes the failure time of the susceptible subject. Assume $X$ and $Z$ are two covariate vectors that describe the covariate effects on the susceptible subjects and cure probability.

%{With cure indicators, the survival time of the population is given by $T= (1-B)\infty +B\times T^*$.}
Under the mixture cure models, the (improper) conditional survival function of %the population 
$T$ is defined as
%To describe the covariate effects on survival time, we assume the overall survival function of $T$ under the mixture cure model is given by
\begin{equation}
    S(t|\bX,\bZ)=1-p(\bX)+p(\bX)S_u(t|\bZ),%(t\ge 0),
    \label{S_p}
\end{equation}
where the incidence component $p(\bX)$ denotes the uncure probability, % of being susceptible, 
and the latency component $S_u(\cdot|\bZ)$ represents the survival function of uncured groups.
%$S_u(t|Z)=P(T^*>t|B=1,Z)=\exp(-\Lambda(t|Z))$ is the conditional survival function for the uncured subjects and  $p(X)$ is the uncured probability of subjects.

In the latency part, we assume the semiparametric transformation model for individuals susceptible to the event, that is 
\begin{equation}
 %S_u(t|\bZ)= \exp\Big\{-G[\int_0^t\exp(\bbeta^{\top}\bZ(s))\,{\rm d}\Lambda(s)]\Big\},
  S_u(t|\bZ)=\exp\left[-G\left\{\exp(\bbeta^{\top}\bZ)\Lambda(t)\right\}\right],
 \label{S_u}
\end{equation}
where $\bbeta$ is a vector of regression parameters, $\Lambda(\cdot)$ is an unknown increasing function with $\Lambda(0)=0$ and $ G(\cdot)$  is a specific monotone transformation function. Model (\ref{S_u}) is a class of flexible semiparametric models, including the linear transformation model \citep{chen2002semiparametric} and the generalized odds rate model \citep{zhou2018computationally}. %\xy{add the motivation for frailty-induced model}
We consider a class of frailty-induced transformation function for $ G(x)$, that is $ G(x)$ $=-\log\left\{\int_0^{\infty}\exp(-x\xi)f(\xi)\,{\rm d}\xi\right\},$
% The cumulative hazard function takes the form
% \begin{equation}
%     \Lambda(t|Z)=G[\int_0^t\exp\{\beta^{\top}Z\}\,{\rm d}\Lambda(s)],\label{e3}
% \end{equation}
% where $\bbeta$ is vector of regression parameters, $\Lambda(\cdot)$ is the cumulative baseline hazard function %an unknown increasing function 
% and $G(\cdot)$ is a specific transformation function that is thrice continuously differentiable and strictly increasing with $G(0)=0$, $G'(0)>0$ and $G(\infty)=\infty$\citep{zeng2006efficient}. The choices of $G(x)=x$ and $G(x)=\log(1+x)$ yield the proportional hazards and proportional odds models, respectively.
%We consider a class of frailty-induced transformations  
%\begin{equation}
%$   G(x)=-\log\left\{\int_0^{\infty}\exp(-x\xi)f(\xi)\,{\rm d}\xi\right\},$
% \label{frailty-induced}
%\end{equation}
 where $f(\xi)$ is the density function of the frailty $\xi$. 
 %\blue{This representation is useful for devising the EM algorithm, as it treats $\xi$ as an unobserved random variable, enabling iterative estimation by alternating between estimating the expectation of the frailty given current parameters and maximizing the complete-data likelihood.}
 % with support $[0,\infty)$. 
%with
The choice of the gamma density  with mean 1 and variance $r$ for $f(\xi)$ yields the class of logarithmic transformations $G(x) = r^{-1}\log(1 + r x)$ with $r \ge 0$,
%By assuming $\xi$ follows a $\textit{Gamma}(1,r)$ distribution, we obtain $G(x) = r^{-1}\log(1 + r x)$, %where $r \ge 0$. This formulation 
%allows Model (\ref{S_u}) to encompass
which includes the proportional hazards model ($r=0$), and the proportional odds model ($r=1$) as special cases \citep{zeng2006efficient,zeng2016maximum}.
%The choice of the gamma density with unit mean and variance $r$ for $f(\xi)$ yields the class of logarithmic transformations, $G(x) = r^{-1}\log(1 + r x) (r \ge 0)$,
%in which $r=0$ and $r=1$ correspond to the proportional hazards and proportional odds models, respectively\citep{zeng2006efficient,zeng2016maximum}. 
%{\color{red} Discuss the relationship between the linear transformation and semiparatric transformation model}
%\blue{Semiparametric linear transformation models include proportional hazards and proportional odds models, but such estimators are computationally demanding and statistically inefficient and can not accommodate time-dependent covariates. While a broad class of transformation models proposed by \cite{zeng2006efficient,zeng2016maximum} allows time-dependent covariates and also includes the proportional hazards and proportional odds models as special cases. }

In the incidence part,
%{\color{red}Apparently, the covariates may} also have certain effects on $B$,
we consider a {semiparametric} single-index model in terms of the susceptible probability %for the incidence component, that is
\begin{equation}
    p(\bX)=%Pr(B=1|X)=
    g(\bgamma^{\top}\bX),
    \label{pi}
\end{equation}
where $\bgamma=(\gamma_1,\ldots,\gamma_{d_1})^{\top}$ is a vector of regression coefficients and $g(\cdot)$ is an unknown smooth link function (monotone or non-monotone) %. The link function $g(\cdot)$ can be any (smooth) function 
with values between 0 and 1. %The choice of $g(x)=x$ and $g(x)=\log [\exp \{\exp (x)\}-1]$ yields the logistic regression model and the complementary log-log model, respectively. 
%{\color{red}we usually using other expression. Disscuss why  we consider the single-index model here, like what king  of model we usually take, and why we not take this commmon model, refer to the paper 2018}
Model (\ref{pi}) provides more flexibility 
compared to conventional parametric models. %and avoids the curse of dimensionality associated with nonparametric models.
Specifically, it allows us to explore the various cure frameworks for incidence, including the logistic regression and complementary log-log model defined by $g(x)=\{1+\exp{(-x)}\}^{-1}$ and $g(x)=\{1-\exp{\{-\exp(x)\}}\}^{-1}$, respectively \citep{%farewell1986mixture, kuk1992mixture,
sy2000estimation,lam2005estimating}. With $\bm X$ being univariate, model (\ref{pi}) reduces to the nonparametric model as described by \cite{xu2014nonparametric} and \cite{lopez2017nonparametric}. 
Furthermore, it avoids the curse of dimensionality associated with completely nonparametric models while maintaining ease of interpretation through its structured approach \citep{horowitz2009semiparametric}.

Assume that $T$ and ($\Tilde{V},K)$ 
%the examination times $V_{ij}$'s and the event time $T_i$ 
are independent conditional on covariates $(\bX,\bZ)$. A random sample of size $n$ comprises
$\mathcal{O}=\{(L_i,R_i,\delta_{L_i},\delta_{I_i},\delta_{R_i},\bX_i,\bZ_i),(i=1,\dots,n)\}$. %{\color{red} correct the notation below}
{Let $\btheta = (\bgamma,\bbeta,g, \Lambda)^\top$} be the set of unknown parameters, then
the likelihood function takes the form 
\begin{equation*}
    \begin{aligned}
L(\btheta|\mathcal{O})&=\prod_{i=1}^n[1-S(R_i|\bX_i,\bZ_i)]^{\delta_{L_i}}[S(L_i|\bX_i,\bZ_i)-S(R_i|\bX_i,\bZ_i)]^{\delta_{I_i}}S(L_i|\bX_i,\bZ_i)^{\delta_{R_i}}.\\
 %    &=\prod_{i=1}^np(\bX_i)^{1-\delta_{R_i}}[1-S_u(R_i|\bZ_i)]^{\delta_{L_i}}[S_u(L_i|\bZ_i)-S_u(R_i|\bZ_i)]^{\delta_{I_i}}[1-p(\bX_i)+p(\bX_i)S_u(L_i|\bZ_i)]^{\delta_{R_i}}.
    \end{aligned}\label{e4}
\end{equation*}

%For the class of frailty-induced transformations described in (\ref{frailty-induced}), (\ref{e4}) can be written as 
{By substituting $S(t|\bX, \bZ )$ with (\ref{S_p}) and
plugging (\ref{S_u}) into the likelihood function,} we have
\begin{equation}
    \begin{aligned}
L(\btheta|\mathcal{O})=&\prod_{i=1}^np(\bX_i)^{1-\delta_{R_i}}\int_{\xi_i}\big[1-\exp\{-\xi_i\exp(\bbeta^{\top}\bZ_i)\Lambda(R_i)\}\big]^{\delta_{L_i}}\\
&\times\big[\exp\{-\xi_i\exp(\bbeta^{\top}\bZ_i)\Lambda(L_i)\}-\exp\{-\xi_i\exp(\bbeta^{\top}\bZ_i)\Lambda(R_i)\}\big]^{\delta_{I_i}}\\
&\times\big[1-p(\bX_i)+p(\bX_i)\exp\{-\xi_i\exp(\bbeta^{\top}\bZ_i)\Lambda(L_i)\}\big]^{\delta_{R_i}}
        f(\xi_i)\,{\rm d}\xi_i.
    \end{aligned}
    \label{L_obs}
\end{equation}

\subsection{Identifiability of the Model}
%{\color{red}An important issue for both the single-index and the mixture cure model is identifiability meaning that}
%\begin{equation*}
 %   \begin{aligned}
  %      L(L,R,\delta,X,Z;g,\gamma,\beta,\Lambda)=L(L,R,\delta,X,Z;\Tilde{g},\Tilde{\gamma},\Tilde{\beta},\Tilde{\Lambda}) a.s.\implies g=\Tilde{g},\gamma=\Tilde{\gamma},\beta=\Tilde{\beta},\Lambda=\Tilde{\Lambda},
%    \end{aligned}
%\end{equation*}
%where {\color{red}$L(L,R,\delta,X,Z;g,\gamma,\beta,\Lambda)$} denotes the likelihood of the model given the parameters. {\color{red}In practice, the examination times are bounded. This makes it impossible for us to observe cured subjects in the data. The solution to this problem is to assume the existence of a so-called "cure threshold" $\alpha<\infty$,  such that $T >\alpha$ implies that $T =\infty $ \citep{taylor1995semi}.}
%To ensure the identification of the model, we require the following set of assumptions
%An important issue for the single-index mixture cure model is identifiability.
{Let $\bgamma_0$, $\bbeta_0$, $g_0$, and $\Lambda_0$ % and $G_0$ 
be the true parameter values. Denote by $\mathcal{X}$ and $\mathcal{Z}$  %denotes 
the support of $\bX$ and  $\bZ$, respectively. % and define %the index $\mathcal{D}_X$ and $\mathcal{D}_Z$ the supports of $\bgamma_0^\top \bX$ and $\bbeta_0^\top \bZ$.
%=\{\bgamma^\top \bX: \bX \in \mathcal{X}\}$ denotes the support of the index $\bgamma^\top\bX$. Similarly, $\mathcal{Z}$ denotes the support of $\bZ$ and $\mathcal{D}_Z=\{\bbeta^\top \bZ: \bZ \in \mathcal{Z}\}$ denotes the support of the index $\bbeta^\top\bZ$.
We impose the following conditions for model identifiability
%Let $\mathcal{S}=\Bbb{R}^{d_1}\times\Bbb{R}^{d_2}$ be the support of $(\bX,\bZ)$, and }
%To ensure the identification of the model, for a continuous random vector $\bX$, we require the following set of assumptions.
\begin{enumerate}
\item [(A1)] The function  $g_0$ is differentiable and not constant on the support of $\bgamma_0^{\top}\bX$.
\item[(A2)] The covariate vector $\bX$ is continuous and does not contain an intercept.  
    The support $\mathcal{X}$ is not contained in any proper linear subspace of $\Bbb{R}^{d_1}$.  
\item[(A3)] %$\bgamma_0=(\gamma_1,\dots,\gamma_{d_1})^\top$, $\gamma_1=1$ or $|| \bgamma_0||=1$ with a fixed sign of $\gamma_1$, where $||\cdot||$ denotes the Euclidean norm. 
{The parameter $\bgamma_0 \in \mathcal{A}_{\bgamma}$, where $\mathcal{A}_{\bgamma}=\{(\gamma_1,\gamma_2,\dots,\gamma_{d_1})|\sum_{j=1}^{d_1}\gamma_j^2=1, \gamma_1>0 \text{ or } \gamma_1=1\}$.}
%,1}\bigcup\mathcal{A}_{\bgamma,2}\bigcup\mathcal{A}_{\bgamma,3}$, where $\mathcal{A}_{\bgamma,1}=\{(\gamma_1,\gamma_2,\dots,\gamma_{d_1})|\sum_{j=1}^{d_1}\gamma_j^2=1, \gamma_1>0\}$, $\mathcal{A}_{\bgamma,2}=\{(\gamma_1,\gamma_2,\dots,\gamma_{d_1})|\sum_{j=1}^{d_1}\gamma_j^2=1, \gamma_1<0\}$ and $\mathcal{A}_{\bgamma,3}=\{(\gamma_1,\gamma_2,\dots,\gamma_{d_1})| \gamma_1=1\}$.}
%\item[(A4)]  The components of $\bX$ are continuously distributed random variables that have a joint probability density function.
%\item[(A5)] For the mixture of continuous and discrete variables $\bX$, varying the values of the discrete components must not divide the support of $\bgamma^{\top}\bX$ into disjoint subsets, and the function g is not periodic.
\item[(A4)] %{The matrix $\textnormal{Var}(\bm Z)$ is full rank.} 
The covariate vector $\bZ$ does not contain an intercept, and the support $\mathcal{Z}$ is not contained in any proper linear subspace of $\Bbb{R}^{d_2}$.
\item[(A5)]  The transformation function $G$ is continuously differentiable on $[0,\infty)$ with $G(0) = 0$, $G'(x)>0$ and $G(\infty) =\infty$.
\item[(A6)] For all $\bX$, $0<g_0(\bgamma_0^{\top}\bX)<1$. $\lim _{t \rightarrow \infty} S_u(t| \bZ)=0 $ for all $\bZ$, indicating the
latency function is proper.
\end{enumerate}
Assumptions (A1)-(A3) are needed to identify the single-index model for the incidence (Theorem 2.1 in \cite{horowitz2009semiparametric} and \cite{amico2019single}). Assumptions (A4)-(A5) are required for the identifiability of the latency model. (A6) %is required for the mixture cure model with interval-censored data.
imposes mild conditions to ensure separate identifiability of the incidence and latency components.
%Assumption (A7) requires that two adjacent monitoring times be separated by at least $\epsilon$; otherwise, the data may contain exact observations, which would entail a different theoretical treatment \citep{zeng2016maximum}.
%Assumption (A8) is a typical assumption needed to identify the mixture cure model with interval censored data \citep{li2001identifiability,ma2009cure,liu2021generalized}. }
\begin{enumerate}
    \item[]{\bf Proposition 1.}
    Under (A1)-(A6), the model given by (\ref{S_p}), (\ref{S_u}), and (\ref{pi}) is identifiable.
\end{enumerate}
The proof of Proposition 1 is given in the \hyperref[A.3]{Appendix}. Note that the above assumptions apply when $\bX$ is a continuous random vector. % a continuous random vector $\bX$.
For $\bX$ that contains both continuous and discrete components, the following additional conditions must hold: 
\begin{enumerate}
    \item[(A7)] Varying the values of the discrete components must not divide the support of $\bgamma_0^{\top}\bX$ into disjoint subsets.
    \item[(A8)] The function $g_0$ is not periodic.
\end{enumerate}}
%(i) varying the values of the discrete components must not divide the support of $\bgamma_0^{\top}\bX$ into disjoint subsets, and (ii) the function g is not periodic.

%{\color{red} add refer to the appendix}{\color{green}done}
%\blue{The aforementioned identifiability results only take into account the case where $\bX$ is a vector of continuous covariates. When $\bX$ encompasses a mix of continuous and discrete variables, Assumption (A9) and (A10) are necessary \citep{horowitz2009semiparametric}.}

% \begin{enumerate}
%     \blue{\item[(A9)]When the values of discrete components vary, the support of $\bgamma^{\top}\bX$ must not be divided into disjoint subsets.
%     \item[(A10)]The link function $g(\cdot)$ is not periodic. }
% \end{enumerate}

\section{Estimation Procedure}\label{sec3}
%Due to the involvement of the integral in the transformation model, directly maximizing the intractable likelihood function  (\ref{L_obs}) is challenging.
The involvement of the integral in the transformation model makes it challenging to directly maximize the intractable likelihood function  (\ref{L_obs}).
 Therefore, we propose a four-layer data augmentation procedure and develop an expectation-maximization (EM) algorithm %that employs four layers of latent random variables to estimate the parameters.
for parameter estimation.
\subsection{Data Augmentation}\label{sec3.1}

The first layer simplifies the likelihood function by introducing a latent cure indicator $B_i $ for subject $i$. %$(i=1,\dots,n)$. %{\color{red}Notation consistency} Specifically,
Assume $B_i \sim \textnormal{Bern}(p(\bX_i))$, where $\textnormal{Bern}(p(\bX_i))$ denotes the Bernoulli distribution with parameter $p(\bX_i)$, representing the probability of being uncured. %Utilizing the relationship  $(1-B_i)\delta_{R_i}= 1-B_i$,
Note that $B_i=1$ when $\delta_{R_i}=0$ for data that are not right-censored. 
For subject $i$ with $\delta_{R_i}=1$, the contribution to the likelihood is {$1-p(\bX_i)$ if $B_i=0$, and $p(\bX_i) S_u(L_i|\bZ_i)$ if $B_i=1$.}
Utilizing the relationship  $(1-B_i)\delta_{R_i}= 1-B_i$, the augmented data likelihood function can be written as
\begin{equation*}
    \begin{aligned}
        L_1(\boldsymbol{\theta}|\mathcal{O},\boldsymbol{B})&=\prod_{i=1}^np(\bX_i)^{B_i}[1-p(\bX_i)]^{1-B_i}\int_{\xi_i}\big[1-\exp\{-\xi_i\exp(\bbeta^{\top}\bZ_i)\Lambda(R_i)\}\big]^{\delta_{L_i}}\\
        &\times\big[\exp\{-\xi_i\exp(\bbeta^{\top}\bZ_i)\Lambda(L_i)\}-\exp\{-\xi_i\exp(\bbeta^{\top}\bZ_i)\Lambda(R_i)\}\big]^{\delta_{I_i}}\\
        &\times\exp\{-\delta_{R_i}B_i\xi_i\exp(\bbeta^{\top}\bZ_i)\Lambda(L_i)\}
        f(\xi_i)\,{\rm d}\xi_i.
    \end{aligned}\label{L1}
\end{equation*}
% The first and second layer introduces the cure indicators $B$ and frailty variable $\xi$, yielding of augmented data likelihood functions:
%and using the fact that $(1-B_i)\delta_{R,i}=1-B_i$, the likelihood function has the form
In the second layer, we introduce the latent variable $\xi_i$. % (i=1,\dots,n)$.
Conditional on $\xi_i$ and $B_i$, the augmented likelihood function then tasks the form 
\begin{equation*}
    \begin{aligned}
L_2(\boldsymbol{\theta}|\mathcal{O},\boldsymbol{B},\boldsymbol{\xi})&=\prod_{i=1}^np(\bX_i)^{B_i}[1-p(\bX_i)]^{1-B_i}\big[1-\exp\{-\xi_i\exp(\bbeta^{\top}\bZ_i)\Lambda(R_i)\}\big]^{\delta_{L_i}}\\
        &\times\big[\exp\{-\xi_i\exp(\bbeta^{\top}\bZ_i)\Lambda(L_i)\}-\exp\{-\xi_i\exp(\bbeta^{\top}\bZ_i)\Lambda(R_i)\}\big]^{\delta_{I_i}}\\
        &\times\exp\{-\delta_{R_i}B_i\xi_i\exp(\bbeta^{\top}\bZ_i)\Lambda(L_i)\}
        f(\xi_i).
    \end{aligned}\label{L2}
\end{equation*}

In the third layer, we introduced nonhomogenous Poisson latent variables to construct the augmented likelihood function. % following the approaches outlined by \cite{zeng2016maximum} and \cite{zhou2018computationally}. 
{The technique leads to %the closed-form expressions of $\eta_l, l = 1, , \dots, k$, 
simplified likelihood framework and is commonly utilized for interval-censored data \citep{zeng2016maximum,zhou2018computationally}.}
Specifically, we define two Poisson random variables $Y_i$ and $W_i$ %(i=1,\dots,n)$ 
with means given by %$\exp(\bbeta^{\top}\bZ_i)[\delta_{L_i}\Lambda(R_i)+\delta_{I_i}\Lambda(L_i)]\xi_i$ and $\exp(\bbeta^{\top}\bZ_i)\{\delta_{I_i}[\Lambda(R_i)-\Lambda(L_i)]+\delta_{R_i}\Lambda(L_i)\}\xi_i$. Let $\lambda_i=\exp(\bbeta^{\top}\bZ_i)[\delta_{L_i}\Lambda(R_i)+\delta_{I_i}\Lambda(L_i)]$ and $\omega_i=\exp(\bbeta^{\top}\bZ_i)\{\delta_{I_i}[\Lambda(R_i)-\Lambda(L_i)]+\delta_{R_i}\Lambda(L_i)\}$. 
$\lambda_i\xi_i$ and $\omega_i\xi_i$,  where $\lambda_i=\exp(\bbeta^{\top}\bZ_i)[\delta_{L_i}\Lambda(R_i)+\delta_{I_i}\Lambda(L_i)]$ and $\omega_i=\exp(\bbeta^{\top}\bZ_i)\{\delta_{I_i}[\Lambda(R_i)-\Lambda(L_i)]+\delta_{R_i}\Lambda(L_i)\}$. {Clearly, $Y_i$ and $W_i$ are independent conditional on $\xi_i$ and $\bZ_i$.}
Define $Y_i\equiv0$ if $Y_i\sim$Possion$(0)$. Within this framework, $Y_i>0$ and $W_i=0$ if $T_i$ is left-censored, $Y_i=0$ and $W_i>0$ if $T_i$ is interval-censored, $Y_i=W_i=0$ if $T_i$ is right-censored. By leveraging these Poisson latent variables, the augmented likelihood function can be formulated as
%{\color{red}We introduce latent variables $Y_i$ and $W_i(i=1,\dots,n)$, which, conditional on $\xi_i$, are independent Possion random variables with means $\lambda_i\xi_i$ and $\omega_i\xi_i$, respectively, where $\lambda_i=\exp(\beta^{\top}Z_i)[\delta_{L_i}\Lambda(R_i)+\delta_{I_i}\Lambda(L_i)]$ and $\omega_i=\exp(\beta^{\top}Z_i)\{\delta_{I_i}[\Lambda(R_i)-\Lambda(L_i)]+\delta_{R_i}\Lambda(L_i)\}$. And we define $Y_i\equiv0$ if $Y_i\sim$Pois$(0)$.} $Y_i$ and $W_i$ are independent of each other given $\xi_i$. In this framework, $Y_i>0$ and $W_i=0$ if $T_i$ is left-censored, $Y_i=0$ and $W_i>0$ if $T_i$ is interval-censored, $Y_i=W_i=0$ if $T_i$ is right-censored. By using the above Poisson latent variables, the likelihood function \ref{ee8} can be rewritten as 
% \begin{equation}
%     \begin{aligned}
%         L_2(\boldsymbol{\theta}|\mathcal{O},\boldsymbol{B,Y,W})&=\prod_{i=1}^np(\bX_i)^{B_i}[1-p(\bX_i)]^{1-B_i}\int_{\xi_i}(1-P(Y_i=0|\xi_i))^{\delta_{L_i}}\\
%         &\times([1-P(W_i=0|\xi_i)]P(Y_i=0|\xi_i))^{\delta_{I_i}}\\
%         &\times(P(Y_i=0|\xi_i)P(W_i=0|\xi_i))^{\delta_{R_i}B_i}
%         f(\xi_i)\,{\rm d}\xi_i,
%     \end{aligned}\label{ee9}
% \end{equation}
% {\color{red}which is the same as \eqref{e5}. Thus, maximization of \eqref{e5} is equivalent to maximum likelihood estimation \eqref{ee9}.
%  We maximize \eqref{ee9} through an EM algorithm by treating $Y_i$, $W_i$ and $\xi_i$ as missing data. }The complete-data likelihood is
\begin{equation*}
    \begin{aligned} L_3(\boldsymbol{\theta}|\mathcal{O},\boldsymbol{B,\xi,Y,W})&=\prod_{i=1}^np(\bX_i)^{B_i}\{1-p(\bX_i)\}^{1-B_i}\left\{\frac{(\lambda_i\xi_i)^{Y_i}}{Y_i!}e^{-\lambda_i\xi_i}\right\}^{\delta_{L_i}}\\
    &\times\left\{\frac{(\omega_i\xi_i)^{W_i}}{W_i!}e^{-(\omega_i+\lambda_i)\xi_i}\right\}^{\delta_{I_i}}
        \exp(-\delta_{R_i}B_i\omega_i\xi_i)f(\xi_i),
    \end{aligned}\label{L3}
\end{equation*}
under the constraint $\delta_{L_i}I(Y_i > 0) + \delta_{I_i}I(W_i > 0) + \delta_{R_i} = 1$ for $i = 1,\dots, n$. %The observed likelihood function (\ref{L_obs}) can be achieved by integrating out the latent variables $B_i$'s, $\xi_i$'s, $Y_i$'s and $W_i$'s in the augmented likelihood function (\ref{L3}).

We utilize the monotone splines technique defined in \cite{Ramsay1988Monotone} to approximate the unspecified increasing function $\Lambda(\cdot)$ in the transformation model. Specifically, we approximate $\Lambda(\cdot)$ with a linear combination of basis I-splines:
%The technique reduces the number of unknown parameters that require estimation while preserving sufficient modelling flexibility.
%The third and fourth layers of data augmentation utilise the spline approximation of the cumulative hazard function and the Poisson process. Specifically, we employ the monotone splines of \cite{Ramsay1988Monotone} to approximate $\Lambda(\cdot)$, that is
%We consider the semiparametric maximum likelihood estimation approach to estimate $\gamma$, $\beta$, and $\Lambda(\cdot)$. 
%Monotone splines are used to model $\Lambda(\cdot)$, which helps reduce the number of unknown parameters that require estimation while preserving sufficient modeling flexibility. Let
\begin{equation}
    \Lambda(t)=\sum_{l=1}^k\eta_lb_l(t),
    \label{Lamb}
\end{equation}
where $\{b_l(t)\}$ is a set of  I-spline basis functions ranging from 0 to 1, $\boldsymbol{\eta}=(\eta_1,\dots,\eta_k)^\top$ are non-negative coefficients to ensure the monotonicity of $\Lambda(\cdot)$, and %{\color{red} discuss this approach} 
$k$ is the number of basis I-splines %The shape of monotone splines is
determined by knot placement and the degrees of splines. % while smoothness is controlled by the degree. The number of basis functions equals the number of knots plus the degree. These functions are piecewise polynomials defined within a finite interval of censoring times in the data. The basis functions are uniquely determined by specifying increasing knot points and selecting a degree.
{This spline-based formulation offers a computationally efficient alternative to step-function approximations \citep{zeng2016maximum} by reducing the parameter dimensionality, while simultaneously providing a smooth estimate of the hazard rate.}
As a general guideline, using an order of 2 (quadratic) or 3 (cubic) with 5 to 15 knots offers adequate smoothness and sufficient flexibility. Knot placement can be either evenly spaced or determined by percentiles, based on our extensive experience.

In the fourth layer, 
%The augmented likelihood (\ref{L3}) is expressed as the multiplication of Poisson probability mass functions. ,
we consider further data augmentation to derive a closed-form update for the coefficients of the monotone spline basis, utilizing the additivity property of Poisson random variables and the linear form of $\Lambda(t)$ in \eqref{Lamb}. Decompose both $Y_{i}$ and $W_i$ as a sum of $k$ independent Poisson random variables, $Y_i=\sum_{l=1}^k Y_{il}$ and $W_i=\sum_{l=1}^k W_{i l}$, where $Y_{il} \sim \mathcal{P}\{ \eta_l\exp(\bbeta^{\top}\bZ_i)[\delta_{L_i}b_l(R_i)+\delta_{I_i}b_l(L_i)]\xi_i  \}$ and $W_{i l} \sim \mathcal{P}\{\eta_l\exp(\bbeta^{\top}\bZ_i)\{\delta_{I_i}[b_l(R_i)-b_l(L_i)]+\delta_{R_i}b_l(L_i)\}\xi_i\}$ for $l=1, \ldots, k$, where $\mathcal{P}$ denotes the Poisson probability mass function. Let $\lambda_{il}=\eta_l\exp(\bbeta^{\top}\bZ_i)[\delta_{L_i}b_l(R_i)+\delta_{I_i}b_l(L_i)]$ and $\omega_{il}=\eta_l\exp(\bbeta^{\top}\bZ_i)\{\delta_{I_i}[b_l(R_i)-b_l(L_i)]+\delta_{R_i}b_l(L_i)\}$.
% To utilize the monotone spline representation of $\Lambda(\cdot)$, we establish that $Y_{il}|\xi_i\sim$Pois$(\lambda_{il}\xi_i)$ and $W_{il}|\xi_i\sim$Pois$(\omega_{il}\xi_i)$, where $\lambda_{il}=\eta_l\exp(\bbeta^{\top}\bZ_i)[\delta_{L_i}b_l(R_i)+\delta_{I_i}b_l(L_i)]$ and $\omega_{il}=\eta_l\exp(\bbeta^{\top}\bZ_i){\delta_{I_i}[b_l(R_i)-b_l(L_i)]+\delta_{R_i}b_l(L_i)}$. Note that $\sum_{l=1}^k\lambda_{il}=\lambda_i$ and $\sum_{l=1}^k\omega_{il}=\omega_i$, $Y_i$ and $\sum_{l=1}^kY_{il}$ have the same distribution.
The final complete likelihood function %of $\boldsymbol{\theta}=(\boldsymbol{\gamma,\beta,\eta})^\top$ 
based on $\xi_i$'s, $B_i$'s, $Y_{il}$'s and $W_{il}$'s has the form
\begin{equation}\label{complete}
    \begin{aligned}        
L_4(\boldsymbol{\theta}|\mathcal{O},\boldsymbol{B,\xi,Y,W})&=\prod_{i=1}^np(\bX_i)^{B_i}\left\{1-p(\bX_i)\right\}^{1-B_i}\left[\prod_{l=1}^k\left\{\frac{(\lambda_{il}\xi_i)^{Y_{il}}}{Y_{il}!}e^{-\lambda_{il}\xi_i}\right\}^{\delta_{L_i}}\right.\\
&\times\left.\left\{\frac{(\omega_{il}\xi_i)^{W_{il}}}{W_{il}!}e^{-(\omega_{il}+\lambda_{il})\xi_i}\right\}^{\delta_{I_i}}
 \exp(-\delta_{R_i}B_i\omega_{il}\xi_i)\right]f(\xi_i),
    \end{aligned}
\end{equation}
under the constraint $\delta_{L_i}I(\sum_{l=1}^kY_{il} > 0) + \delta_{I_i}I(\sum_{l=1}^kW_{il} > 0) + \delta_{R_i} = 1$ for $i = 1,\dots, n$. The observed likelihood function \eqref{L_obs} can be achieved by integrating out the latent variables $\boldsymbol{B}$, $\boldsymbol{\xi}$, $Y_l$'s and $W_l$'s in the complete likelihood function \eqref{complete}. {The deduction of equation \eqref{complete} to \eqref{L_obs} is given in the \hyperref[A.1]{Appendix}.}

\subsection{Nonparametric regression for Incidence}

%There are two ways to approximate the nonparametric link function $g$. 
For the nonparametric link function $g$, two popular techniques, spline and kernel methods, can be utilized. 

The spline-based sieve approach has been extensively studied in the semiparametric models of survival data \citep{ding2011sieve,zhao2017sieve,liu2021generalized}.
In the single-index model, the nonparametric function $g$ involves the parametric component $\bgamma^{\top}\bX$, and thus it requires constructing the splines over intervals for a given $\bgamma$. Specifically, let $a^{\gamma}=\min _{1 \leq i \leq n}\{ \bgamma^{\top}\bX_i\}$ and $ b^\gamma= \max _{1 \leq i \leq n}\{\bgamma^{\top}\bX_i\}$. We approximate the link function using B-spline functions on $[a^{\gamma},b^{\gamma}]$ for a fixed $\bgamma$:
\begin{equation}
    {g}({\bgamma}^{\top}\bX)=\sum_{j=1}^m\psi_jb_j^*({\bgamma}^{\top}\bX),\label{apg1}
\end{equation}
where $\{b_j^*(\cdot)\}$ represents a set of B-spline basis functions, and $\boldsymbol{\psi}=(\psi_1,\dots,\psi_m)$ denotes the spline coefficients.
 Note that the end points of such intervals vary with different values of $\bgamma$, yielding more challenges in the estimation procedure.

Alternative nonparametric techniques, such as the kernel techique, could also be used for the estimation of $g$. 
The kernel approach has been investigated for the nonparametric covariate functions \citep{amico2019single,wang2021kernel,wang2022kernel}.
 For a given $\boldsymbol{\gamma}$, the link function $g(\cdot)$ can be obtained by the (unfeasible) leave-one-out kernel estimator \citep{amico2019single}
\begin{equation}
    g_{-i}(\bgamma^{\top}\bX_i)=\frac{\sum_{j\ne i}^n K(\frac{\bgamma^{\top}\bX_i-\bgamma^{\top}\bX_j}{h})B_j}{\sum_{l\ne i}^n K(\frac{\bgamma^{\top}\bX_i-\bgamma^{\top}\bX_l}{h})},
    \label{apg2}
\end{equation}
where $ K(\cdot) $ is a kernel function, $h$ is a one-dimensional bandwidth  and $\boldsymbol{B}$ is the latent cure status, which can be replaced by its expectation in the calculation. {The bandwidth $h$ is selected via likelihood cross-validation, with the complete procedure detailed in the simulation studies (Section 4).}
For an illustration,
we compared the kernel estimator with the cubic B-splines estimator for the incidence part in
the simulation studies.

\subsection{EM Algorithm}\label{sec3.2}
%The EM algorithm involves maximizing the likelihood of the complete data by iteratively alternating between an Expectation (E) step and a Maximization (M) step until convergence.In the E-step, we take
The expectation step involves calculating  $E\{\log L_c(\boldsymbol{\theta}|\mathcal{O},\boldsymbol{B,Y,W})\}$ with respect to all latent variables %with respect to all latent variables 
$\boldsymbol{Y}$, $\boldsymbol{W}$, $\boldsymbol{B}$ and $\boldsymbol{\xi}$ %in the log-likelihood function
conditional on the observed data $\mathcal{O}$ and the current parameter $\boldsymbol{\theta}^{(d)}=(\boldsymbol{\gamma}^{(d)},\boldsymbol{\beta}^{(d)},\boldsymbol{\eta}^{(d)})$. 
%For ease of exposition, we suppress all conditional arguments in the expectations henceforth. 
 We omit all conditional arguments in the expectations henceforth for the sake of clarity; for instance, we write $E(B_{i})$ instead of $E(B_{i}| \mathcal{O},\boldsymbol{\theta}^{(d)})$.
%This yields the objective function % This yields $Q(\boldsymbol{\theta},\boldsymbol{\theta}^{(d)})=E[\log{L_c(\boldsymbol{\theta})}|\boldsymbol{\theta}^{(d)}]$, with  $\boldsymbol{\theta}^{(d)}=(\boldsymbol{\gamma}^{(d)'},\boldsymbol{\beta}^{(d)'},\boldsymbol{\eta}^{(d)'})'$, and $Q(\boldsymbol{\theta},\boldsymbol{\theta}^{(d)})$ 
The objective function can be written as the sum of two parts
\begin{equation*}
Q(\boldsymbol{\theta},\boldsymbol{\theta}^{(d)})=Q_1(\boldsymbol{\gamma},\boldsymbol{\gamma}^{(d)})+Q_2(\boldsymbol{\beta,\eta};\boldsymbol{\beta}^{(d)},\boldsymbol{\eta}^{(d)}),
\end{equation*}
where{\begin{align}    Q_1(\boldsymbol{\gamma},\boldsymbol{\gamma}^{(d)})=&\sum_{i=1}^nE(B_i)\log[g(\bgamma^{\top}\bX_i)]+[1-E(B_i)]\log[1-g(\bgamma^{\top}\bX_i)],\label{eQ1}\\
Q_2(\boldsymbol{\beta,\eta};\boldsymbol{\beta}^{(d)},\boldsymbol{\eta}^{(d)})=&\sum_{i=1}^n\sum_{l=1}^k\bigg\{[\delta_{L_i}E(Y_{il})+\delta_{I_i}E(W_{il})](\log\eta_l+\bbeta^{\top}\bZ_i)\nonumber\\
    &-\eta_l\exp(\bbeta^{\top}\bZ_i)E(\xi_iB_i)[(1-\delta_{R_i})b_l(R_i)+\delta_{R_i}b_l(L_i)]\bigg\},\label{eQ2}
\end{align} }
%where $l(\boldsymbol{\theta}^{(d)})$ is a function of $\boldsymbol{\theta}^{(d)}$, and is free of $\boldsymbol{\theta}$.
after omitting additive terms that are free of $\boldsymbol{\theta}$.

We show that the conditional expectations of the latent variables are given as:
\setlength{\jot}{12pt}
\iffalse
\begin{gather*}
E(Y_{il})=\frac{\eta_l^{(d)}b_l(R_i)E(Y_i|\boldsymbol{\theta}^{(d)},\mcO)}{\sum_{l=1}^k\eta_l^{(d)}b_l(R_i)},\\%\label{eeyil}; 
E(Y_i)=\frac{\delta_{L_i}N_{i2}^{(d)}\int_{\xi_i}\xi_if(\xi_i){\rm d}\xi_i}{1-\exp\{-G(N_{i2}^{(d)})\}},\\
E(W_{il})=\frac{\eta_l^{(d)}[b_l(R_i)-b_l(L_i)]E(W_i|\boldsymbol{\theta}^{(d)},\mcO)}{\sum_{l=1}^k\eta_l^{(d)}[b_l(R_i)-b_l(L_i)]},\\
E(W_i)=\frac{\delta_{I_i}(N_{i2}^{(d)}-N_{i1}^{(d)})\exp[-G(N_{i1}^{(d)})]G'(N_{i1}^{(d)})}{\exp\{-G(N_{i1}^{(d)})\}-\exp\{-G(N_{i2}^{(d)})\}},\\
E(B_{i})=\delta_{L_i}+\delta_{I_i}+\frac{\delta_{R_i}p^{(d)}(\bX_i)\exp\{-G(N_{i1}^{(d)})\}}{1-p^{(d)}(\bX_i)+p^{(d)}(\bX_i)\exp\{-G(N_{i1}^{(d)})\}}%\label{eebi}
\end{gather*} 
\fi
\begin{align}
E(Y_{il})&=\frac{\eta_l^{(d)}b_l(R_i)E(Y_i)}{\sum_{l=1}^k\eta_l^{(d)}b_l(R_i)},\label{eeyil} 
\quad
E(Y_i)=\frac{\delta_{L_i}N_{i2}^{(d)}\int_{\xi_i}\xi_if(\xi_i){\rm d}\xi_i}{1-\exp\{-G(N_{i2}^{(d)})\}},\\
E(W_{il})&=\frac{\eta_l^{(d)}[b_l(R_i)-b_l(L_i)]E(W_i)}{\sum_{l=1}^k\eta_l^{(d)}[b_l(R_i)-b_l(L_i)]},\\
E(W_i)&=\frac{\delta_{I_i}(N_{i2}^{(d)}-N_{i1}^{(d)})\exp\{-G(N_{i1}^{(d)})\}G'(N_{i1}^{(d)})}{\exp\{-G(N_{i1}^{(d)})\}-\exp\{-G(N_{i2}^{(d)})\}},\\
E(B_{i})&=\delta_{L_i}+\delta_{I_i}+\frac{\delta_{R_i}p^{(d)}(\bX_i)\exp\{-G(N_{i1}^{(d)})\}}{1-p^{(d)}(\bX_i)+p^{(d)}(\bX_i)\exp\{-G(N_{i1}^{(d)})\}},\label{eebi}
\end{align} 
and 
\begin{align}
E(\xi_iB_i)&=\delta_{L_i}\frac{\int_{\xi_i}\xi_if(\xi_i){\rm d}\xi_i-\exp\{-G(N_{i2}^{(d)})\}G'(N_{i2}^{(d)})}{1-\exp\{-G(N_{i2}^{(d)})\}}\nonumber\\
    &+\delta_{I_i}\frac{\exp\{-G(N_{i1}^{(d)})\}G'(N_{i1}^{(d)})-\exp\{-G(N_{i2}^{(d)})\}G'(N_{i2}^{(d)})}{\exp\{-G(N_{i1}^{(d)})\}-\exp\{-G(N_{i2}^{(d)})\}}\nonumber\\
    &+\delta_{R_i}\frac{p^{(d)}(\bX_i)\exp\{-G(N_{i1}^{(d)})\}G'(N_{i1}^{(d)})}{1-p^{(d)}(\bX_i)+p^{(d)}(\bX_i)\exp\{-G(N_{i1}^{(d)})\}},\label{eexiibi}   
\end{align}
where $N_{i1}^{(d)}=\exp(\bbeta^{(d)\top}Z_i)\sum_{l=1}^k\eta_l^{(d)}b_l(L_i)$, $N_{i2}^{(d)}=\exp(\bbeta^{(d)\top}Z_i)\sum_{l=1}^k\eta_l^{(d)}b_l(R_i)$. The detailed derivations of the
above conditional expectations are presented in the \hyperref[A.2]{Appendix}.

The maximization step finds $\boldsymbol{\theta}^{(d+1)}=\arg\max Q(\boldsymbol{\theta},\boldsymbol{\theta}^{(d)})$, where $Q(\boldsymbol{\theta},\boldsymbol{\theta}^{(d)})=Q_1(\boldsymbol{\gamma},\boldsymbol{\gamma}^{(d)})+Q_2(\boldsymbol{\beta,\eta};\boldsymbol{\beta}^{(d)},\boldsymbol{\eta}^{(d)})$.

We update $\boldsymbol{\gamma}^{(d+1)}$ by maximizing $Q_1(\boldsymbol{\gamma},\boldsymbol{\gamma}^{(d)})$ described in Eq (\ref{eQ1}). For %a given $\boldsymbol{\gamma}$, 
the kernel-based approach, the link function $g(\cdot)$ can be obtained by the leave-one-out kernel estimator \citep{amico2019single}
 \begin{equation}
     g_{-i}(\bgamma^{\top}\bX_i)=\frac{\sum_{j\ne i}^n K(\frac{\bgamma^{\top}\bX_i-\bgamma^{\top}\bX_j}{h})E(B_j)}{\sum_{l\ne i}^n K(\frac{\bgamma^{\top}\bX_i-\bgamma^{\top}\bX_l}{h})},\label{apg3}
\end{equation}
where $h$ is a one-dimensional bandwidth. 
Substituting expression \eqref{apg3} into Eq \eqref{eQ1} and maximizing it, we obtain the updated estimate $\bgamma^{(d+1)}$. Using $\bgamma^{(d+1)}$, we 
compute ${g}_{-i}^{(d+1)}(({\bgamma}^{(d+1)})^{\top}\bX_i)$ with the estimator provided in Equation \eqref{apg3}. {The sieve estimator of $g$ can be obtained in a similar way.}

Setting the derivative  $\frac{\partial Q_2(\boldsymbol{\beta,\eta};\boldsymbol{\beta}^{(d)},\boldsymbol{\eta}^{(d)})}{\partial \eta_l}=0$, for $l=1,\dots,k$,  we obtain the closed-form solution for $\eta_l$ based on $\bbeta$:
 \begin{equation}
\eta_l(\bbeta)=\frac{\sum_{i=1}^n\{\delta_{L_i}E(Y_{il})+\delta_{I_i}E(W_{il})\}}{\sum_{i=1}^n \exp(\bbeta^{\top}\bZ_i)E(\xi_iB_i)\{(1-\delta_{R_i})b_l(R_i)+\delta_{R_i}b_l(L_i)\}}, \quad l=1,\dots,k.
 \label{est_gamma}
    \end{equation}
Substituting  $\eta_l(\bbeta)$ into Eq (\ref{eQ2}) and %let $\frac{\partial Q_2(\boldsymbol{\beta,\eta};\boldsymbol{\beta}^{(d)},\boldsymbol{\eta}^{(d)})}{\partial \boldsymbol{\beta}}=0$, 
differentiating with respect to $\boldsymbol{\beta}$, we can obtain $\boldsymbol{\beta}^{(d+1)}$ by solving the score function:
\begin{equation}
    \sum_{i=1}^n\sum_{l=1}^k\left[\{\delta_{L_i}E(Y_{il})+\delta_{I_i}E(W_{il})\}-\eta_l(\bbeta)\exp(\bbeta^{\top}\bZ_i)E(\xi_iB_i)\{(1-\delta_{R_i})b_l(R_i)+\delta_{R_i}b_l(L_i)\}\right]\bZ_i=0.
\label{est_beta}
\end{equation}
%Given $\boldsymbol{\beta}^{(d+1)}$, updated $\eta^{(d+1)}_l$ according to Eq \eqref{est_gamma}.
The updated $\eta^{(d+1)}_l = \eta_l(\bbeta^{(d+1)})$ for $l = 1, 2, \dots k$ is determined by Eq (\ref{est_gamma}).

\iffalse
consider the partial derivatives of $Q_2(\boldsymbol{\beta,\eta};\boldsymbol{\beta}^{(d)},\boldsymbol{\eta}^{(d)})$ with respect to $(\boldsymbol{\beta,\eta})$, which are given by
\begin{align}
    \frac{\partial Q_2(\boldsymbol{\beta,\eta};\boldsymbol{\beta}^{(d)},\boldsymbol{\eta}^{(d)})}{\partial \boldsymbol{\beta}}=&\sum_{i=1}^n\sum_{l=1}^k\Big\{[\delta_{L_i}E(Y_{il})+\delta_{I_i}E(W_{il})]\nonumber\\
    &-\exp(\beta^{\top}Z_i)E(\xi_iB_i)[(1-\delta_{R_i})b_l(R_i)+\delta_{R_i}b_l(L_i)]\Big\}Z_i\\
    \frac{\partial Q_2(\boldsymbol{\beta,\eta};\boldsymbol{\beta}^{(d)},\boldsymbol{\eta}^{(d)})}{\partial \eta_l}=&\sum_{i=1}^n\Big\{\eta_l^{-1}[\delta_{L_i}E(Y_{il})+\delta_{I_i}E(W_{il})]\nonumber\\
&-\exp(\beta^{\top}Z_i)E(\xi_iB_i)[(1-\delta_{R_i})b_l(R_i)+\delta_{R_i}b_l(L_i)]\Big\}
\end{align}
Obviously, $(\boldsymbol{\beta}^{(d+1)},\boldsymbol{\eta}^{(d+1)})$ is a solution to the system of equations given by $\frac{\partial Q_2(\boldsymbol{\beta,\eta};\boldsymbol{\beta}^{(d)},\boldsymbol{\eta}^{(d)})}{\partial \boldsymbol{\beta}}=0$ and $\frac{\partial Q_2(\boldsymbol{\beta,\eta};\boldsymbol{\beta}^{(d)},\boldsymbol{\eta}^{(d)})}{\partial \eta_l}=0$, for $l=1,\dots,k$.
%Once the conditional expectations of observed likelihood is obtained, the maximization can be realized as described in subsection 3.4
\fi
%\subsection{Algorithm}
We summarise the proposed EM algorithm for calculating estimates for  $\btheta$ as follows:

\begin{description}
\item[Step 1.]Set initial values $\boldsymbol{\theta}^{(0)}=(\boldsymbol{\gamma}^{(0)},\boldsymbol{\beta}^{(0)},\boldsymbol{\eta}^{(0)})$, where $\boldsymbol{\beta}^{(0)}=\boldsymbol{0} $, $\boldsymbol{\eta}^{(0)}=\boldsymbol{1}$, and $\boldsymbol{\gamma}^{(0)}$ is obtained by fitting the generalized additive model using pseudo values $1-\delta_{R_{i}}$. %{\color{blue}$g^{(0)}=g_{-i}\{(\bgamma^{0})^{\top}\bX_i\}$.} {\color{red}there is no initial values of $g$}
%In our numerical studies, we advocate the use of gam() function in R package mgcv to get the initial values of $\bgamma^{(0)}$ and set the initial values of the parameters $(\boldsymbol{\beta}^{(0)},\boldsymbol{\eta}^{(0)})$ as $(\boldsymbol{0},\boldsymbol{1})$;
\item[Step 2.] In the $(d+1)$th iteration, calculate the expectations in Eqs \eqref{eeyil} -\eqref{eexiibi} based on $(\boldsymbol{\theta}^{(d)},\mcO)$.
%\item[Step 3.] {\color{blue} Obtain bandwidth $h^{(d)}$ by maximizing Eq \eqref{cvh} on the interval $[0.1,0.5]$.}
\item[Step 3.] Obtain $\boldsymbol{\gamma}^{(d+1)}$ by maximizing $Q_1(\boldsymbol{\gamma},\boldsymbol{\gamma}^{(d)})$ given in Eq (\ref{eQ1}), %via Newton–Raphson algorithm. %with replacing $p(X_i)$ with estimator \eqref{epg} in $Q_1(\boldsymbol{\theta},\boldsymbol{\theta}^{d)})$, 
and calculate the Nadaraya-Watson estimator $g(\bgamma^{\top}X_i)^{(d+1)}$ %is also obtained, the new estimator is denoted 
by $g_{-i}^{(d+1)}((\bgamma^{(d+1)})^{\top}X_i)$.
\item[Step 4.] Obtain $\bbeta^{(d+1)}$ by calculating the estimating
equation Eq (\ref{est_beta}).
%$\bbeta^{(d+1)}$ by maximizing $Q_2(\boldsymbol{\beta,\eta(\beta)};\boldsymbol{\beta}^{(d)},\boldsymbol{\eta}^{(d)})$ %where
%\begin{equation}
%    \begin{aligned}
%\eta_l(\bbeta)=\frac{\sum_{i=1}^n[\delta_{L_i}E(Y_{il})+\delta_{I_i}E(W_{il})]}{\sum_{i=1}^n \exp(\bbeta^{\top}\bZ_i)E(\xi_iB_i)[(1-\delta_{R_i})b_l(R_i)+\delta_{R_i}b_l(L_i)]}
%  \end{aligned}
% \end{equation}
%for $l=1,\dots,k$;
\item[Step 5.] Obtain  $\eta^{(d+1)}_l$ by Eq (\ref{est_gamma}) based on $\bbeta^{(d+1)}$.%$\eta^{(d+1)}_l=\eta_l(\beta^{(d+1)})$ for $l=1,\dots,k$;
\item[Step 6.]  Repeat Steps 2 to 5 until the convergence criterion is satisfied, obtaining the final iterative values $\boldsymbol{\hat{\theta}}=(\boldsymbol{\hat{\gamma}},\boldsymbol{\hat{\beta}},\boldsymbol{\hat{\eta}})$.
\end{description}
The initial values of $\boldsymbol{\gamma}$ are obtained using the gam function from R package mgcv, and the Newton–Raphson algorithm is employed to maximise the objective function.

{We utilize the bootstrap method to estimate the asymptotic variance of the estimator for practical use. {This method further facilitates the variance estimation of the cumulative baseline hazard function
$\Lambda(t)$, at any fixed time point $t$.} 
Alternatively, %other techniques such as 
the profile likelihood technique~\citep{murphy2000profile} and
the resampling approach that generalizes the perturbing technique of \cite{jin2003rank}
can also be employed. }% to estimate the standard deviation. The jackknife method~\citep{shao2012jackknife} is another resampling approach that can be used for variability assessment. These methods provide different perspectives and can be chosen based on the specific characteristics of the problem.%} %Additionally, the profile likelihood technique \citep{murphy2000profile} can be used for variance estimation.

\section{Simulation}\label{sec4}
%The simulation setting needs to be clarified first, these include two partsgenerating data and estimation procedures. 

%For generating data, one needs to first generate a cure indicator (similar to \cite{amico2019single} ). For those uncured subjects, generate failure time from conditional survival function.

{Extensive simulation studies are conducted to evaluate the finite sample performance of the proposed estimators, including the SMCI model with kernel estimator (SMCI-K) and the SMCI model with spline estimator (SMCI-S)}. For comparison, we employ the generalized odds rate mixture cure (GORMC) model proposed by \cite{zhou2018computationally}, which assumes a logistic regression for the incidence and a semiparametric transformation model for the latency. 

For the incidence component, we consider three different scenarios. Scenario 1 employs a logistic regression model. Scenario 2 utilizes a non-logistic but monotonic link function defined as  $g(u) = (1+\tanh(1.5u^5))/2$. The third scenario posits a non-logistic model for incidence with a non-monotonic shape, expressed as $g(u) = \exp(4.8u^3 - 8u^2 + 3.2u + 0.85) / [1 + \exp(4.8u^3 - 8u^2 + 3.2u + 0.85)]$. In each scenario, we consider three covariates:  $X_1 \sim \textrm{Unif}(-1,2) $, $X_2 \sim \textrm{N}(0,1)$ and $X_3 \sim \textrm{Bern}(0.5) $, and the true parameters $\bgamma = %(0,1/\sqrt3,-1/\sqrt3,1/\sqrt3)$
(1/\sqrt3,-1/\sqrt3,1/\sqrt3)^{\top}$, respectively. %{\color{red} add the cure proportion}
%The average cure proportion is approximately 35\%. 
Given the incidence component  $p(\bX)$, the uncured status $B$ is generated from a Bernoulli distribution with the parameter equal to $p(\bX)$. %Corresponding to the cure proportions approximately equals to 35\%.
Depending on the scenario, the cure proportion ranged from 22.6\% (in Scenario  2 with $r=2$) to 38\% (in Scenario 1 with $r=1$).

% % Suppose the data are generated from the mixture cure model \eqref{S_p}. Initially,
% %The uncured indicator  $B$ is generated from a Bernoulli distribution with the probability specified by the single-index model (\ref{pi}). 
% We assume three different scenarios for the incidence.
% Scenario 1 assumes a logistic regression model for the incidence. Scenario 2 considers the non-logistic but monotone link function $g(u) = (1+\tanh(1.5u^5))/2$. 
% The third scenario posits a nonlogistic model for incidence with a non monotone shape $g(u) = \exp(4.8u^3 - 8u^2 + 3.2u + 0.85) / [1 + \exp(4.8u^3 - 8u^2 + 3.2u + 0.85)]$.
% In each scenario, we consider three-dimensional time-independent covariates: $X_1 \sim U(-1,2) $, $X_2 \sim N(0,1)$ and $X_3 \sim Ber(0.5) $, and the true parameters set %and set the true parameters 
% $\bgamma = %(0,1/\sqrt3,-1/\sqrt3,1/\sqrt3)$
% (1/\sqrt3,-1/\sqrt3,1/\sqrt3)$, corresponding to the cure proportion equals to ., , ,  {\color{red} add the cure proportion} respectively. %Three different scenarios for the incidence component are explored.
% Given the incidence component  $p(\bX)$, the uncured status $B$ is generated from a Bernoulli distribution with the parameter equal to $p(\bX)$.

For the latency component,  we %consider three-dimensional covariates,
generate the failure times of susceptible subjects from the transformation model %(\ref{S_u})
{\begin{equation*}
%\Lambda_u(t|\bZ)=G[\int_0^t\exp(\bbeta^{\top}\bZ(s))\,{\rm d}\Lambda(s)],
\Lambda_u(t|\bZ)=G\left\{\exp(\bbeta^{\top}\bZ)\Lambda(t)\right\},
\end{equation*}}
where $G(x)=r^{-1}\log(1+rx)$ and $\Lambda(t) = 0.5\log(1 + t) +0.5t^{1.5}+0.5 t^3$.  We 
consider covariates $Z_1 \sim \textnormal{Unif}(0,2) $, $Z_2 \sim \textnormal{N}(0,1)$ and $Z_3 \sim \textnormal{Bern}(0.5) $, with parameters set as $\bbeta = (1,-1,1)^{\top}$.  
We investigate three models for the survival function of susceptible subjects by specifying $r=0$, 1 or 2, corresponding to the proportional hazards model, proportional odds model and the generalized odds rate model, respectively.
The interval-censored data are then generated in the following way similar to that in \cite{zhou2018computationally}. {On average, there were $10$–$15\%$ left-censored observations and $40$–$45\%$ right-censored ones.} Two sample sizes $n=200$ and $n=500$ are considered, and a total of 200 replicated datasets are generated for each sample size. {The simulation setting is detailed in Appendix Table \ref{tab:Simulation setting}.}
% The survival times for the uncured observations are generated according to $S_u(t|Z)=\int_0^{\infty}\exp(-\xi\exp(\beta^{\top}Z)\Lambda(t))f(\xi)\,{\rm d}\xi$ for given choice of $\beta$, $\Lambda(t)$ and $f(t)$. For three experiments, the baseline transformation function has the form of $\Lambda(t) = 0.5\log(1 + t) +0.5t^{1.5}+0.5 t^3$. $f(t)$ is gamma density with unit mean and variance $r$, which is used to consider the proportional hazards (PH) model and the generalized odds ratio (GOR) model. Different models with $r = 0, 0.5, 1$ and $2$ are considered.
% {\color{red} consider the PH(r=0), PO(r=1) and generalized odds ratio (r>0)}

%We further generate the interval-censored data as follows: each subject will have a sequence of ordered visit numbers $0=V_{i0} <\dots< V_{iv_i}$, where the total number of observations $v_i$ for each subject is generated from 1 plus a Poisson random variable with mean $\zeta$. The lengths of intervals between adjacent visits $\tau_{ij} = V_{ij} - V_{i(j-1)}$ for $j=1,\dots,v_i$ follow independent exponential distributions with mean $1/\kappa$. $\kappa$ and $\zeta$ will be adjusted according to different models to achieve different censoring rate. And the coefficients are set to be $\gamma=(0,1/\sqrt{3},-1/\sqrt{3},1/\sqrt{3})$ and $\beta=(1,-1,1)$. 

%{\color{red}The choices for the parameters $\kappa$ and $\zeta$ are given in Table \ref{tab:Simulation setting} as well as the averages, over the 200 simulated datasets for each setting, of the cure proportion and the censoring rate.??}

According to \cite{wang2016flexible}, we utilize 5 equally spaced knots at the percentiles with the degree of 3 for the monotone splines.  We choose the Epanechnikov kernel 
\begin{equation*}
    K(u)=\frac{(3-0.6u^2)I(u^2\leq 5)}{4\sqrt{5}}
\end{equation*}
for the SMCI-K model,
%proposed model, 
and select bandwidth by the likelihood cross-validation criterion. The criterion is defined as 
%Before incidence estimation, we use the likelihood cross-validation procedure method to select bandwidth. The selected bandwidth $h^{(d+1)}$ is given by maximizing the $Q_1(\bgamma,\bgamma^{(d)})$ evaluated at $\hat{\bgamma}^{(d)}$, that is, 
$\textnormal{CV}^{(d+1)}(h)=\sum_1^{n}E(B_i)\log\{\hat{g}_{h,-i}^{(d)}(\hat{\bgamma}^{(d)\top}\bX_i)\}+(1-E(B_i))\log\{1-\hat{g}_{h,-i}^{(d)}(\hat{\bgamma}^{(d)\top}\bX_i)\}$, where $\hat{g}_{h,-i}^{(d)}(\hat{\bgamma}^{(d)\top}\bX_i)$ is the leave-one-out kernel estimator \eqref{apg3} with $\bgamma$ replaced by $\hat{\bgamma}^{(d)}$ and based on a bandwidth $h$. We choose the optimal bandwidth on the interval $[0.1,0.5]$ based on $h^{(d+1)} = \arg\min_h \textnormal{CV}^{(d+1)}(h)$. % on the interval [0.1,0.5]. 
{%We compared the kernel estimator with the Spline estimator. In Scenarios 1, 2, and 3, 
For the SMCI-S approach,
we utilise the cubic-spline, with the number of knots set to 3, 8, and 10 in Scenarios 1,2, and 3, respectively.}

To evaluate the performance of these three estimators for incidence, the average squared error (ASE) is considered as a criterion, given by  $\textnormal{ASE}(\hat{p})=M^{-1}\sum_{i=1}^M\big\{\hat{g}(\hat{\bgamma}^{\top}\bX_i)-g(\bgamma^{\top}\bX_i)\big\}^2$.
 It is computed over a grid of points $(\boldsymbol{x}_j)_j=(\{x_{j 1}, x_{j 2}, x_{j 3}, \})_j,$ $j = 1, ..., M$, {where $M$ is the number of value sets.}
 For $X_{1}$ and $X_{2}$, we use a grid of size 0.1 over $[-1, 2]$ and $[-1.5, 1.5]$ respectively, while  $X_{ 3}$ takes values in $\{0, 1\}$, resulting in $M=$ 1992.
 %{\color{red}specify the value of $M$} %{\color{red} specified the number of M}.
%For the latency component, we compute bias (Bias), empirical estimated standard error (ESE), empirical standard deviation (ESD), and 95\% empirical coverage probability (CP) of $\hat{\bbeta}$ for each model.
{For the latency component, we compute the following metrics of $\hat\beta$, over $R$
simulation replicates:
%For the latency component, we compute theperformance metrics of $\hat\beta$ over $R$ simulation replicates, including the Bias: 
the Bias, $$\operatorname{Bias}(\hat{\beta})=\frac{1}{R} \sum_{r=1}^R(\hat{\beta}^{(r)}-\beta);$$ the  empirical estimated standard error (ESE), $$\operatorname{ESE}(\hat{\beta})=\frac{1}{R} \sum_{r=1}^R\widehat{SE}^{(r)};$$ the empirical standard deviation (ESD): $$\operatorname{ESD}(\hat{\beta})= \sqrt{\frac{1}{R-1} \sum_{r=1}^R(\hat{\beta}^{(r)}-\overline{\hat{\beta}})^2}, \textnormal{ where } \overline{\hat{\beta}}=\frac{1}{R} \sum_{r=1}^R \hat{\beta}^{(r)};$$ and
95\% empirical coverage probability (CP), defined as the proportion of replicates in which the nominal $95 \%$ confidence interval, $\hat{\beta}^{(r)} \pm 1.96 \times \widehat{S E}(\hat{\beta}^{(r)})$, contains the true parameter $\beta$.}

\begin{figure}[H]
\centering  
\begin{subfigure}[b]{0.32\textwidth}
  \includegraphics[width=\textwidth]{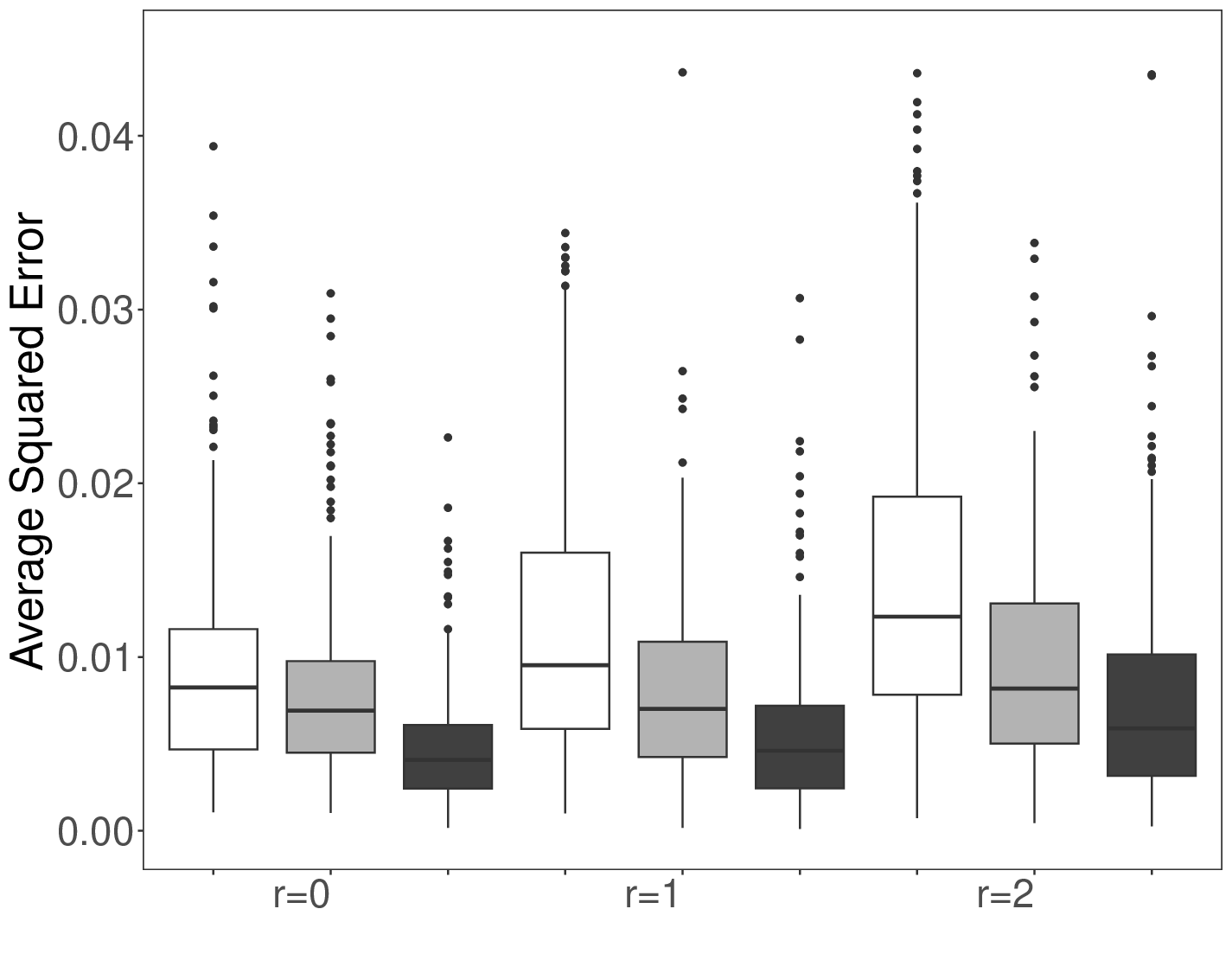}
\caption{Scenario 1; $n=200$}\label{ASEa}
\end{subfigure}
\begin{subfigure}[b]{0.32\textwidth}
  \includegraphics[width=\textwidth]{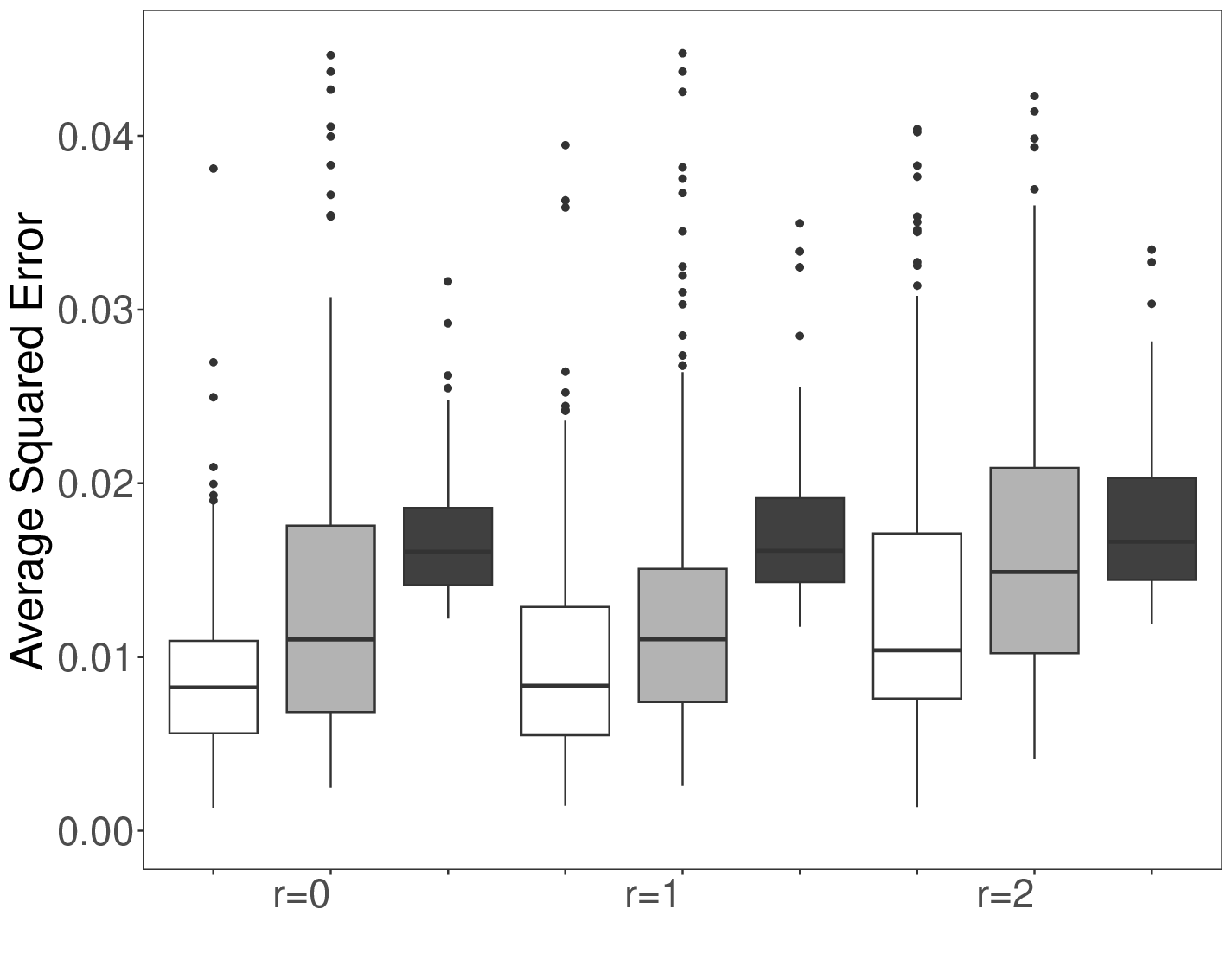}
  \caption{Scenario 2; $n=200$}\label{ASEb}
\end{subfigure}
\begin{subfigure}[b]{0.32\textwidth}
  \includegraphics[width=\textwidth]{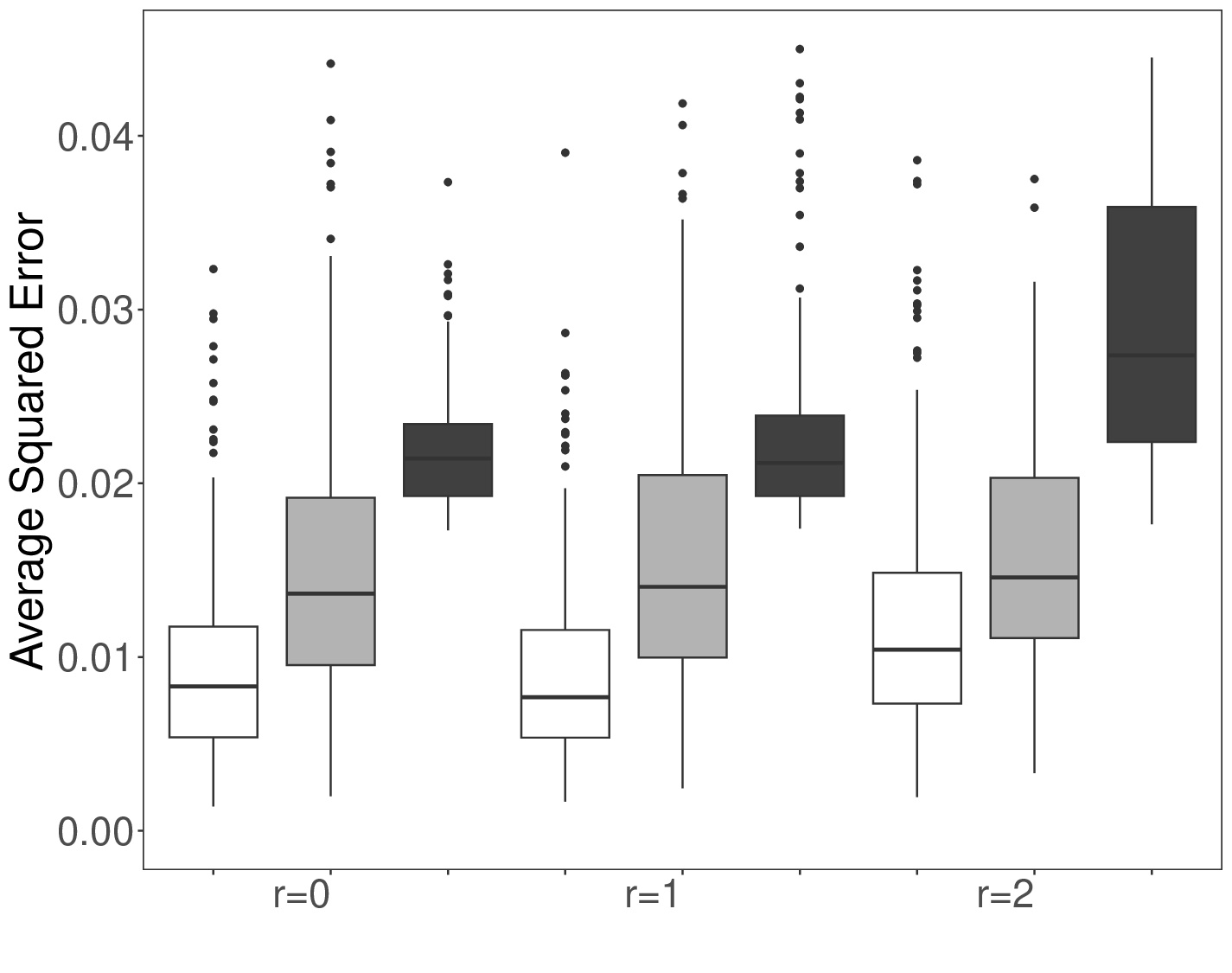}
  \caption{Scenario 3; $n=200$}\label{ASEc}
\end{subfigure}
\begin{subfigure}[b]{0.32\textwidth}
  \includegraphics[width=\textwidth]{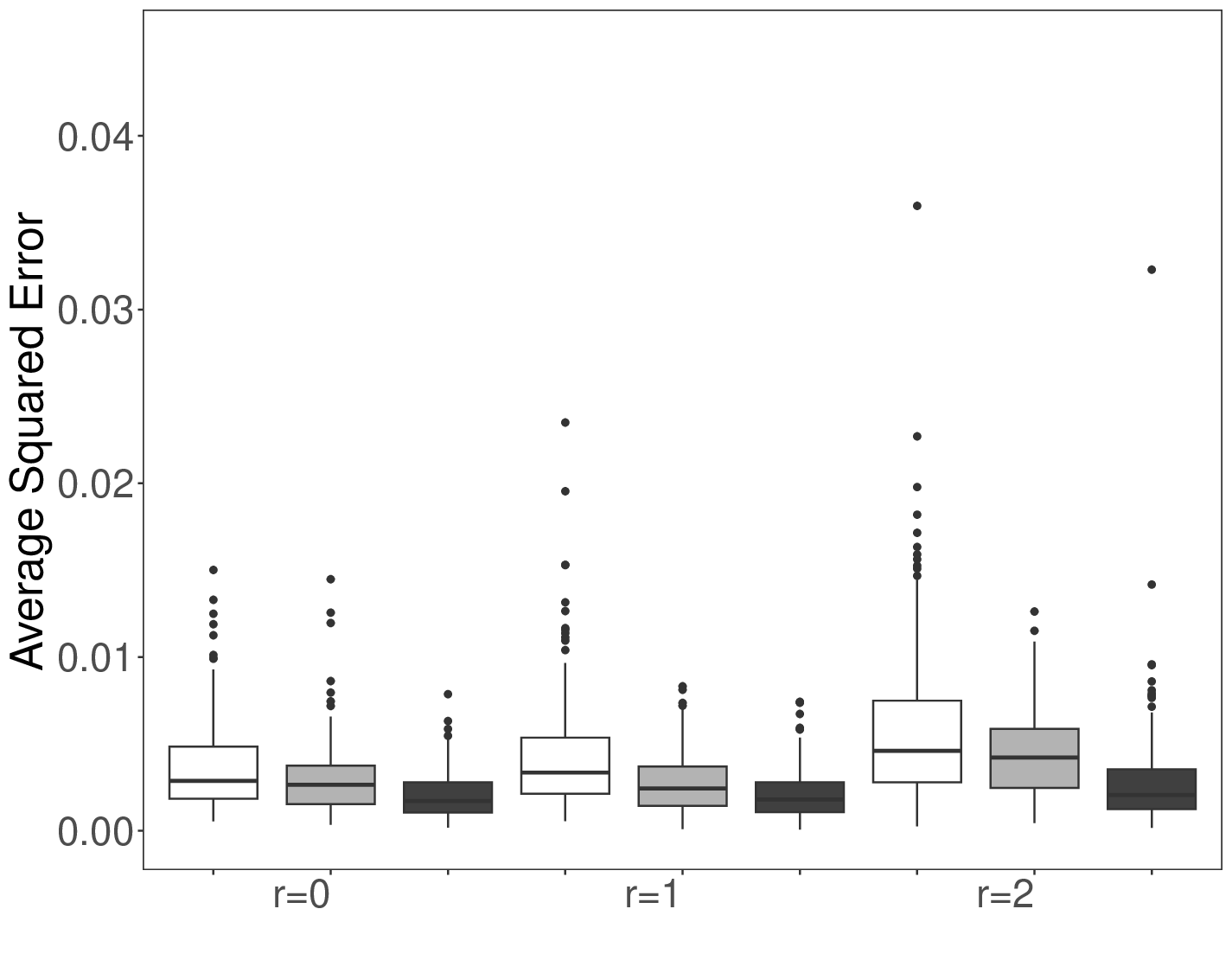}
  \caption{Scenario 1; $n=500$}\label{ASEd}
\end{subfigure}
\begin{subfigure}[b]{0.32\textwidth}
  \includegraphics[width=\textwidth]{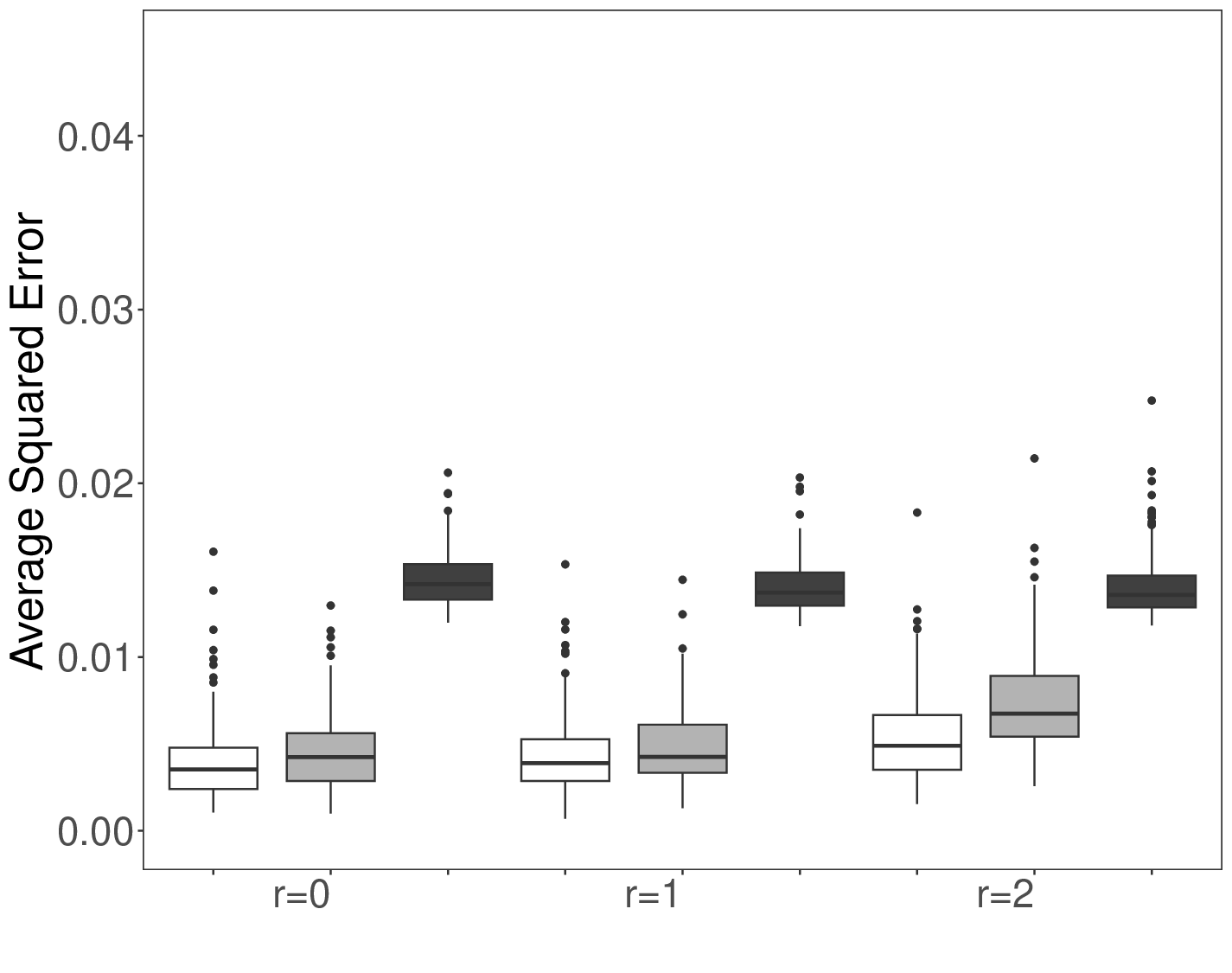}
  \caption{Scenario 2; $n=500$}\label{ASEe}
\end{subfigure}
\begin{subfigure}[b]{0.32\textwidth}
  \includegraphics[width=\textwidth]{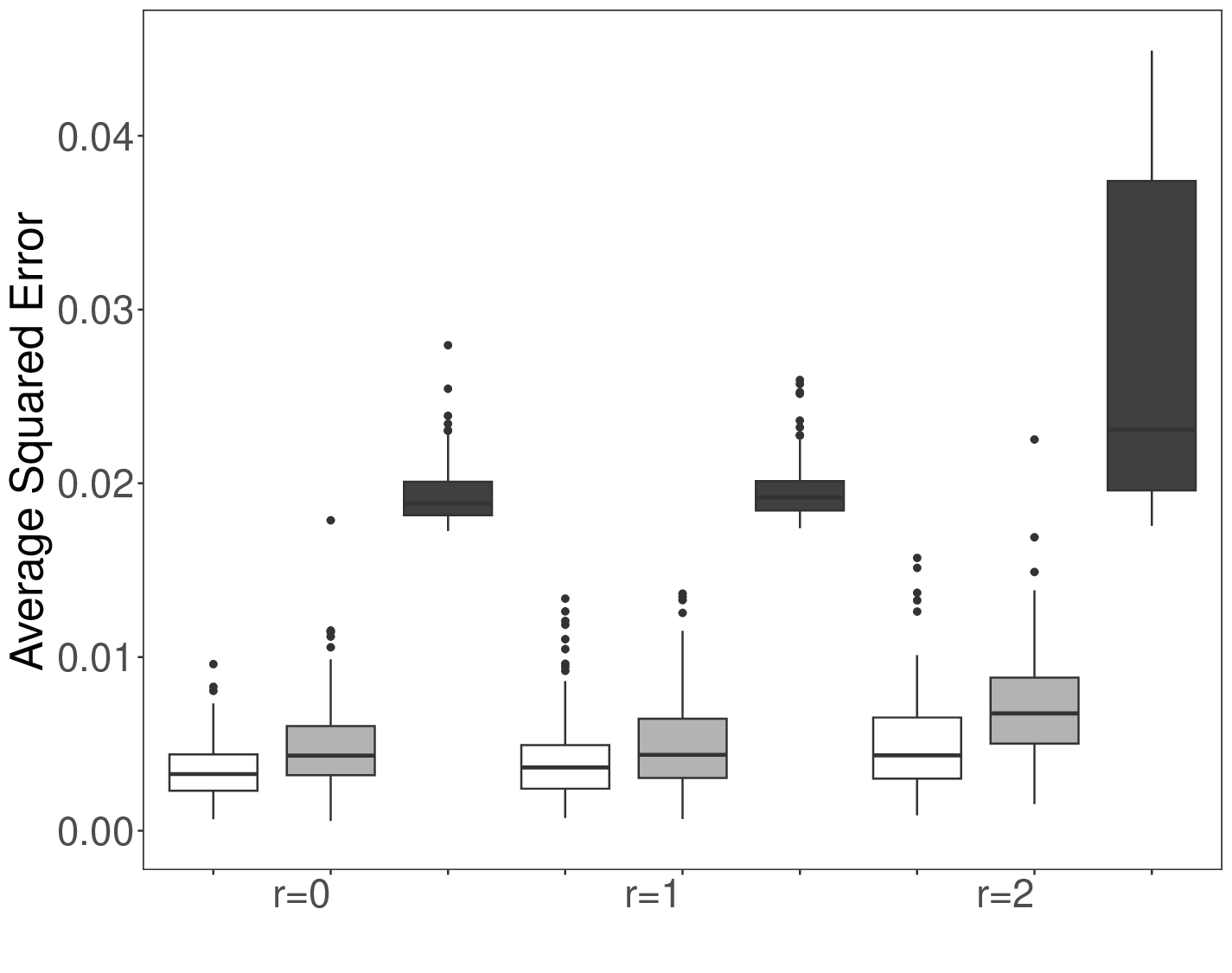}
  \caption{Scenario 3; $n=500$}\label{ASEf}
\end{subfigure}
\caption{{Boxplots of ASE for the SMCI-K (white boxplots), SMCI-S (grey boxplots) and GORMC (dark boxplots) models with $n=200$ and 500. }}
\label{ASE}
\end{figure}

\begin{figure}[H]
    \centering
    \begin{subfigure}[b]{0.32\textwidth}
        \includegraphics[width=\textwidth]{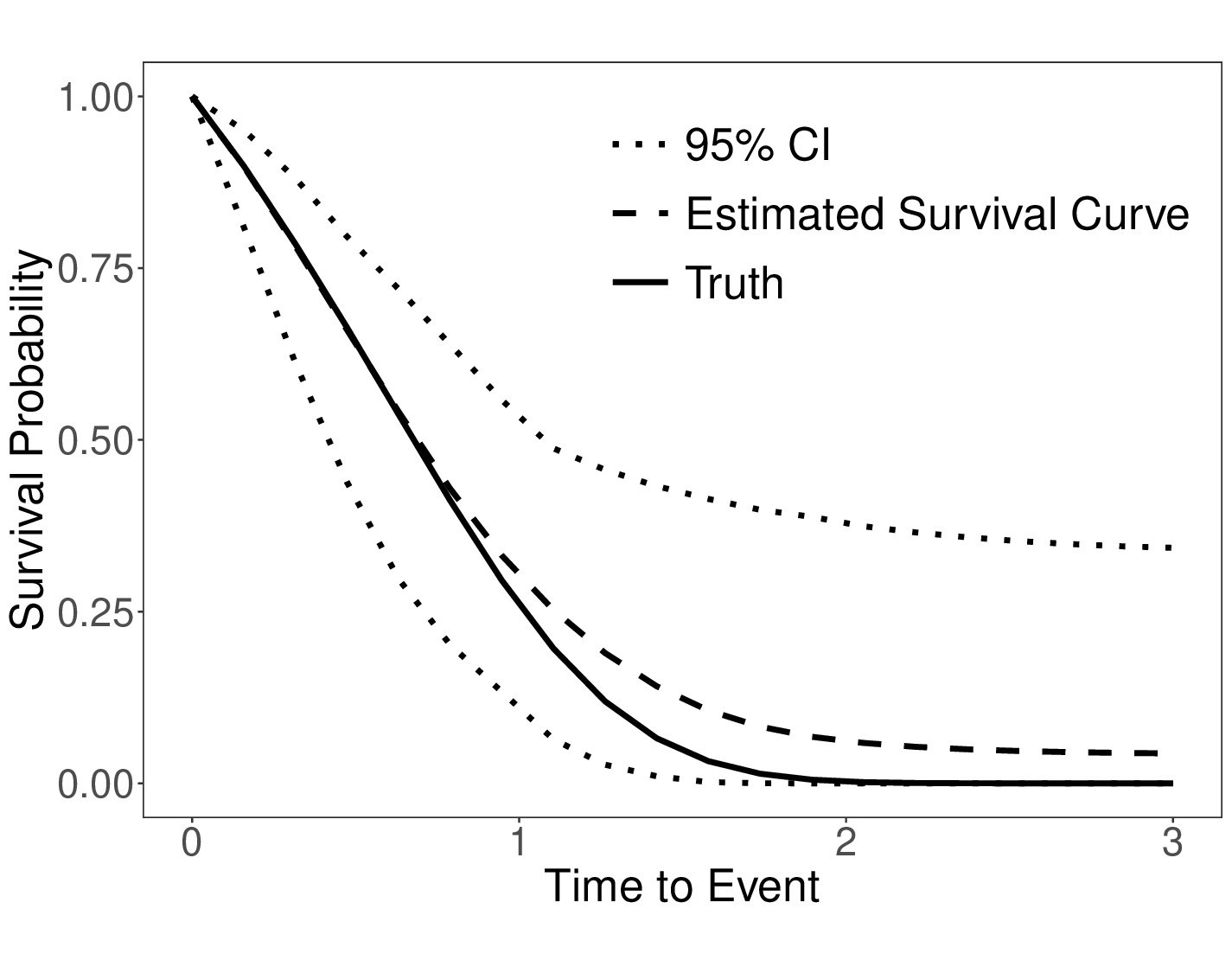}
        \caption{$n=200$; $r=0$}
    \end{subfigure}
    \begin{subfigure}[b]{0.32\textwidth}
        \includegraphics[width=\textwidth]{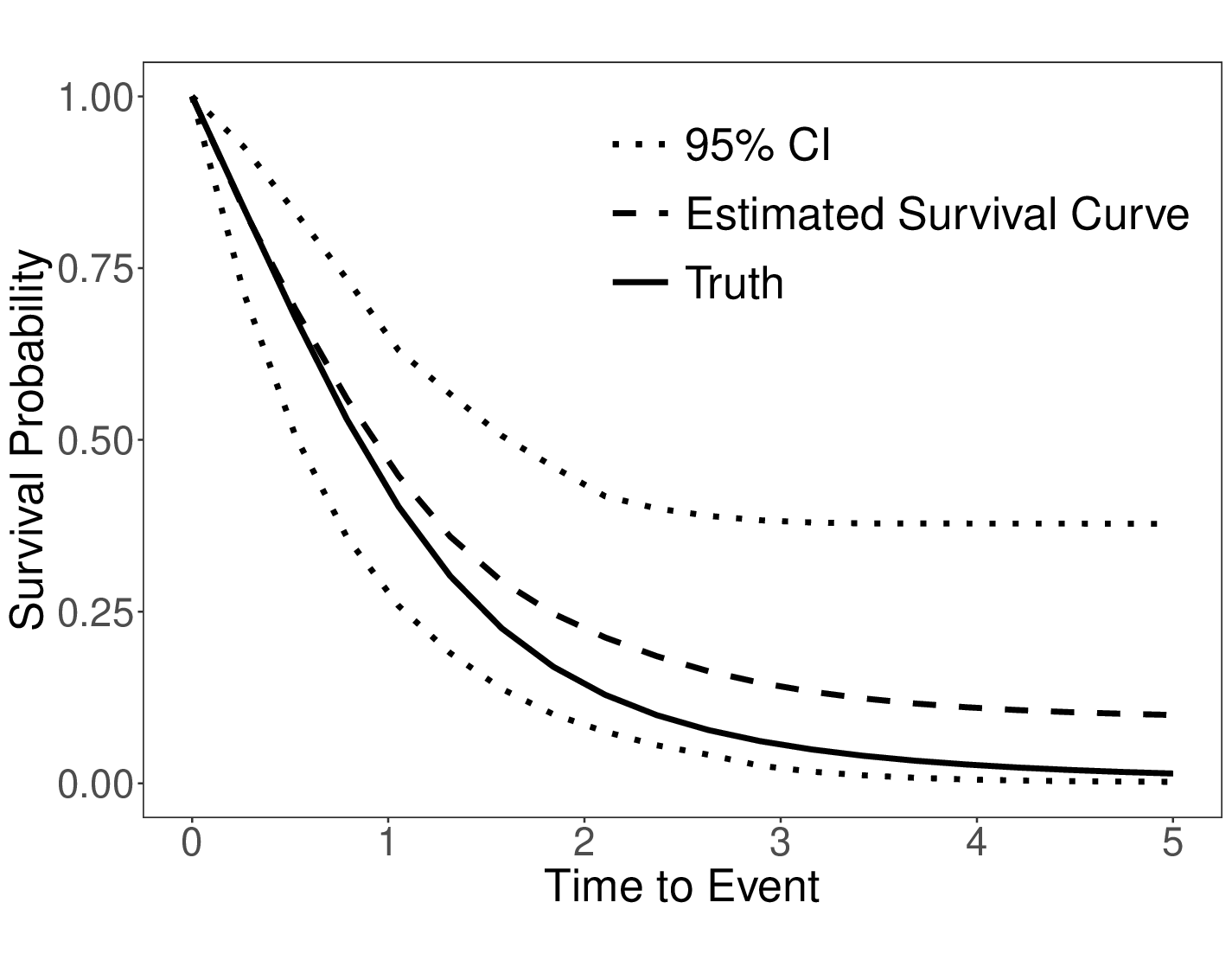}
        \caption{$n=200$; $r=1$}
    \end{subfigure}
    \begin{subfigure}[b]{0.32\textwidth}
        \includegraphics[width=\textwidth]{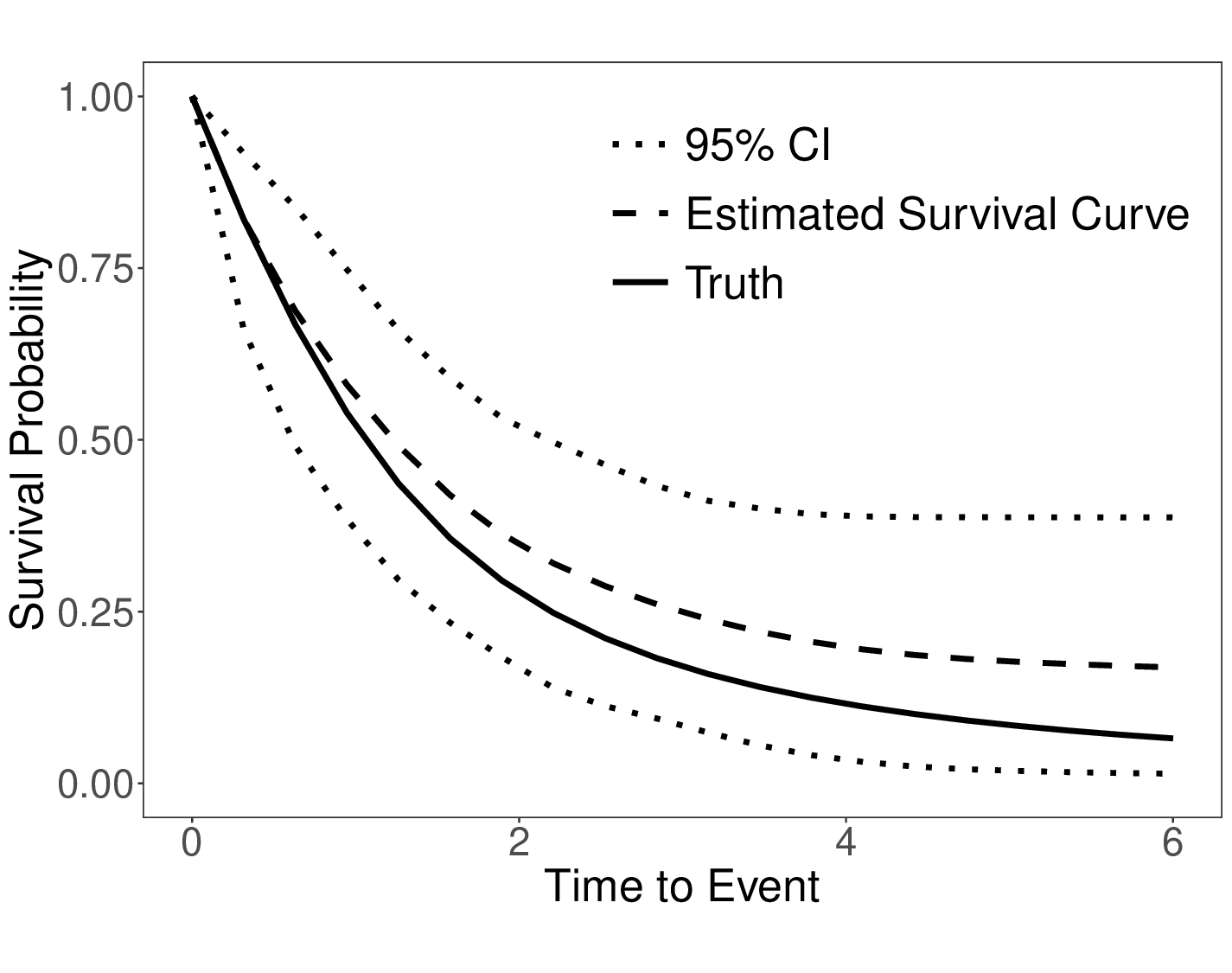}
        \caption{$n=200$; $r=2$}
    \end{subfigure}
    \begin{subfigure}[b]{0.32\textwidth}
        \includegraphics[width=\textwidth]{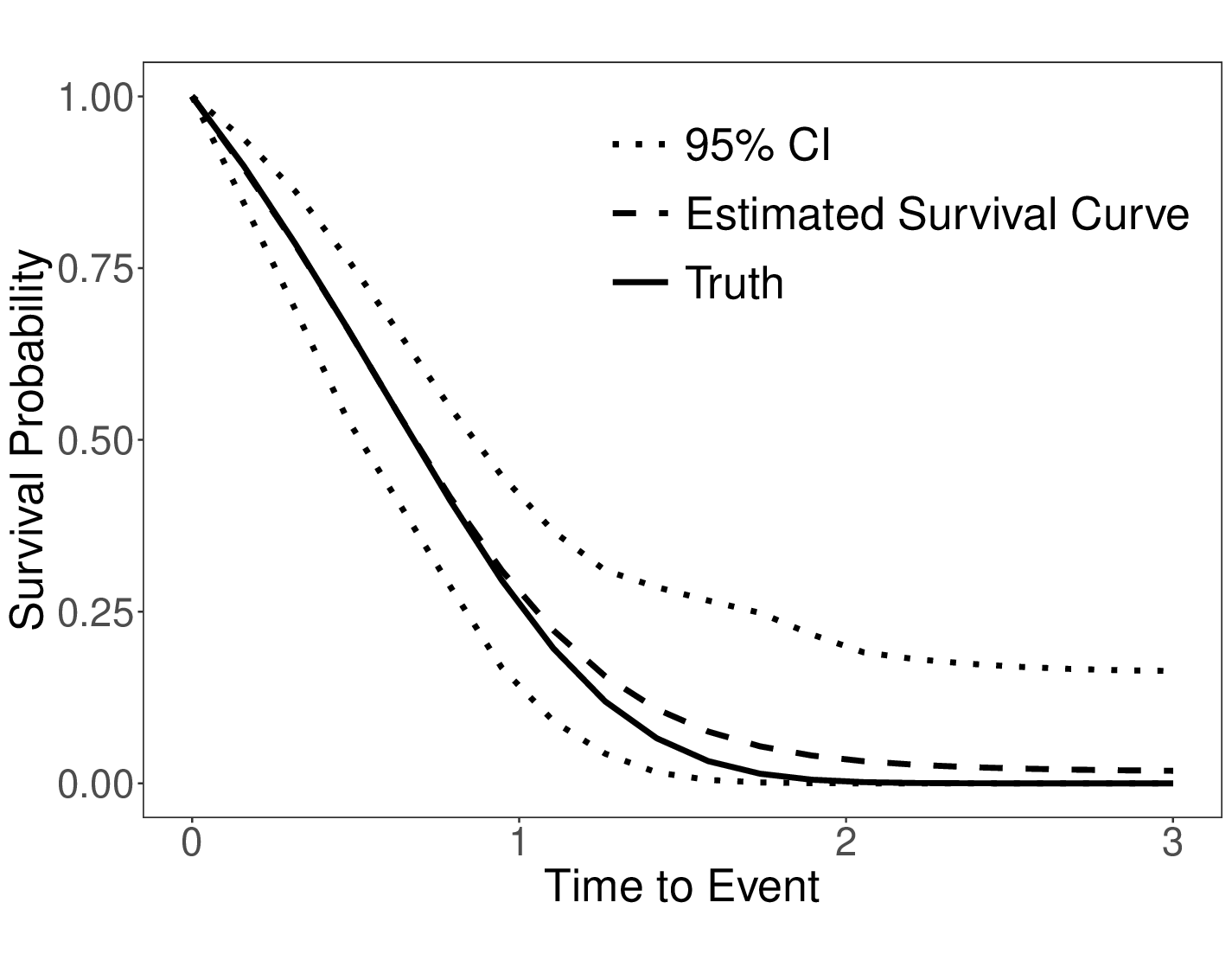}
        \caption{$n=500$; $r=0$}
    \end{subfigure}
    \begin{subfigure}[b]{0.32\textwidth}
        \includegraphics[width=\textwidth]{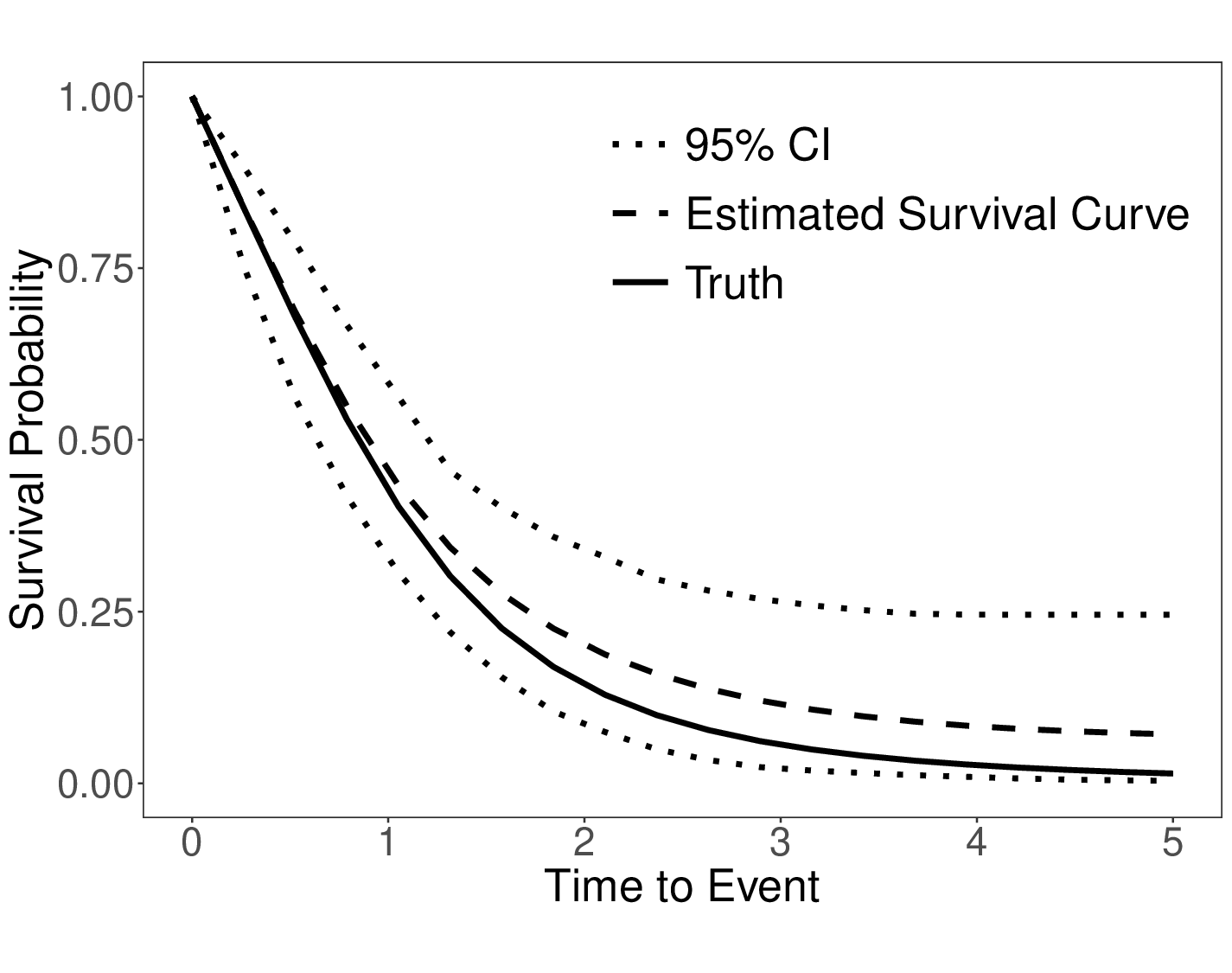}
        \caption{$n=500$; $r=1$}
    \end{subfigure}
    \begin{subfigure}[b]{0.32\textwidth}
        \includegraphics[width=\textwidth]{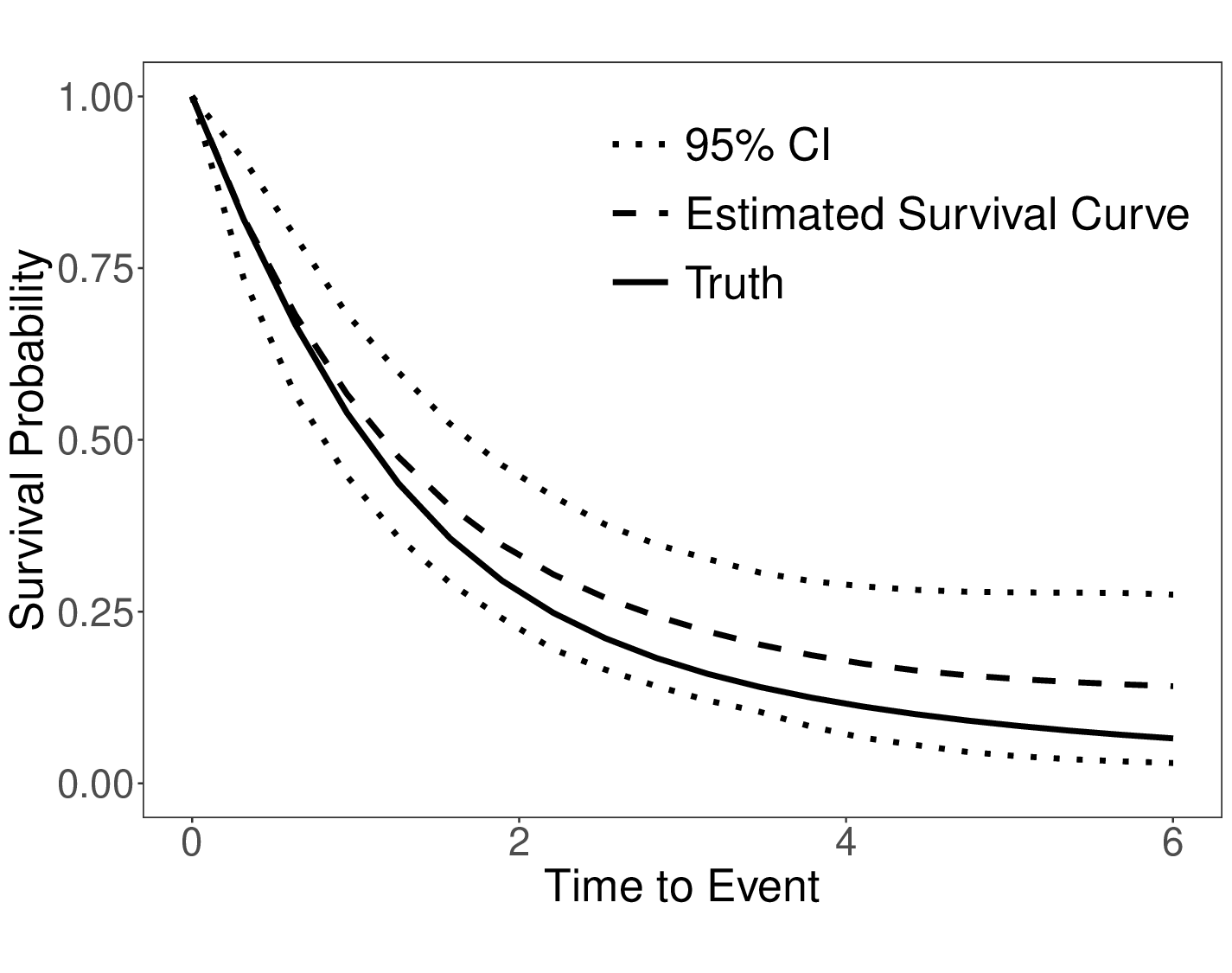}
        \caption{$n=500$; $r=2$}
    \end{subfigure}
    \caption{Estimated baseline survival functions $S_u(\cdot|\bZ=0)$ for Scenario 1. The solid curves show the true values, the dashed curves pertain to averaged estimates, and the dotted curves represent the upper and lower 2.5\% quantiles of the estimates. %\blue{2.5\% and 97.5\% quantiles.}
   % \xy{how are the confidence bands constructed?} 
    }%{\color{red}add explanation of dotted curves}}
    \label{Estimated baseline survival functions for Scenario 1 with n = 500.}
\end{figure}

%{\color{red} Discuss as the following order: Scenario 1(n=200, n=500), the performance in the incidence evaluated by the ASE and figure, the performance in latency. Scenario 2 and 3}

{For the incidence, Figure \ref{ASE} displays the boxplots of the average squared errors (ASEs) of $\hat{p}(\bX)$ obtained from the proposed SMCI-K, SMCI-S, and the GORMC methods.
%For the incidence, Figure \ref{ASE} displays the boxplots of the ASEs of $\hat{p}(\bX)$ obtained from all models. 
It shows that the proposed SMCI-K and SMCI-S methods demonstrate outstanding performance compared to the GORMC in some cases. %, with the kernel estimator showing better performance than the spline estimator. 
The ASEs for SMCI-K and SMCI-S are generally smaller than those from the GORMC method, regardless of the value of $r$.  An exception occurs in Scenario 1, where the underlying model is correctly specified, and all three methods exhibit small ASEs.
The SMCI-K approach achieves lower ASEs than SMCI-S in small-sample settings, with comparable performance at large sample sizes.
Both of them produce more outliers compared to the GORMC approach, due to their sensitivity to small changes or noise in the data, especially with small sample sizes.
%Overall, the SMCI-K method shows stable performance across all scenarios, consistently outperforming the SMCI-S and the GORMC in non-logistic scenarios, making it a robust choice when the true underlying model of the data is unknown or non-logistic.}
Overall, the results indicate that the proposed methods typically perform well and are preferable to the GORMC method when the true underlying model of the data is unknown. }

{For the latency component, Tables \ref{Simulation result of Scenario 1}-\ref{Simulation result of Scenario 3} present the summary statistics of $\hat{\bbeta}$ in three scenarios for the three models. % It shows that the proposed SMCI performs
%When the true model for the incidence is a logistic model, the proposed approach as well as the GORMC method for the estimation of $\bbeta$  with small biases. The estimated standard errors are comparable with the empirical approach and the CPs are close to the nominal level of 0.95. 
The SMCI-K and SMCI-S approaches yield smaller biases for
$\bbeta$ relative to the GORMC method except in Scenario 1. 
A significant advantage is observed in the case $r=2$ and $n=500$, which may 
arise from the fact that the inaccurate estimate of the \blue{incidence} %probability of being uncured
has a greater impact on the estimate of $\bbeta$ when the number of censored observations is large.
%As illustrated in Figure \ref{Estimated baseline survival functions for Scenario 1 with n = 500.},
Figures \ref{Estimated baseline survival functions for Scenario 1 with n = 500.}-\ref{Estimated baseline survival functions for Scenario 3 with n = 500.} exhibit the estimated baseline survival curves, which slightly deviate from the true curves when $p(\bX)$ is correctly specified and reveal negligible bias otherwise.} %Figure \ref{ASE} displayed the ASE of these two models in various settings.

\begin{figure}[H]
    \centering
    \begin{subfigure}[b]{0.32\textwidth}
        \includegraphics[width=\textwidth]{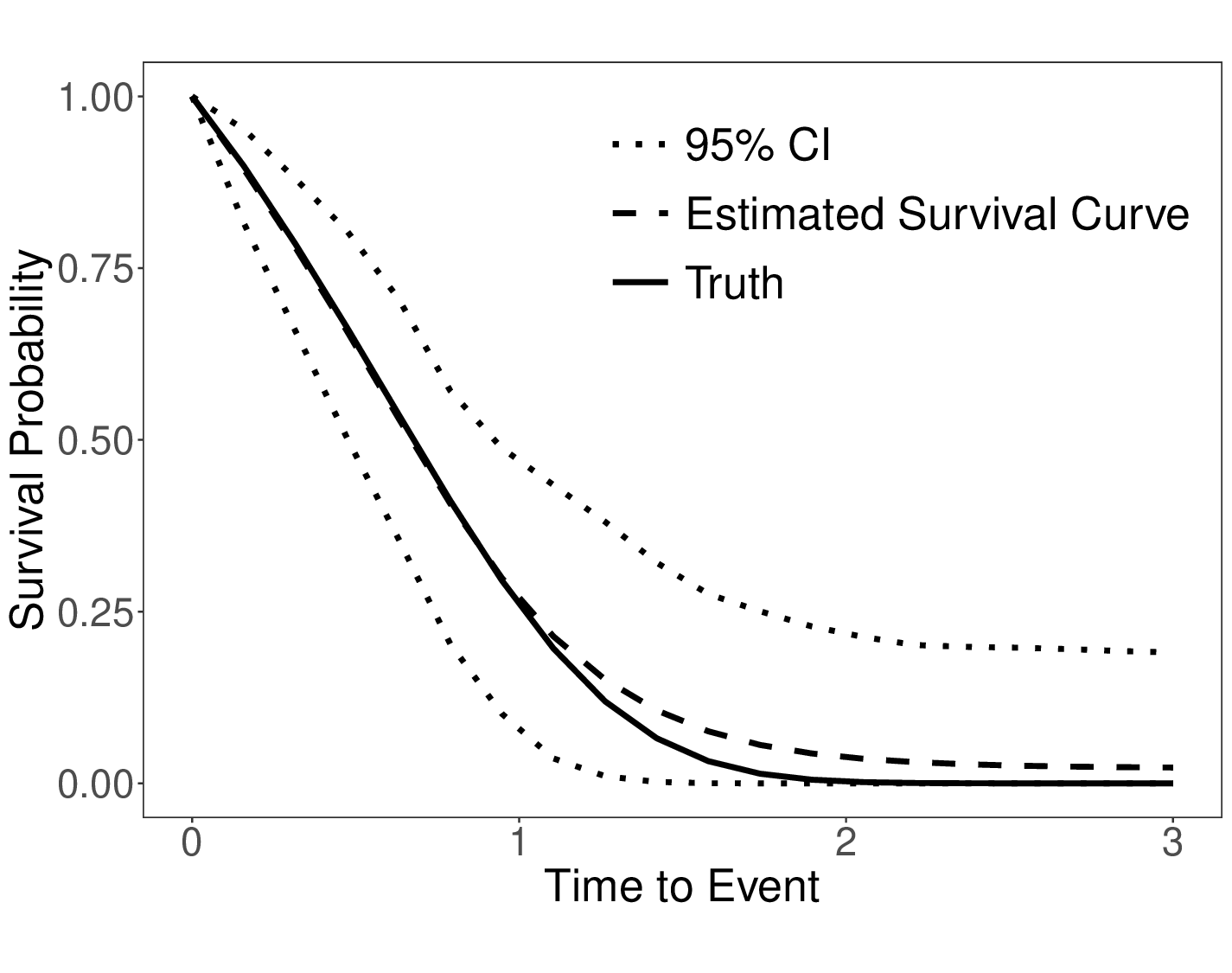}
        \caption{$n=200$; $r=0$}
    \end{subfigure}
    \begin{subfigure}[b]{0.32\textwidth}
        \includegraphics[width=\textwidth]{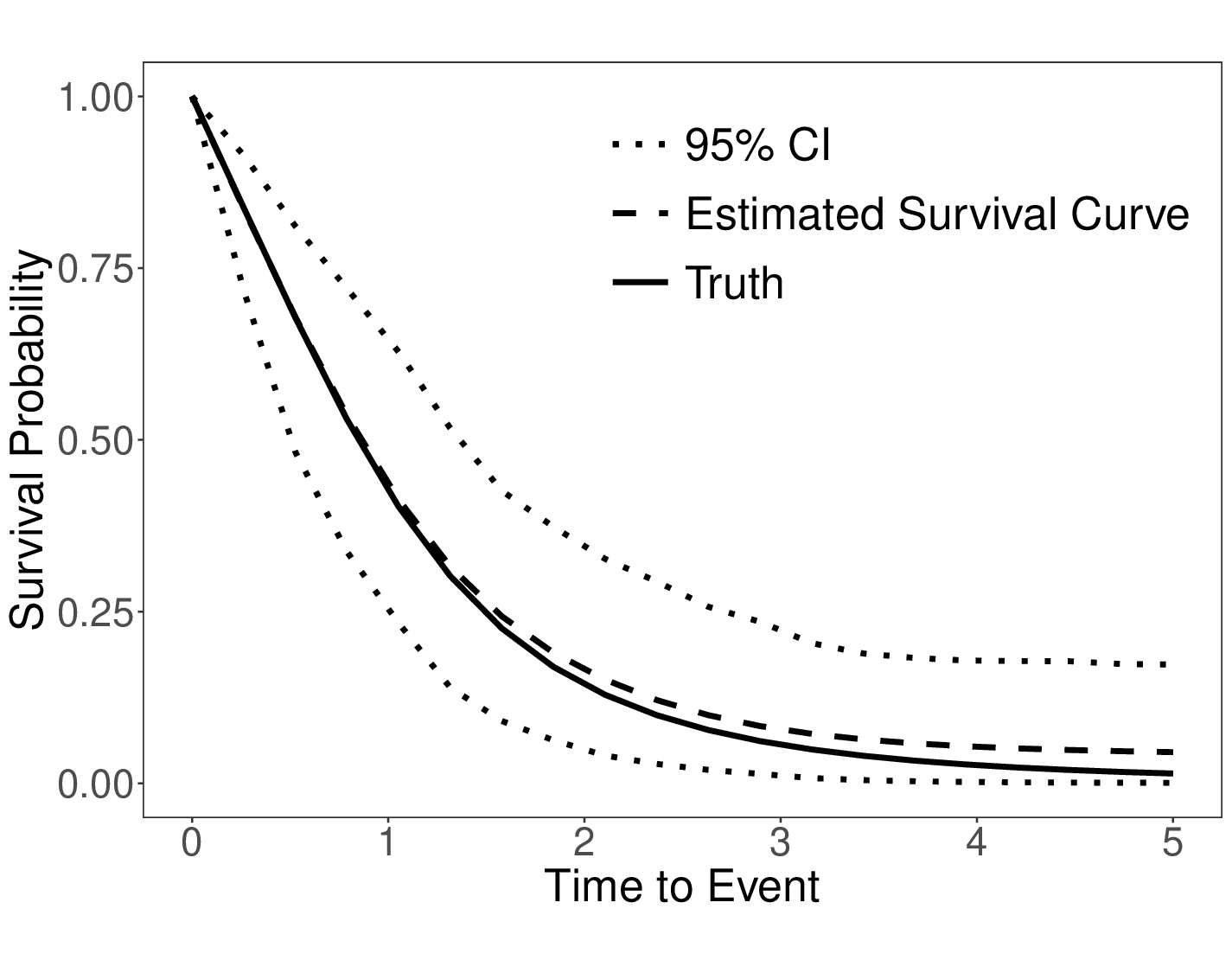}
        \caption{$n=200$; $r=1$}
    \end{subfigure}
    \begin{subfigure}[b]{0.32\textwidth}
        \includegraphics[width=\textwidth]{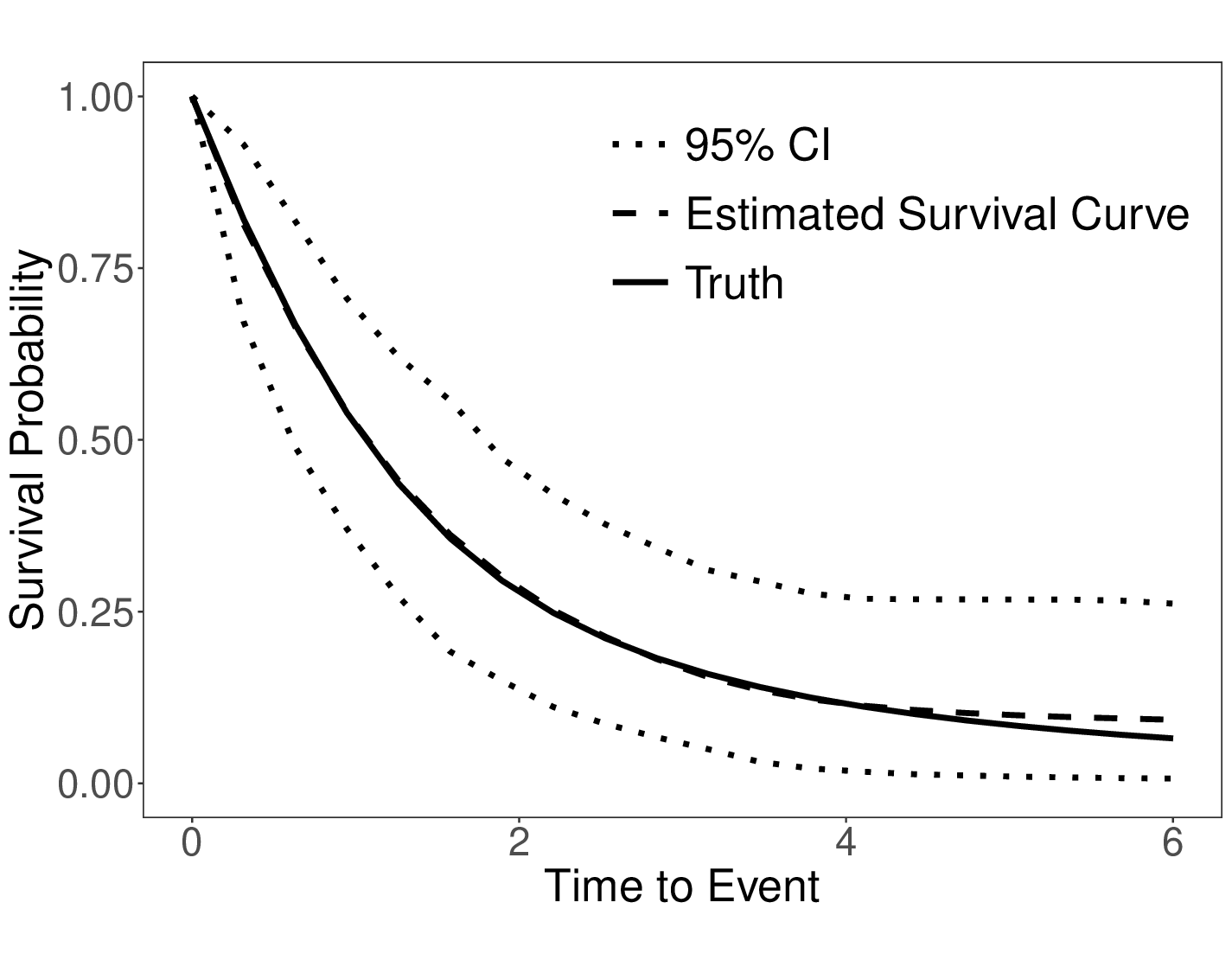}
        \caption{$n=200$; $r=2$}
    \end{subfigure}
    \begin{subfigure}[b]{0.32\textwidth}
        \includegraphics[width=\textwidth]{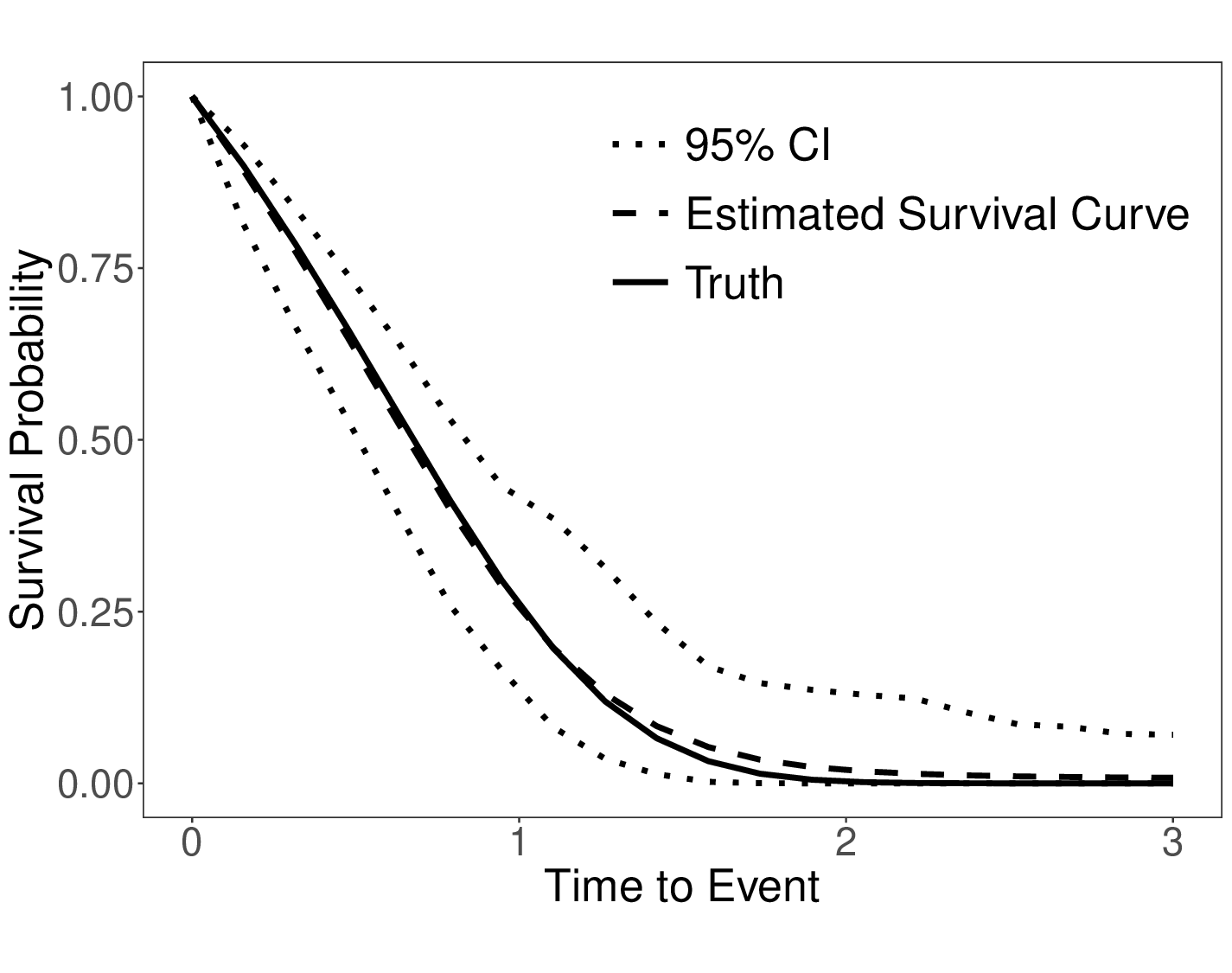}
        \caption{$n=500$; $r=0$}
    \end{subfigure}
    \begin{subfigure}[b]{0.32\textwidth}
        \includegraphics[width=\textwidth]{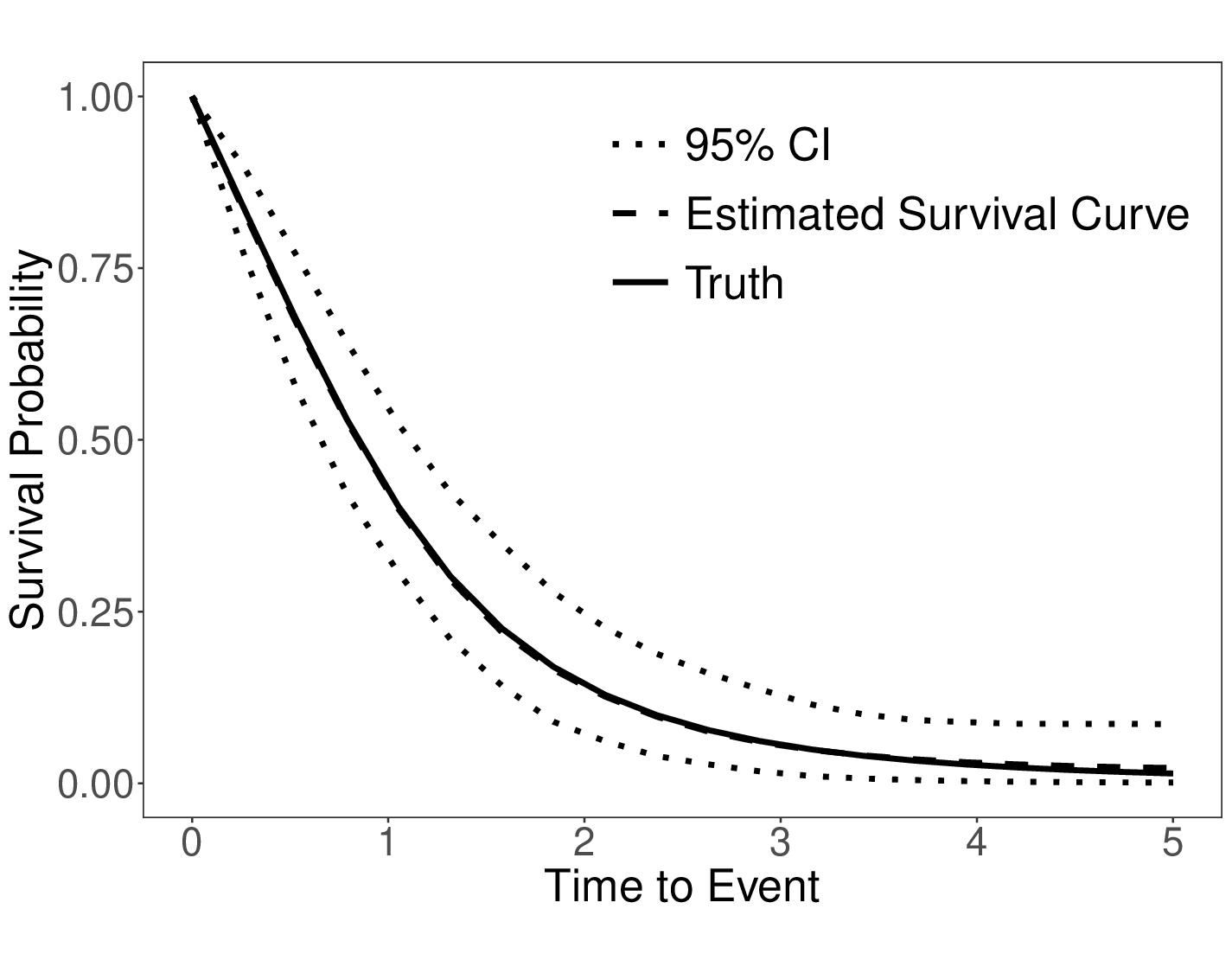}
        \caption{$n=500$; $r=1$}
    \end{subfigure}
    \begin{subfigure}[b]{0.32\textwidth}
        \includegraphics[width=\textwidth]{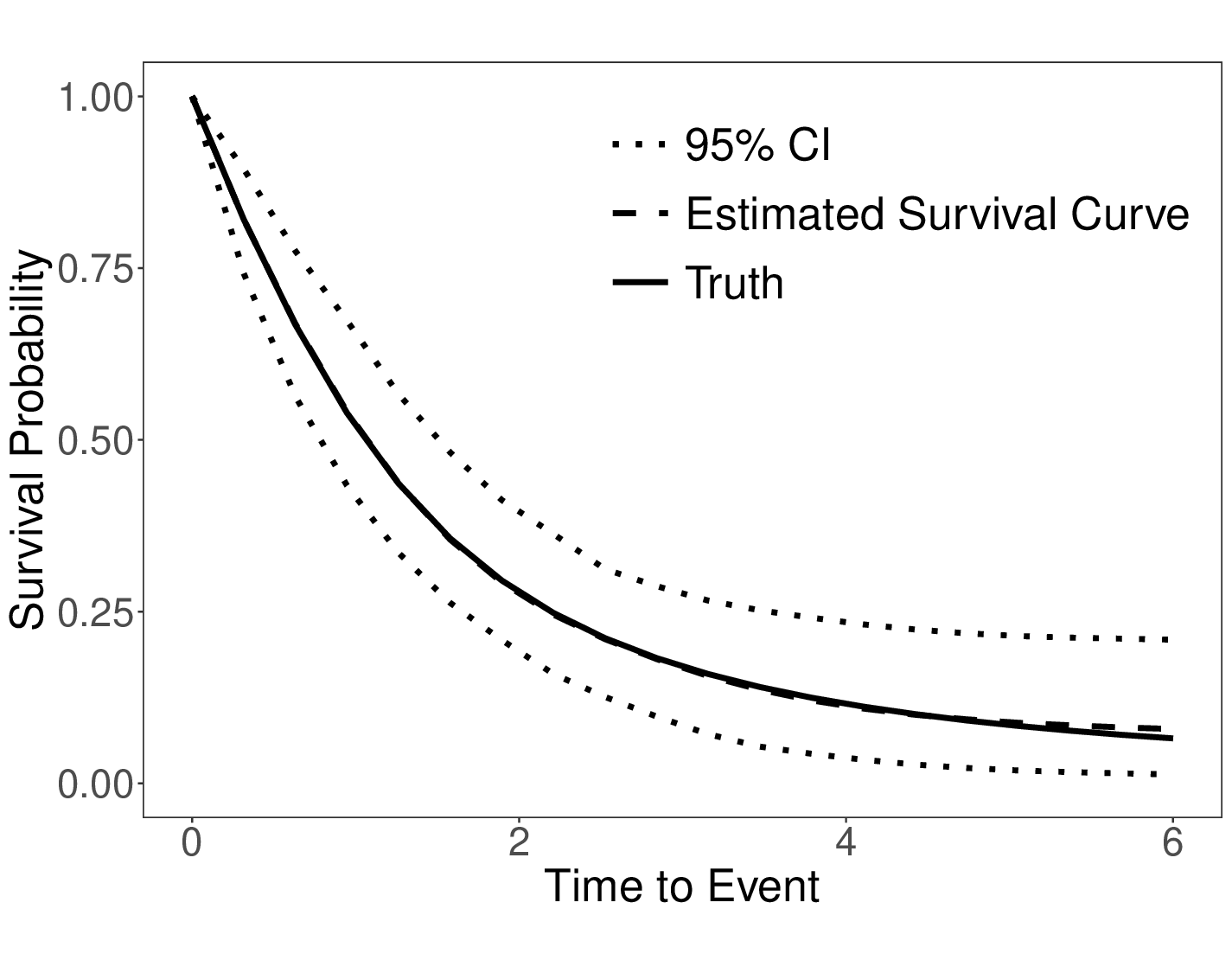}
        \caption{$n=500$; $r=2$}
    \end{subfigure}
    \caption{Estimated baseline survival functions $S_u(\cdot|\bZ=0)$ for Scenario 2. The solid curves show the true values, the dashed curves pertain to averaged estimates, and the dotted curves represent the upper and lower 2.5\% quantiles of the estimates. 
    }
    \label{Estimated baseline survival functions for Scenario 2 with n = 500.}
\end{figure}

\begin{figure}[H]
    \centering
    \begin{subfigure}[b]{0.32\textwidth}
        \includegraphics[width=\textwidth]{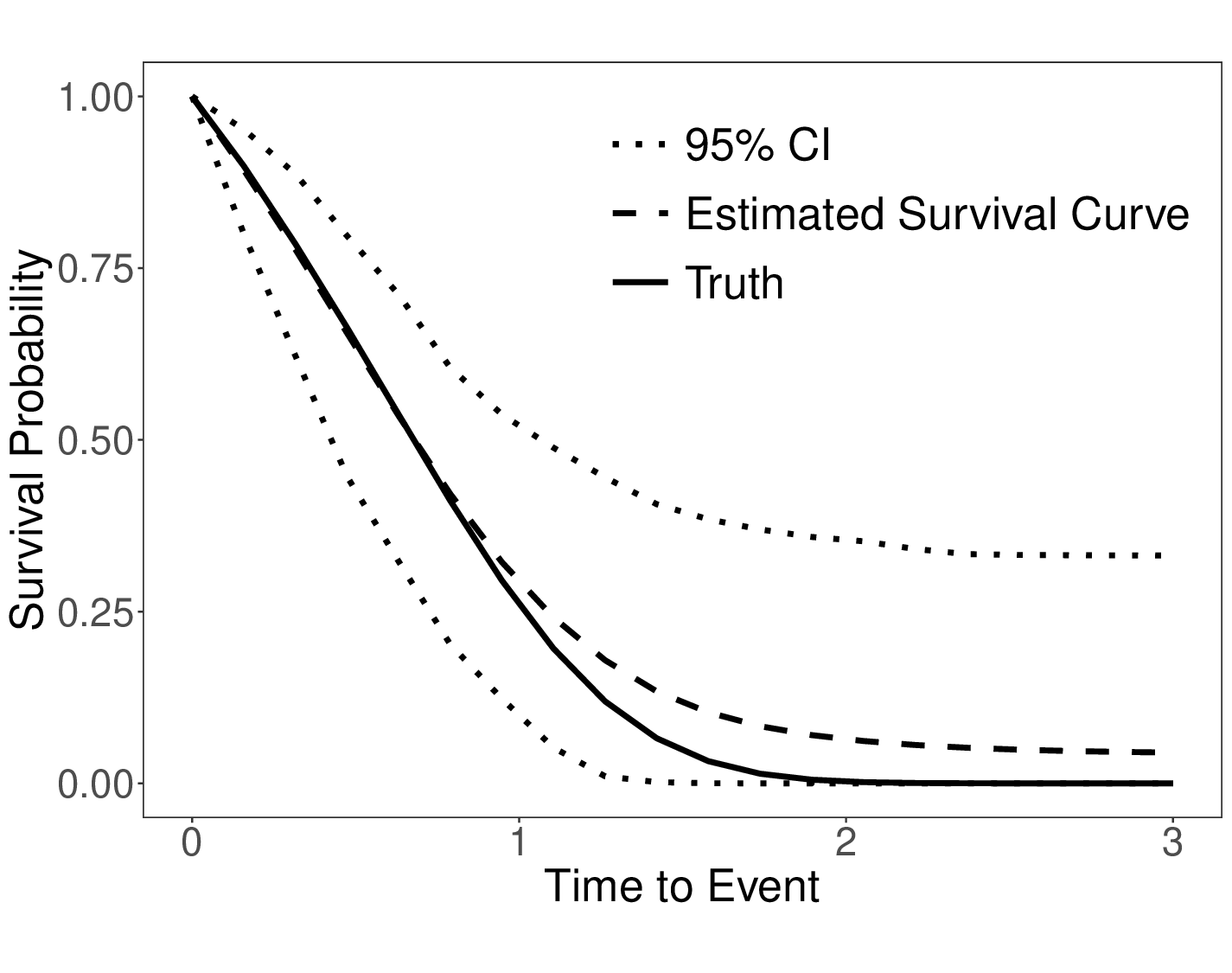}
        \caption{$n=200$; $r=0$}
    \end{subfigure}
    \begin{subfigure}[b]{0.32\textwidth}
        \includegraphics[width=\textwidth]{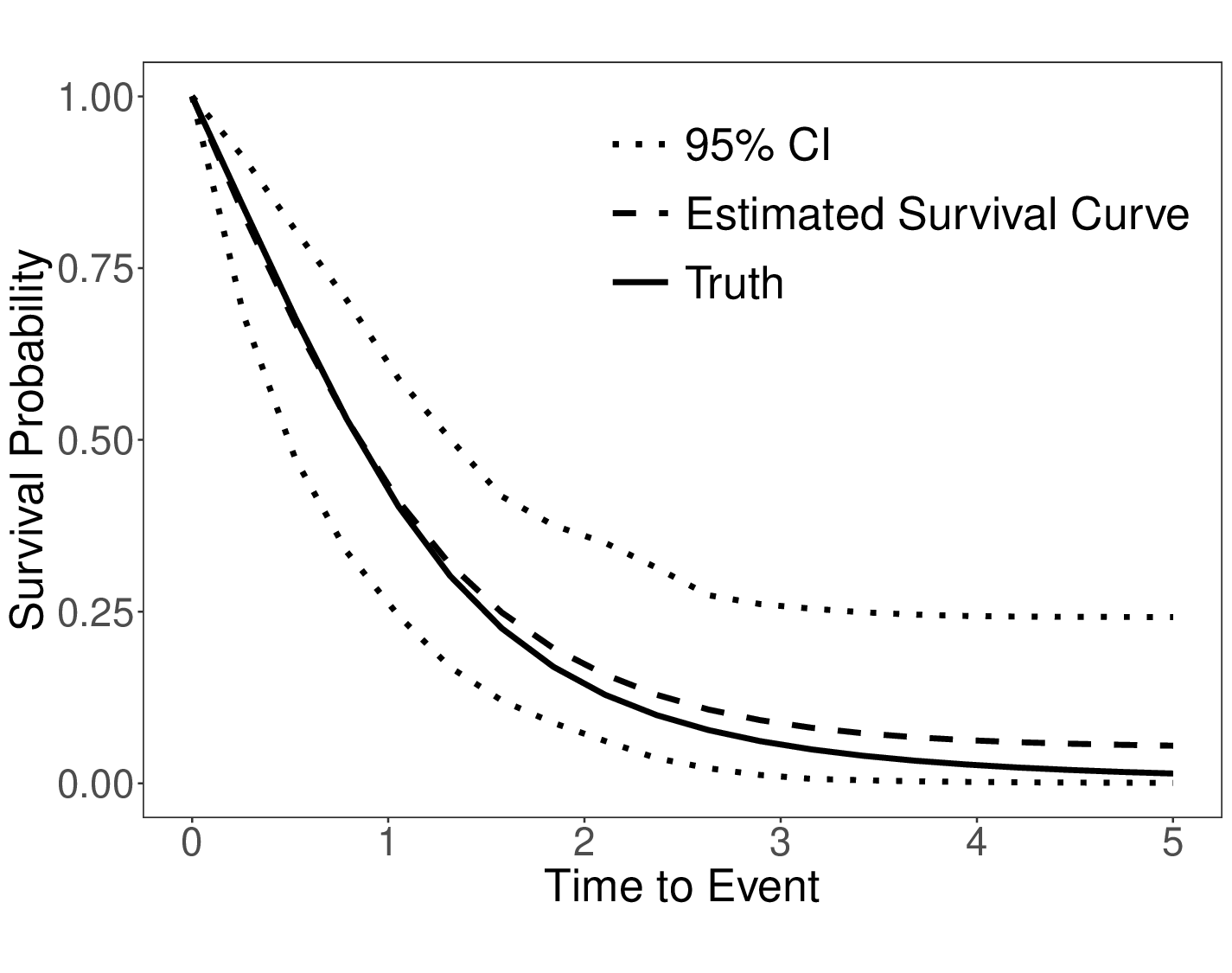}
        \caption{$n=200$; $r=1$}
    \end{subfigure}
    \begin{subfigure}[b]{0.32\textwidth}
        \includegraphics[width=\textwidth]{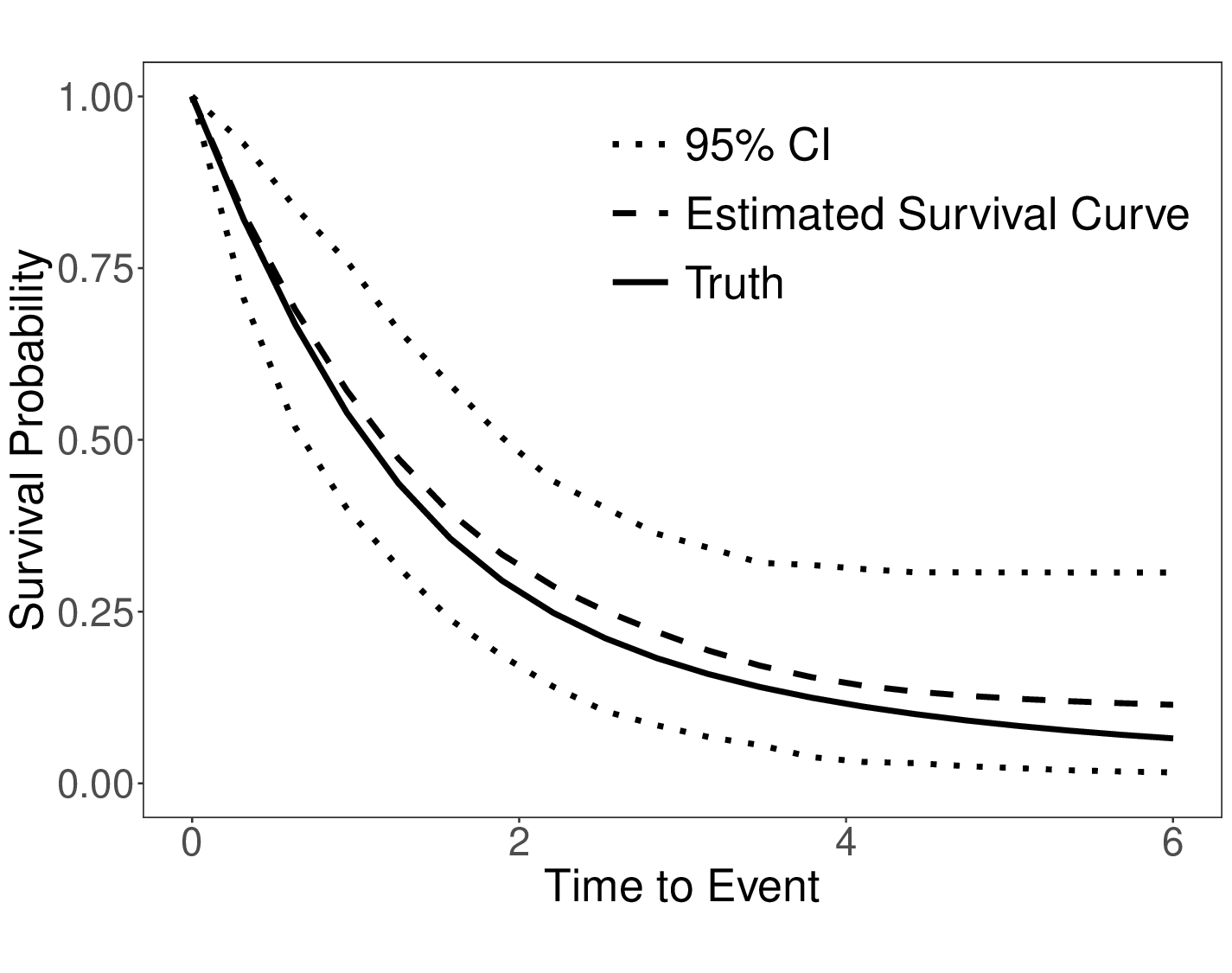}
        \caption{$n=200$; $r=2$}
    \end{subfigure}
    \begin{subfigure}[b]{0.32\textwidth}
        \includegraphics[width=\textwidth]{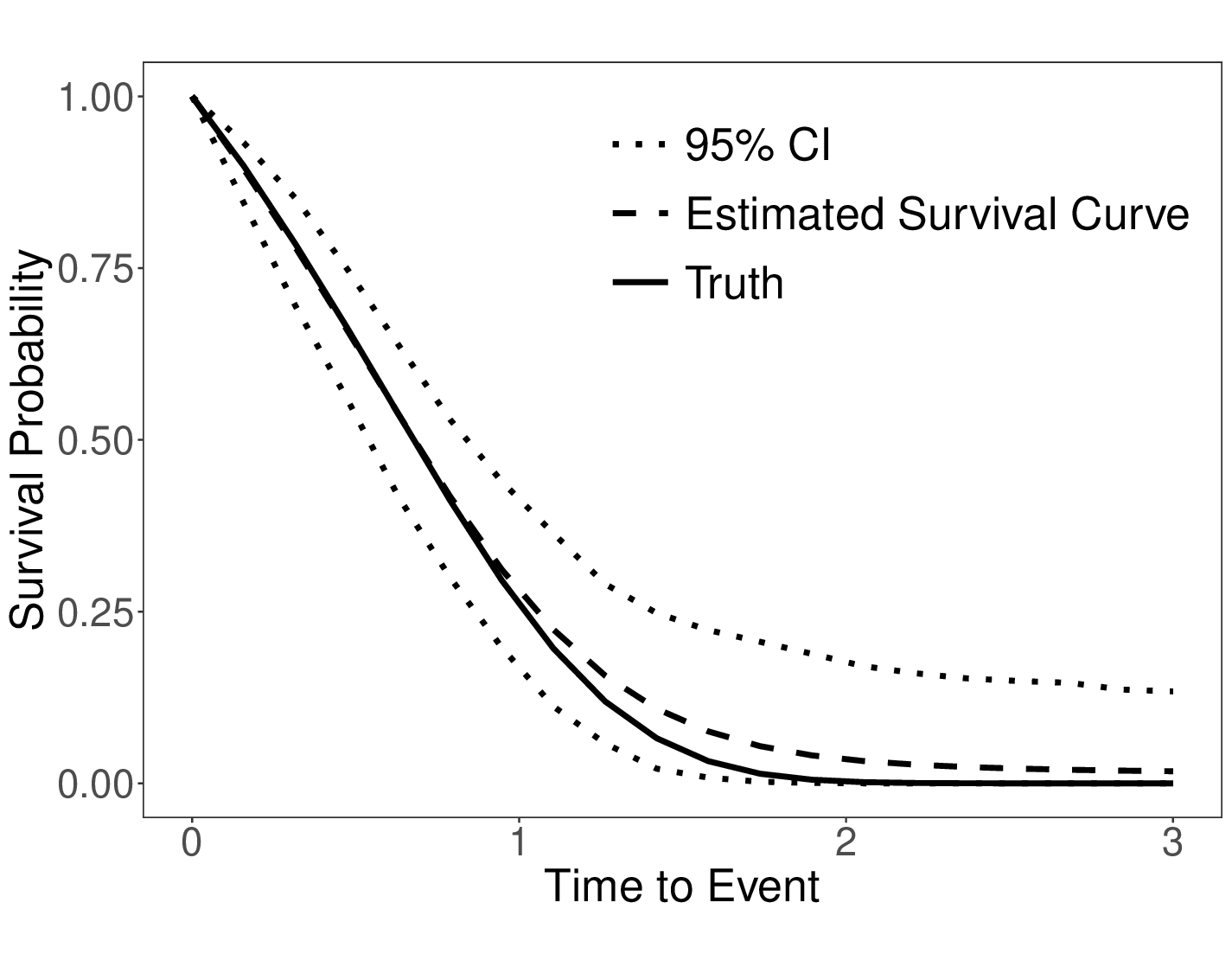}
        \caption{$n=500$; $r=0$}
    \end{subfigure}
    \begin{subfigure}[b]{0.32\textwidth}
        \includegraphics[width=\textwidth]{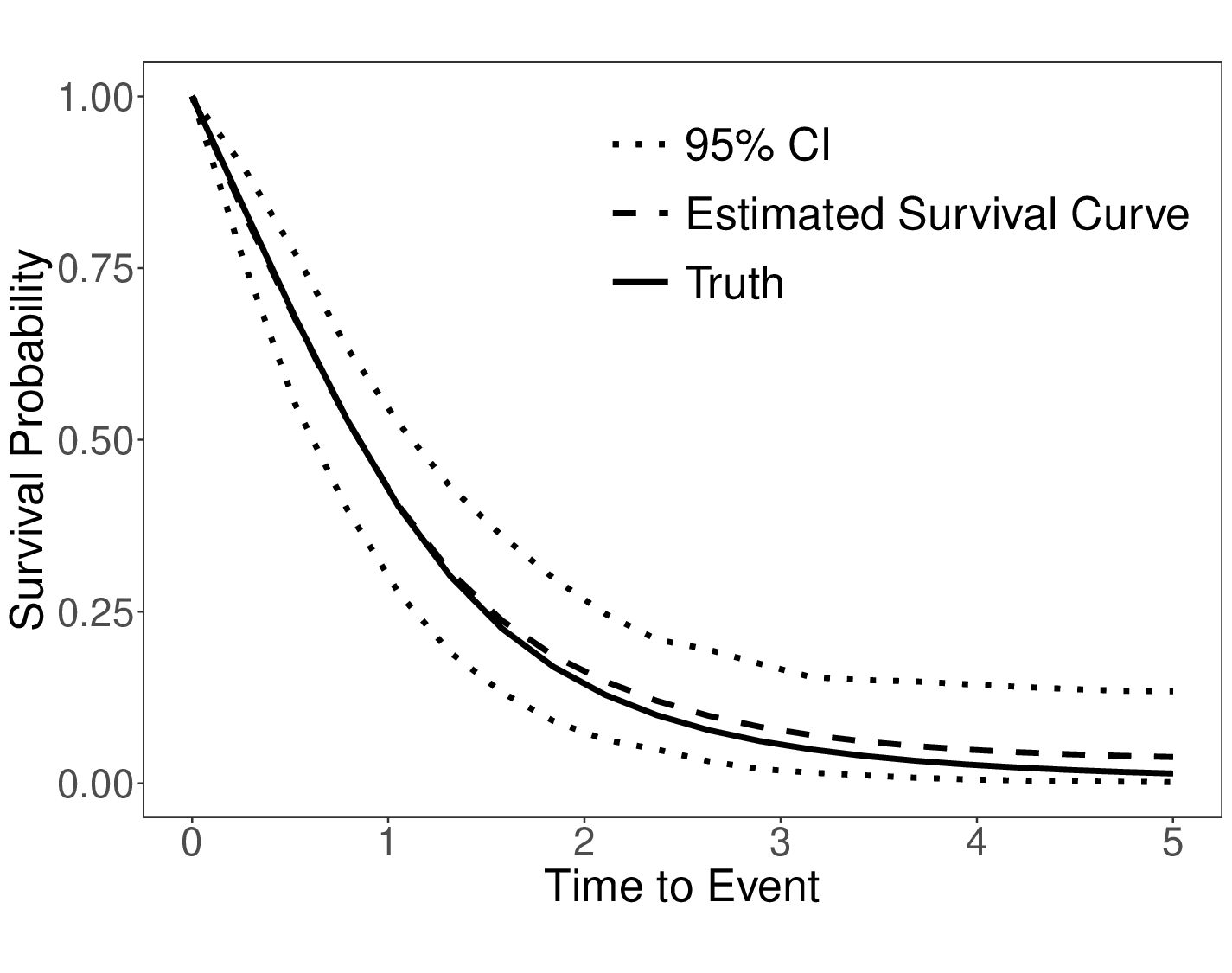}
        \caption{$n=500$; $r=1$}
    \end{subfigure}
    \begin{subfigure}[b]{0.32\textwidth}
        \includegraphics[width=\textwidth]{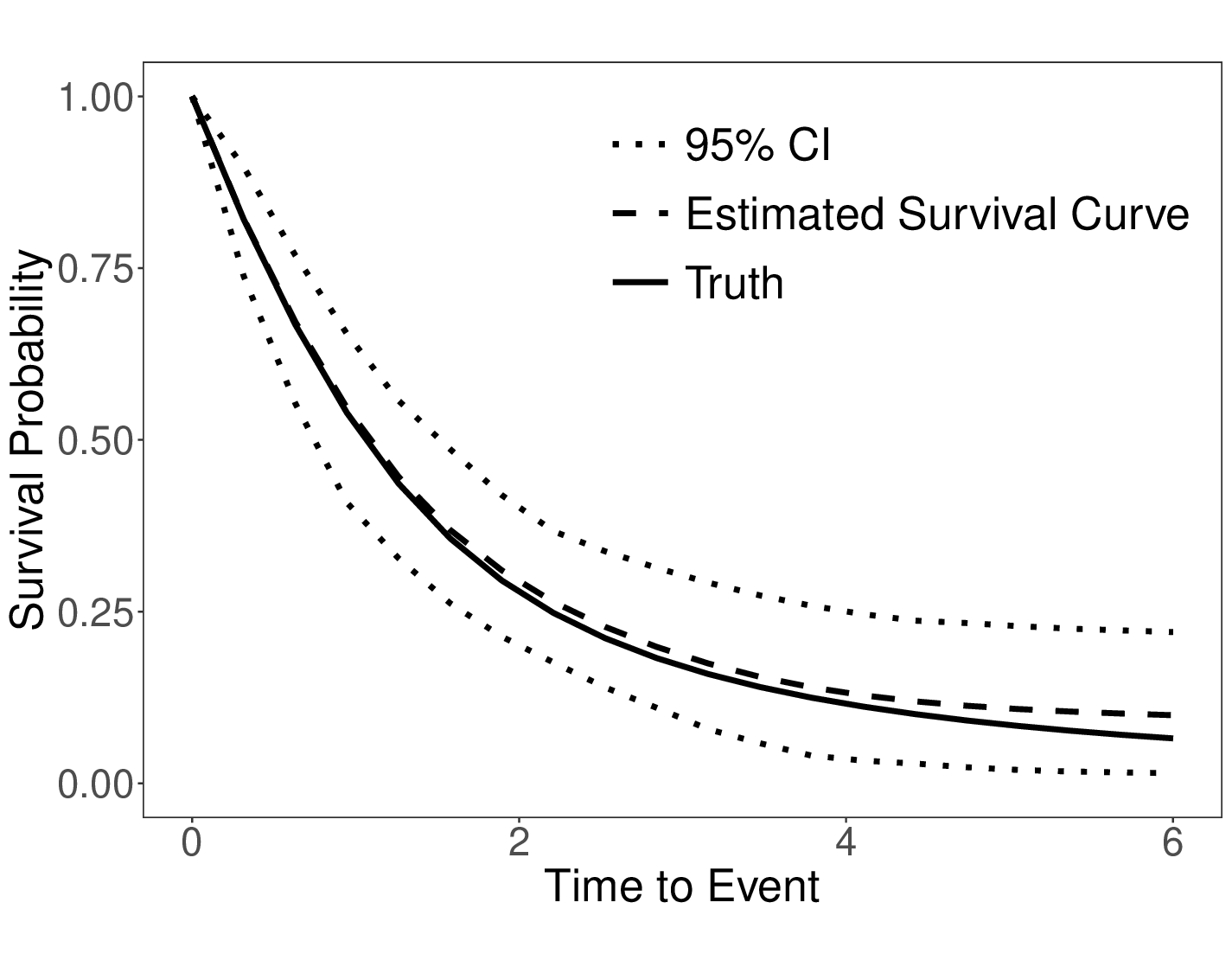}
        \caption{$n=500$; $r=2$}
    \end{subfigure}
    \caption{Estimated baseline survival functions $S_u(\cdot|\bZ=0)$ for Scenario 3. The solid curves show the true values, the dashed curves pertain to averaged estimates, and the dotted curves represent the upper and lower 2.5\% quantiles of the estimates.  }
    \label{Estimated baseline survival functions for Scenario 3 with n = 500.}
\end{figure}

\begin{table}[htp]
\centering
\caption{Simulation result of Scenario 1 with $n = 200$ and 500. Bias: the estimated bias; ESD: empirical standard deviation; ESE: empirical standard error estimate; CP: the 95\% empirical coverage probabilities.}
\begin{tabular}{ccccccccccccccccc}
\hline\hline
&& \multicolumn{4}{c}{SMCI-K} && \multicolumn{4}{c}{SMCI-S}&& \multicolumn{4}{c}{GORMC} \\%& \multicolumn{4}{c}{SIC} \\
    %    \cline{2-13}
\cline{3-6} \cline{8-11} \cline{13-16}
$r$ &Par&Bias &ESD &ESE &CP &&Bias &ESD &ESE &CP &&Bias &ESD &ESE &CP \\
\hline
\multicolumn{2}{c}{ $n=200$} \\
\multirow{3}{*}{$0$}&$\beta_1$ &0.02 &0.26 &0.26 &0.94 && 0.07 & 0.24 & 0.25 & 0.94 &&0.06 &0.24 &0.24 &0.95 \\
&$\beta_2$ &0.00 &0.16 &0.18 &0.98 &&-0.05 & 0.18 & 0.17 & 0.94 &&-0.05 &0.14 &0.17 &0.97   \\
&$\beta_3$ &0.01 &0.30 &0.30 &0.94 && 0.07 & 0.32 & 0.28 & 0.92 &&0.06 &0.27 &0.27 &0.96   \\ 
% &\multicolumn{1}{c}{$r=1$}\\
\multirow{3}{*}{$1$}&$\beta_1$ &-0.03 &0.33 &0.34 &0.94 && 0.01 & 0.34 & 0.35 & 0.96 &&0.00 &0.34 &0.33 &0.93  \\
&$\beta_2$ &0.05 &0.20 &0.21 &0.96 &&-0.02 & 0.20 & 0.22 & 0.94 &&0.02 &0.22 &0.21 &0.94  \\
&$\beta_3$ &0.01 &0.35 &0.39 &0.98 && 0.04 & 0.37 & 0.40 & 0.95 &&0.04 &0.36 &0.38 &0.97 \\
%&\multicolumn{1}{c}{{$r=2$}}\\
\multirow{3}{*}{$2$}&$\beta_1$ &-0.06 &0.41 &0.41 &0.94 && 0.03 & 0.41 & 0.43 & 0.96 &&-0.01 &0.43 &0.41 &0.94  \\
&$\beta_2$ &0.03 &0.22 &0.25 &0.96 &&-0.01 & 0.25 & 0.27 & 0.95 &&-0.02 &0.24 &0.25 &0.96   \\
&$\beta_3$ &-0.06 &0.47 &0.48 &0.94 &&-0.02 & 0.47 & 0.50 & 0.96 &&-0.03 &0.48 &0.47 &0.93 \\
%\hline
\multicolumn{2}{c}{ $n=500$}\\
\multirow{3}{*}{$0$}
&$\beta_1$ &0.01 &0.14 &0.16 &0.96 && 0.03 & 0.16 & 0.15 & 0.93 &&0.02 &0.14 &0.15 &0.94  \\
&$\beta_2$ &0.01 &0.09 &0.11 &0.96 &&-0.01 & 0.11 & 0.10 & 0.94 &&-0.01 &0.09 &0.10 &0.98  \\
&$\beta_3$ &-0.01 &0.17 &0.17 &0.95 && 0.03 & 0.17 & 0.16 & 0.95 &&0.00 &0.16 &0.16 &0.96  \\  
\multirow{3}{*}{$1$}
& $\beta_1$ &0.00 &0.20 &0.21 &0.96 &&-0.03 & 0.20 & 0.21 & 0.96 &&0.01 &0.21 &0.20 &0.96  \\
&$\beta_2$ &0.03 &0.12 &0.13 &0.96 && 0.00 & 0.13 & 0.13 & 0.93 &&0.01 &0.12 &0.13 &0.94\\
&$\beta_3$ &-0.01 &0.23 &0.24 &0.92&&-0.02 & 0.24 & 0.24 & 0.94 & &0.00 &0.23 &0.23 &0.92 \\
\multirow{3}{*}{$2$}
&$\beta_1$ &-0.06 &0.26 &0.25 &0.93 &&-0.03 & 0.26 & 0.26 & 0.96 &&-0.02 &0.26 &0.26 &0.93  \\
&$\beta_2$ &0.05 &0.15 &0.15 &0.94 && 0.02 & 0.16 & 0.16 & 0.95 &&0.01 &0.16 &0.16 &0.95 \\
&$\beta_3$ &-0.03 &0.29 &0.29 &0.95 &&-0.01 & 0.29 & 0.30 & 0.97 &&0.00 &0.30 &0.29 &0.95 \\
\hline
    \end{tabular}
\label{Simulation result of Scenario 1}%n=200,nsimu=200
\end{table}

\begin{table}[htp]
\centering
\caption{Simulation result of Scenario 2 with $n = 200$ and 500. Bias: the estimated bias; ESD: empirical standard deviation; ESE: empirical standard error estimate; CP: the 95\% empirical coverage probabilities.}
\begin{tabular}{ccccccccccccccccc}
\hline\hline
&& \multicolumn{4}{c}{SMCI-K} && \multicolumn{4}{c}{SMCI-S}&& \multicolumn{4}{c}{GORMC} \\%& \multicolumn{4}{c}{SIC} \\
    %    \cline{2-13}
\cline{3-6} \cline{8-11} \cline{13-16}
$r$ &Par&Bias &ESD &ESE &CP &&Bias &ESD &ESE &CP &&Bias &ESD &ESE &CP \\
\hline
\multicolumn{2}{c}{ $n=200$} \\
\multirow{3}{*}{$0$}&$\beta_1$ &0.01 &0.22 &0.24 &0.96 && 0.03 & 0.22 & 0.23 & 0.96 &&0.03 &0.23 &0.24 &0.95 \\
&$\beta_2$ &-0.01 &0.16 &0.17 &0.94 &&-0.03 & 0.16 & 0.16 & 0.94 &&-0.03 &0.16 &0.16 &0.93 \\ 
&$\beta_3$ &0.00 &0.27 &0.27 &0.95 && 0.03 & 0.26 & 0.26 & 0.97 &&0.03 &0.27 &0.26 &0.92 \\   
 \multirow{3}{*}{$1$}&
$\beta_1$ &-0.01 &0.27 &0.32 &0.96 && 0.02 & 0.33 & 0.33 & 0.96 &&0.01 &0.28 &0.31 &0.98  \\
&$\beta_2$ &0.02 &0.19 &0.19 &0.96 &&-0.01 & 0.21 & 0.20 & 0.94 &&-0.01 &0.20 &0.20 &0.96\\
&$\beta_3$ &0.01 &0.34 &0.36 &0.95 && 0.06 & 0.34 & 0.37 & 0.96 &&0.04 &0.35 &0.36 &0.95\\
\multirow{3}{*}{$2$}
&$\beta_1$ &0.00 &0.41 &0.40 &0.95 && 0.04 & 0.43 & 0.42 & 0.94 &&0.04 &0.42 &0.4 &0.94 \\
&$\beta_2$ &0.01 &0.23 &0.24 &0.94 &&-0.02 & 0.24 & 0.26 & 0.95 &&-0.03 &0.24 &0.24 &0.94 \\
&$\beta_3$ &0.01 &0.46 &0.46 &0.96 && 0.00 & 0.49 & 0.48 & 0.94 &&0.04 &0.48 &0.45 &0.94 \\
\multicolumn{2}{c}{$n=500$}\\
\multirow{3}{*}{$0$}
&$\beta_1$ &-0.03 &0.15 &0.15 &0.94 && 0.01 & 0.14 & 0.14 & 0.95 &&0.00 &0.14 &0.14 &0.96 \\ 
&$\beta_2$ &0.01 &0.11 &0.11 &0.95 && 0.00 & 0.10 & 0.09 & 0.96 &&-0.01 &0.09 &0.09 &0.96 \\ 
&$\beta_3$ &-0.01 &0.17 &0.16 &0.94 && 0.02 & 0.15 & 0.15 & 0.94 &&0.02 &0.16 &0.15 &0.95 \\  
\multirow{3}{*}{$1$}
&$\beta_1$ &-0.01 &0.18 &0.19 &0.98 &&-0.01 & 0.19 & 0.20 & 0.95 &&0.01 &0.19 &0.19 &0.98 \\
&$\beta_2$ &0.00 &0.12 &0.12 &0.96 && 0.01 & 0.12 & 0.12 & 0.96 &&-0.02 &0.12 &0.12 &0.96 \\
&$\beta_3$ &0.01 &0.20 &0.22 &0.96 &&-0.03 & 0.23 & 0.22 & 0.94 &&0.02 &0.20 &0.22 &0.96\\
\multirow{3}{*}{$2$}&
$\beta_1$ &0.00 &0.24 &0.24 &0.96 && 0.00 & 0.25 & 0.25 & 0.95 &&0.03 &0.25 &0.24 &0.96 \\
&$\beta_2$ &0.00 &0.14 &0.14 &0.96 && 0.02 & 0.15 & 0.16 & 0.96 &&-0.03 &0.14 &0.15 &0.97\\
&$\beta_3$ &0.01 &0.25 &0.28 &0.97 && 0.01 & 0.28 & 0.29 & 0.96 &&0.04 &0.27 &0.28 &0.96\\
\hline
\end{tabular}
\label{Simulation result of Scenario 2}%n=200,nsimu=200
\end{table}

\begin{table}[htp]
\centering
\caption{Simulation result of Scenario 3 with $n = 200$ and 500. Bias: the estimated bias; ESD: empirical standard deviation; ESE: empirical standard error estimate; CP: the 95\% empirical coverage probabilities.}
\begin{tabular}{ccccccccccccccccc}
\hline\hline
&& \multicolumn{4}{c}{SMCI-K} && \multicolumn{4}{c}{SMCI-S}&& \multicolumn{4}{c}{GORMC} \\%& \multicolumn{4}{c}{SIC} \\
    %    \cline{2-13}
\cline{3-6} \cline{8-11} \cline{13-16}
$r$ &Par&Bias &ESD &ESE &CP &&Bias &ESD &ESE &CP &&Bias &ESD &ESE &CP \\
\hline
\multicolumn{2}{c}{ $n=200$} \\
\multirow{3}{*}{$0$}&
$\beta_1$ &0.00 &0.23 &0.25 &0.94 && 0.03 & 0.24 & 0.24 & 0.94 &&0.05 &0.23 &0.23 &0.95 \\
&$\beta_2$ &0.03 &0.17 &0.17 &0.96 &&-0.03 & 0.14 & 0.16 & 0.98 &&-0.03 &0.16 &0.16 &0.95 \\
&$\beta_3$ &-0.04 &0.26 &0.28 &0.96 && 0.03 & 0.24 & 0.26 & 0.99 &&0.01 &0.25 &0.25 &0.96 \\
\multirow{3}{*}{$1$}&
$\beta_1$ &-0.03 &0.33 &0.33 &0.94 && 0.09 & 0.33 & 0.34 & 0.93 &&-0.02 &0.35 &0.31 &0.92 \\
&$\beta_2$ &-0.02 &0.19 &0.20 &0.95 &&-0.04 & 0.20 & 0.21 & 0.96 &&-0.02 &0.21 &0.20 &0.92 \\
&$\beta_3$ &-0.05 &0.35 &0.37 &0.96 && 0.02 & 0.38 & 0.38 & 0.96 &&-0.04 &0.37 &0.35 &0.92  \\
\multirow{3}{*}{$2$}&
$\beta_1$ &0.03 &0.39 &0.40 &0.96 && 0.01 & 0.44 & 0.42 & 0.93 &&-0.04 &0.40 &0.36 &0.92 \\
&$\beta_2$ &0.02 &0.23 &0.24 &0.95 &&-0.02 & 0.25 & 0.26 & 0.95 &&0.08 &0.27 &0.23 &0.89 \\
&$\beta_3$ &0.05 &0.47 &0.46 &0.94 && 0.03 & 0.45 & 0.48 & 0.96 &&-0.02 &0.46 &0.42 &0.92 \\
\multicolumn{2}{c}{$n=500$}\\
\multirow{3}{*}{$0$}&$\beta_1$ &-0.01 &0.14 &0.15 &0.96 && 0.00 & 0.14 & 0.14 & 0.98 &&0.02 &0.12 &0.14 &0.98  \\
&$\beta_2$ &0.02 &0.10 &0.11 &0.97 && 0.01 & 0.09 & 0.09 & 0.97 &&0.00 &0.09 &0.09 &0.97  \\
&$\beta_3$ &-0.02 &0.18 &0.17 &0.94 &&-0.03 & 0.15 & 0.16 & 0.98 &&0.02 &0.16 &0.16 &0.93  \\ 
\multirow{3}{*}{$1$}
&$\beta_1$ &-0.01 &0.19 &0.20 &0.98 &&-0.01 & 0.20 & 0.20 & 0.95 &&-0.02 &0.19 &0.19 &0.96  \\
&$\beta_2$ &0.01 &0.12 &0.12 &0.94 &&-0.01 & 0.13 & 0.13 & 0.96 &&0.02 &0.13 &0.12 &0.93 \\
&$\beta_3$ &-0.03 &0.22 &0.22 &0.95 &&-0.02 & 0.23 & 0.23 & 0.92 &&-0.04 &0.23 &0.22 &0.95 \\
\multirow{3}{*}{$2$}
&$\beta_1$ &-0.04 &0.24 &0.24 &0.93 && 0.00 & 0.23 & 0.25 & 0.96 &&-0.14 &0.26 &0.22 &0.83 \\
&$\beta_2$ &0.02 &0.16 &0.14 &0.92 && 0.02 & 0.14 & 0.16 & 0.96& &0.12 &0.18 &0.14 &0.78 \\
&$\beta_3$ &0.01 &0.26 &0.28 &0.96 && 0.02 & 0.31 & 0.29 & 0.93 &&-0.09 &0.26 &0.26 &0.91  \\
\hline
\end{tabular}
\label{Simulation result of Scenario 3}%n=200,nsimu=200
\end{table}

We also examine the performance of the estimated incidence link functions. % with the curves presented in the \hyperref[A.4]{Appendix}.
As expected, the estimated averages closely align with the true values, suggesting that the proposed SMCI-K and SMCI-S models provide accurate approximations (Appendix Figures \ref{Estimated incidence link functions under n=200.}-\ref{Estimated incidence link functions under n=500.}).

%We performed a second simulation study using a larger number of covariates and variable pool sizes. This study, summarized in the Supplementary Material, revealed the same findings.

{We perform extensive simulations to further evaluate the method's numerical robustness and computational performance. Robustness checks against varying spline orders and knot specifications confirm stable estimation accuracy (Appendix Tables \ref{tab:spline_order}-\ref{tab:knot} and Figures \ref{fig:ASE_order_knot}-\ref{fig:baseline_order}). Assessment of scalability with increasing covariate dimensions demonstrates numerical stability and linear time complexity, validating computational efficiency in large dimensional settings (Appendix Table \ref{tab:com_scala} and Figures \ref{fig:ASE_pq}-\ref{SC_pq}). Finally, detailed diagnostics of the EM algorithm, encompassing iteration plots and sensitivity to diverse initial values, verify its reliable convergence (typically within 50–150 iterations) and insensitivity to initialization (Appendix Table \ref{tab:intial values} and Figures \ref{log-lik-K}-\ref{fig2: initial values}). All supporting results are provided in Appendix A.4.
}

% The parameter estimates are highly consistent and exhibit negligible bias and
% proper coverage probabilities across all initializations. This consistency strongly suggests that
% our algorithm is insensitive to initial values.

% As the sample size
% increases, the estimated standard errors and sample standard deviations reduce significantly,
% while no clear decreasing trend is observed for bias. The high censorship scenario yields results
% comparable to those previously discussed. The proposed method performs well with varying the number and location of knots results in reasonable estimates
% across all configurations considered

%We also plot the estimated  survival curves of uncured groups, and present the mean of estimated baseline survival curves along with 2.5\% and 97.5\% quantiles under all settings in Figure \ref{Estimated baseline survival functions.}. The estimated baseline survival curves have negligible bias, they are close to the truth in all cases.

\section{Real Data Analysis}\label{sec5}

%The ADNI data has been investigated by the mixture cure model 
%\citep{zhou2021mediation,shi2022functional}\cite{scolas2018diagnostic}

As an illustration, we apply our approach to 
 the analysis of the motivating data
from %analyze the dataset related to a study on Alzheimer’s disease,
the Alzheimer’s Disease Neuroimaging Initiative (ADNI) %which was launched in 2004, spearheaded by Dr. Michael W. Weiner 
\citep{weiner20152014,weiner2017alzheimer,weiner2017recent}. 
%ADNI is a multi-center, longitudinal observational study aimed at developing biomarkers for Alzheimer's disease, enhancing the understanding of its pathophysiology, improving diagnostic methods for early detection, and refining the design of clinical trials. 
MCI, regarded as a transitional stage between normal cognition and more severe AD, is a potential target for identifying individuals at risk of developing AD.
The main objective of this study is to identify the risk factors associated with the conversion of time from MCI to the onset of AD in elderly patients. {In the dataset, individuals were in states of normal cognitive function, MCI, or AD, with direct transitions from MCI to death being rare; consequently, mortality had a negligible influence as a competing risk in our analysis.}
Additional information about the dataset can be found on the official website (\href{https://adni.loni.usc.edu}{\textcolor{blue}{https://adni.loni.usc.edu}}).
%This study recorded the date of each follow-up and diagnostic results of participants. We are interested in the duration between the diagnosis of MCI and AD, which is subject to interval censoring. 
The ADNI dataset consists of 950 patients, and 63\% of them had never developed severe AD by the end of the study. Figure \ref{Analysis of the ADNI study}(a) presents the nonparametric maximum likelihood estimator of the survival function. The estimated survival curve plateaus around 0.4, indicating the existence of a proportion of unsusceptible subjects in the population. 

\begin{figure}[H]
\centering  
\begin{subfigure}[b]{0.4\textwidth}
  \includegraphics[angle=270,width=\textwidth]{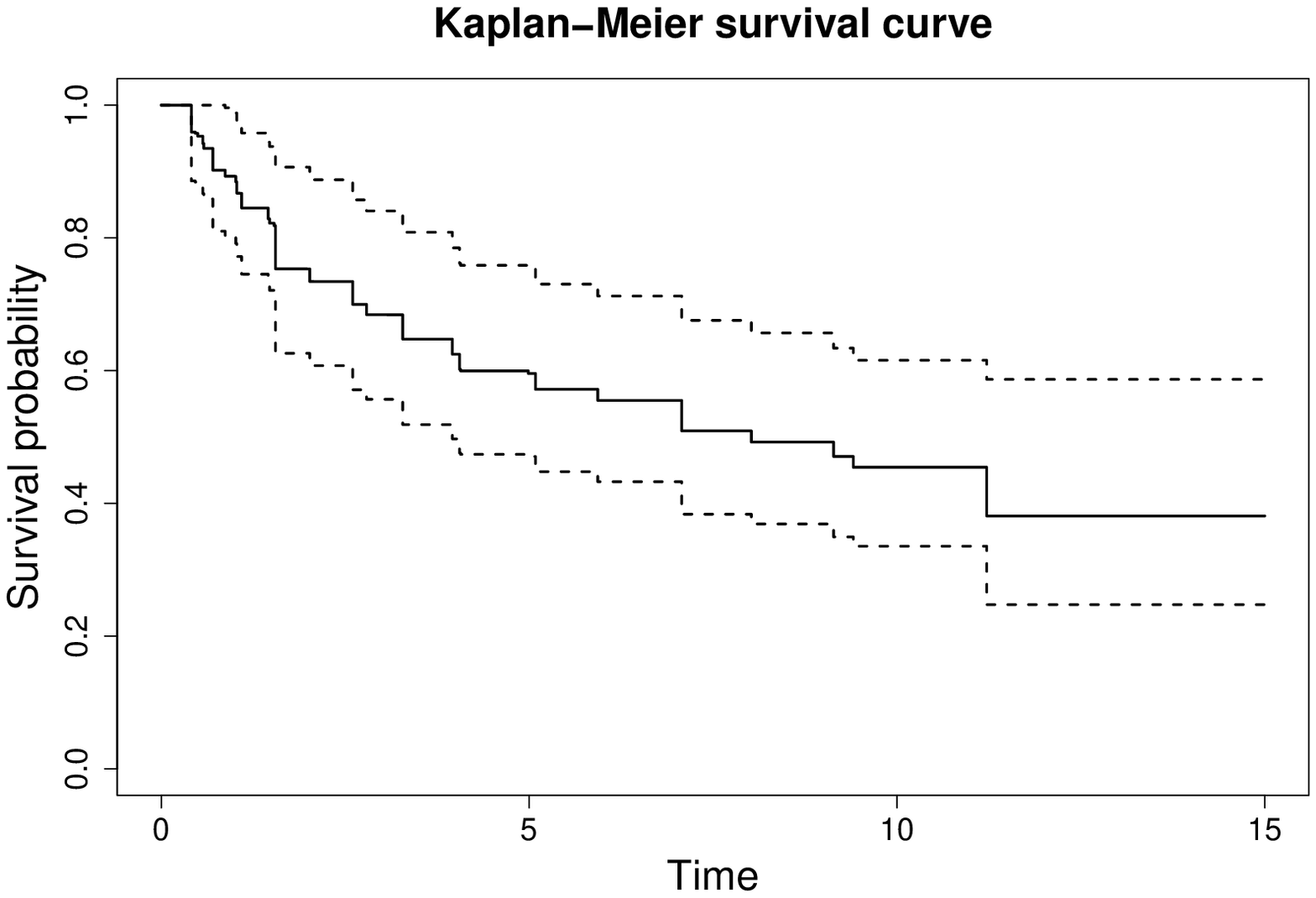}
  \caption{Kaplan-Meier survival curve}\label{KMsurvival curve}
\end{subfigure}
\begin{subfigure}[b]{0.4\textwidth}
  \includegraphics[angle=270,width=\textwidth]{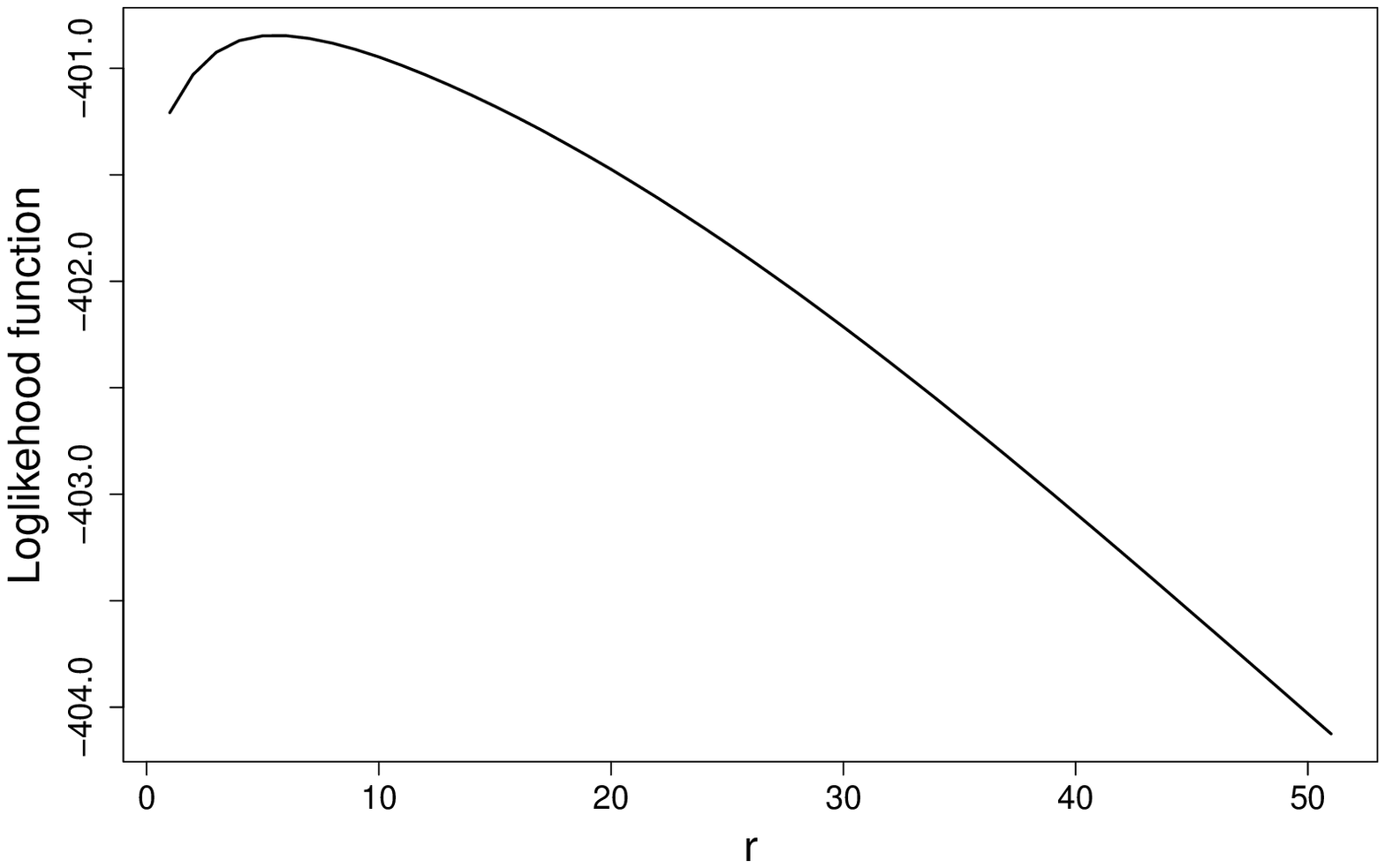}
  \caption{loglikelihood}\label{loglikelihood}
\end{subfigure}
\caption{Analysis of the ADNI study: (a) The Kaplan-Meier survival curve; (b) the loglikelihood at the nonparametric maximum likelihood estimates plotted as a function of $r$ in the logarithmic transformations.}
\label{Analysis of the ADNI study}
\end{figure}

In our analysis, %four prognostic factors were included in the model, both in the latency and the incidence part: 
we include four baseline prognostic factors in the model for both the latency and incidence components: age (ranged from 55 to 91), gender (0 for female and 1 for male), years of education (ranged from 4 to 20), and APOE4 status (0 for carrying no APOE alleles, 1 for carrying 1 APOE allele and 2 for carrying 2 APOE alleles). 
We fit model (\ref{S_p}) with the class of logarithmic transformations $G(x) = r ^{-1} \log(1 + rx)(r\ge 0)$. 
The dataset is partitioned into a training ($n=633$) and a validation ($n=317$) sets. 
The training set is utilised for model estimation, while the validation set evaluates log-likelihood to select the tuning parameter $r$ for SMCI-K, SMCI-S and GORMC models. The turning parameter $r$ is selected via grid search over the interval [0,5] with a step size of 0.1.
{As depicted in Figure \ref{Analysis of the ADNI study}(b), the log-likelihood for the SMCI-K model is maximized at $r=0.3$, yielding a maximum value of $-400.95$. Similarly, for the SMCI-S model, the log-likelihood is maximized at $r=0$, corresponding to a value of $-403.25$. For the GORMC model, the optimal value is $r=0$, corresponding to a log-likelihood of $-410.94$. }
%the log-likelihood was maximized at $r=0.3$, yielding a maximum value of −400.85. For the GORMC model, the optimal value is $r=0$, corresponding to a log-likelihood of −410.94.%with the log-likelihood value reaching −410.94 when $r=0$.}
%The parameter $r$ was determined by the grid search with $r$ ranging from $0$ to $5$ by $0.1$. %Figure \ref{loglikelihood} shows that the loglikelihood is maximized at $r = 0.3$. Therefore, the final model was chosen with $r = 0.3$, which yielded an observed log-likelihood of $-1142.789$.
 %Figure \ref{Analysis of the ADNI study}(b) illustrates that the log-likelihood is maximized at $r = 0.3 $. Consequently, the final model was selected with $r = 0.3$, yielding an observed log-likelihood of $-400.85$. %{\color{red} how to determine the value of r in the GOMRC model?} {\color{green}done} {\color{green} We have obtained, using the same method, that the likelihood function of the GORMC model attains its maximum value of $-1156.506$ when $r=0.5$. Meaning that the SMCI model performs better than the GORMC model in predicting the survival probability. To understand why the SMCI model performs better, we compare the estimates of the two models in the incidence and latency.}
 %Similar approach has been applied to the GORMC model, with the log-likelihood value reaching to $-410.94$ when $r=0$.
 %\xy{add the description of train and test procedure.}
 
Table \ref{tab:Analysis results for ADNI}  summarises the estimation results for regression coefficients, standard errors and p-values obtained from these three methods. {The larger log-likelihood value and the smaller AIC and BIC values indicate that the SMCI-K performs better than the SMCI-S and GORMC models in predicting survival probability.}
%The bigger log-likelihood value indicates that our proposal performs better than  
%SMCI model performs better than 
%the GORMC model in predicting the survival probability. %{\color{red} should be bigger?}
%under the SMCI model, including parameter estimates, standard errors and corresponding p-values.% and the standard errors computed by the bootstrap approach with 200 resamples and P-values. For comparison, Table \ref{tab:Analysis results for ADNI} also shows the results of the GORMC model and SIC model.
%Additionally, for comparison, Table \ref{tab:Analysis results for ADNI} includes the results from the GORMC model and SMCI model. 

\begin{table}[H]
    \centering
    \captionsetup{skip=0pt}
    {\caption{Analysis Results for ADNI Data under the SMCI and GORMC models.}
    \label{tab:Analysis results for ADNI}
    \begin{tabular}{ccccccccccccc}
    \hline\hline
    & \multicolumn{3}{c}{SMCI-K} && \multicolumn{3}{c}{SMCI-S} && \multicolumn{3}{c}{GORMC}\\
        \cline{2-4}\cline{6-8}\cline{10-12}
         &Est. &SE &p-value&&Est. &SE &p-value&&Est. &SE &p-value\\
         \hline
        \textbf{ Incidence}\\
         (intercept)&- &- &- &&-&-&-&&-0.09 &0.27 &0.75  \\
         Age&0.59 &0.20 &$0.00^*$ &&0.51 &0.14 &$0.00^*$&&0.74 &0.16 &$0.00^*$ \\
         Gender&0.07 &0.34 &0.83&&-0.07&0.26&0.80 &&0.06 &0.29 &0.84 \\
         Edu&-0.09 &0.22 &0.67 &&-0.09&0.13&0.47& &-0.04 &0.15 &0.78\\
         APOE4&0.80 &0.33 &$0.02^*$&&0.85&0.11&$0.00^*$& &1.53 &0.29 &$0.00^*$ \\
         \textbf{Latency}\\
         Age&-0.03 &0.17 &0.85&&-0.06 &0.11    &0.57 &&-0.19 &0.17 &0.28 \\
         Gender&-0.30 &0.22 &0.17 &&-0.31 &0.21  &  0.14 &&-0.21 &0.32 &0.52\\
         Edu&0.08 &0.11 &0.43 &&0.09& 0.10  &  0.34 &&0.03 &0.15 &0.83\\
         APOE4&0.41 &0.19 &$0.03^*$ &&0.28& 0.14  &  $0.05^*$ &&0.35 &0.23 &0.13\\
         \textbf{log-likelihood}&\multicolumn{3}{c}{-400.95} && \multicolumn{3}{c}{-403.25} && \multicolumn{3}{c}{-410.94} \\
         \textbf{AIC}&\multicolumn{3}{c}{1563.43} && \multicolumn{3}{c}{1581.96 } && \multicolumn{3}{c}{1570.59} \\
         \textbf{BIC}&\multicolumn{3}{c}{1634.64} && \multicolumn{3}{c}{1653.16 } && \multicolumn{3}{c}{1646.25} \\
         \hline
         $^*$ p-value$<=0.05$
    \end{tabular}}
\end{table}

\begin{figure}[H]
\centering  
\begin{subfigure}[b]{0.32\textwidth}
  \includegraphics[angle=270,width=\textwidth]{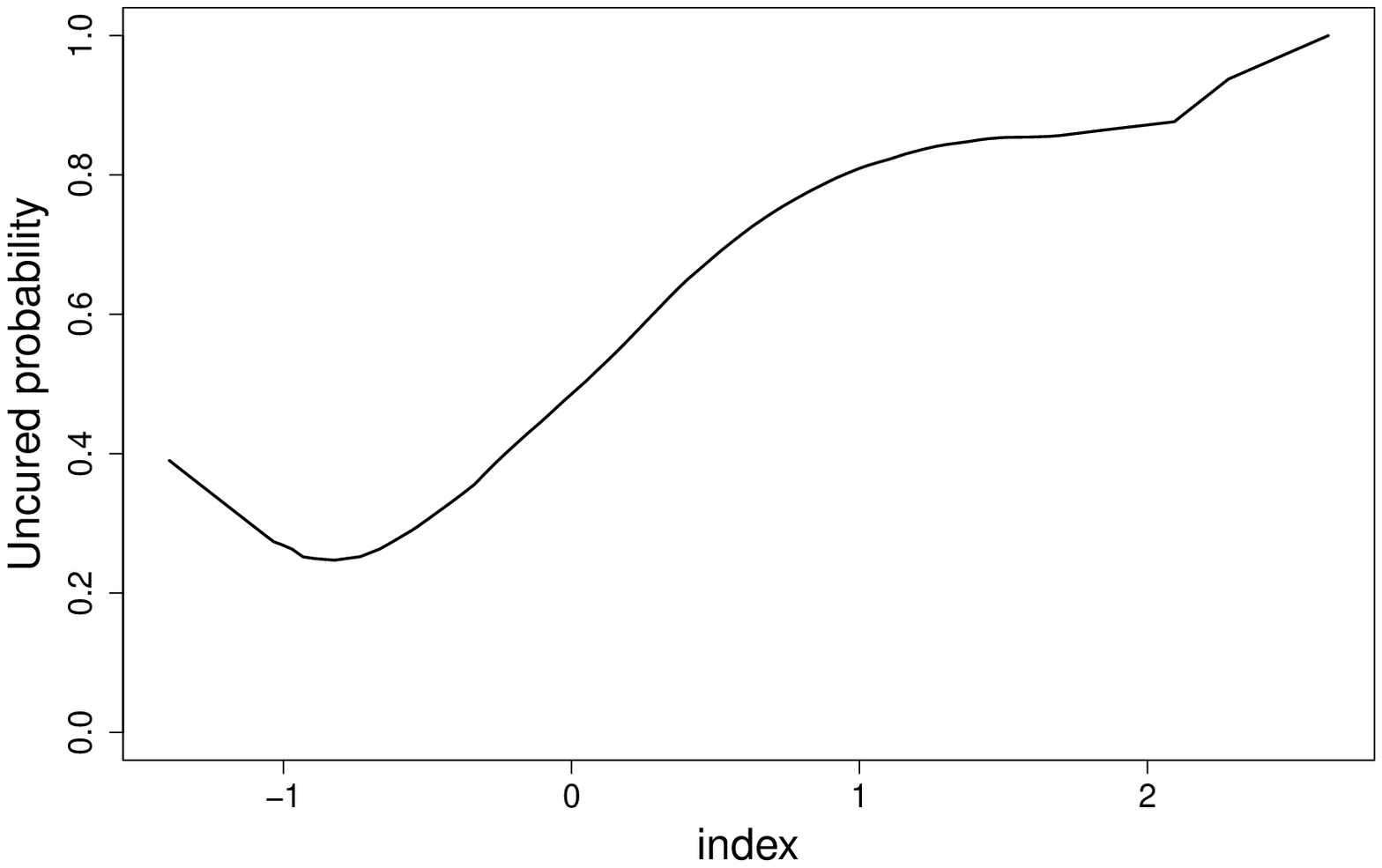}
  \caption{SMCI-K}\label{SMCIlink1}
\end{subfigure}
\begin{subfigure}[b]{0.32\textwidth}
  \includegraphics[angle=270,width=\textwidth]{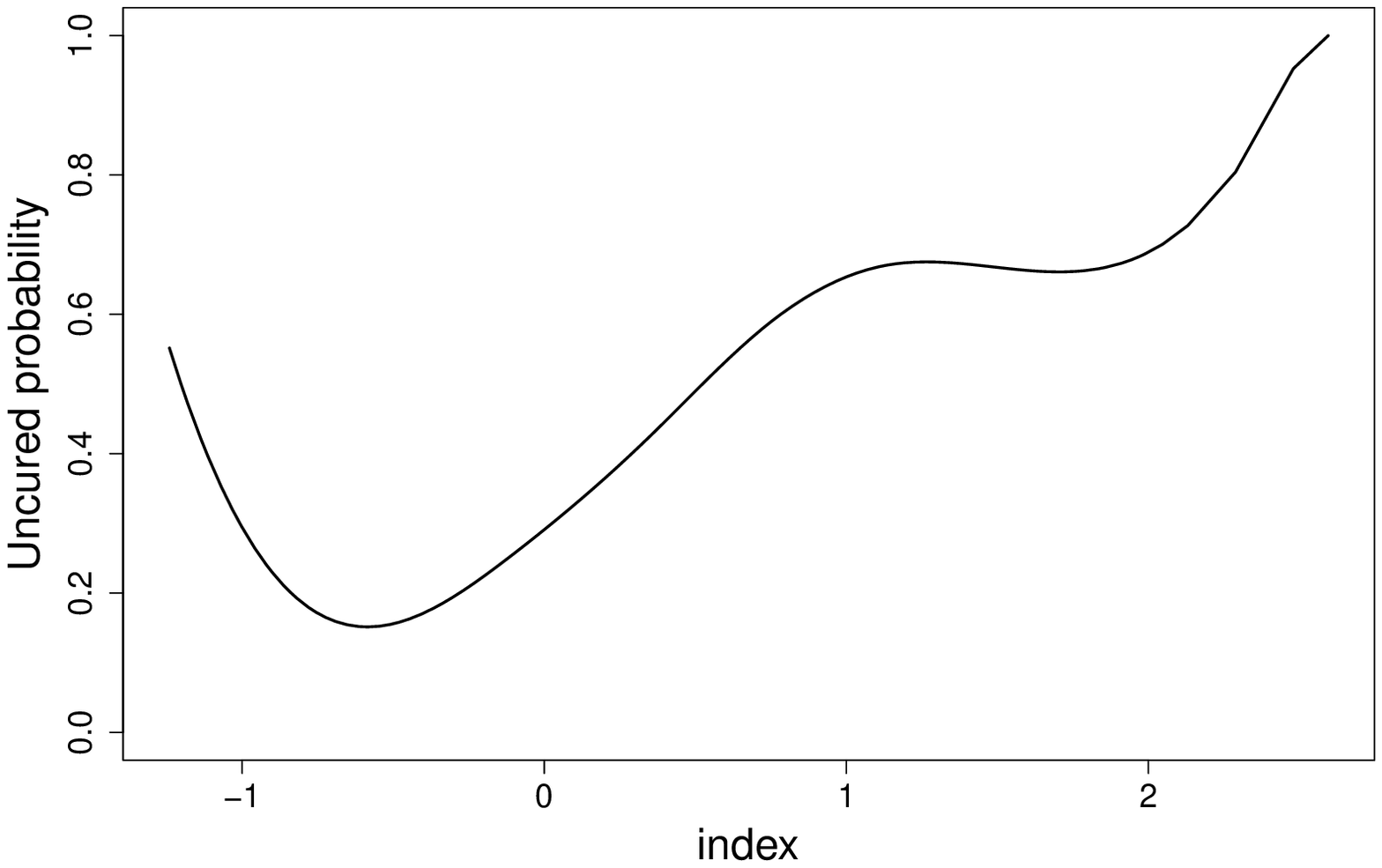}
  \caption{SMCI-S}\label{SMCIlinksp1}
\end{subfigure}
\begin{subfigure}[b]{0.32\textwidth}
  \includegraphics[angle=270,width=\textwidth]{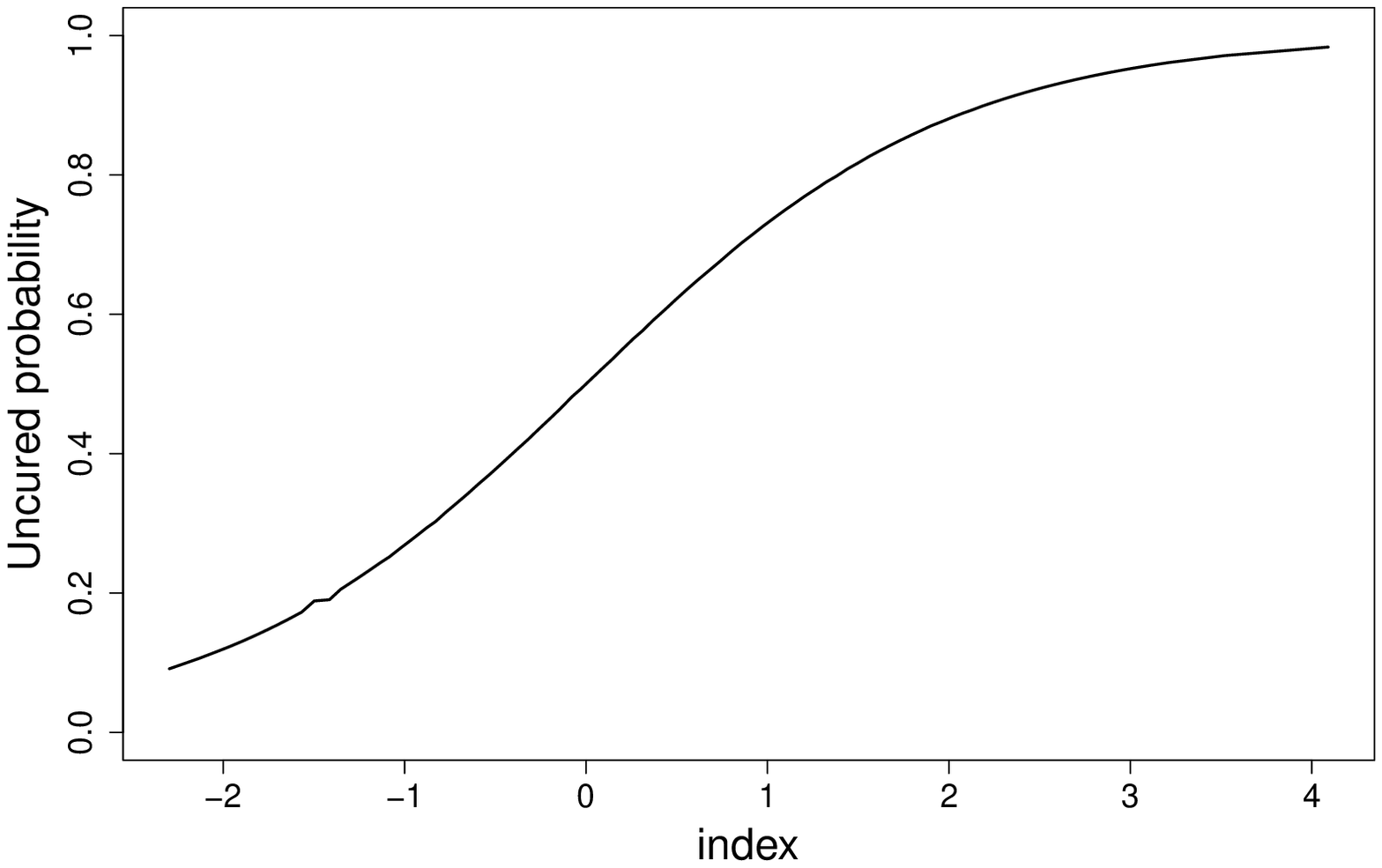}
  \caption{GORMC}\label{GORlink1}
\end{subfigure}
\begin{subfigure}[b]{0.32\textwidth}
  \includegraphics[angle=270,width=\textwidth]{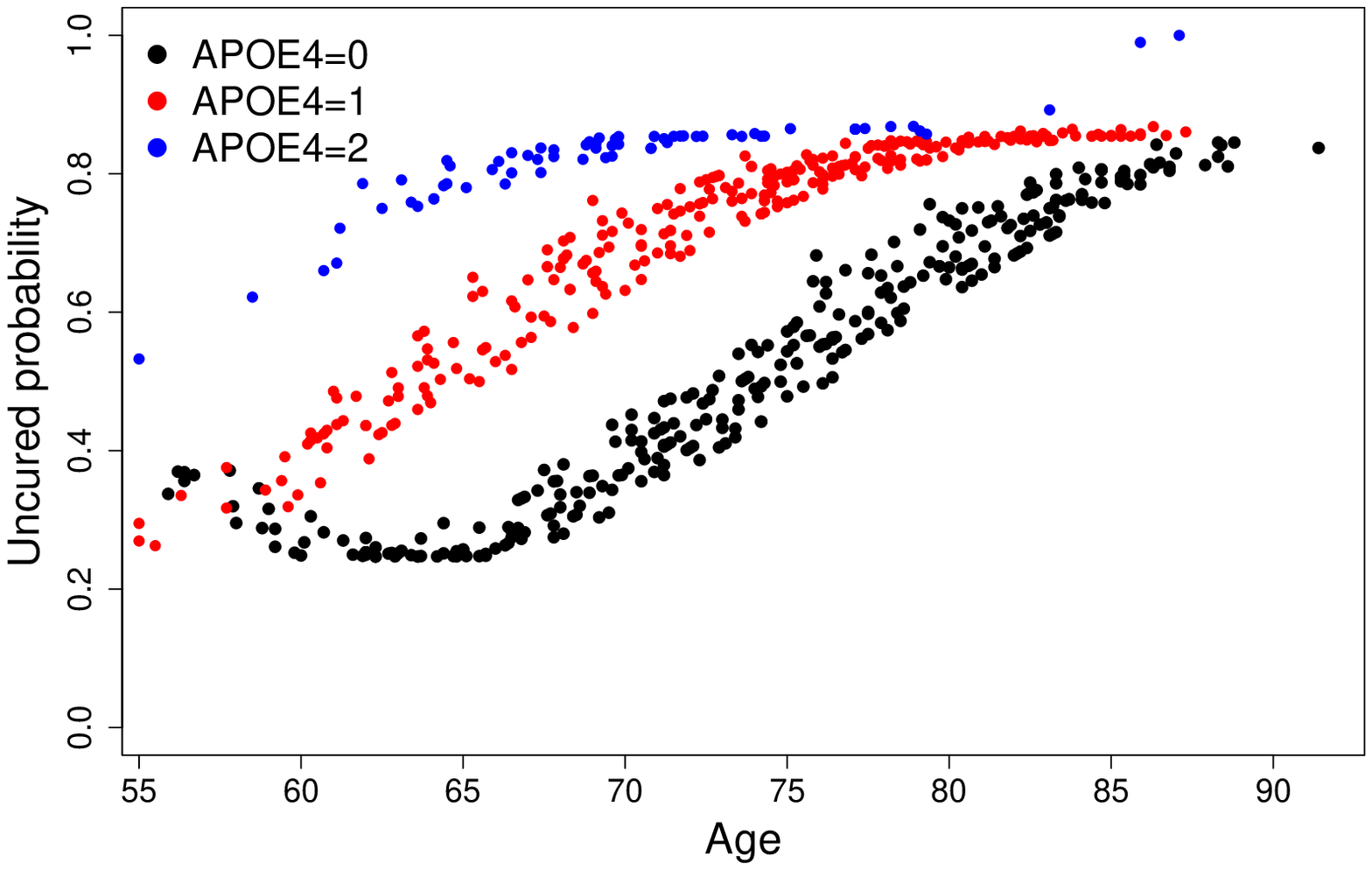}
  \caption{SMCI-K}\label{SMCIlink2}
\end{subfigure}
\begin{subfigure}[b]{0.32\textwidth}
  \includegraphics[angle=270,width=\textwidth]{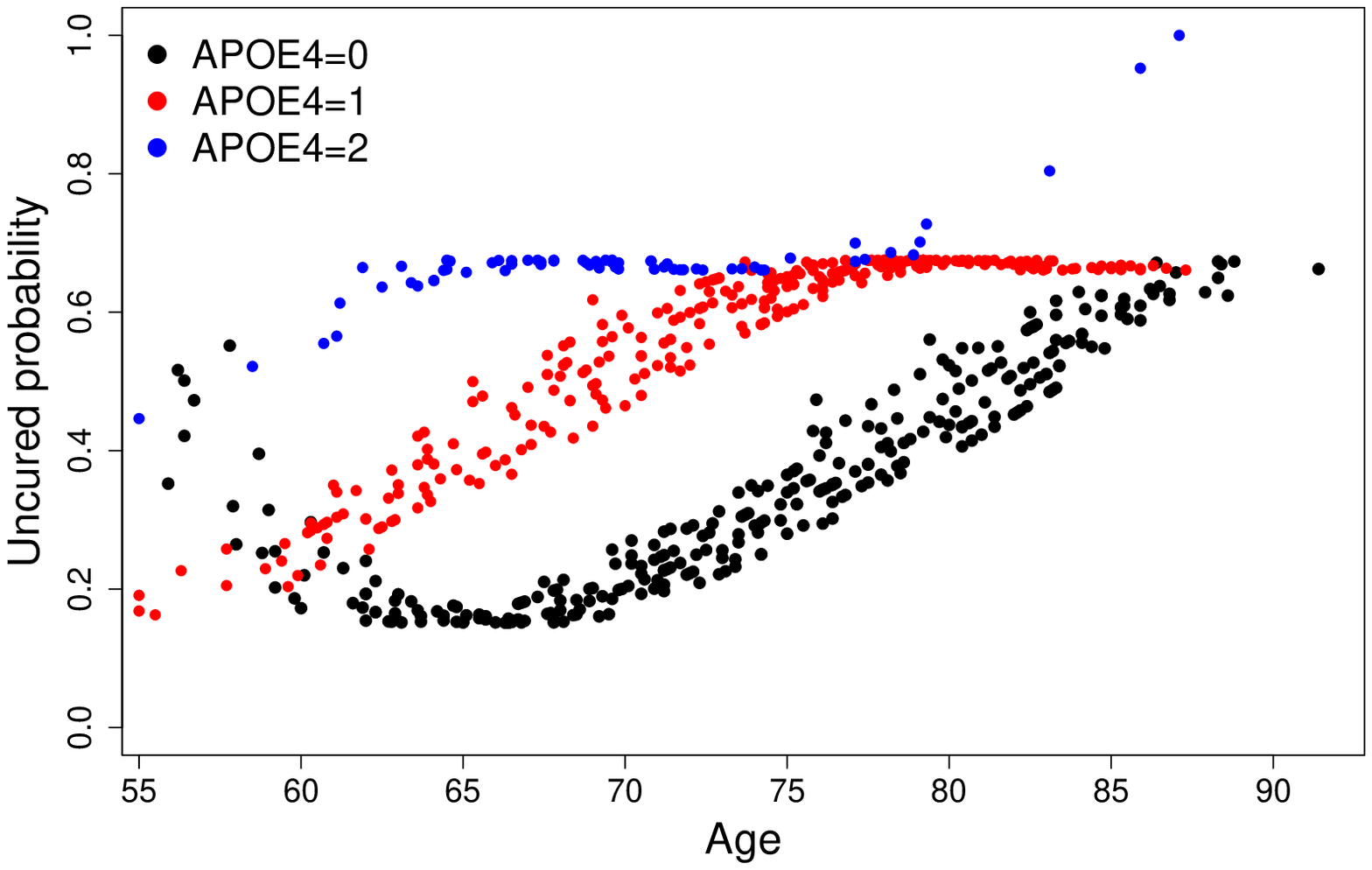}
  \caption{SMCI-S}\label{SMCIlinksp2}
\end{subfigure}
\begin{subfigure}[b]{0.32\textwidth}
  \includegraphics[angle=270,width=\textwidth]{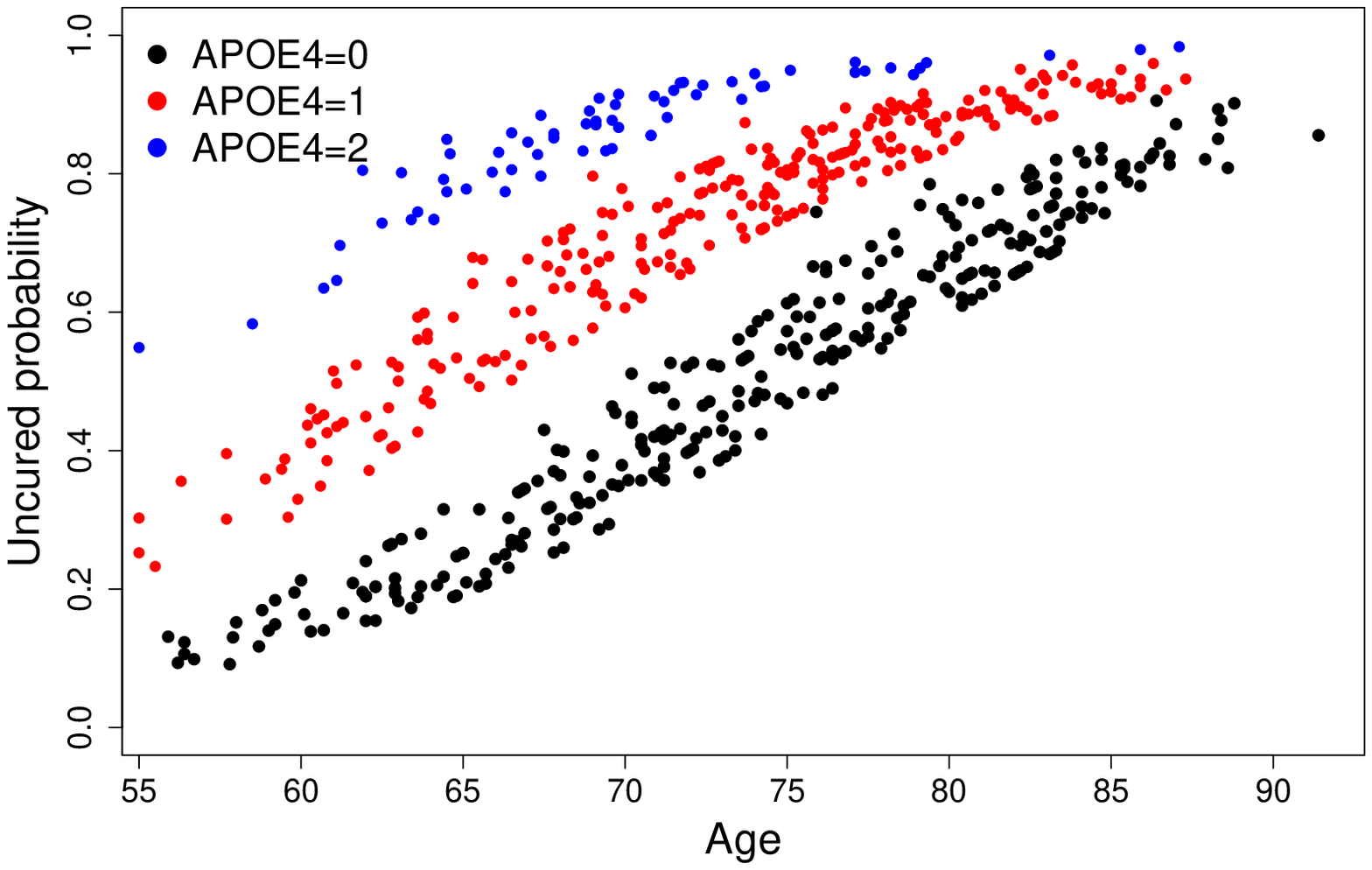}
  \caption{GORMC}\label{GORlink2}
\end{subfigure}
%\begin{subfigure}[b]{0.32\textwidth}
 % \includegraphics[angle=270,width=\textwidth]{image/SICLink_estimates.eps}
  %\caption{SIC}
%\end{subfigure}
\caption{Estimated link functions for (a) the SMCI-K model, (b) the SMCI-S model, (c) the GORMC model, and plots of age effects %of the effects of age
on the uncure probability 
for (d) the SMCI-K model, (e) the SMCI-S model and (f) the GORMC model.}
\label{Estimated link functions}
\end{figure}

For the incidence component, %the 
results indicate that 
%the covariates have consistent effects in both models, with 
both age and APOE4 status have a significant effect %on the %uncured probability \blue{incidence} 
(at the 5\% level) across all three models. %significance level for all three models.
%{\color{blue}For the incidence, Table \ref{tab:Analysis results for ADNI} shows that the covariates' impacts have the same direction for the SMCI and GORMC models. Both models identify age and APOE4 status as significant predictors of uncured probability at the 5\% level. Covariates such as gender and education do not show significant associations with uncured probability in either model. 
Figures \ref{Estimated link functions}(a)-(c) illustrate the estimated link functions. % for the three models, with 
The GORMC model %showing a clear increasing trend, 
consistently exhibits a clear increasing trend, whereas the SMCI-K and SMCI-S methods reveal a non-monotonic pattern % display a non-monotonic pattern featuring
with both decreasing and increasing trends. To exhibit the covariates effects on the uncured probability, % of covariates, 
%The estimated non-monotonic link function prevents the straightforward interpretation of the parameters. 
%To further explore these findings, we analyze the link functions for both models, as shown in Figure \ref{Estimated link functions}(a)-(b). The GORMC model assumes a monotonic logistic link function, which might oversimplify these relationships. In contrast, the link estimate of SMCI is different, there is a non-monotone trend in Figure \ref{Estimated link functions}(a).
%As proposed by \cite{li1989regression}, within a broader context, even when two models utilize distinct linking functions for estimation, the relative importance of parameter estimates remains comparable, thereby enabling an assessment of which variable impacts most of the outcomes.
%To directly interpret the parameter estimates, we plot the effect of age on uncured probability for the two models in
Figures \ref{Estimated link functions}(d)-(f) present age effects for different APOE4 statuses, as modeled by the three methods. %the effects of age on uncured probability 
%\blue{incidence} 
%given the biomarker AOPE4 for three methods. %The links of the SMCI-K and SMCI-S methods have a similar shape. The estimated link function from GORMC differs considerably from the other two methods.
In the GORMC model,  the lines representing APOE4 groups do not cross, preserving their relative order. %the lines do not cross, and the order of the APOE4 groups is preserved.
In contrast, the SMCI-K and SMCI-S approaches reveal %that the effects of age vary across different APOE4 statuses. 
that the age-related effects on the uncured probability vary across different APOE4 statuses. 
{Specifically, individuals homozygous for APOE4 (carrying two alleles) % carrying two APOE4 alleles 
consistently exhibit the highest uncured probability across all ages, which aligns with evidence that APOE4 homozygosity accelerates amyloid-$\beta$ accumulation and sustains elevated risk \citep{safieh2019apoe4,emrani2020apoe4}. %With advancing age, the non‑cure probability also rises among individuals with a single APOE4 allele. In contrast, 
For individuals carrying a single APOE4 allele, the uncured probability also demonstrates an increasing trend with age. Among non-carriers, the uncured probability initially decreases with age until approximately 65 years, then subsequently increases. This non-linear effect likely reflects cumulative neurodegenerative and vascular aging processes.}

%The uncured probability also increases with age among individuals carrying a single APOE4 allele. Among non-carriers, the uncured probability decreases with age until approximately 65 and subsequently increases, reflecting a nonlinear effect likely driven by cumulative neurodegenerative and vascular ageing processes.}

{For the latency component, Table \ref{tab:Analysis results for ADNI} indicates that APOE4 status is the only significant predictor of the conditional survival probability among susceptible subjects under the SMCI-K and SMCI-S models. Conversely, no covariates achieve statistical significance in the GORMC model. %whereas no covariate is significant in the GORMC model.
In the SMCI-K and SMCI-S models, individuals with one or two APOE4 alleles progress from MCI to AD more rapidly than non-carriers, exhibiting a clear dose–response pattern in progression risk. The estimated survival curves from the SMCI-K model (Figure 7) provide consistent visual evidence, showing lower survival probabilities for APOE4 carriers.}

\begin{figure}[H]
\centering  
\begin{subfigure}[b]{0.7\textwidth}
  \includegraphics[angle=270,width=\textwidth]{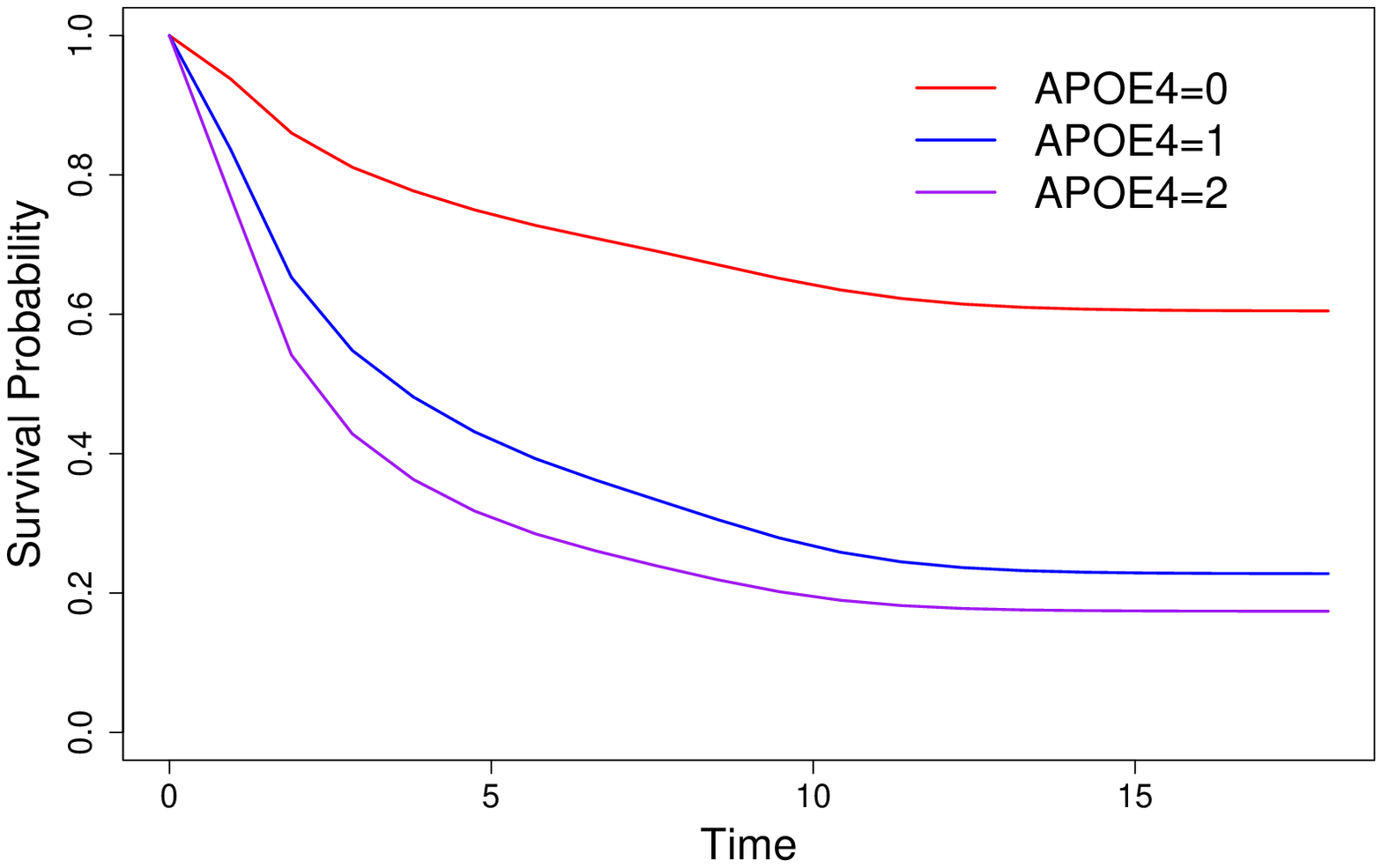}
\end{subfigure}
\caption{Estimated survival curves in the SMCI-K model for the average level and three different APOE4 status.}
\label{estimated survival curves}
\end{figure}

\section{Conclusion}\label{sec6}

%This paper proposes a class of semi-parametric models 

We proposed a class of flexible single-index semiparametric transformation
cure models for analyzing interval-censored data with a cured subgroup, %This class of regression models includes the commonly used proportional hazards mixture cure models, proportional odds mixture cure models, and generalized proportional odds mixture cure models. 
which includes a flexible single-index model for incidence and a semiparametric transformation model for latency. 
%The single-index framework 
Our proposal allows {interpretability} and avoids the potential {curse-of-dimensionality} issue for large dimension covariates in the incidence and includes the proportional hazard cure, proportional odds cure and generalized odds cure models as special cases.
We introduce the four-layer data augmentation to facilitate the computation of estimation and obtain the maximum likelihood estimation via the EM algorithm.
%We used the EM algorithm to calculate the parameter maximum likelihood estimator for this model type. Moreover, we have demonstrated the identifiability of the proposed model. Based on a numerical study, we demonstrate the strong performance of the method in estimating the uncure probability, and its ability to outperform logistic models when the real link function is not a logistic function.
{The numerical study demonstrates
that %our proposal 
both the SMCI-K and
SMCI-S methods provide more accurate estimations than the logistic-based transformation cure method unless the model is correctly specified, %in general, 
which is robust to the model assumptions.
We recommend the utilization of the kernel-based approach in the small-sample setting.}

{
%\noindent\textbf{Remark.}
Note that we employed the log-likelihood values, AIC and BIC as the criterion to evaluate model performance. %The test for the single-index model with an unknown link function in the cure rate remains unestablished.
To check the existence of nonlinear effects in the mixture cure model, one can utilize the graphical procedure \citep{scolas2018diagnostic}. %proposed by \cite{scolas2018diagnostic} can be utilized to conduct diagnostic checks on mixture cure models with interval-censored data. By plotting deviance residuals against individual covariates, one may assess the linearity assumption; the presence of systematic features in the plot generally indicates a nonlinear effect.
Recently, \cite{muller2019goodness} proposed tests for the parametric form of the cure rate function with a single covariate. However, the test for the single-index model in the incidence components, which allows for multidimensional covariates, remains to be investigated in the future.
}

{A promising direction for future research is to establish the large-sample properties of the proposed estimator. Although extensive numerical studies provide empirical evidence for the consistency of the estimators, a rigorous formal theoretical proof of consistency and asymptotic normality remains a significant open challenge.
%, the method appears to yield consistent estimation of the coefficients in the survival and cure models. Deriving the consistency and asymptotic normality of parameters would be of interest in such a model. 
%There have been some 
Despite recent developments %have been made on 
in the theoretical foundation of the single-index model with censored data %proportional hazards model 
\citep{groeneboom2018current,li2023corporate}, %lu2006class,li2023corporate},
the literature on the theoretical properties of single-index mixture cure models remains relatively limited.
\cite{musta2024single} investigated the consistency of estimators for the single-index mixture cure model with right censored data under monotonicity constraints. A potential direction of research would be on the large-sample properties of estimators for a single-index mixture cure model with interval censoring.}
% This study does not exhaust the notable issues in cure survival data with interval censoring.
% The implicit characterization of the single-index function and involvement of finite and infinite parameters in the estimation %the parameters contains not only the finite parameters $gamma$ and $beta$ but also the infinite parameters $\Lambda$ 
% pose challenges to establishing the theoretical properties of the estimators. \cite{musta2024single} explored the consistency of the estimator for the single-index mixture cure model without monotonicity constraints. 
Moreover, our proposal assumes the conditional independence of the failure time and the observational process. More complex censoring mechanisms, including informative censoring \citep{yu2022regression,wang2016regression} and bivariate interval-censored data \citep{zhou2017sieve}, could be explored. 
In addition, our method only incorporates nonparametric effects of risk factors on the incidence part while presuming a linear constriction on the latency. The more flexible double semiparametric models should be considered \citep{lee2024partly}. 
%Last, many literatures have studied the goodness-of-fit test for the mixture cure model \citep{muller2019goodness, geng2023goodness} and single-index model \citep{xia2004goodness}, but none have incorporated multidimensional covariates in interval-censored data, especially goodness-of-fit tests for the cure rate in a mixture cure model. A reasonable next step would be to target tests with multi-dimensional covariates.
These extensions raise theoretical and computational challenges, and further investigation is required.
%For partly interval-censored data, some subjects can be exactly observed, but extending the proposed algorithm is difficult. Further research is expected in the future for partly interval-censored data. 

%Proving the consistency of a single-index mixture cure model is challenging, and an extension of our work can be to demonstrate the consistency of a single-index mixture cure model. \cite{musta2024single} has already demonstrated the consistency of a single index mixture cure model in right-censored data when the link function is monotonic. The model proposed in this article allows for time-independent covariates, which can be developed into time-dependent covariates in subsequent work.

\section*{Acknowledgment }
%{We are grateful to the editor, associate editor and three referees for their valuable comments and suggestions, which have greatly improved the manuscript.}
This research was supported by the National Social Science Fund of China (24CTJ035). %andthe Basic Research Program of Guangzhou Municipal Science and Technology Bureau(Grant No, SL2022A04J01237).

\begin{appendices}
\section*{Appendix}
\label{Appendix}

{The Appendix includes the deduction of the complete likelihood function, derivations of the conditional expectations in the expectation-maximization algorithm, the proofs of identifiability, and results from additional simulation studies. %Users can access our R code at the corresponding author’s GitHub page \hyperlink{https://github.com/liuxystat/SMCI}{https://github.com/liuxystat/SMCI}.
}

\subsection*{A.1 Deduction of equation \eqref{complete} to \eqref{L_obs} in Section \ref{sec3.1}}\label{A.1}
%How to derive eq (\ref{L3}) to eq (\ref{complete})\\
%{\color{red} comments:  eq \ref{complete}  can be derived by integrating out $Y$ and $W$ from  eq \ref{L3}, specifically, consider$\delta_{L_i}$=1, $\delta_{I_i}=1$ and $\delta_{R_i}=1$  respectively}\\

We derive the equation \eqref{complete} to \eqref{L_obs} given in section \ref{sec3.1} of the manuscript. In \eqref{complete}, $B_i$ is the latent cure indicator, $Y_{il}$s and $W_{il}$s are independent Poisson random variables conditional on $\xi_i$ with the mean parameters $\lambda_{il}\xi_i$ and $\omega_{il}\xi_i$, respectively, where 
\begin{equation*}
\begin{aligned}
    \lambda_{il}=&\eta_l\exp(\bbeta^{\top}\bZ_i)\{\delta_{L_i}b_l(R_i)+\delta_{I_i}b_l(L_i)\},\\
    \omega_{il}=&\eta_l\exp(\bbeta^{\top}\bZ_i)[{\delta_{I_i}\{b_l(R_i)-b_l(L_i)\}+\delta_{R_i}b_l(L_i)}].
\end{aligned}
\end{equation*}
Under the above construction, we can consider $(\boldsymbol{Y}_l,\boldsymbol{W}_l,\boldsymbol{\xi},\boldsymbol{B})$ as missing data. % \eqref{complete} also has this form
%$$
%\begin{aligned}
 %   L_4=&\prod_{i=1}^np(\bX_i)^{B_i}(1-p(\bX))^{1-B_i}\prod_{l=1}^kP(Y_{il}|\xi_i)P(W_{il}|\xi_i)^{\delta_{I_i}+\delta_{R_i}}\Big[\delta_{L_i}I(\sum_{l=1}^kY_{il}>0)\\
%    &+\delta_{I_i}I(\sum_{l=1}^kY_{il}=0,\sum_{l=1}^k
    %W_{il}>0)+\delta_{R_i}B_iI(\sum_{l=1}^kY_{il}=0,\sum_{l=1}^kW_{il}=0)\Big]f(\xi_i){\rm d}\xi_i.
%\end{aligned}
%$$
The %observed-data 
complete-data likelihood contributed for the individual $i=1,\dots,n$ is
\begin{equation}
    \begin{aligned}
        L^{(i)}=&p(\bX_i)^{B_i}\{1-p(\bX_i)\}^{1-B_i}f(\xi_i)
        \prod_{l=1}^k\left\{\frac{(\lambda_{il}\xi_i)^{Y_{il}}}{Y_{il}!}e^{-\lambda_{il}\xi_i}\right\}^{\delta_{L_i}}\times\left\{\frac{(\omega_{il}\xi_i)^{W_{il}}}{W_{il}!}e^{-(\omega_{il}+\lambda_{il})\xi_i}\right\}^{\delta_{I_i}}\\
 &\times\exp(-\delta_{R_i}B_i\omega_{il}\xi_i)\label{Li}
    \end{aligned}
\end{equation}
under the constraint $\delta_{L_i}I(\sum_{l=1}^kY_{il} > 0) + \delta_{I_i}I(\sum_{l=1}^kW_{il} > 0) + \delta_{R_i} = 1$.

Firstly, for $\delta_{L_i}=1$, we have $B_i=1$, $P(B_i=1)=1$ and $W_{il} \equiv 0$ since $\omega_{il}=0$. Equation \eqref{Li} can be rewritten as
$L^{(i)}=p(\bX_i)f(\xi_i)\left\{I(\sum_{l=1}^kY_{il}>0)\prod_{l=1}^k\frac{(\lambda_{il}\xi_i)^{Y_{il}}}{Y_{il}!}e^{-\lambda_{il}\xi_i}\right\}$. Integrating $Y_{il}$, $W_{il}$, $\xi_i$ and $B_i$ out of $L^{(i)}$, we obtain
\begin{equation}
    \begin{aligned}
    &\int_{\xi_i}\sum_{B_{i}}\sum_{Y_{il}}\sum_{W_{il}}p(\bX_i)f(\xi_i)\left\{I(\sum_{l=1}^kY_{il}>0)\prod_{l=1}^k\frac{(\lambda_{il}\xi_i)^{Y_{il}}}{Y_{il}!}e^{-\lambda_{il}\xi_i}\right\}{\rm d}W_{il}{\rm d}Y_{il}{\rm d}B_{i}{\rm d}\xi_i\\
    %=&\int_{\xi_i}\sum_{W_{il}}p(\bX_i)f(\xi_i)P(\sum_{l=1}^kY_{il}>0|\xi_i){\rm d}W_{il}{\rm d}\xi_i\\
    %=&p(\bX_i)\int_{\xi_i}f(\xi_i)[1-P(\sum_{l=1}^kY_{il}=0|\xi_i)]{\rm d}\xi_i\\
    %=&p(\bX_i)\int_{\xi_i}f(\xi_i)[1-\prod_{l=1}^kP(Y_{il}=0|\xi_i)]{\rm d}\xi_i\\
    %=&p(\bX_i)\int_{\xi_i}f(\xi_i)[1-\exp(-\sum_{l=1}^k\lambda_{il}\xi_i)]{\rm d}\xi_i\\
    =&p(\bX_i)\int_{\xi_i}f(\xi_i)\left[1-\exp\{-\xi_i\exp(\bbeta^{\top}\bZ_i)\Lambda(R_i)\}\right]{\rm d}\xi_i.\label{LideltaL}
\end{aligned}
\end{equation}

Similarly, if $\delta_{I_i}=1$, we have $B_i=1$ and $P(B_i=1)=1$. Equation 
\eqref{Li} can be rewritten as $L^{(i)}=p(\bX_i)f(\xi_i)\left\{I(\sum_{l=1}^kY_{il}=0,\sum_{l=1}^kW_{il}>0)\prod_{l=1}^k\frac{(\omega_{il}\xi_i)^{W_{il}}}{W_{il}!}e^{-(\lambda_{il}+\omega_{il})\xi_i}\right\}$. Integrating $Y_{il}$, $W_{il}$, $\xi_i$ and $B_i$ out of $L^{(i)}$, we have
\begin{equation}
\begin{aligned}
    &\int_{\xi_i}\sum_{B_{i}}\sum_{Y_{il}}\sum_{W_{il}}p(\bX_i)f(\xi_i)\left\{I(\sum_{l=1}^kY_{il}=0,\sum_{l=1}^kW_{il}>0)\frac{(\omega_i\xi_i)^{(W_{il})}}{W_{il}!}e^{-(\lambda_{il}+\omega_{il})\xi_i}\right\}{\rm d}W_{il}{\rm d}Y_{il}{\rm d}B_{i}{\rm d}\xi_i\\
    %=&\int_{\xi_i}p(\bX_i)f(\xi_i)\sum_{Y_{il}=0}e^{-\lambda_{il}\xi_i}\sum_{W_{il}>0}\frac{(\omega_{il}\xi_i)^{W_{il}}}{W_{il}!}e^{-\omega_{il}\xi_i}{\rm d}W_{il}{\rm d}Y_{il}{\rm d}\xi_i\\
    %=&p(\bX_i)\int_{\xi_i}f(\xi_i)P(\sum_{l=1}^kY_{il}=0|\xi_i)[1-P(\sum_{l=1}^kW_{il}=0|\xi_i)]{\rm d}\xi_i\\
    %=&p(\bX_i)\int_{\xi_i}f(\xi_i)\prod_{l=1}^kP(Y_{il}=0|\xi_i)[1-\prod_{l=1}^kP(W_{il}=0|\xi_i)]{\rm d}\xi_i\\
    %=&p(\bX_i)\int_{\xi_i}f(\xi_i)\exp(-\sum_{l=1}^k\lambda_{il}\xi_i)[1-\exp(-\sum_{l=1}^k\omega_{il}\xi_i)]{\rm d}\xi_i\\
    =&p(\bX_i)\int_{\xi_i}f(\xi_i)\big[\exp\{-\xi_i\exp(\bbeta^{\top}\bZ_i)\Lambda(L_i)\}-\exp\{-\xi_i\exp(\bbeta^{\top}\bZ_i)\Lambda(R_i)\}\big]{\rm d}\xi_i.\label{LideltaI}
\end{aligned}    
\end{equation}

Finally, if $\delta_{R_i}=1$, we have $Y_{il}\equiv0$ since $\lambda_{il}=0$. Equation 
\eqref{Li} can be rewritten as $L^{(i)}=p(\bX_i)^{B_i}\{1-p(\bX_i)\}^{1-B_i}f(\xi_i)\exp\{-I(\sum_{l=1}^kY_{il}=0,\sum_{l=1}^kW_{il}=0)B_i\omega_i\xi_i\}$. Note that $P(B_i=1)=p(\bX_i)$ is defined at the beginning of Section \ref{sec3.1}. Integrating $Y_{il}$, $W_{il}$, $\xi_i$ and $B_i$ out of $L^{(i)}$, it follows that
\begin{equation}
\begin{aligned}
    &\int_{\xi_i}\sum_{B_{i}}\sum_{Y_{il}}\sum_{W_{il}}p(\bX_i)^{B_i}\{1-p(\bX_i)\}^{1-B_i}f(\xi_i)\prod_{l=1}^k\exp\{-I(\sum_{l=1}^kW_{il}=0)B_i\omega_{il}\xi_i\}{\rm d}W_{il}{\rm d}Y_{il}{\rm d}B_{i}{\rm d}\xi_i\\
    %=&\int_{\xi_i}\sum_{B_{i}}p(\bX_i)^{B_i}(1-p(\bX_i))^{1-B_i}f(\xi_i)\sum_{W_{il}=0}\prod_{l=1}^k\exp(-B_i\omega_{il}\xi_i){\rm d}W_{il}{\rm d}B_{i}{\rm d}\xi_i\\
    %=&\int_{\xi_i}f(\xi_i)\sum_{B_{i}}p(\bX_i)^{B_i}(1-p(\bX_i))^{1-B_i}P(\sum_{l=1}^kW_{il}=0|\xi_i)^{B_i}{\rm d}B_{i}{\rm d}\xi_i\\
    %=&\int_{\xi_i}f(\xi_i)\sum_{B_{i}}\big[I(B_i=0)(1-p(\bX_i))+I(B_i=1)p(\bX_i)P(\sum_{l=1}^kW_{il}=0|\xi_i)\big]{\rm d}B_{i}{\rm d}\xi_i\\
    %=&\int_{\xi_i}f(\xi_i)\big[1-p(\bX_i)+p(\bX_i)P(\sum_{l=1}^kW_{il}=0|\xi_i)\big]{\rm d}\xi_i\\
    =&\int_{\xi_i}f(\xi_i)\big[1-p(\bX_i)+p(\bX_i)\exp\{-\xi_i\exp(\bbeta^{\top}\bZ_i)\Lambda(L_i)\}\big]{\rm d}\xi_i.\label{LideltaR}
\end{aligned}
\end{equation}

Combining \eqref{LideltaL}, \eqref{LideltaI} and \eqref{LideltaR} gives
\begin{equation*}
    \begin{aligned}        
&\int_{\xi}\sum_{B_{i}}\sum_{Y_{il}}\sum_{W_{il}}L_4(\boldsymbol{\theta}|\mathcal{O},\boldsymbol{B,\xi,Y,W}){\rm d}W_{il}{\rm d}Y_{il}{\rm d}B_{i}{\rm d}\xi\\
=&\prod_{i=1}^np(\bX_i)^{1-\delta_{R_i}}\int_{\xi_i}\big[1-\exp\{-\xi_i\exp(\bbeta^{\top}\bZ_i)\Lambda(R_i)\}\big]^{\delta_{L_i}}\\
        &\times\big[\exp\{-\xi_i\exp(\bbeta^{\top}\bZ_i)\Lambda(L_i)\}-\exp\{-\xi_i\exp(\bbeta^{\top}\bZ_i)\Lambda(R_i)\}\big]^{\delta_{I_i}}\\
        &\times\big[1-p(\bX_i)+p(\bX_i)\exp\{-\xi_i\exp(\bbeta^{\top}\bZ_i)\Lambda(L_i)\}\big]^{\delta_{R_i}}
        f(\xi_i)\,{\rm d}\xi_i\\
        =&L(\btheta|\mathcal{O}_o).
    \end{aligned}
\end{equation*}
Therefore, \eqref{L_obs} can be derived by integrating out $\boldsymbol{Y}_l$s, $\boldsymbol{W}_l$s $\boldsymbol{\xi}$ and $\boldsymbol{B}$ in \eqref{complete}.

\subsection*{A.2 Proof of Conditional expectations}\label{A.2}
We derive the expressions for expectations given in Section \ref{sec3.2} of the manuscript. As discussed in Section \ref{sec3.1}, $Y_i$ and $W_i$ conditionally follow the truncated Poisson distribution with the mean parameter $\lambda_i^{(d)}\xi_i$ and $\omega_i^{(d)}\xi_i$, respectively, where 
\begin{equation*}
    \begin{aligned}
        \lambda_i^{(d)}=&\exp(\bbeta^{(d)\top}\bZ_i)\{\delta_{L_i}\sum_{l=1}^k\eta_l^{(d)}b_l(R_i)+\delta_{I_i}\sum_{l=1}^k\eta_l^{(d)}b_l(L_i)\}\\
        \omega_i^{(d)}=&\exp(\bbeta^{(d)\top}\bZ_i)\delta_{I_i}[\{\sum_{l=1}^k\eta_l^{(d)}b_l(R_i)-\sum_{l=1}^k\eta_l^{(d)}b_l(L_i)\}+\delta_{R_i}\sum_{l=1}^k\eta_l^{(d)}b_l(L_i)].
    \end{aligned}
\end{equation*}
The posterior density function of $\xi_i$ given the observed data is proportional to $p(\bX_i)\{1-S_u(R_i|\bZ_i)\}^{\delta_{L_i}}\{S_u(L_i|\bZ_i)-S_u(R_i|\bZ_i)\}^{\delta_{I_i}}S_u(L_i|\bZ_i)^{\delta_{R_i}}$, where $$S_u(T_i|\bZ_i)=\exp\{-\xi_i\exp(\bbeta^{(d)\top}\bZ_i)\sum_{l=1}^k\eta_l^{(d)}b_l(T_i)\}.$$ % $p(\bX_i)[\exp(-\xi_i\exp(\bbeta^{(d)\top}\bZ_i)\Lambda^{(d)}(L_i))-\exp(-\xi_i\exp(\bbeta^{(d)\top}\bZ_i)\Lambda^{(d)}(R_i))]f(\xi_i)$. 
We first consider $E(Y_i|\btheta^{(d)},\mcO)$. For $i=1,\dots,n$, we have
\begin{equation*}
    \begin{aligned}
        E(Y_i|\btheta^{(d)},\mcO)&=E(Y_i|\btheta^{(d)},\mcO,Y_i>0)P(Y_i>0|\btheta^{(d)},\mcO)\\
        &=E(E(Y_i|\xi_i,\btheta^{(d)},\mcO,Y_i>0)|\btheta^{(d)},\mcO,Y_i>0)P(Y_i>0|\btheta^{(d)},\mcO),
        %E(Y_i|\btheta^{(d)},\mcO)=&E_{\xi_i}(E(Y_i|\xi_i,\btheta^{(d)},\mcO,Y_i>0)|Y_i>0,W_i=0)=\frac{E_{\xi_i}(E(Y_i|\xi_i,\btheta^{(d)},\mcO,Y_i>0))}{P(Y_i>0,W_i=0)},
        %E(W_i|\btheta^{(d)},\mcO)=&E_{\xi_i}(E(W_i|\xi_i,\btheta^{(d)},\mcO,W_i>0)|Y_i=0,W_i>0)=\frac{E_{\xi_i}(E(W_i|\xi_i,\btheta^{(d)},\mcO,W_i>0))}{P(Y_i=0,W_i>0)},
    \end{aligned}
\end{equation*}
where the outer expectation above is taken with respect to the distribution of $\xi_i$ given $\btheta^{(d)}$, $\mcO$ and $Y_i>0$. It follows that
\begin{equation}
E(Y_i|\xi_i,\btheta^{(d)},\mcO,Y_i>0)=
\begin{cases} 
\frac{\xi_iN_{i2}^{(d)}}{1-\exp(-\xi_iN_{i2}^{(d)})},  & \text{if } \delta_{L_i}=1, \\
0, & \text{if } \delta_{I_i}=1 \text{ or } \delta_{R_i}=1,
\end{cases}
\end{equation}
where $N_{i2}^{(d)}=\exp(\bbeta^{(d)\top}\bZ_i)\sum_{l=1}^k\eta_l^{(d)}b_l(R_i)$. This can be rewritten as
\begin{equation*}
    \begin{aligned}
        E(Y_i|\xi_i,\btheta^{(d)},\mcO,Y_i>0)=\frac{\delta_{L_i}\xi_iN_{i2}^{(d)}}{1-\exp(-\xi_iN_{i2}^{(d)})},
    \end{aligned}
\end{equation*}
and hence
\begin{equation}
    \begin{aligned}
        E(Y_i|\btheta^{(d)},\mcO,Y_i>0)=\frac{\delta_{L_i}N_{i2}^{(d)}\int_{\xi_i}\xi_if(\xi_i){\rm d}\xi_i}{1-\exp\{-G(N_{i2}^{(d)})\}}.\label{aeyi}
    \end{aligned}
\end{equation}
Similarly, for $E(W_i|\btheta^{(d)},\mcO)$, we have
\begin{equation*}
    \begin{aligned}
        E(W_i|\btheta^{(d)},\mcO)&=E(W_i|\btheta^{(d)},\mcO,W_i>0)P(W_i>0|\btheta^{(d)},\mcO)\\
        &=E(E(W_i|\xi_i,\btheta^{(d)},\mcO,W_i>0)|\btheta^{(d)},\mcO,W_i>0)P(W_i>0|\btheta^{(d)},\mcO),
    \end{aligned}
\end{equation*}
where the outer expectation above is taken with respect to the distribution of $\xi_i$ given $\btheta^{(d)}$, $\mcO$ and $W_i>0$. It follows that
\begin{equation*}
E(W_i|\xi_i,\btheta^{(d)},\mcO,W_i>0)=
\begin{cases} 
\frac{\xi_i(N_{i2}^{(d)}-N_{i1}^{(d)})}{1-\exp\{-\xi_i(N_{i2}^{(d)}-N_{i1}^{(d)})\}},  & \text{if } \delta_{I_i}=1, \\
0, & \text{if } \delta_{L_i}=1 \text{ or } \delta_{R_i}=1,
\end{cases}
\end{equation*}
where $N_{i1}^{(d)}=\exp(\bbeta^{(d)\top}\bZ_i)\sum_{l=1}^k\eta_l^{(d)}b_l(L_i)$. This can be rewritten as 
\begin{equation*}
    \begin{aligned}
        E(W_i|\xi_i,\btheta^{(d)},\mcO,W_i>0)=\frac{\delta_{I_i}\xi_i(N_{i2}^{(d)}-N_{i1}^{(d)})}{1-\exp\{-\xi_i(N_{i2}^{(d)}-N_{i1}^{(d)})\}},
    \end{aligned}
\end{equation*}
and hence
\begin{equation}
    \begin{aligned}
        E(W_i|\btheta^{(d)},\mcO)=\frac{\delta_{I_i}(N_{i2}^{(d)}-N_{i1}^{(d)})\exp\{-G(N_{i1}^{(d)})\}G'(N_{i1}^{(d)})}{\exp\{-G(N_{i1}^{(d)})\}-\exp\{-G(N_{i2}^{(d)})\}}.
    \end{aligned}\label{aewi}
\end{equation}
Then we drive the expectations for $E(Y_{il}|\btheta^{(d)},\mcO)$ and $(W_{il}|\btheta^{(d)},\mcO)$. Noting that the distribution of $Y_{il}$ given $Y_i$ and the distribution of $W_{il}$ given $W_i$ are both binomial. Thus,
\begin{align}
    E(Y_{il}|\boldsymbol{\theta}^{(d)},\mcO)=&\frac{\eta_l^{(d)}b_l(R_i)E(Y_i|\boldsymbol{\theta}^{(d)},\mcO)}{\sum_{l=1}^k\eta_l^{(d)}b_l(R_i)},\label{aeyil}\\
    E(W_{il}|\boldsymbol{\theta}^{(d)},\mcO)=&\frac{\eta_l^{(d)}\{b_l(R_i)-b_l(L_i)\}E(W_i|\boldsymbol{\theta}^{(d)},\mcO)}{\sum_{l=1}^k\eta_l^{(d)}\{b_l(R_i)-b_l(L_i)\}}.\label{aewil}
\end{align}

We now drive $E(B_i|\boldsymbol{\theta}^{(d)},\mcO)$. The conditional expectation of $B_i$ is the probability of being cured for the right-censored subject, which is equal to 1 when the failure time is left-censored or interval-censored. Consequently,
\begin{equation}
    \begin{aligned}
    E(B_{i}|\boldsymbol{\theta}^{(d)},\mcO)=&\delta_{L_i}+\delta_{I_i}+\frac{\delta_{R_i}p^{(d)}(X_i)\exp\{-G(N_{i1}^{(d)})\}}{1-p^{(d)}(X_i)+p^{(d)}(X_i)\exp\{-G(N_{i1}^{(d)})\}}.\label{aebi}
\end{aligned}
\end{equation}
For $E(\xi_iB_i|\boldsymbol{\theta}^{(d)},\mcO)$, we have $E(\xi_iB_i|\boldsymbol{\theta}^{(d)},\mcO)=E(\xi_iB_i|\boldsymbol{\theta}^{(d)},\mcO,B_i=1)P(B_i=1|\btheta^{(d)},\mcO)$. For left-censored subject, {$P(\mcO)=P(B_i=1,\btheta^{(d)},\mcO)=p^{(d)}(X_i)(1-\exp\{-G(N_{i2}^{(d)})\})$,} then 
\begin{equation*}
    \begin{aligned}
        E(\xi_iB_i|\boldsymbol{\theta}^{(d)},\mcO)&=\frac{\int_{\xi_i}\xi_ip^{(d)}(X_i)\{1-\exp(-\xi_iN_{i2}^{(d)})\}f(\xi_i){\rm d}\xi_i}{P(\mcO)}\\
        &=\frac{\int_{\xi_i}\xi_if(\xi_i){\rm d}\xi_i-\exp\{-G(N_{i2}^{(d)})\}G'(N_{i2}^{(d)})}{1-\exp\{-G(N_{i2}^{(d)})\}}.\label{exibl}
    \end{aligned}
\end{equation*}
Similarly, for interval-censored subject, $P(\mcO)=P(B_i=1,\btheta^{(d)},\mcO)=p^{(d)}(X_i)[\exp\{-G(N_{i1}^{(d)})\}-\exp\{-G(N_{i2}^{(d)})\}]$. Consequently,
\begin{equation*}
    \begin{aligned}
        E(\xi_iB_i|\boldsymbol{\theta}^{(d)},\mcO)&=\frac{\int_{\xi_i}\xi_ip^{(d)}(X_i)\{\exp(-\xi_iN_{i1}^{(d)})-\exp(-\xi_iN_{i2}^{(d)})\}f(\xi_i){\rm d}\xi_i}{P(\mcO)}\\
        &=\frac{\exp\{-G(N_{i1}^{(d)})\}G'(N_{i1}^{(d)})-\exp\{-G(N_{i2}^{(d)})\}G'(N_{i2}^{(d)})}{\exp\{-G(N_{i1}^{(d)})\}-\exp\{-G(N_{i2}^{(d)})\}}.\label{exibi}
    \end{aligned}
\end{equation*}
We now consider right-censored subject. If $\delta_{R_i}=1$, then $P(\mcO)=1-p^{(d)}(X_i)+p^{(d)}(X_i)\exp\{-G(N_{i1}^{(d)})\}$ and $P(\mcO)=P(B_i=1,\btheta^{(d)},\mcO)=p^{(d)}(X_i)\exp\{-G(N_{i1}^{(d)})\}$. It follows that
\begin{equation*}
    \begin{aligned}
        E(\xi_iB_i|\boldsymbol{\theta}^{(d)},\mcO)&=\frac{\int_{\xi_i}\xi_ip^{(d)}(X_i)\exp(-\xi_iN_{i1}^{(d)})f(\xi_i){\rm d}\xi_i}{P(\mcO)}\\
        &=\frac{p^{(d)}(X_i)\exp\{-G(N_{i1}^{(d)})\}G'(N_{i1}^{(d)})}{1-p^{(d)}(X_i)+p^{(d)}(X_i)\exp\{-G(N_{i1}^{(d)})\}}.\label{exibr}
    \end{aligned}
\end{equation*}
Combining these results gives
\begin{equation}
    \begin{aligned}
    E(\xi_iB_i|\boldsymbol{\theta}^{(d)})=&\delta_{L_i}\frac{\int_{\xi_i}\xi_if(\xi_i){\rm d}\xi_i-\exp\{-G(N_{i2}^{(d)})\}G'(N_{i2}^{(d)})}{1-\exp\{-G(N_{i2}^{(d)})\}}\\
    &+\delta_{I_i}\frac{\exp\{-G(N_{i1}^{(d)})\}G'(N_{i1}^{(d)})-\exp\{-G(N_{i2}^{(d)})\}G'(N_{i2}^{(d)})}{\exp\{-G(N_{i1}^{(d)})\}-\exp\{-G(N_{i2}^{(d)})\}}\\
    &+\delta_{R_i}\frac{p^{(d)}(X_i)\exp\{-G(N_{i1}^{(d)})\}G'(N_{i1}^{(d)})}{1-p^{(d)}(X_i)+p^{(d)}(X_i)\exp\{-G(N_{i1}^{(d)})\}}.\label{aexibi}
\end{aligned}
\end{equation}

From Equations \eqref{aeyi}, \eqref{aewi}, \eqref{aeyil}, \eqref{aewil}, \eqref{aebi} and \eqref{aexibi}, the conditional expectation in Section \ref{sec3.2} are
\begin{align}
E(Y_{il})&=\frac{\eta_l^{(d)}b_l(R_i)E(Y_i)}{\sum_{l=1}^k\eta_l^{(d)}b_l(R_i)}, 
\quad
E(Y_i)=\frac{\delta_{L_i}N_{i2}^{(d)}\int_{\xi_i}\xi_if(\xi_i){\rm d}\xi_i}{1-\exp\{-G(N_{i2}^{(d)})\}},\nonumber\\
E(W_{il})&=\frac{\eta_l^{(d)}\{b_l(R_i)-b_l(L_i)\}E(W_i)}{\sum_{l=1}^k\eta_l^{(d)}\{b_l(R_i)-b_l(L_i)\}},\nonumber\\
E(W_i)&=\frac{\delta_{I_i}(N_{i2}^{(d)}-N_{i1}^{(d)})\exp\{-G(N_{i1}^{(d)})\}G'(N_{i1}^{(d)})}{\exp\{-G(N_{i1}^{(d)})\}-\exp\{-G(N_{i2}^{(d)})\}},\nonumber\\
E(B_{i})&=\delta_{L_i}+\delta_{I_i}+\frac{\delta_{R_i}p^{(d)}(\bX_i)\exp\{-G(N_{i1}^{(d)})\}}{1-p^{(d)}(\bX_i)+p^{(d)}(\bX_i)\exp\{-G(N_{i1}^{(d)})\}},\nonumber
\end{align} 
and 
\begin{align}
E(\xi_iB_i)&=\delta_{L_i}\frac{\int_{\xi_i}\xi_if(\xi_i){\rm d}\xi_i-\exp\{-G(N_{i2}^{(d)})\}G'(N_{i2}^{(d)})}{1-\exp\{-G(N_{i2}^{(d)})\}}\nonumber\\
    &+\delta_{I_i}\frac{\exp\{-G(N_{i1}^{(d)})\}G'(N_{i1}^{(d)})-\exp\{-G(N_{i2}^{(d)})\}G'(N_{i2}^{(d)})}{\exp\{-G(N_{i1}^{(d)})\}-\exp\{-G(N_{i2}^{(d)})\}}\nonumber\\
    &+\delta_{R_i}\frac{p^{(d)}(\bX_i)\exp\{-G(N_{i1}^{(d)})\}G'(N_{i1}^{(d)})}{1-p^{(d)}(\bX_i)+p^{(d)}(\bX_i)\exp\{-G(N_{i1}^{(d)})\}}.\nonumber   
\end{align}

\subsection*{A.3 Proof of identifiablity}\label{A.3}
{%To show that the model is identifiable, 
To establish model identifiability, suppose that there exist two parameter sets $(g, \bgamma,\Lambda, \bbeta)$ and $(\Tilde{g}, \Tilde{\bgamma},\Tilde{\Lambda}, \Tilde{\bbeta})$ such that for all $( \bX, \bZ)$ and $t\in [0,\infty)$,
\begin{equation}
    \begin{aligned}
        &1-g(\bgamma^{\top}\bX)+g(\bgamma^{\top}\bX)S_u(t|\bZ)=1-\Tilde{g}(\Tilde{\bgamma}^{\top}\bX)+\Tilde{g}(\Tilde{\gamma}^{\top}\bX)\Tilde{S}_u(t|\bZ),
    \end{aligned}\label{ident.}
\end{equation}
% \begin{equation}
%     \begin{aligned}
%         &g(\bgamma^{\top}\bX)^{1-\delta_R}\{1-S_u(R|\bZ)\}^{\delta_L}\{S_u(L|\bZ)-S_u(R|\bZ)\}^{\delta_I}\{1-g(\bgamma^{\top}\bX)+g(\bgamma^{\top}\bX)S_u(L|\bZ)\}^{\delta_R}\\
%         =&\Tilde{g}(\Tilde{\bgamma}^{\top}\bX)^{1-\delta_R}\{1-\Tilde{S}_u(R|\bZ)\}^{\delta_L}\{\Tilde{S}_u(L|\bZ)-\Tilde{S}_u(R|\bZ)\}^{\delta_I}\{1-\Tilde{g}(\Tilde{\bgamma}^{\top}\bX)+\Tilde{g}(\Tilde{\gamma}^{\top}\bX)\Tilde{S}_u(L|\bZ)\}^{\delta_R},
%     \end{aligned}\label{ident.}
% \end{equation}
 where $S_u(t|\bZ)=\exp[-G\{\exp(\bbeta^{\top}\bZ)\Lambda(t)\}]$, and $\Tilde{S}_u(t|\bZ)=\exp[-G\{\exp(\Tilde{\bbeta}^{\top}\bZ)\Tilde{\Lambda}(t)\}]$. %Then we need to show 
 We demonstrate identifiability by showing $g=\Tilde{g}$, $\bgamma=\Tilde{\bgamma}$, $\Lambda=\Tilde{\Lambda}$, $\bbeta=\Tilde{\bbeta}$.}

From equation \eqref{ident.}, we obtain %leads to 
$$\frac{g(\bgamma^{\top}\bX)}{\Tilde{g}(\Tilde{\bgamma}^{\top}\bX)}=\frac{1-\Tilde{S}_u(t|\bZ)}{1-S_u(t|\bZ)}=c(\bX),$$
where $c(\bX)$ is a positive function of $\bX$ %not depending on 
independent of $t$.
Rearranging %the equation
yields 
$\Tilde{S}_u(t|\bZ) = c(\bX)S_u(t|\bZ) +1-c(\bX).$
By assumption A(6), 
$\lim _{t \rightarrow \infty} S_u(t| \bZ)=\lim _{t \rightarrow \infty} \Tilde{S}_u(t| \bZ)=0 $. Hence $c(\bX)=1$, implying
%Based on assumption (A7) and the fact that $ \Lambda(t) $ is an increasing function, it follows that as $t$ approaches infinity, both $\Lambda(t)$ and $\Tilde{\Lambda}(t)$ tend to infinity. Consequently, as $t\to \infty$, we observe $\lim_{t\to\infty}\exp\{-G(\exp(\bbeta^{\top}\bZ)\Lambda(t))\}=0$ and $\lim_{t\to\infty}\exp\{-G(\exp(\Tilde{\bbeta}^{\top}\bZ)\Tilde{\Lambda}(t))\}=0$. This leads to the conclusion that $c(\bX)=1$, which in turn implies
$g(\bgamma^{\top}\bX)=\Tilde{g}(\Tilde{\bgamma})$ and $S_u(t|\bZ)=\Tilde{S}_u(t|\bZ)$.

From the equality of survival functions and the assumption that $G$ is strictly increasing, we have
%\blue{For $S_u(t|Z)=\Tilde{S}_u(t|Z)$, we have $G\{\exp(\bbeta^{\top}\bZ)\Lambda(t)\}=G\{\exp(\Tilde{\bbeta}^{\top}\bZ)\Tilde{\Lambda}(t)\}$.Since assumption (A3) states $G$ is strictly increasing, we can deduce that 
$$\exp(\bbeta^{\top}\bZ)\Lambda(t)=\exp(\Tilde{\bbeta}^{\top}\bZ)\Tilde{\Lambda}(t).$$
%which can be rewritten as 
Rewriting this relationship yields
$
    \frac{\exp(\bbeta^{\top}\bZ)}{\exp(\Tilde{\bbeta}^{\top}\bZ)}=\frac{\Tilde{\Lambda}(t)}{\Lambda(t)}=c_1, \label{c_1}
$
where $c_1$ is a positive constant independent of $\bZ$ and $t$. %This implies that $(\bbeta-\Tilde{\bbeta})^{\top}\bZ=\log(c_1).$
%If $ \bbeta-\Tilde{\bbeta} \ne 0$, assumption (A5) guarantees that $\textnormal{Var}\{(\bbeta-\Tilde{\bbeta})^{\top}\bZ\}= (\bbeta-\Tilde{\bbeta})^{\top} \textnormal{Var}(\bZ) (\bbeta-\Tilde{\bbeta})> 0$, %But since $\log(c_1)$ is a constant, its variance 
%while $\textnormal{Var}\{\log(c_1)\}=0$, creating a contradiction. %Thus, 
Based on assumption (A4), we have % must be zero, which contradicts the previous statement. Hence, it must be that
$\bbeta=\Tilde{\bbeta}$, and $c_1=1$, which further implies % along with 
$\Lambda(t)=\Tilde{\Lambda}(t)$.%}

%\blue{For $g(\bgamma^\top\bX)=\Tilde{g}(\Tilde{\bgamma}^\top\bX)$, if $\bX$ is a vector of continuous covariates, then satisfying assumption (A1) implies that $g=\Tilde{g}$ and $\bgamma=\Tilde{\bgamma}$. When $\bX$ is a mixture of continuous and discrete variables, satisfying assumptions (A1) and (A2) ensures that $g=\Tilde{g}$ and $\bgamma=\Tilde{\bgamma}$. These results follow from Theorem 2.1 in \cite{horowitz2009semiparametric} and \cite{amico2019single}.}

For the equality $g(\bgamma^\top\bX)=\Tilde{g}(\Tilde{\bgamma}^\top\bX)$, the identifiability of parameters can be established based on Theorem 2.1 in \cite{horowitz2009semiparametric}.

\subsection*{A.4 More simulation results}\label{A.4}

\begin{figure}[H]
    \centering
    \begin{subfigure}[b]{0.32\textwidth}
        \includegraphics[width=\textwidth]{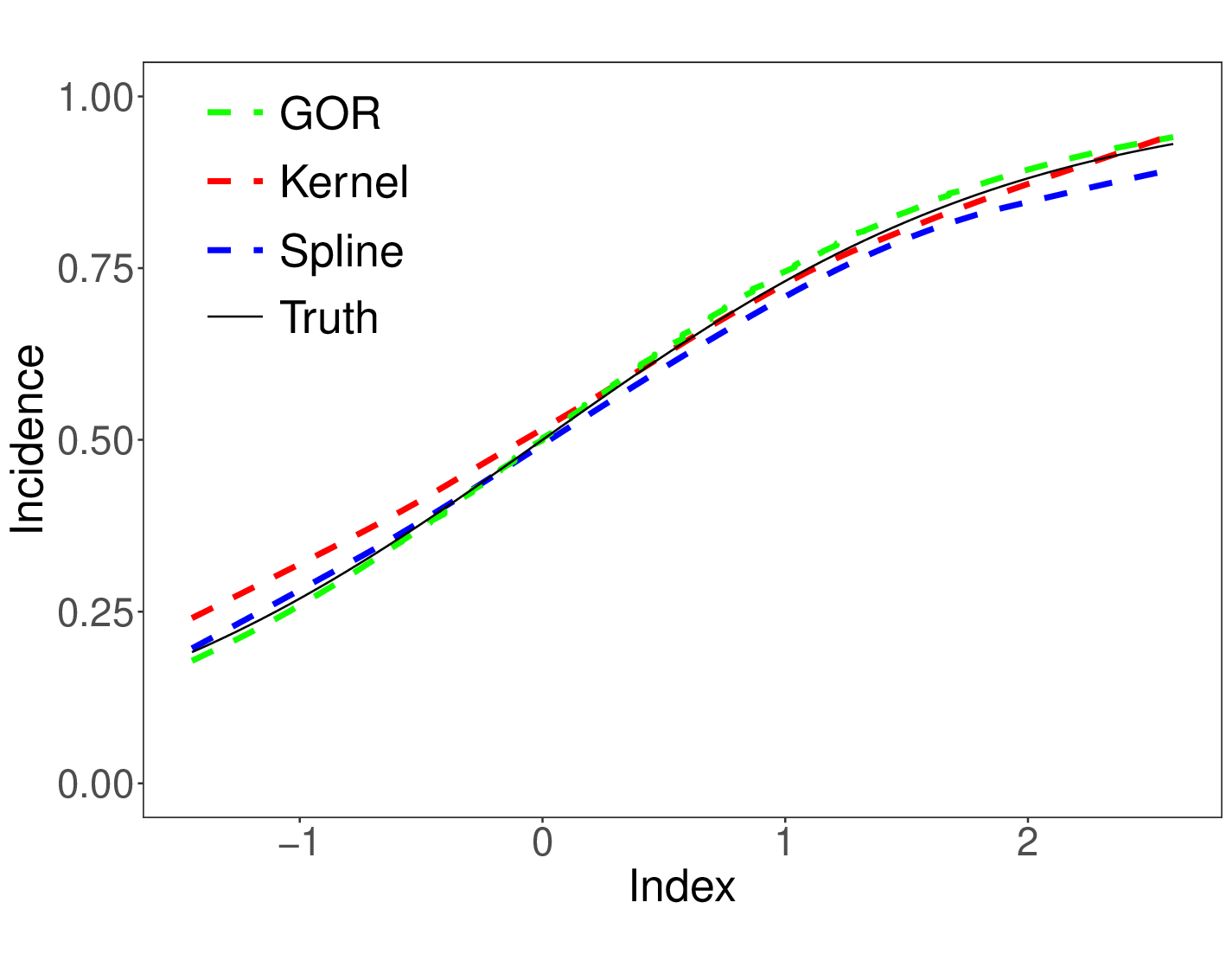}
        \caption{Scenario 1; r=0}
    \end{subfigure}
    \begin{subfigure}[b]{0.32\textwidth}
        \includegraphics[width=\textwidth]{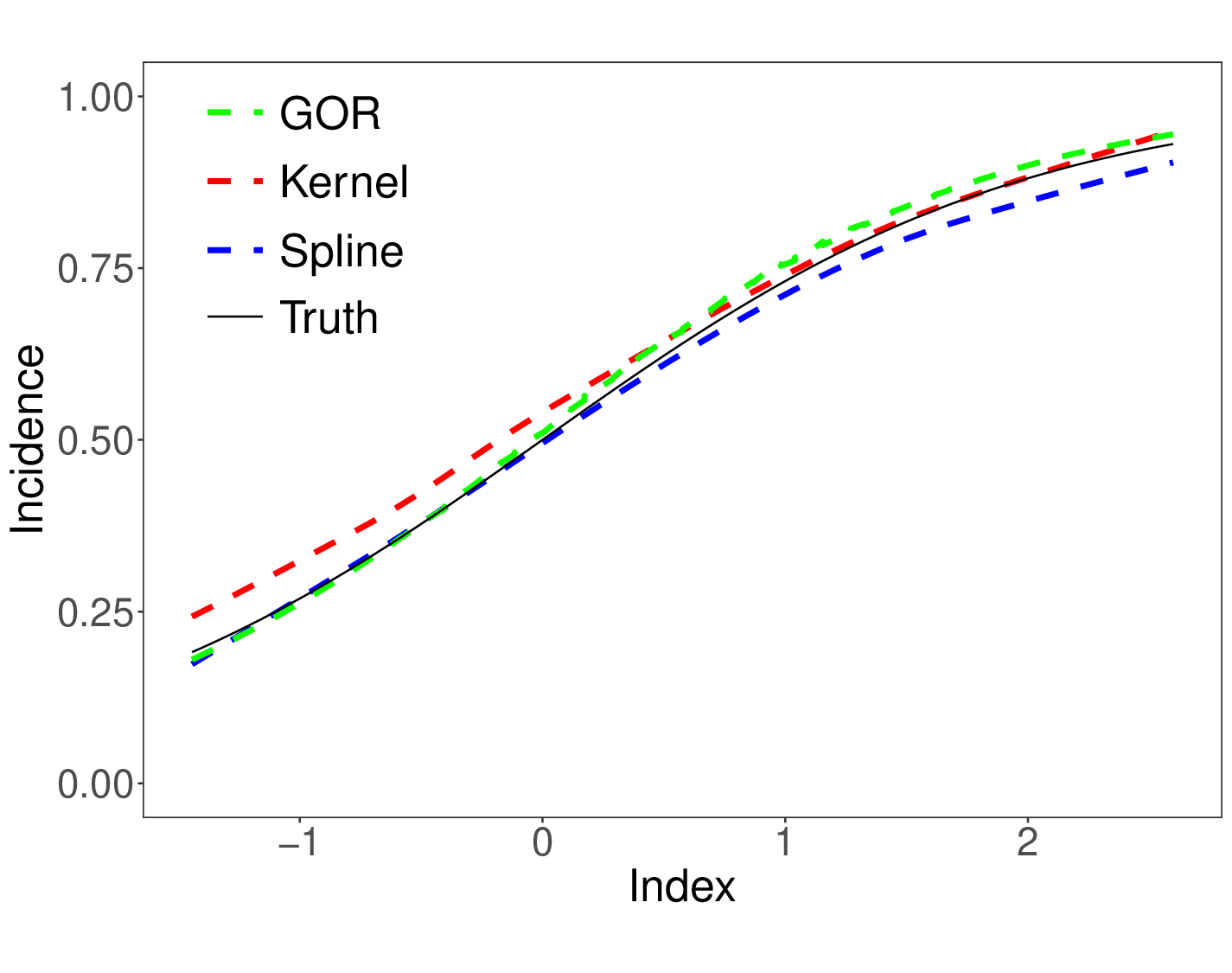}
        \caption{Scenario 1; r=1}
    \end{subfigure}
    \begin{subfigure}[b]{0.32\textwidth}
        \includegraphics[width=\textwidth]{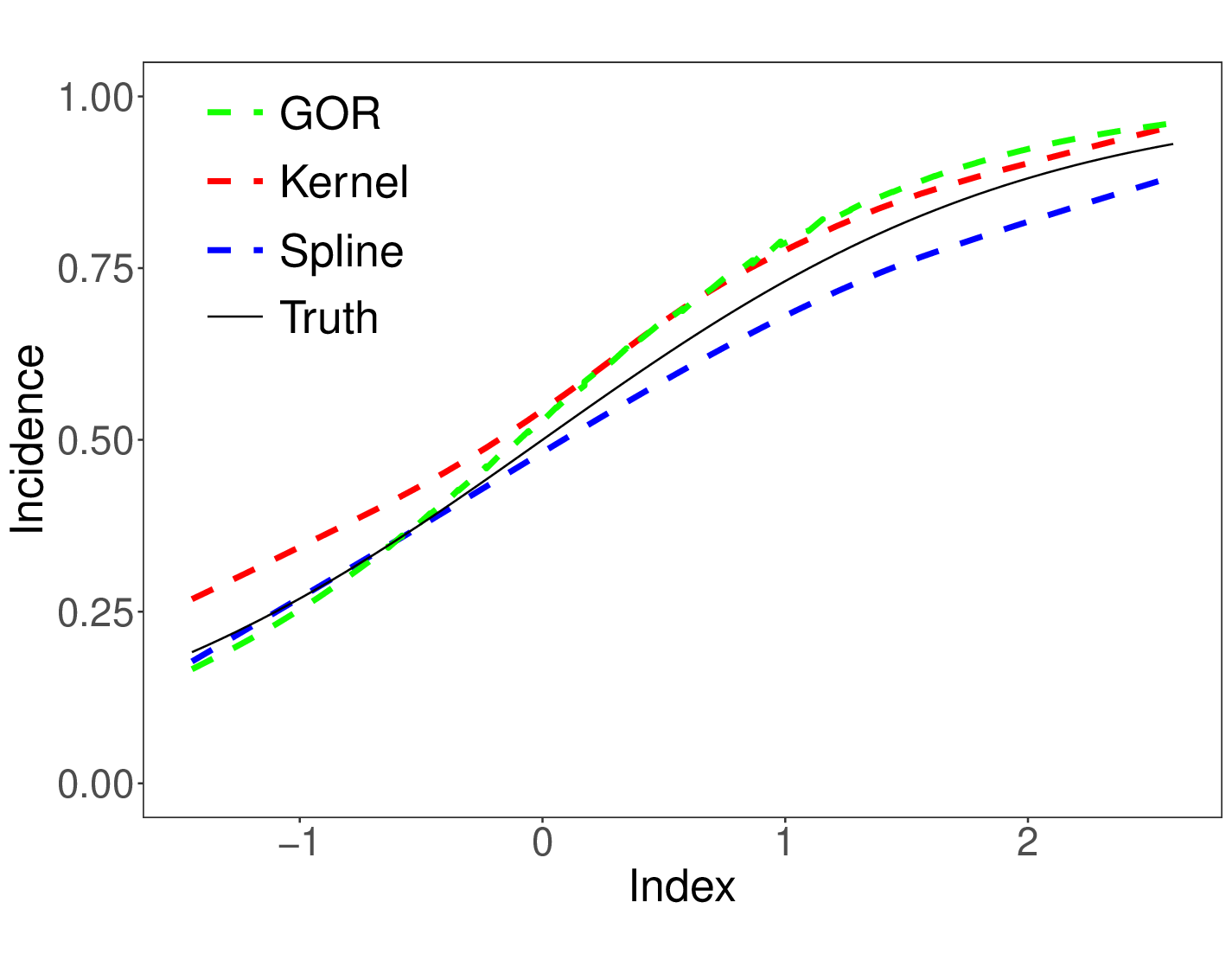}
        \caption{Scenario 1; r=2}
    \end{subfigure}
    \begin{subfigure}[b]{0.32\textwidth}
        \includegraphics[width=\textwidth]{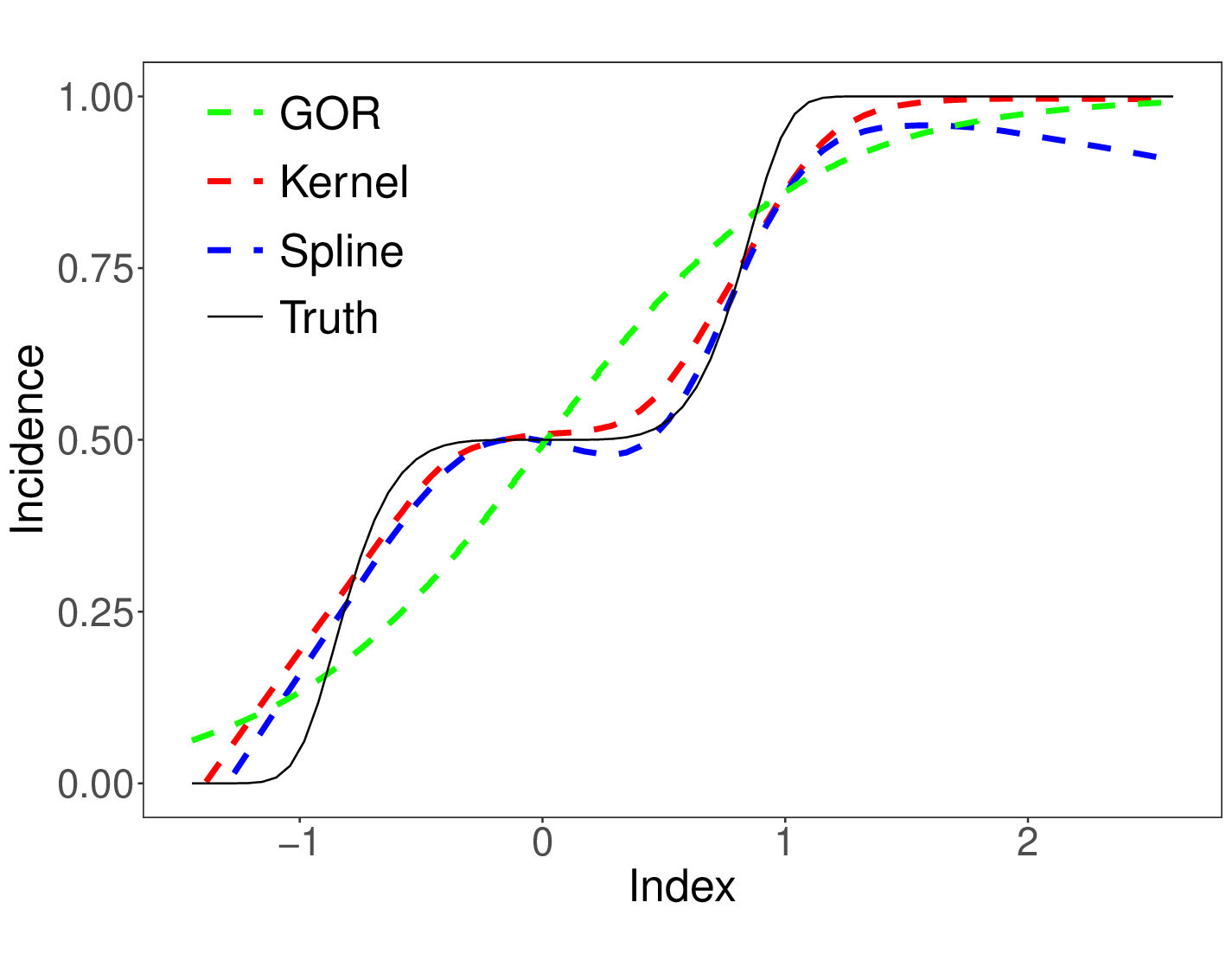}
        \caption{Scenario 2; r=0}
    \end{subfigure}
    \begin{subfigure}[b]{0.32\textwidth}
        \includegraphics[width=\textwidth]{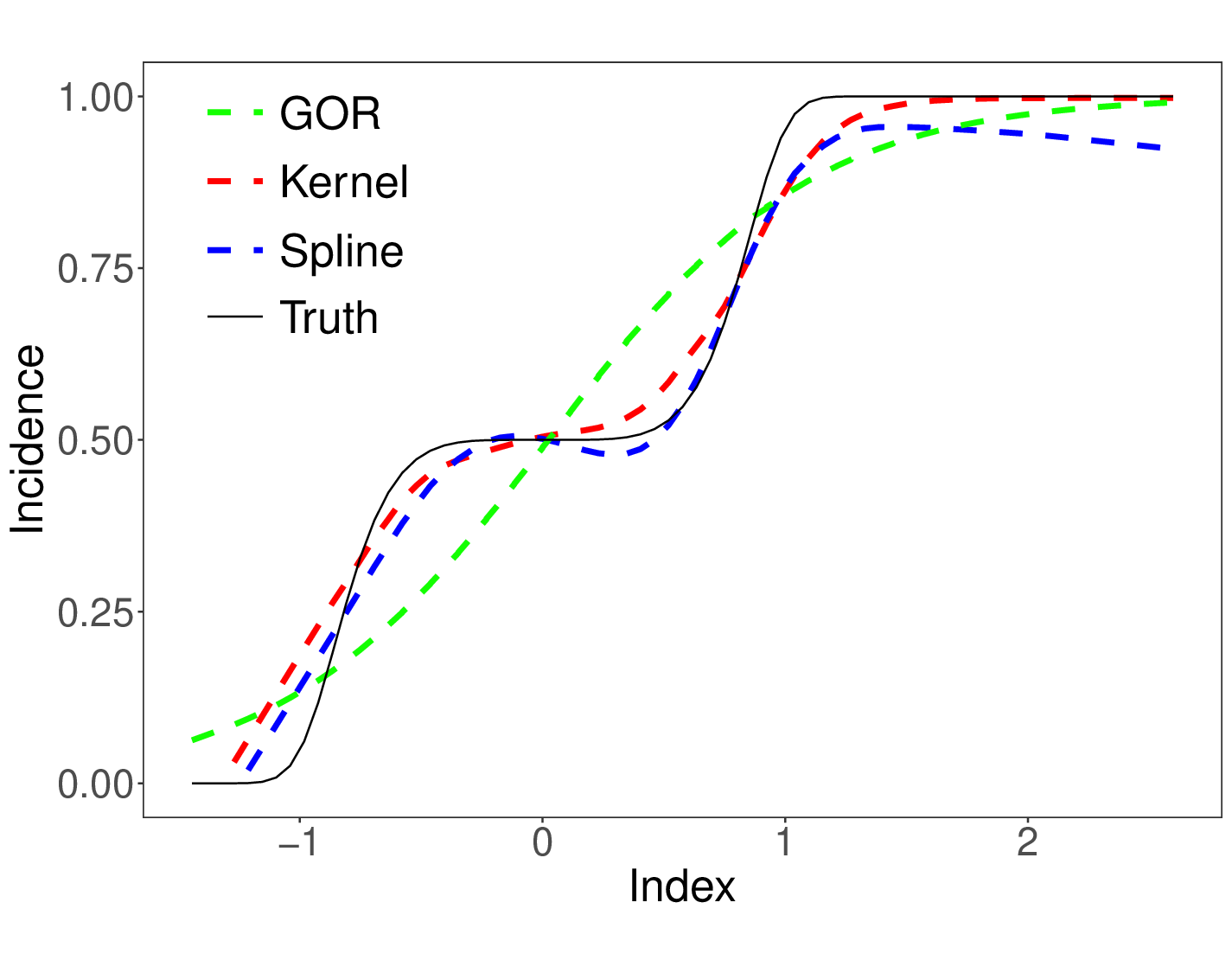}
        \caption{Scenario 2; r=1}
    \end{subfigure}
    \begin{subfigure}[b]{0.32\textwidth}
        \includegraphics[width=\textwidth]{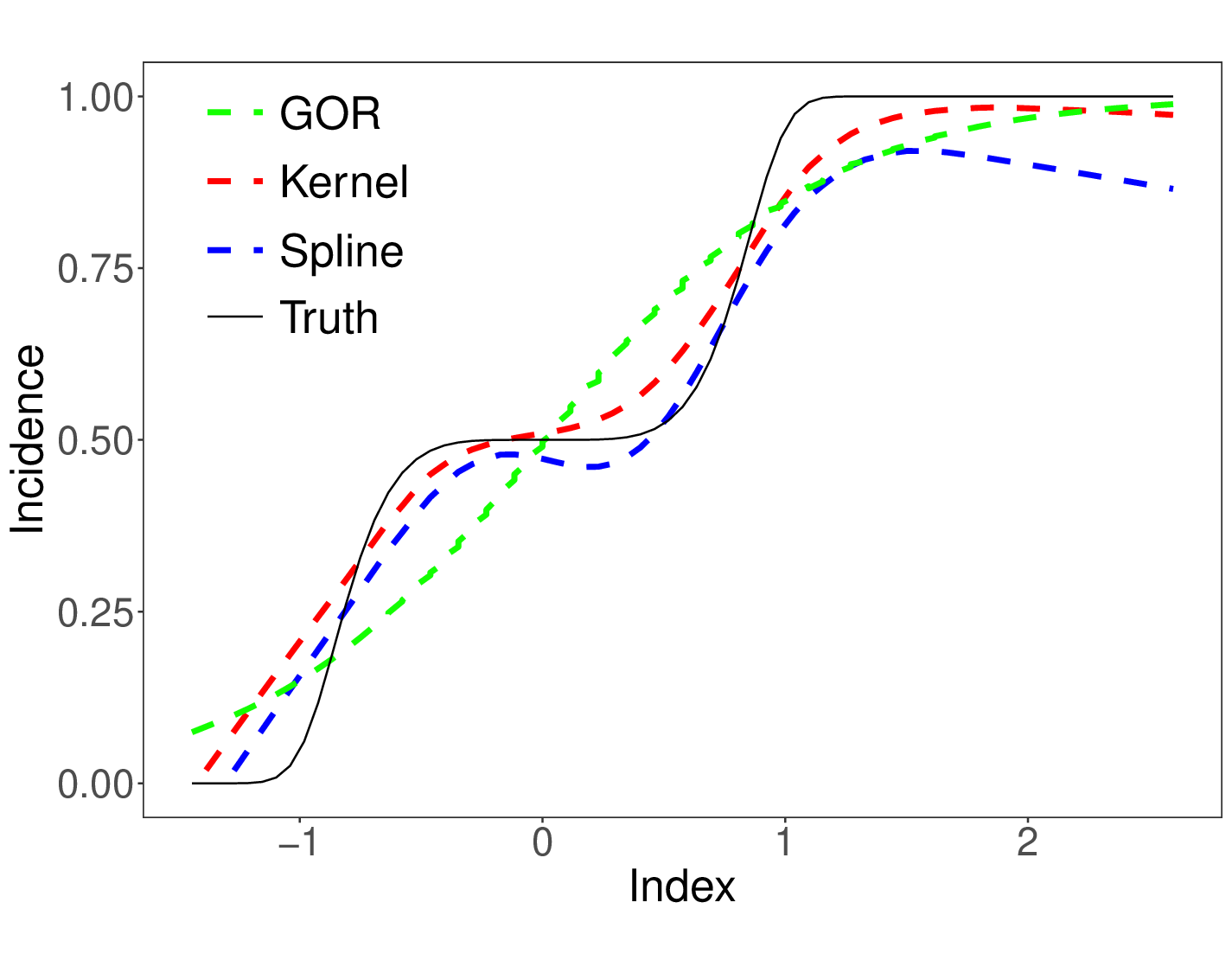}
        \caption{Scenario 2; r=2}
    \end{subfigure}
    \begin{subfigure}[b]{0.32\textwidth}
         \includegraphics[width=\textwidth]{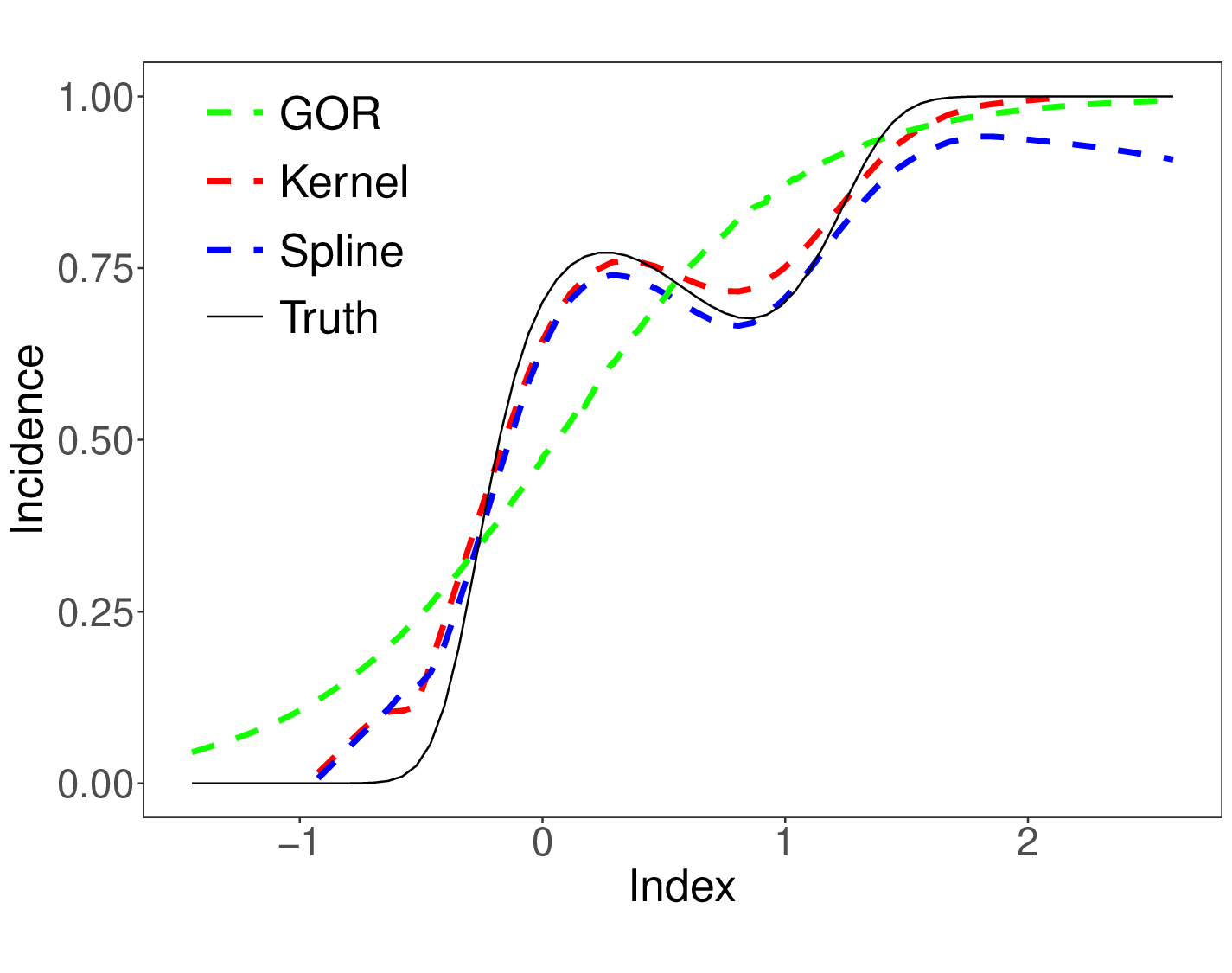}
        \caption{Scenario 3; r=0}
    \end{subfigure}
    \begin{subfigure}[b]{0.32\textwidth}
        \includegraphics[width=\textwidth]{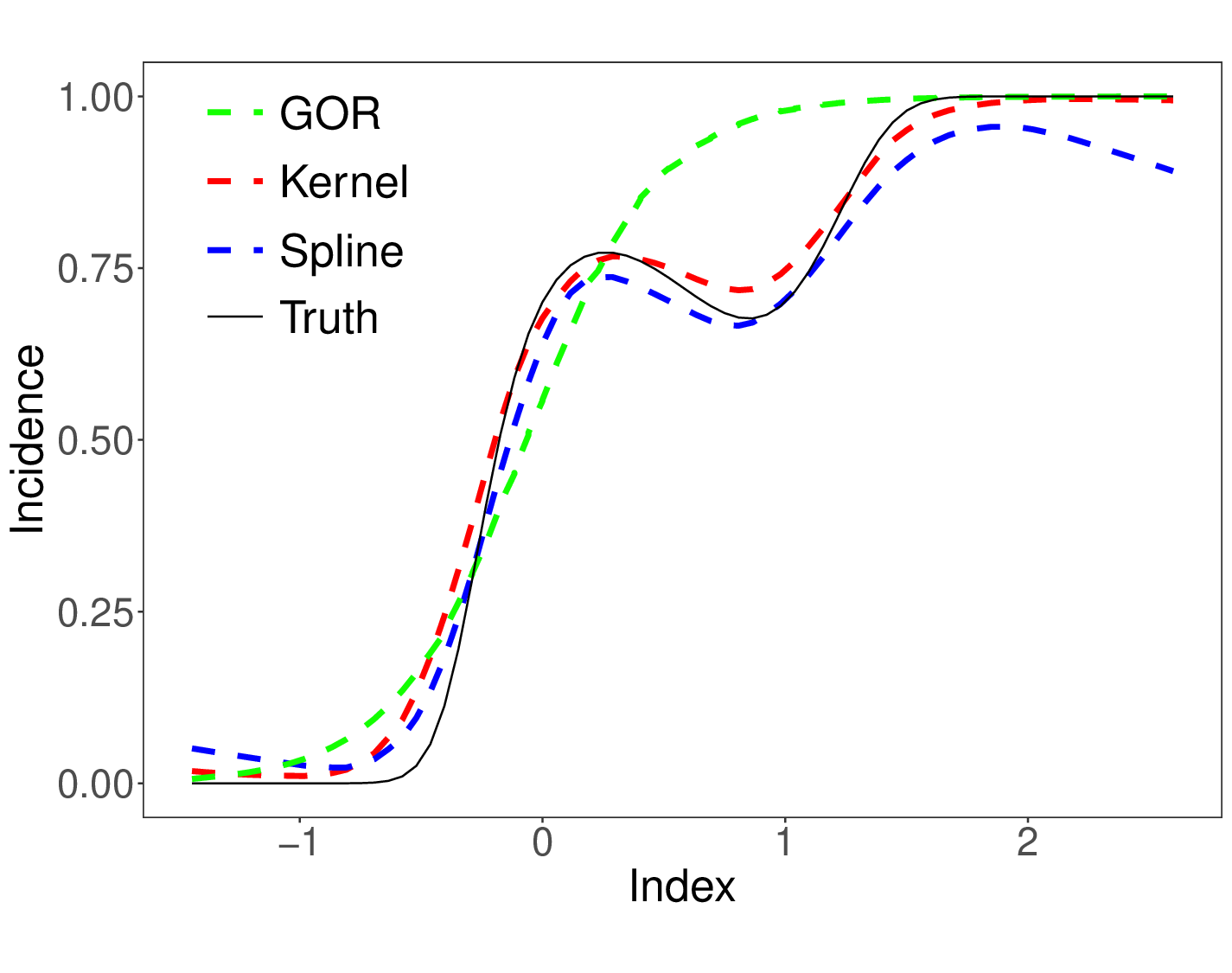}
        \caption{Scenario 3; r=1}
    \end{subfigure}
    \begin{subfigure}[b]{0.32\textwidth}
        \includegraphics[width=\textwidth]{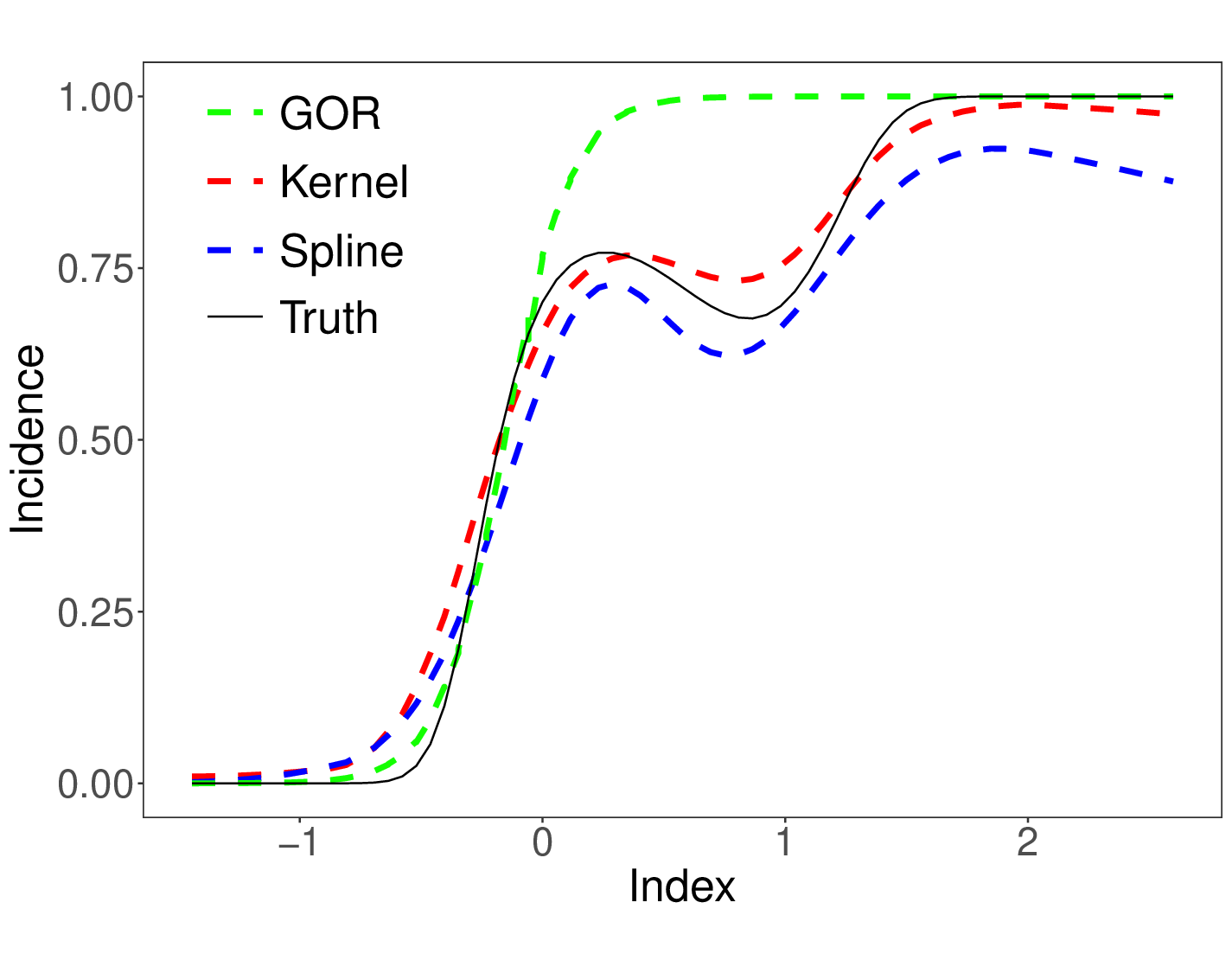}
        \caption{Scenario 3; r=2}
    \end{subfigure}
    \caption{{%Average estimates of incidence link functions with $n=200$. The black curves represent the true values. The red curves show the average estimates using the kernel method. The blue curves show the average estimates using the spline method. The green curves show the logistic estimates. 
    Estimated incidence link functions($n=200$). True values are shown in black, with method-specific estimates displayed as: kernel method (red), spline regression (blue), and logistic modeling (green). }}
    \label{Estimated incidence link functions under n=200.}
\end{figure}

\begin{figure}[H]
    \centering
    \begin{subfigure}[b]{0.32\textwidth}
        \includegraphics[width=\textwidth]{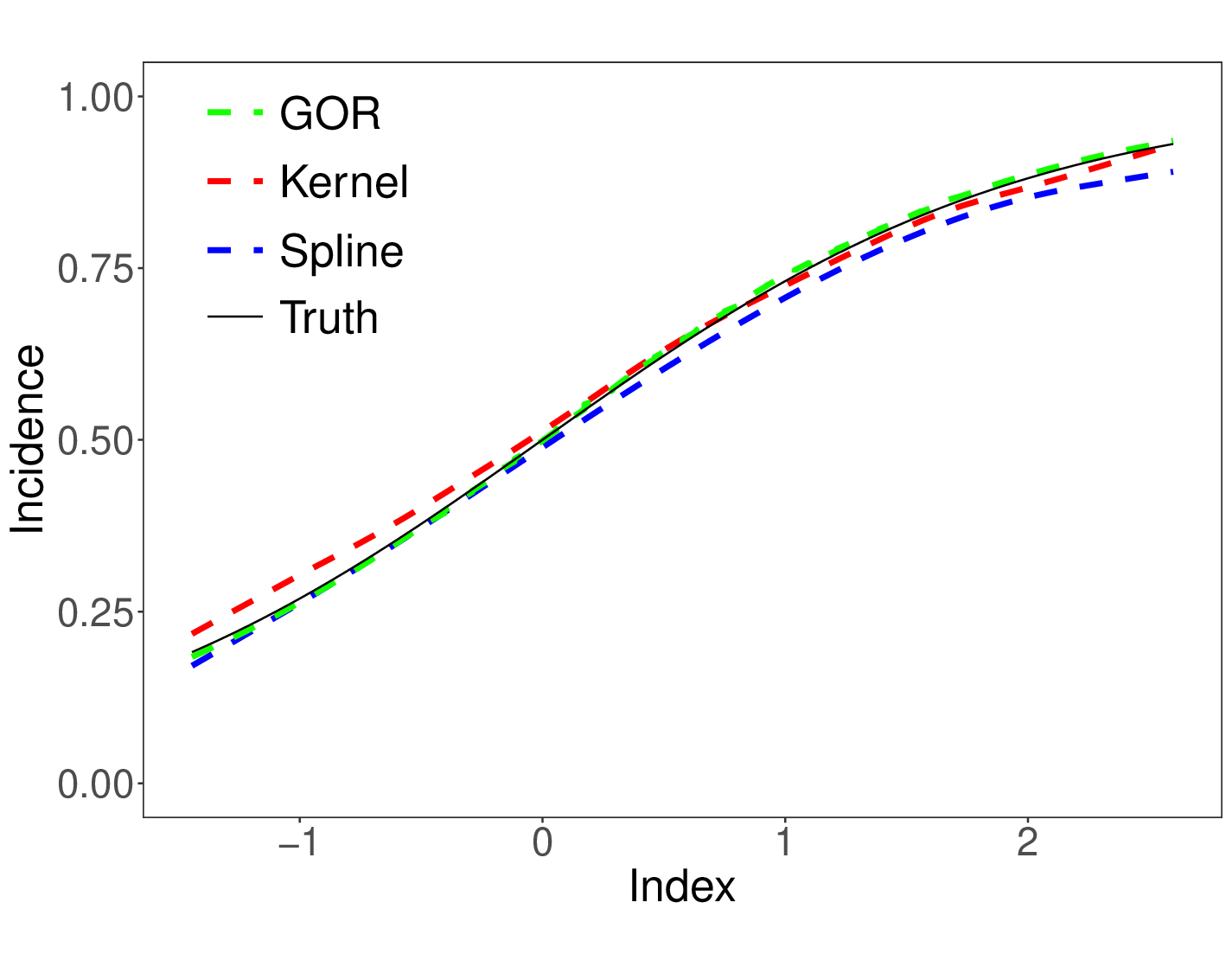}
        \caption{Scenario 1; r=0}
    \end{subfigure}
    \begin{subfigure}[b]{0.32\textwidth}
        \includegraphics[width=\textwidth]{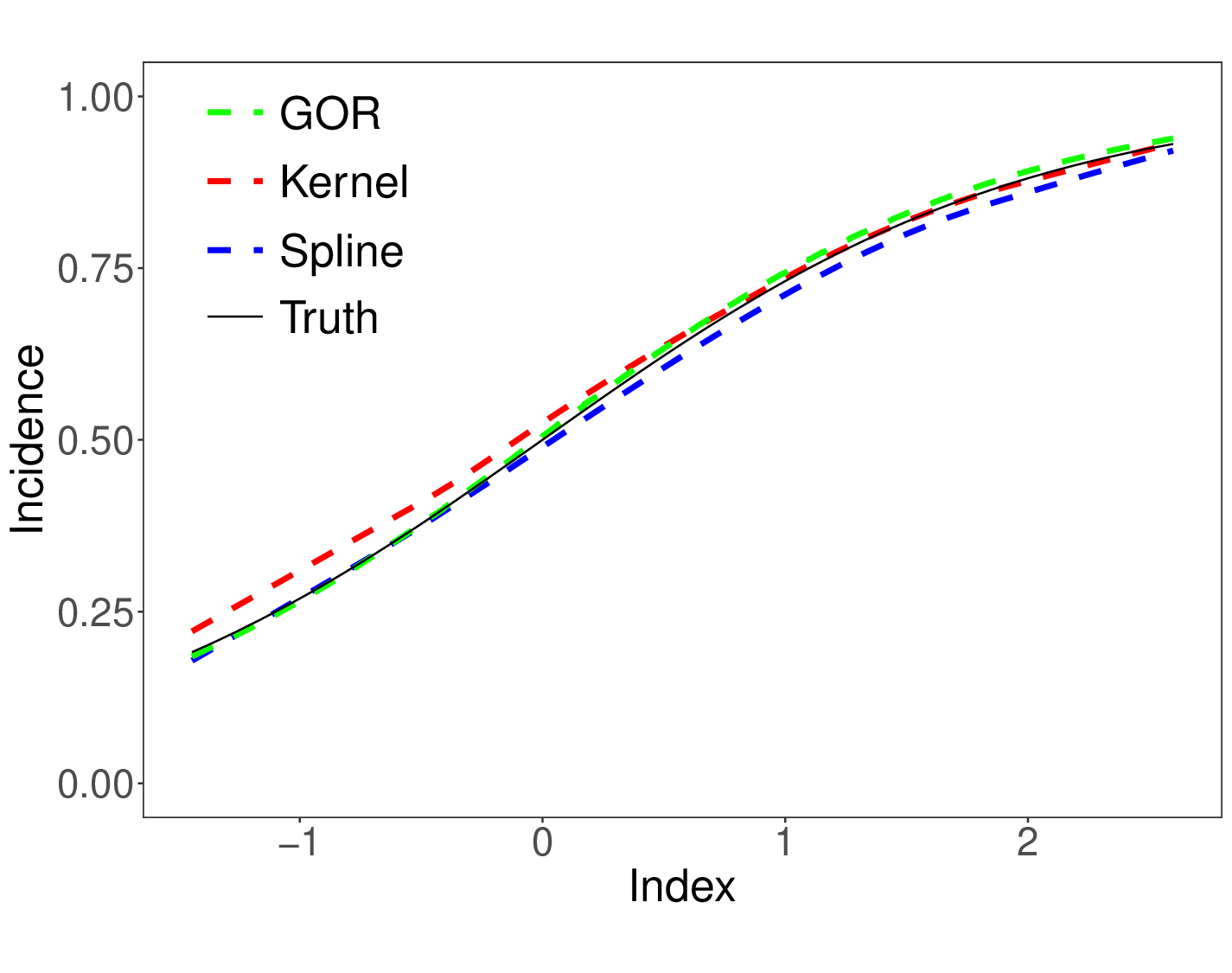}
        \caption{Scenario 1; r=1}
    \end{subfigure}
    \begin{subfigure}[b]{0.32\textwidth}
        \includegraphics[width=\textwidth]{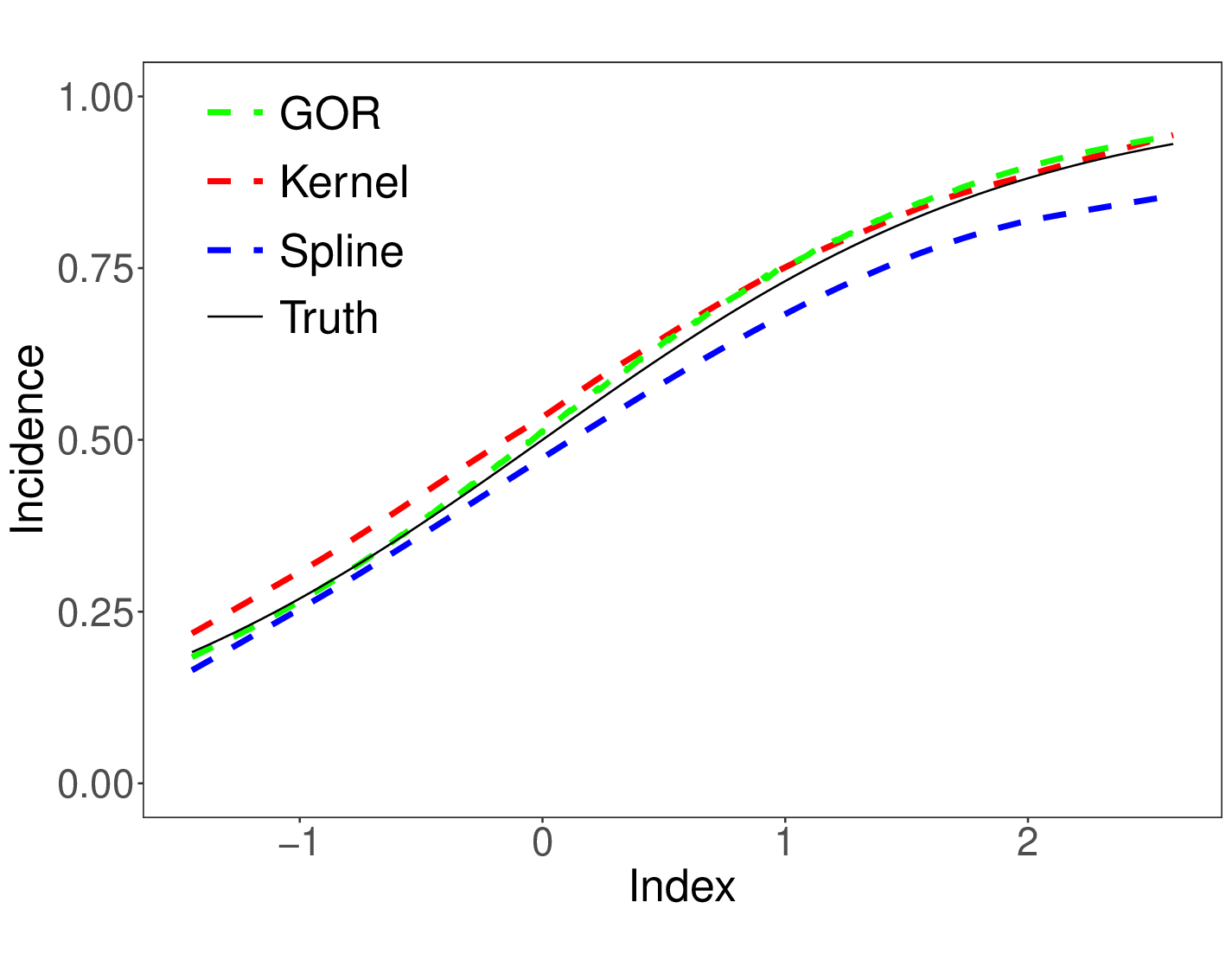}
        \caption{Scenario 1; r=2}
    \end{subfigure}
    \begin{subfigure}[b]{0.32\textwidth}
        \includegraphics[width=\textwidth]{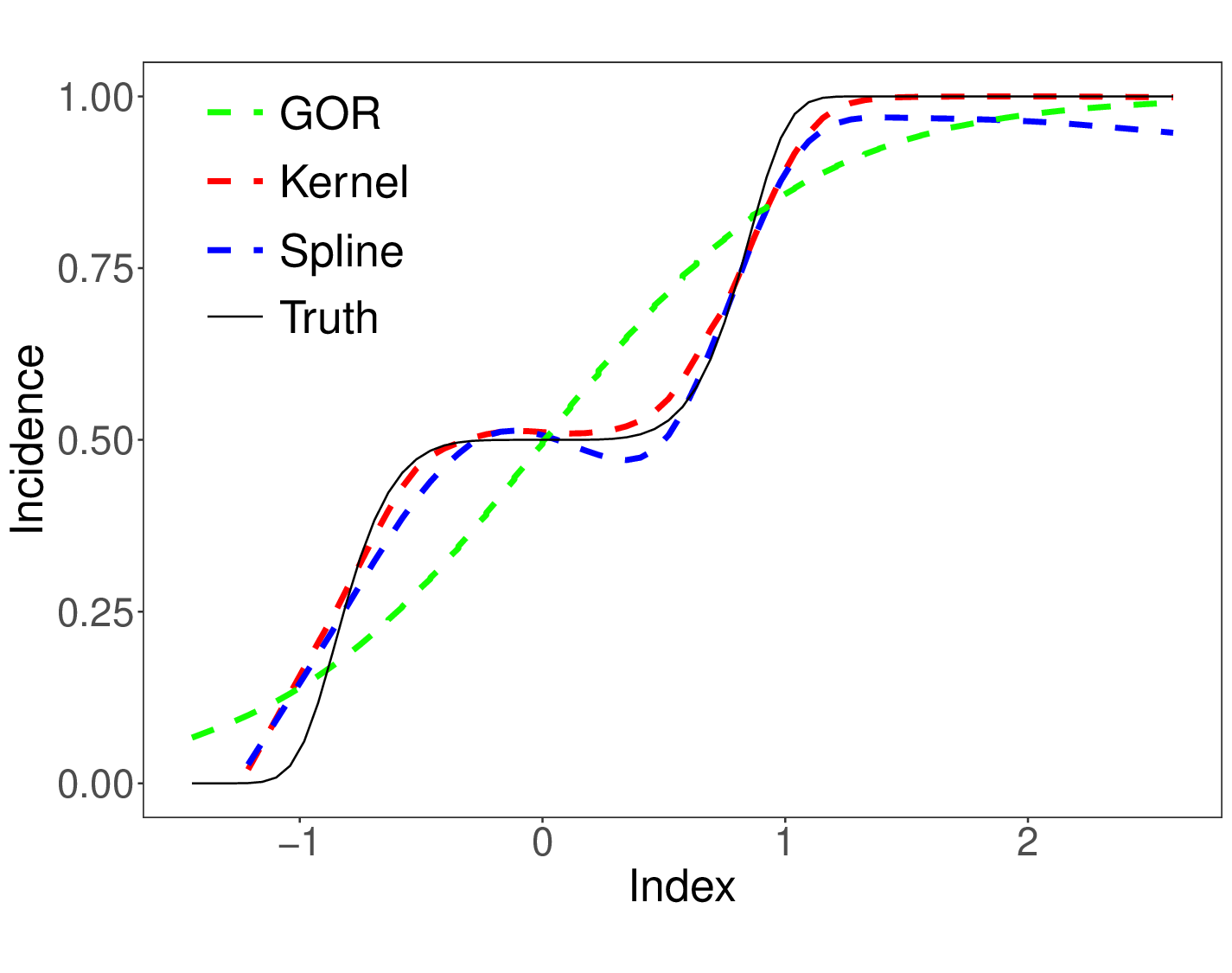}
        \caption{Scenario 2; r=0}
    \end{subfigure}
    \begin{subfigure}[b]{0.32\textwidth}
        \includegraphics[width=\textwidth]{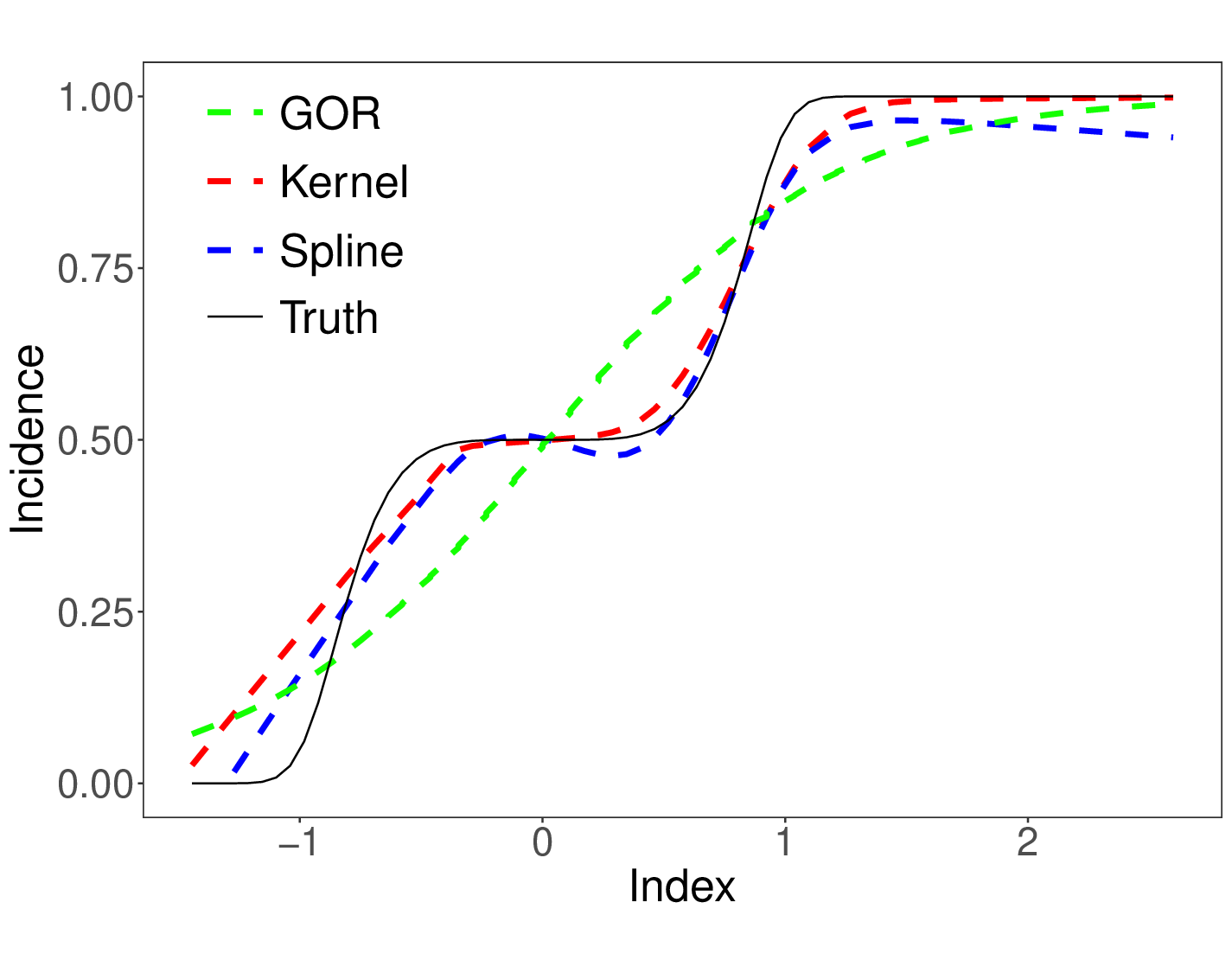}
        \caption{Scenario 2; r=1}
    \end{subfigure}
    \begin{subfigure}[b]{0.32\textwidth}
        \includegraphics[width=\textwidth]{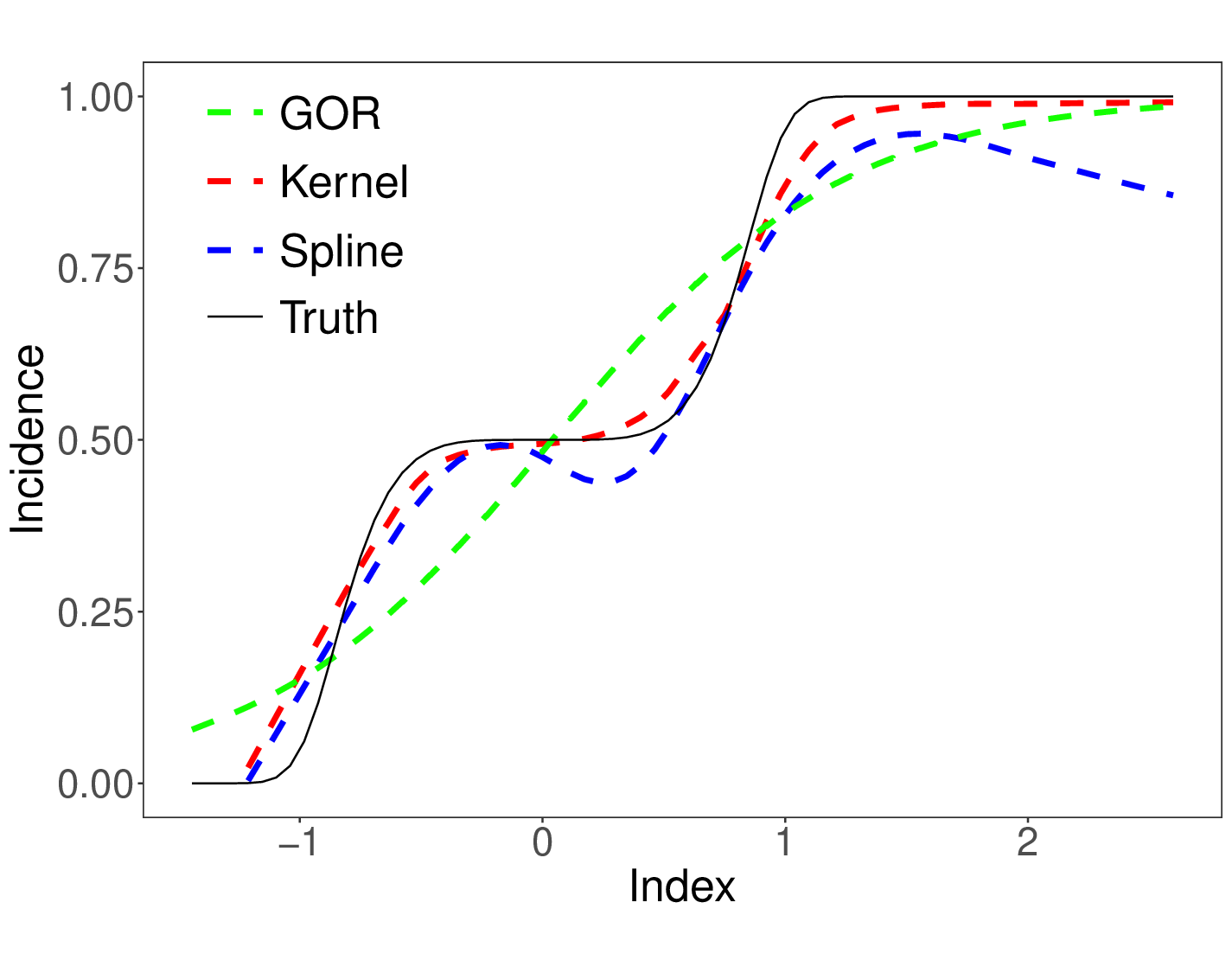}
        \caption{Scenario 2; r=2}
    \end{subfigure}
    \begin{subfigure}[b]{0.32\textwidth}
         \includegraphics[width=\textwidth]{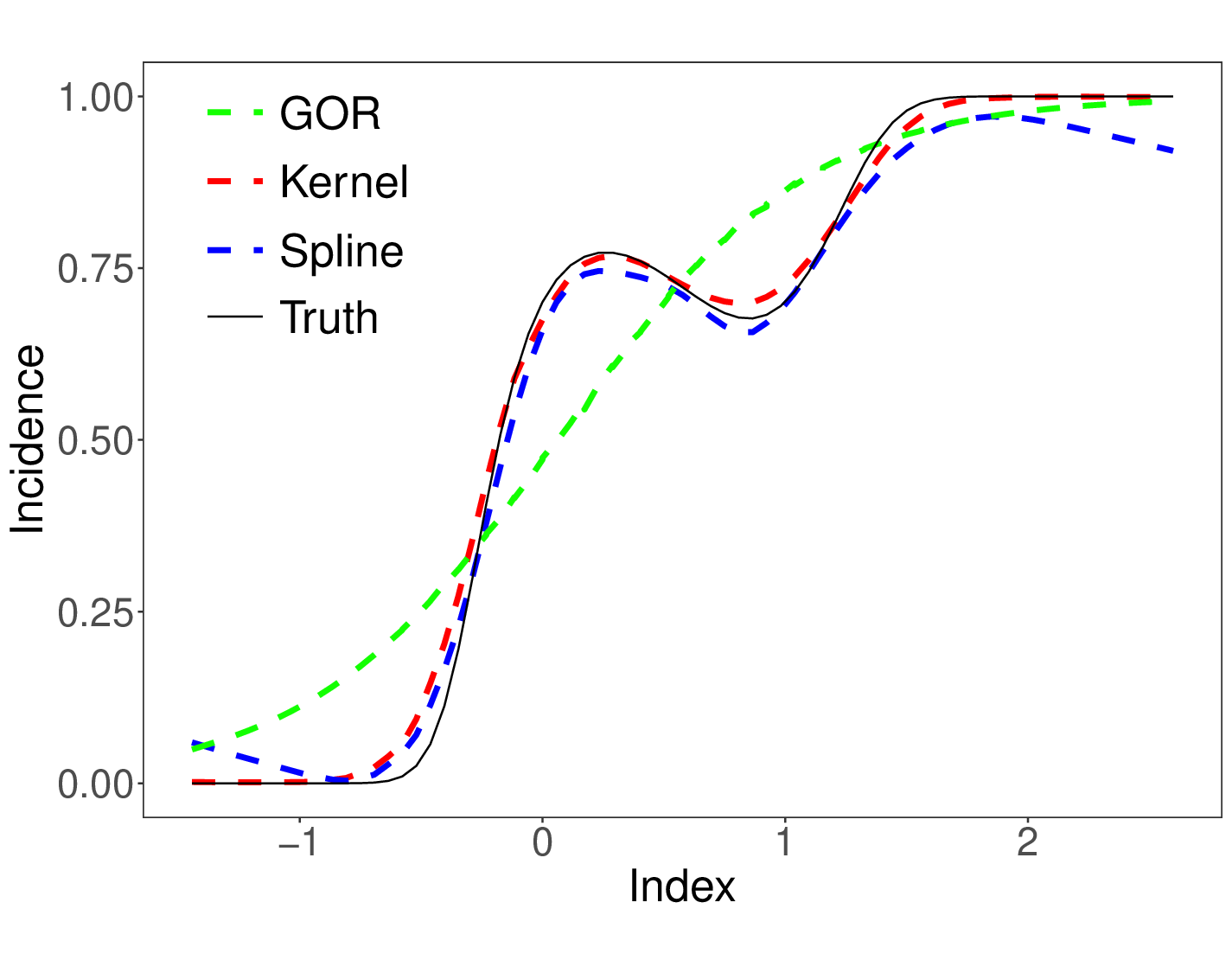}
        \caption{Scenario 3; r=0}
    \end{subfigure}
    \begin{subfigure}[b]{0.32\textwidth}
        \includegraphics[width=\textwidth]{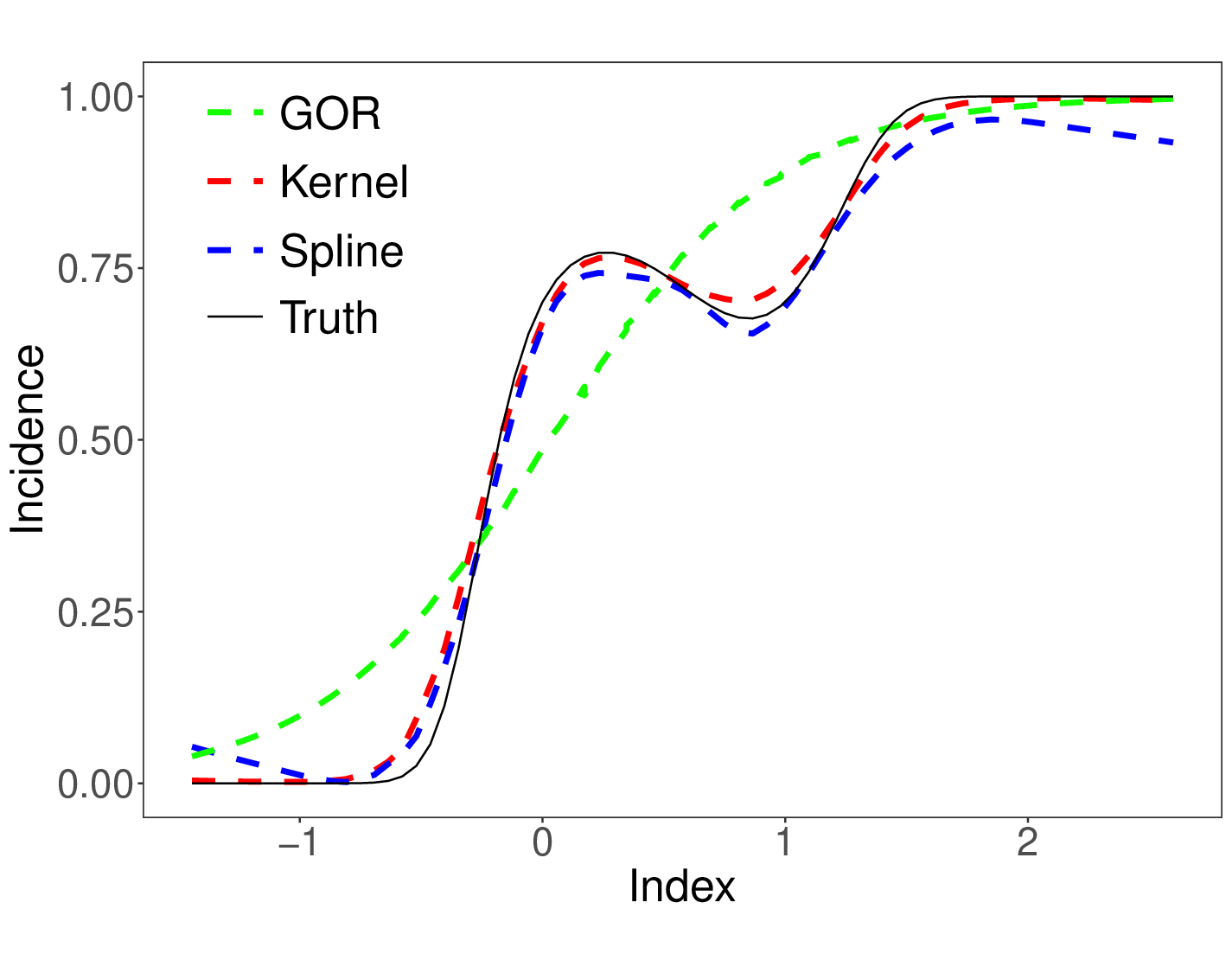}
        \caption{Scenario 3; r=1}
    \end{subfigure}
    \begin{subfigure}[b]{0.32\textwidth}
        \includegraphics[width=\textwidth]{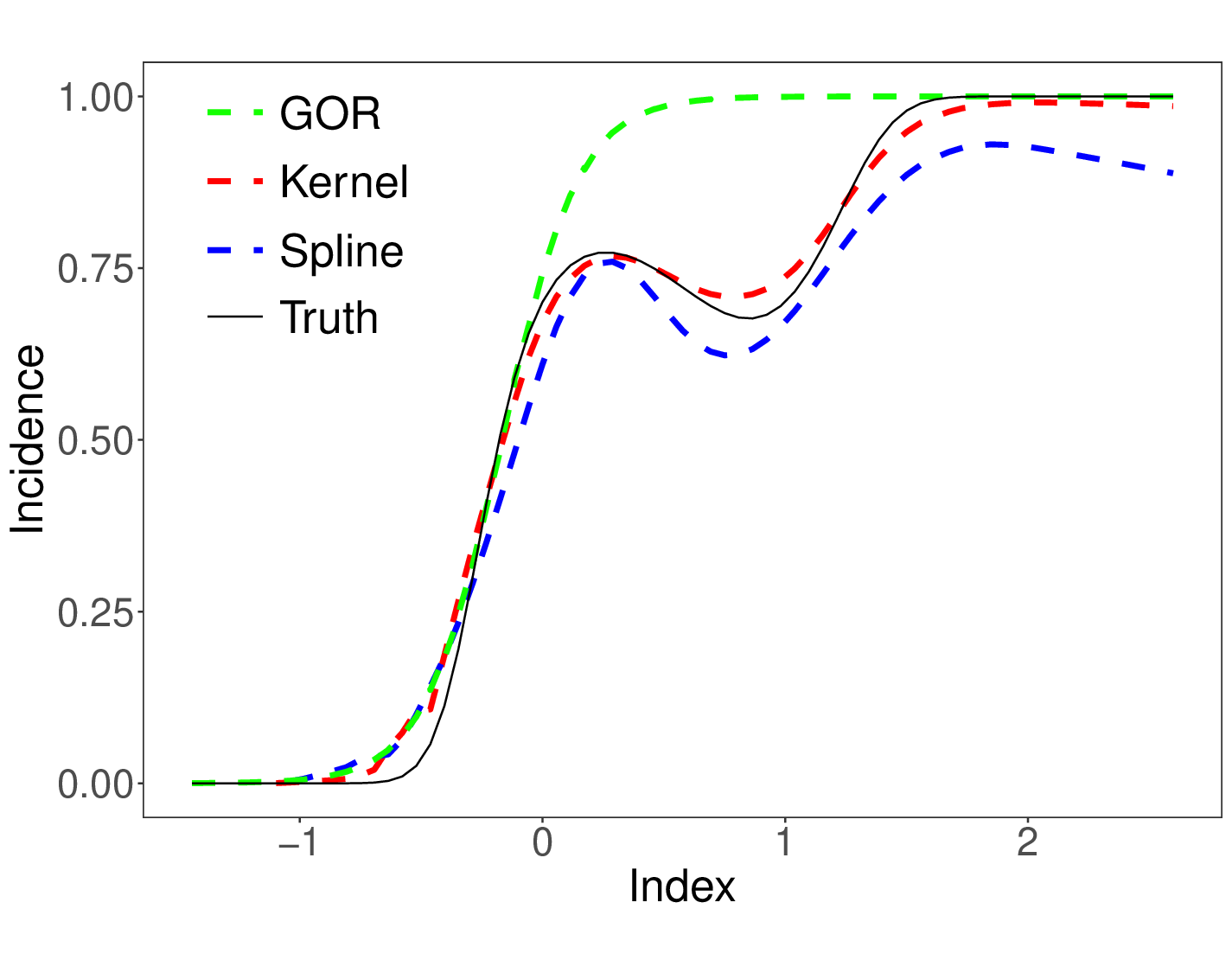}
        \caption{Scenario 3; r=2}
    \end{subfigure}
    \caption{{%Average estimates of incidence link functions. The black curves represent the true values. The red curves show the average estimates using the kernel method. The blue curves show the average estimates using the spline method. The green curves show the logistic estimates for $n = 500$.
     Estimated incidence link functions($n=500$). True values are shown in black, with method-specific estimates displayed as: kernel method (red), spline regression (blue), and logistic modeling (green).} }
    \label{Estimated incidence link functions under n=500.}
\end{figure}

\begin{table}[H]
    \centering
    \captionsetup{skip=0pt}
    \caption{Simulation setting}
    \label{tab:Simulation setting}
    \begin{tabular}{ccccccccccc}
    \hline\hline
    &&& \multicolumn{3}{c}{{$n=200$}} && \multicolumn{3}{c}{{$n=500$}} \\
\cline{5-7} \cline{9-11}
         Scenario& $r$  &$\kappa$ &$\zeta$& $p_{cure}$ & $p_L$ & $p_R$& &$p_{cure}$ & $p_L$ & $p_R$ \\
         \hline
         &	0&	3&	5&0.38&0.13&0.45&&	0.38&	0.12&	0.46\\
         1&	1&	5&	5&0.38&0.13&0.44&&	0.38&	0.13&	0.45\\
         &	2&	6&	8&0.38&0.12&0.46&&	0.38&	0.12&	0.46\\
         
         &	0&	3&	5&0.33&0.14&0.41&&	0.32&	0.14&	0.41\\
         2&	1&	5&	8&0.33&0.14&0.39&&	0.33&	0.14&	0.39\\
         &	2&	6&	8&0.32&0.13&0.42&&	0.32&	0.13&	0.42\\
        
         &	0&	3&	5&0.33&0.13&0.42&&	0.34&	0.13&	0.42\\
         3&	1&	5&	5&0.33&0.14&0.40&&	0.34&	0.14&	0.46\\
         &	2&	6&	8&0.33&0.12&0.42&&	0.34&	0.13&	0.45\\
         \hline
    \end{tabular}
\end{table}

\subsubsection*{A.4.1 Sensitivity Analysis: Spline Orders and Knots}

{To evaluate the impact of spline order and knot specification
on model performance, we have conducted additional simulations under Scenario 3 with 200 sample sizes and $r=1$, varying the spline order (quadratic, cubic, or quartic) and the number of knots (6–8 quantile-based). % to evaluate the sensitivity of model performance to knot specification.
The results, summarized in Tables \ref{tab:spline_order}-\ref{tab:knot} and Figures \ref{fig:ASE_order_knot}-\ref{fig:baseline_order},
demonstrate that the estimation performance for both parametric and nonparametric components remains stable and accurate across all tested specifications.}

\blue{
\begin{table}[htpb!]
\centering
\renewcommand{\arraystretch}{1.3}
%\resizebox{0.65\linewidth}{!}{
\begin{tabular}{cccccccc}
  \hline
  $n=200$ & True & Bias & SSD & SE & CP& \\ 
    \hline
\multicolumn{7}{l}{Order 2 (quadratic)}\\
$\beta_1 $&1 &0.00 &0.32 &0.34 &0.95\\
$\beta_2$  &-1 &0.01 &0.20 &0.21 &0.96\\
$\beta_3$  &1 &0.01 &0.33 &0.38 &0.98\\
\multicolumn{7}{l}{Order 3 (cubic)}\\
$\beta_1 $&1 &0.01 &0.33 &0.34 &0.94\\
$\beta_2$  &-1 &0.00 &0.21 &0.21 &0.96\\
$\beta_3$  &1 &0.01 &0.33 &0.38 &0.98\\
\multicolumn{7}{l}{Order 4 (quartic)}\\
$\beta_1 $&1 &0.01 &0.33 &0.33 &0.94\\
$\beta_2$  &-1 &-0.01 &0.21 &0.21 &0.94\\
$\beta_3$  &1 &0.02 &0.33 &0.38 &0.97\\
\hline
\end{tabular}
%}
\caption{
Simulation results for Scenario 3  with 
$n=200$ and $r=1$ using quadratic, cubic, or quartic I-splines. % with 5 quantile-based knots.
Bias: the estimated bias; ESD:
empirical standard deviation; ESE: empirical standard error estimate; CP: the 95\% empirical
coverage probabilities.}
\label{tab:spline_order}
\end{table}
}

\blue{
\begin{table}[htpb!]
\centering
\renewcommand{\arraystretch}{1.3}
%\resizebox{0.65\linewidth}{!}{
\begin{tabular}{cccccccc}
  \hline
  $n=200$ & True & Bias & SSD & SE & CP& \\ 
    \hline
\multicolumn{7}{l}{6 quantile-based knots}\\
$\beta_1 $& 1 &0.08 &0.30 &0.33 &0.96\\
$\beta_2$  &-1 &0.00 &0.21 &0.21 &0.93\\
$\beta_3$  &1&0.00 &0.36 &0.38 &0.97\\
\multicolumn{7}{l}{7 quantile-based knots}\\
$\beta_1 $&1 &0.08 &0.30 &0.33 &0.96\\
$\beta_2$  &-1 &-0.01 &0.21 &0.21 &0.93\\
$\beta_3$  &1 &0.00 &0.36 &0.38 &0.96\\
\multicolumn{7}{l}{8 quantile-based knots}\\
$\beta_1 $& 1 &0.06 &0.32 &0.33 &0.96\\
$\beta_2$  &-1 &0.01 &0.24 &0.38 &0.95\\
$\beta_3$  &1 &-0.01 &0.38 &0.38 &0.95\\
\hline
\end{tabular}
%}
\caption{
Simulation results of Scenario 3 with $n = 200$ and $r=1$ using 6-8 quantile-based knots. % based on 200 replicates. 
Bias: the estimated bias; ESD:
empirical standard deviation; ESE: empirical standard error estimate; CP: the 95\% empirical
coverage probabilities.}
\label{tab:knot}
\end{table}
}

\begin{figure}[H]
    \centering
    \begin{subfigure}[b]{0.47\textwidth}
        \includegraphics[width=0.7\textwidth, angle=-90]{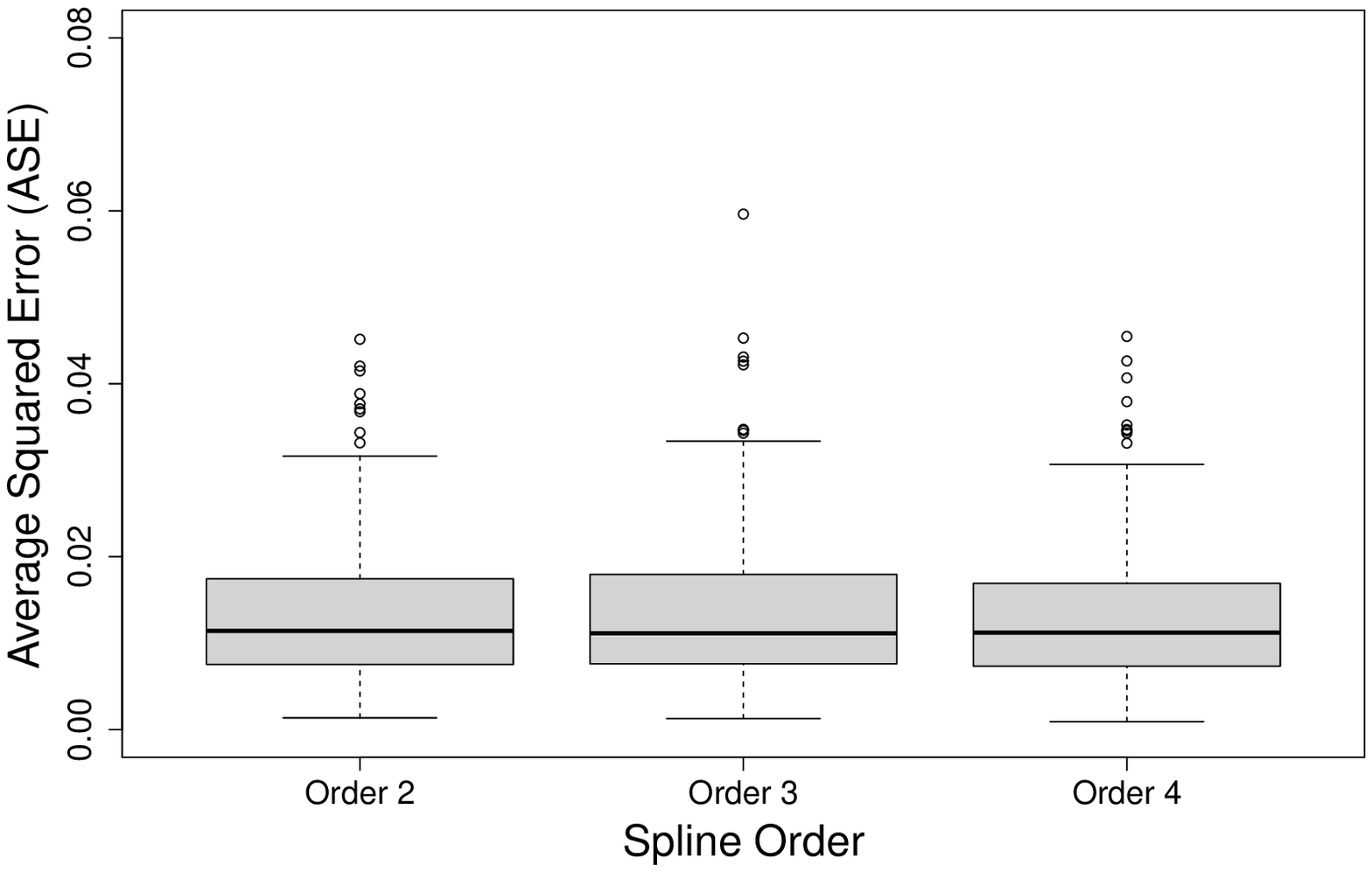}
        \caption{ASE by Spline Order}
    \end{subfigure}
    \begin{subfigure}[b]{0.47\textwidth}
        \includegraphics[width=0.7\textwidth, angle=-90]{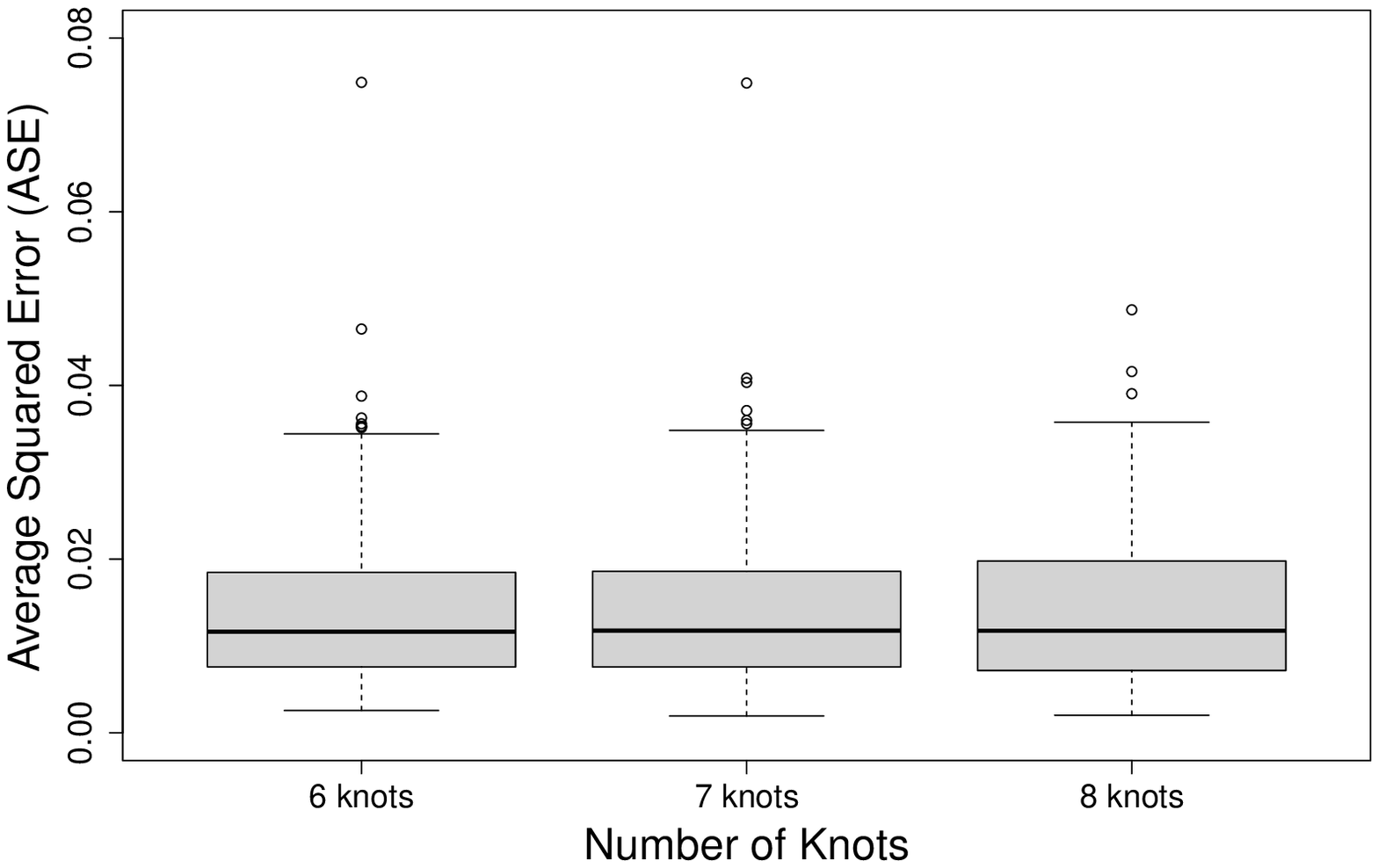}
        \caption{ASE by Knot Number}
    \end{subfigure}
    \caption{Boxplots of %the average squared error (ASE) 
    ASE for the incidence link function in Scenario 3 with $n=200$ and $r=1$, employing quadratic, cubic, or quartic I-splines (left panel) and 6-8 quantile-based knots (right panel).% (a) ASE distribution across spline orders 2, 3, and 4. (b) ASE distribution across knot numbers 6, 7, 8, and 9. The results are based on 200 replications under the setting $n=200$, Scenario 3, $r=1$.
    }
    \label{fig:ASE_order_knot}
\end{figure}

% Additionally, we examined the sensitivity of the estimated baseline survival function $S_0(t)$ to these tuning parameters. Figure~\ref{fig:baseline_order} displays the estimated baseline survival curves obtained under different spline orders (with 7 knots) and knot numbers (with cubic splines). The curves exhibit excellent agreement with the true baseline survival function.
\begin{figure}[H]
    \centering
    \begin{subfigure}[b]{0.32\textwidth}
        \includegraphics[width=\textwidth]{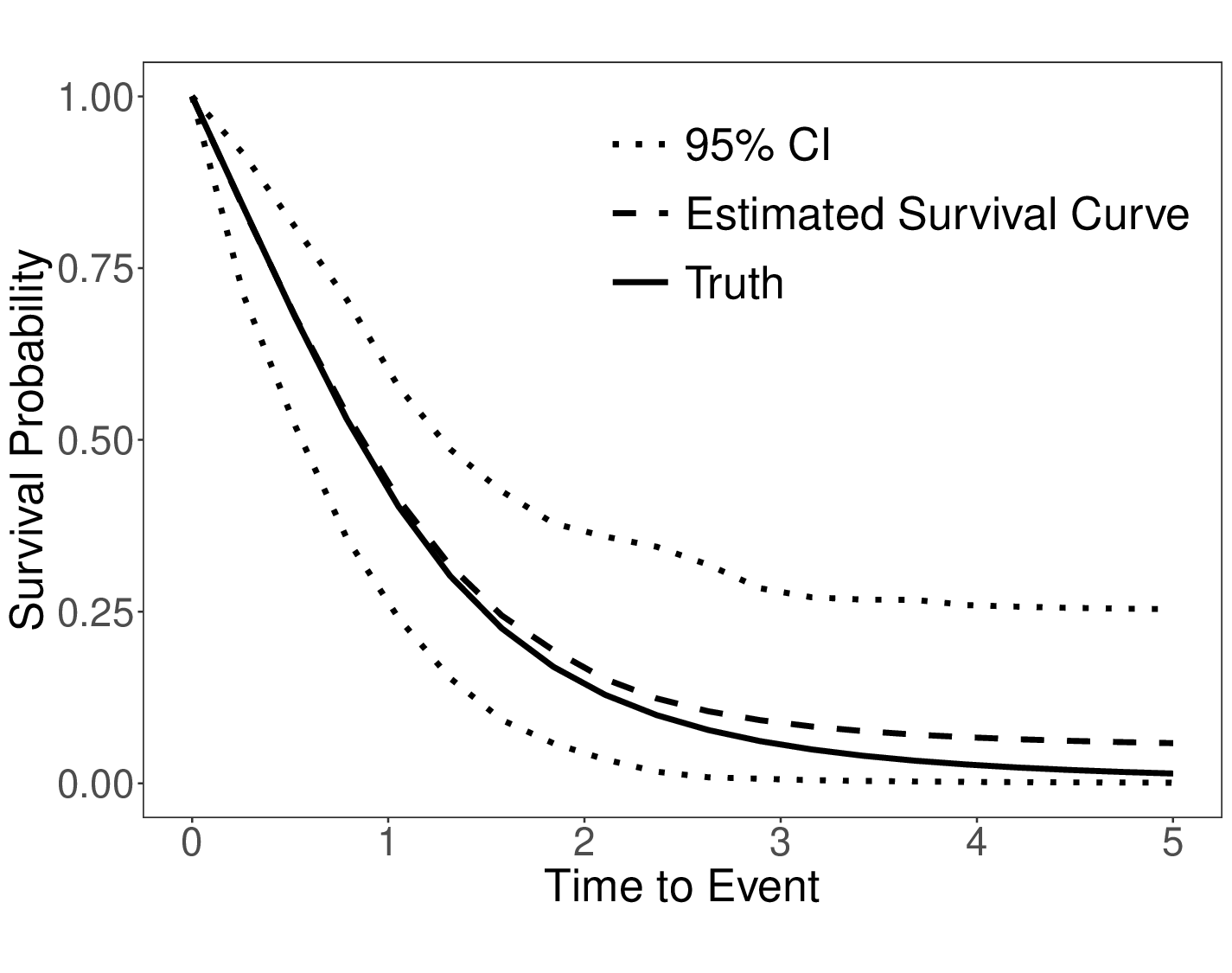}
        \caption{Order 2 (quadratic)}
    \end{subfigure}
    \begin{subfigure}[b]{0.32\textwidth}
        \includegraphics[width=\textwidth]{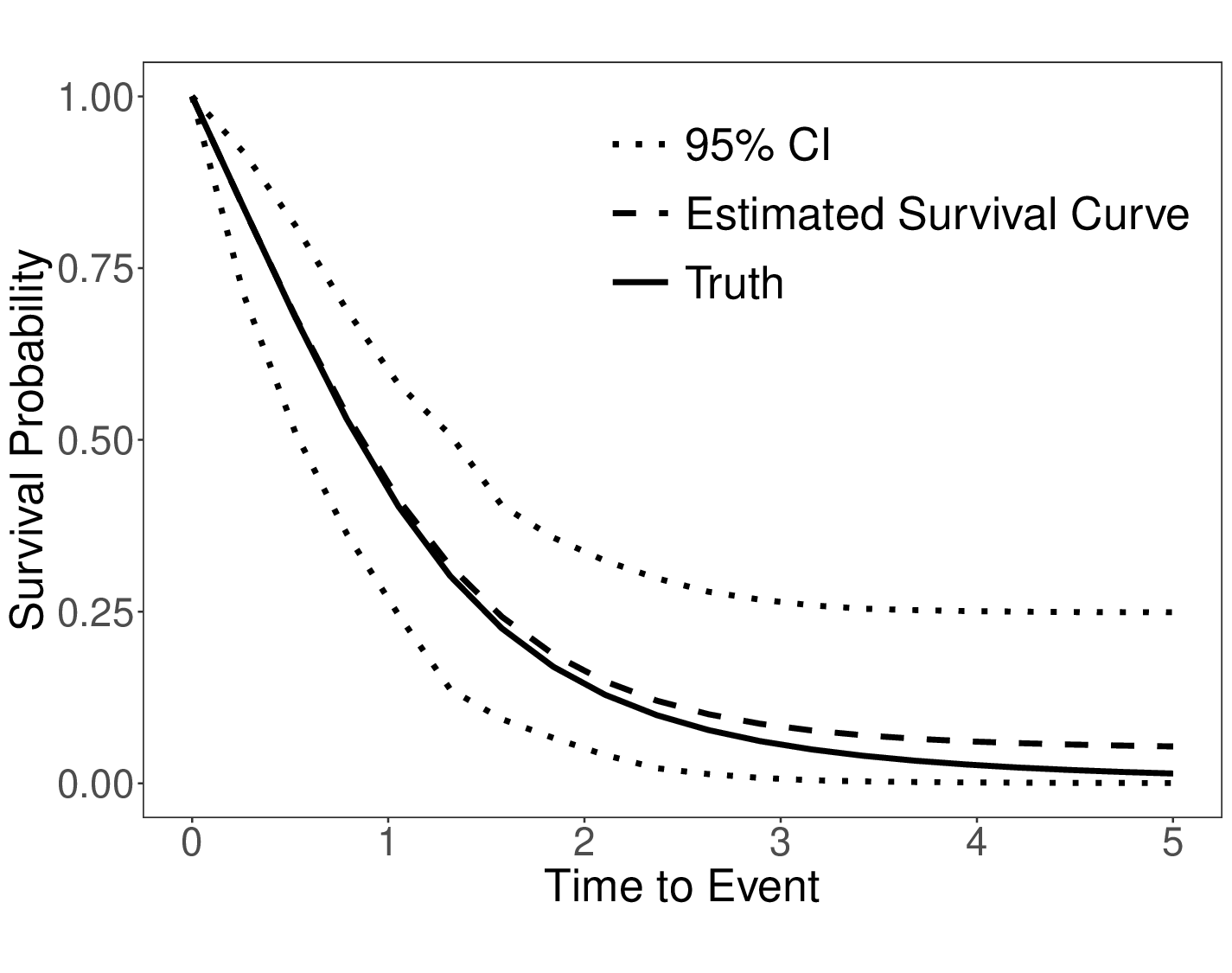}
        \caption{Order 3 (cubic)}
    \end{subfigure}
    \begin{subfigure}[b]{0.32\textwidth}
        \includegraphics[width=\textwidth]{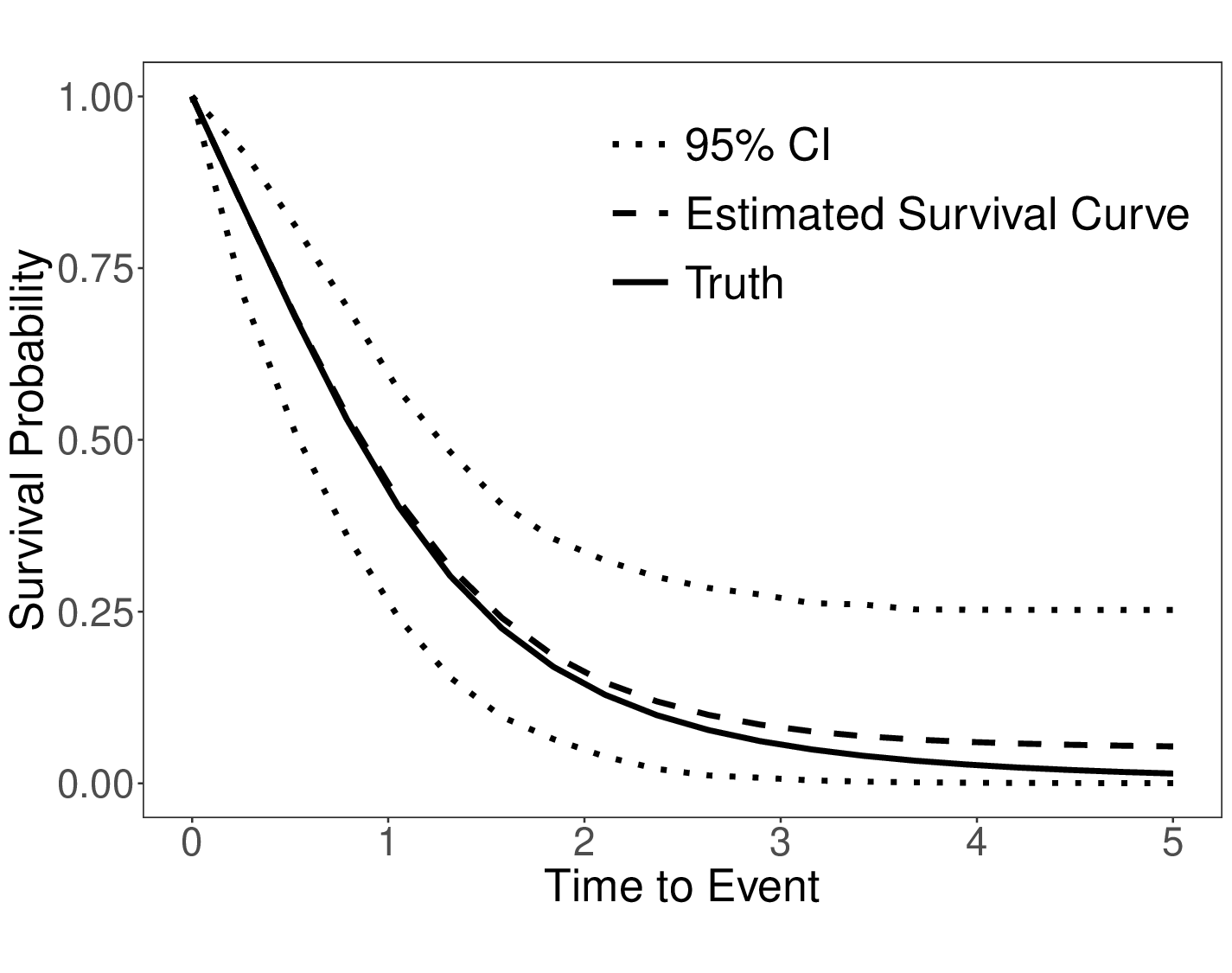}
        \caption{Order 4 (quartic)}
    \end{subfigure}
    \begin{subfigure}[b]{0.32\textwidth}
        \includegraphics[width=\textwidth]{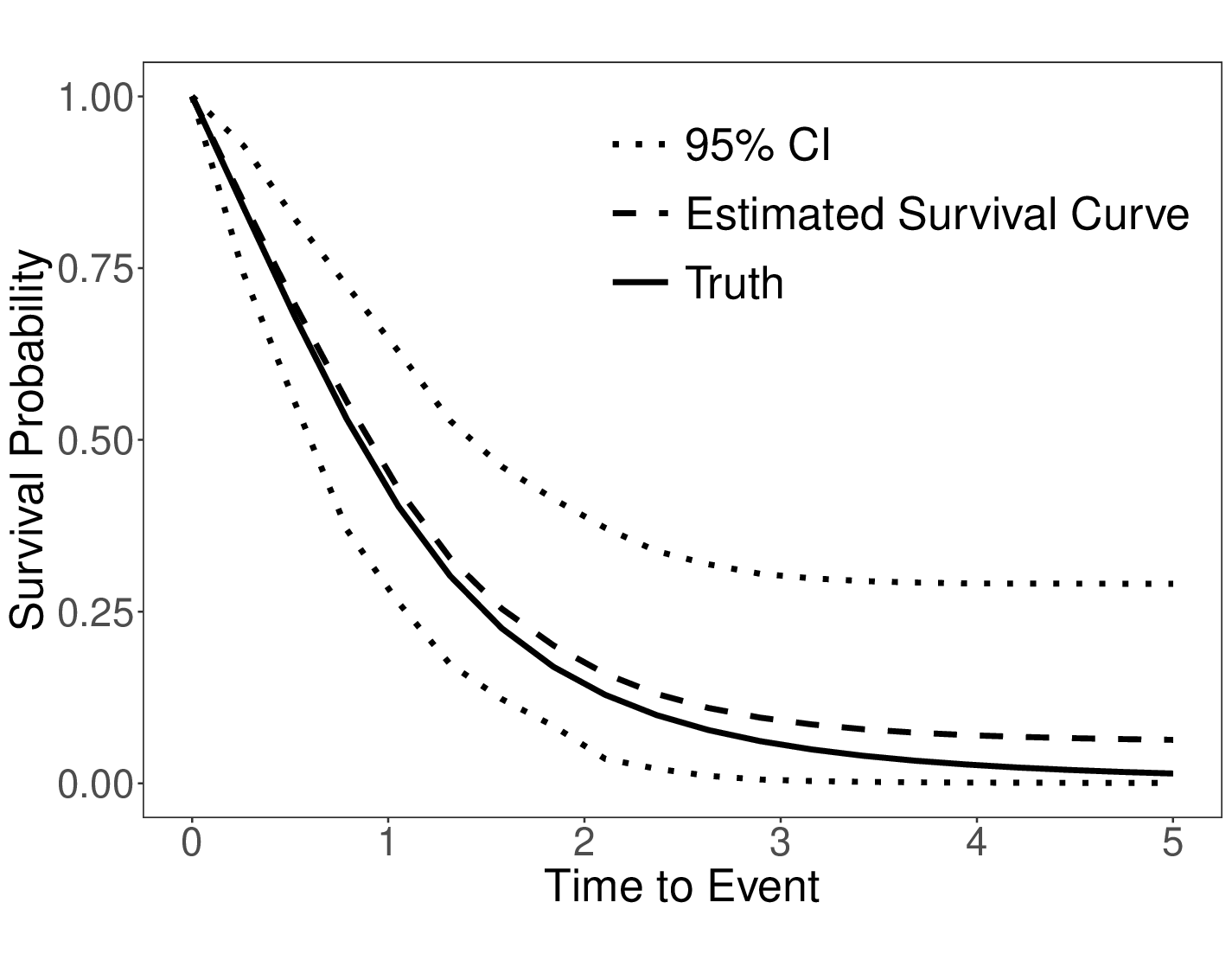}
        \caption{6 knots}
    \end{subfigure}
    \begin{subfigure}[b]{0.32\textwidth}
        \includegraphics[width=\textwidth]{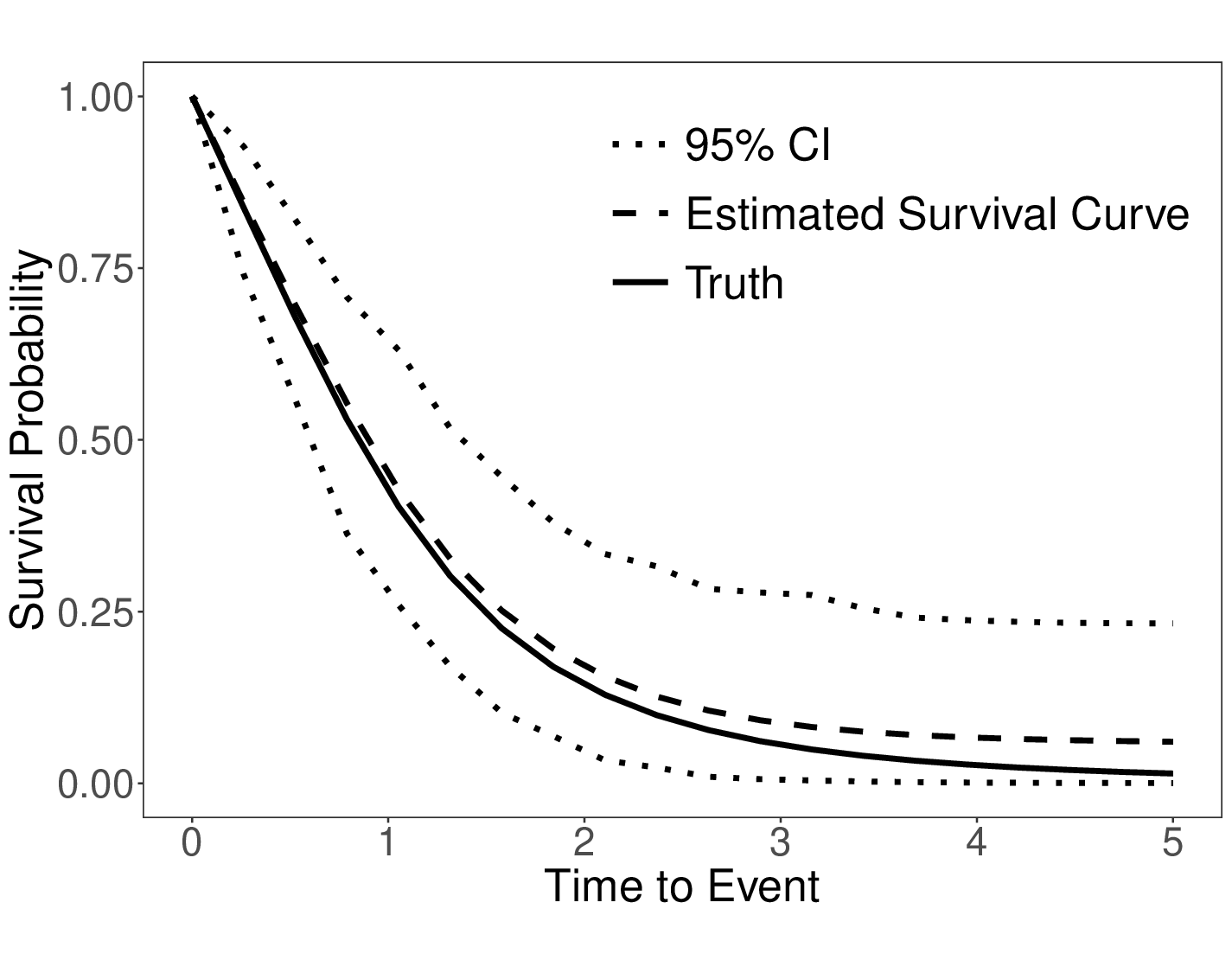}
        \caption{7 knots}
    \end{subfigure}
    \begin{subfigure}[b]{0.32\textwidth}
        \includegraphics[width=\textwidth]{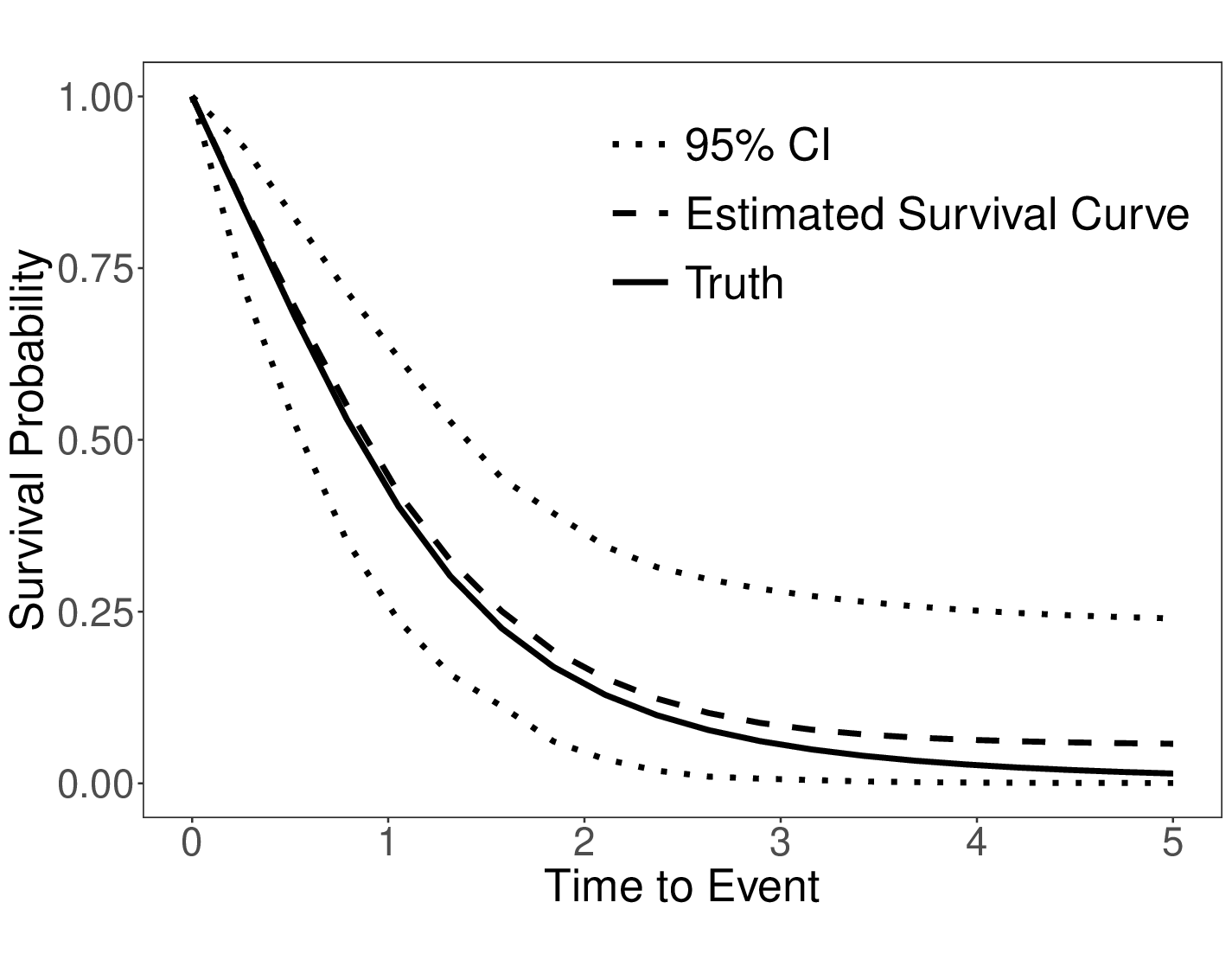}
        \caption{8 knots}
    \end{subfigure}
    \caption{%Estimated baseline survival functions using different spline orders (with 7 knots) and different numbers of knots (with cubic splines). (a) Quadratic splines (order 2); (b) Cubic splines (order 3); (c) Quartic splines (order 4); (d) 6 knots; (e) 7 knots; (f) 8 knots. The results are based on 200 replications under the setting $n=200$, Scenario 3, $r=1$.
    Estimated baseline survival functions of Scenario 3 with $n=200$ and $r=1$, employing quadratic, cubic, or quartic I-splines (upper penal) and 6-8 quantile-based knots (lower panel).
    }
    \label{fig:baseline_order}
\end{figure}

\subsubsection*{A.4.2 Computational Efficiency and Scalability}

{To empirically assess scalability, we conducted additional simulations with varying covariate dimensions for the incidence ($d_1$) and latency ($d_2$) components: $(d_1,d_2)=(5,5), (10,10)$. 
As summarized in Table \ref{tab:com_scala} and Figures \ref{fig:ASE_pq}-\ref{SC_pq}, the results indicate that 1) the running time increases approximately linearly with the total number of covariates. This suggests that the EM algorithm remains efficient for moderate-dimensional settings; and 2)  the estimation performance for both parametric and nonparametric components remains stable across dimensions, demonstrating numerical robustness. }

\begin{table}[htpb!]
\centering
\renewcommand{\arraystretch}{1.3}
%\resizebox{0.65\linewidth}{!}{
\begin{tabular}{cccccccc}
  \hline
 ($n,d_1,d_2$) & 
& Average Time(second) & Avg. Bias & MSE &\\ 
    \hline
%\multicolumn{7}{l}{Covariate Dimension}\\
SMCI-K\\
(200,5,5)&&64.44 &0.01 &0.57\\
(200,10,10)&&88.18 &-0.04 &0.53\\
%(20,20)&&146.92 &-0.09 &0.34\\
SMCI-S\\
(200,5,5)&&14.20 &0.01 &0.56\\
(200,10,10)&&23.60 &-0.04 &0.52\\
%(20,20)&&66.30 &-0.09 &0.34\\
% &$\beta_1 $& & & & &\\
% (5,5)&$\beta_2$& & & \\
%  &$\beta_3$ & & & & &\\
%  \hline
% %\multicolumn{7}{l}{Covariate Dimension}\\
% &$\beta_1 $& & & & &\\
% (10,10)&$\beta_2$& & & &\\
%  &$\beta_3$ & & & & &\\
%   \hline
% %\multicolumn{7}{l}{Order 4 (quartic)}\\
% &$\beta_1 $& & & & &\\
% (20,20)&$\beta_2$& & & & &\\
%  &$\beta_3$ & & & & &\\
\hline
\end{tabular}
%}
\caption{
%Simulation results for Scenario 3  with $n=200$ and $r=1$ with different dimensions of covariates.Bias: the estimated bias; ESD:empirical standard deviation; ESE: empirical standard error estimate; CP: the 95\% empirical coverage probabilities.
Computational time and estimation accuracy under varying covariate dimension $d_1$ (incidence) and $d_2$ (latency). % Average Time: time per iteration; Avg. Bias = $d_2\sum_{j=1}^{d_2}(\hat{\beta}_j-{\beta}_j)$; MSE = 
}
\label{tab:com_scala}
\end{table}

\begin{figure}[H]
    \centering
    \begin{subfigure}[b]{0.47\textwidth}
        \includegraphics[width=0.7\textwidth, angle=-90]{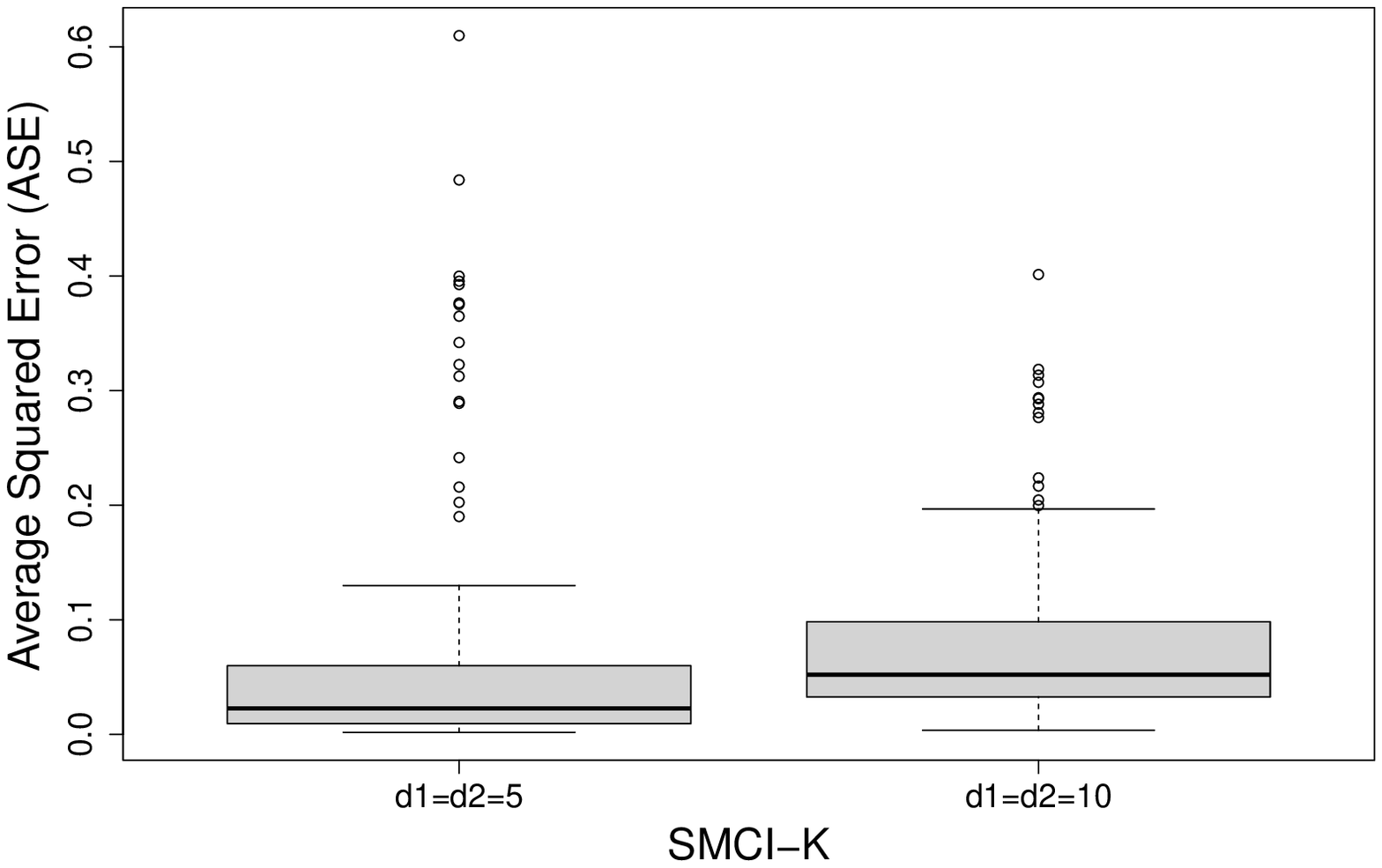}
        \caption{ASE by SMCI-K}
    \end{subfigure}
    \begin{subfigure}[b]{0.47\textwidth}
        \includegraphics[width=0.7\textwidth, angle=-90]{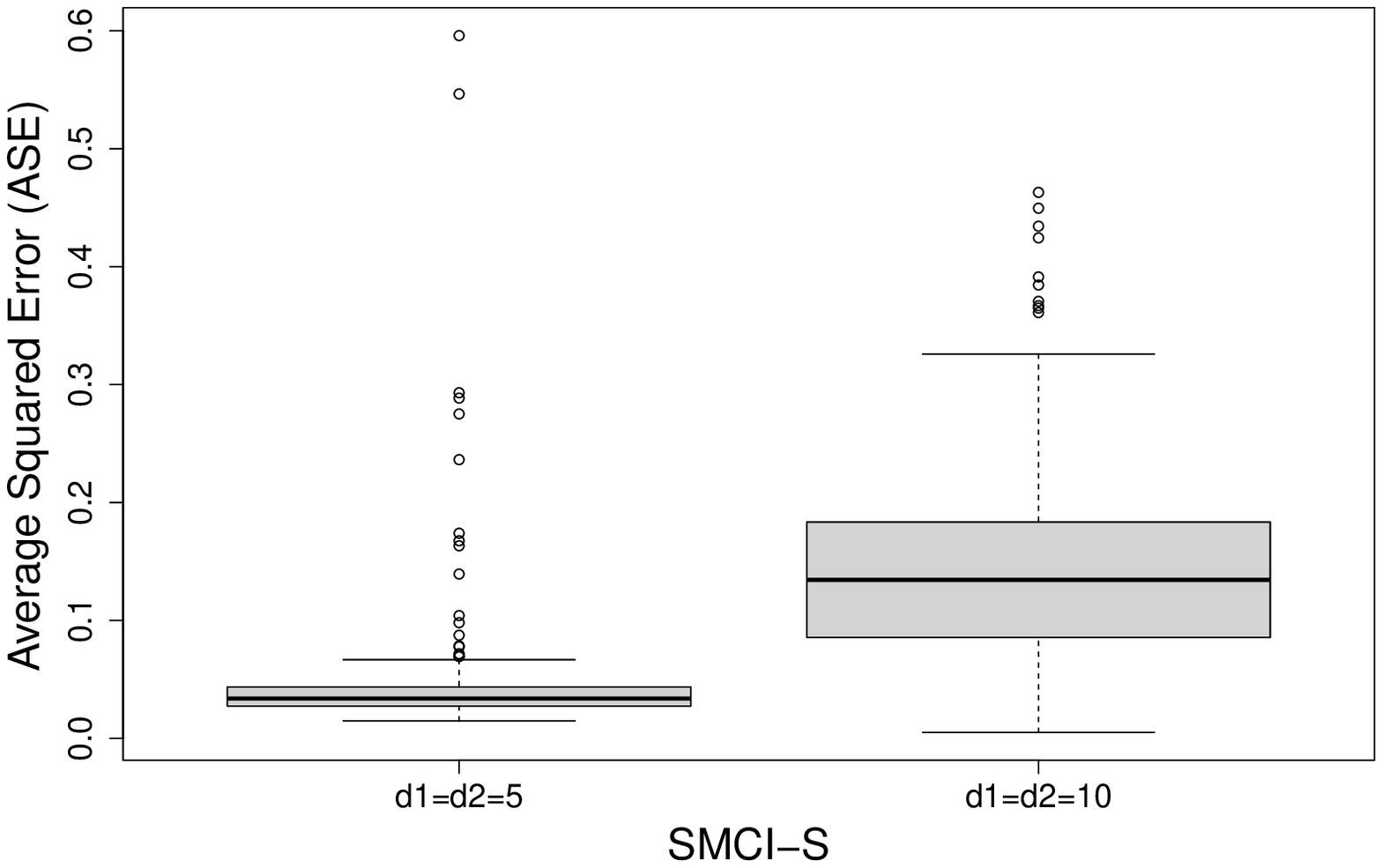}
        \caption{ASE by SMCI-S}
    \end{subfigure}
    \caption{Boxplots of ASE for the incidence link function in Scenario 3 with $n=500$ and $r=1$ under varying covariate dimension $d_1$ (incidence) and $d_2$ (latency).
    }
    \label{fig:ASE_pq}
\end{figure}

\begin{figure}[H]
    \centering
        \begin{subfigure}[b]{0.4\textwidth}
        \includegraphics[width=\textwidth]{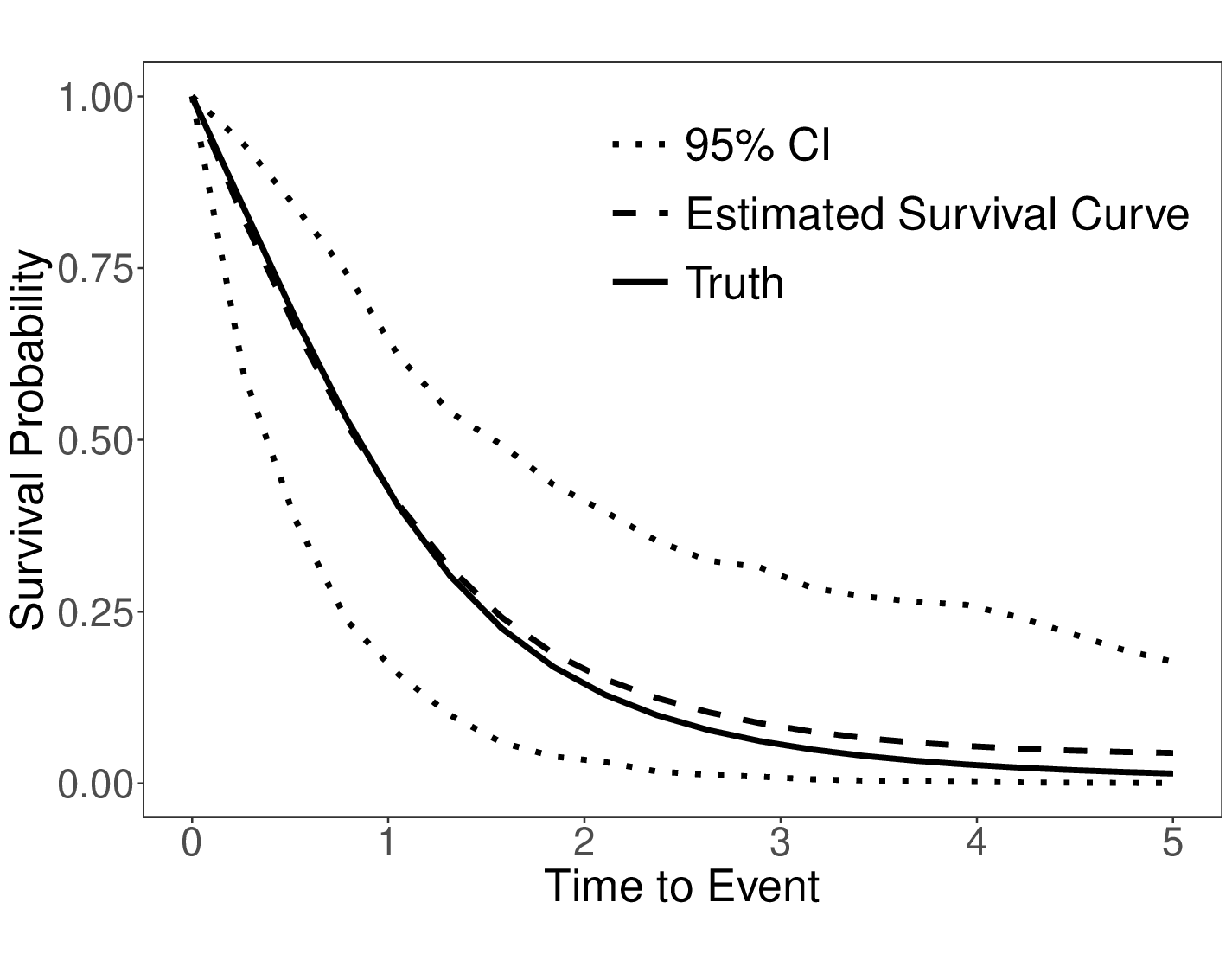}
        \caption{$d_1=d_2=5$, SMCI-K}
    \end{subfigure}
    \begin{subfigure}[b]{0.4\textwidth}
        \includegraphics[width=\textwidth]{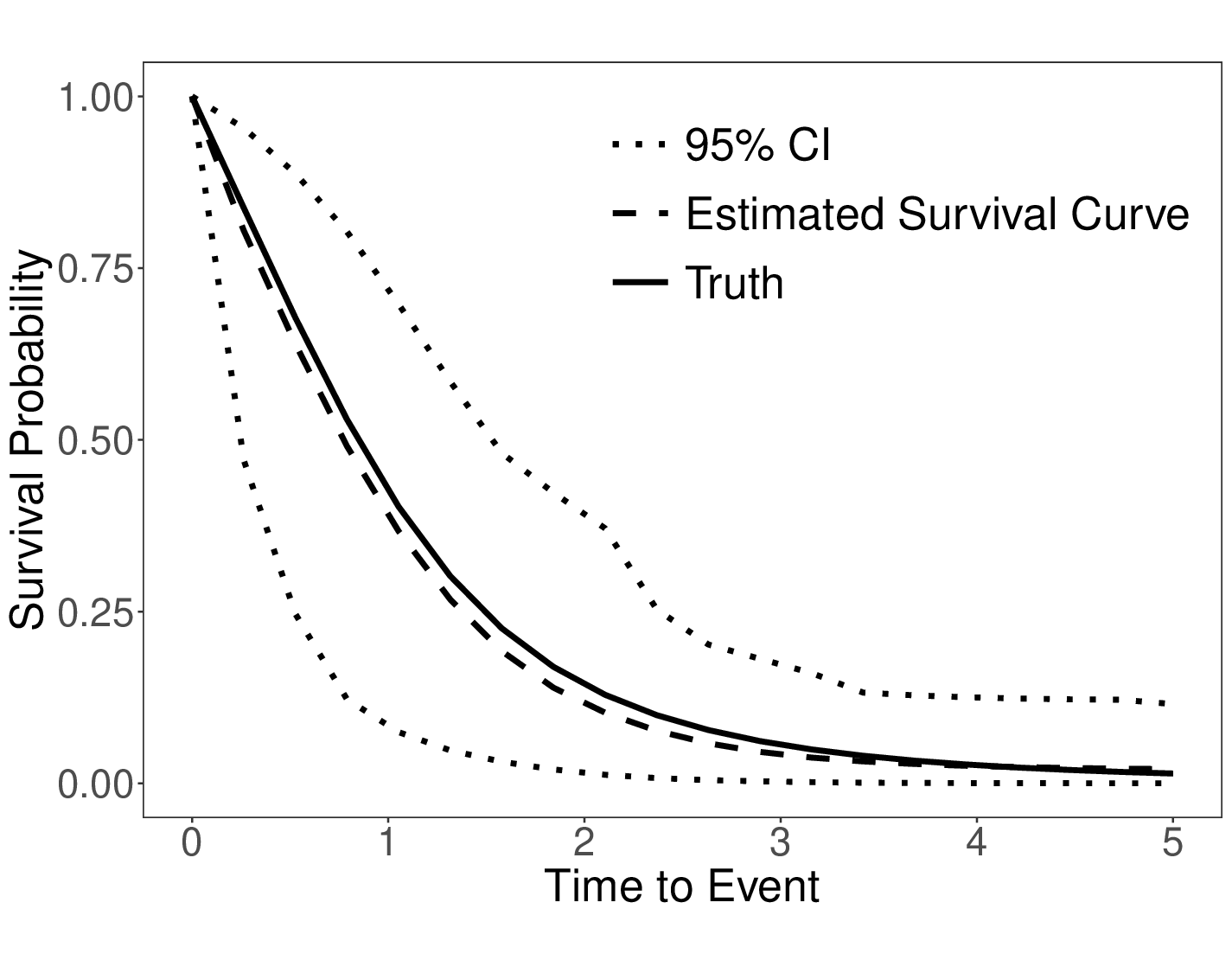}
        \caption{$d_1=d_2=5$, SMCI-S}
    \end{subfigure}
    \begin{subfigure}[b]{0.4\textwidth}
        \includegraphics[width=\textwidth]{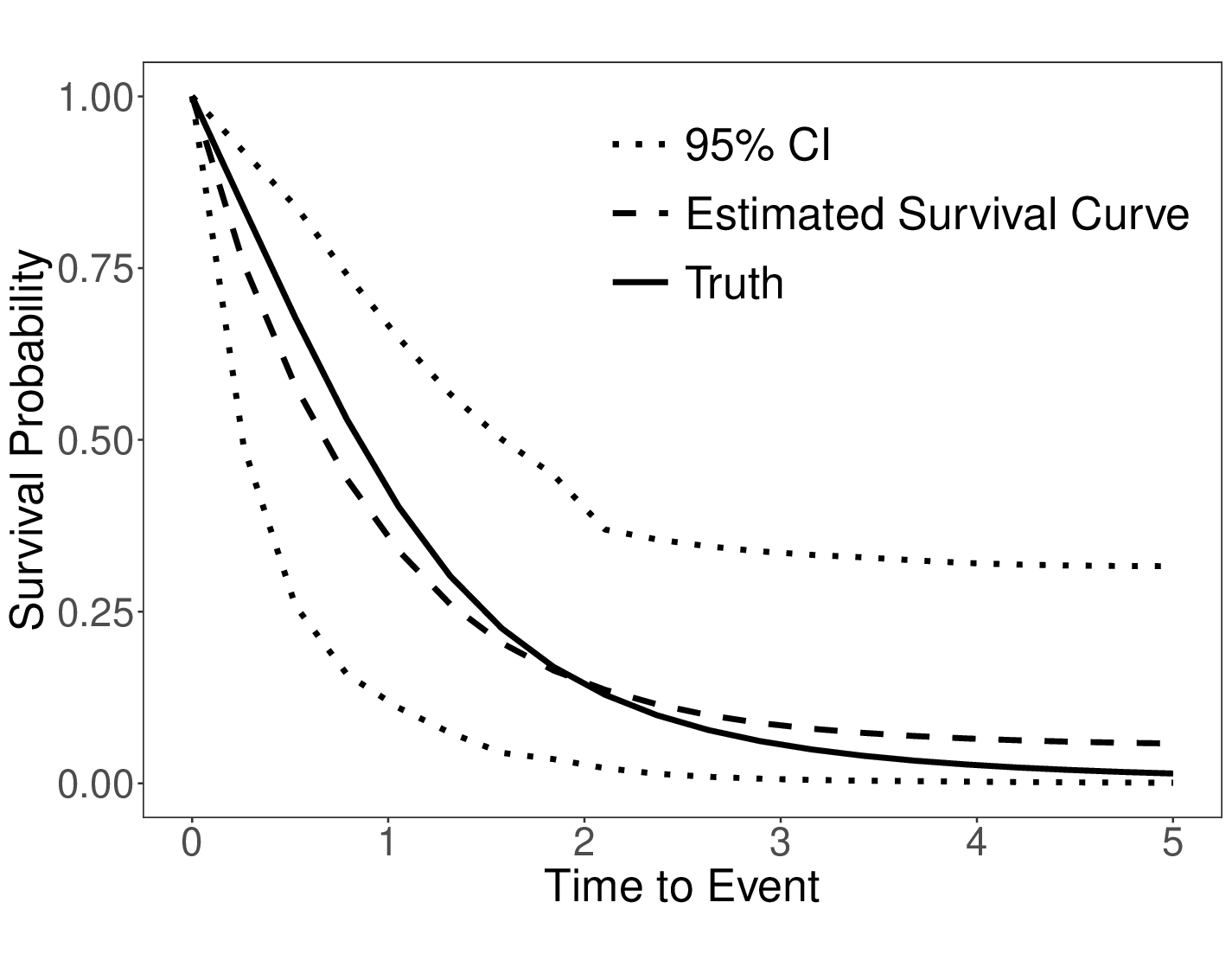}
        \caption{$d_1=d_2=10$, SMCI-K}
    \end{subfigure}
    \begin{subfigure}[b]{0.4\textwidth}
        \includegraphics[width=\textwidth]{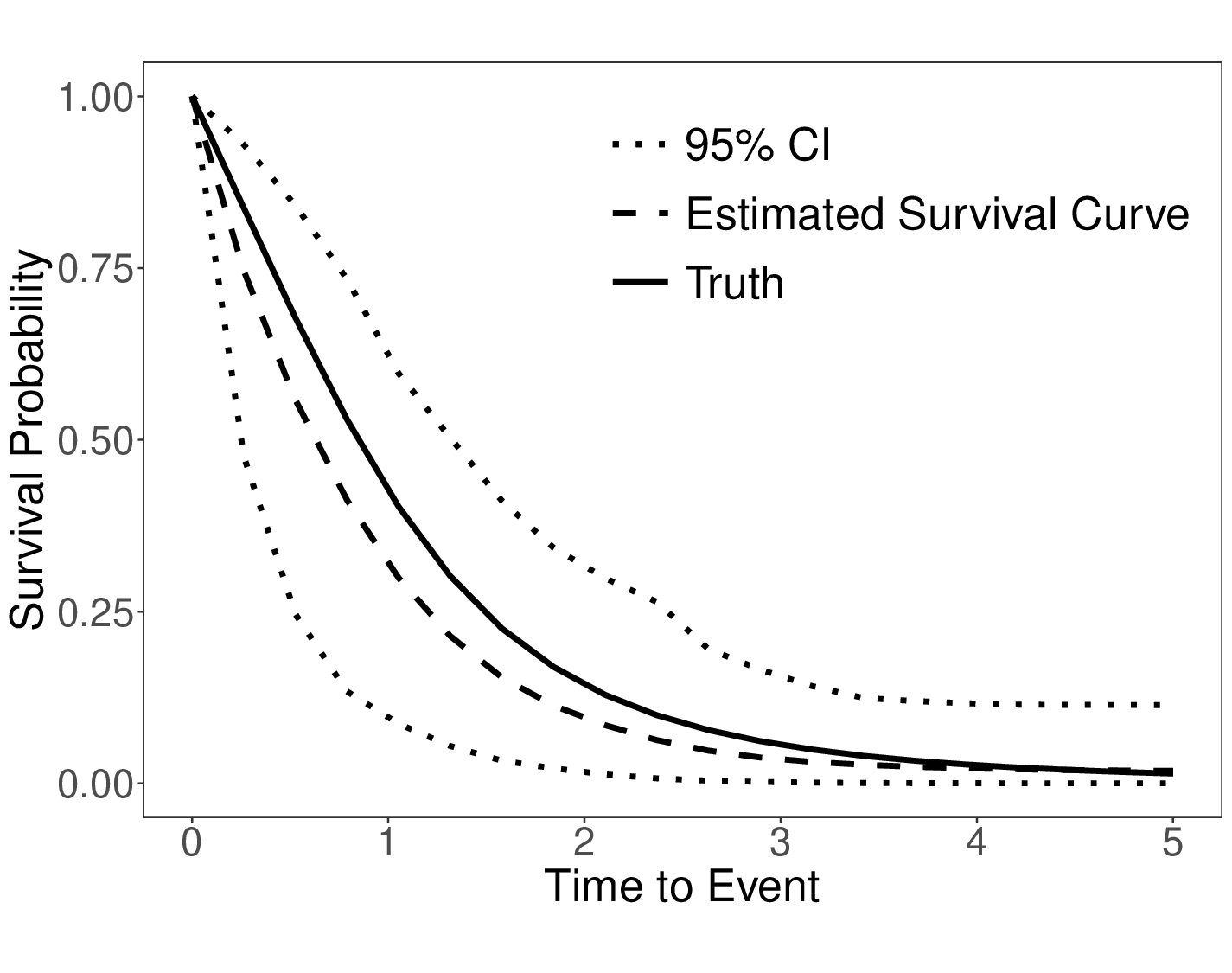}
        \caption{$d_1=d_2=10$, SMCI-S}
    \end{subfigure}
    \caption{Estimated baseline survival functions of Scenario 3 with $n=500$ and $r=1$, under varying covariate dimension $d_1$ (incidence) and $d_2$ (latency).
    }
    \label{SC_pq}
\end{figure}

\subsubsection*{A.4.3 EM Algorithm Diagnostics}

For convergence diagnostics, we provide iteration plots showing the evolution of the 
log-likelihood across EM iterations for proposed models (SMCI-K and SMCI-S). 
As demonstrated in Figures \ref{log-lik-K}-\ref{log-lik-S},
the EM algorithm consistently converged within a reasonable range, typically requiring around 50-150 iterations.

\begin{figure}[H]
    \centering
        \begin{subfigure}[b]{0.32\textwidth}
        \includegraphics[width=\textwidth]{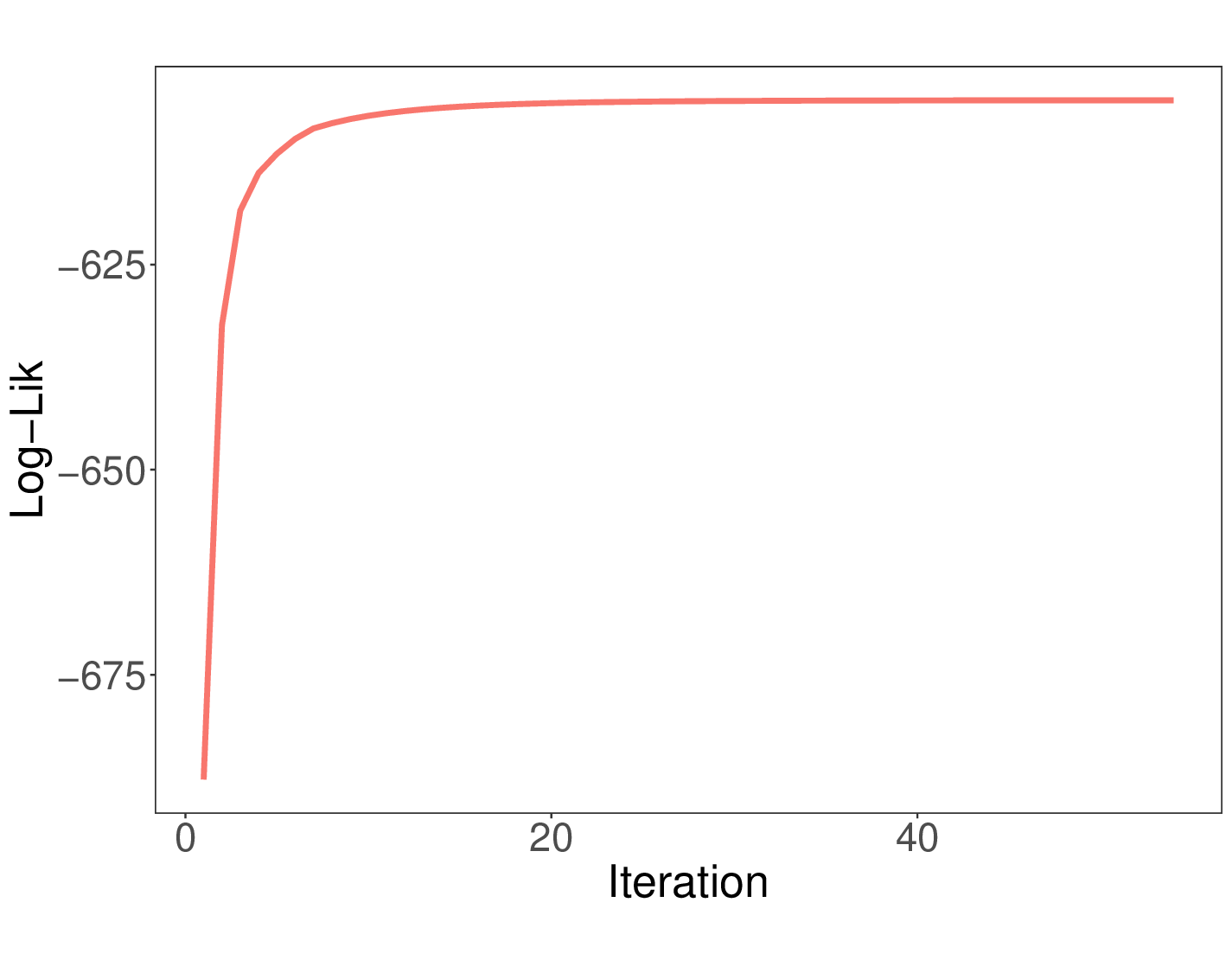}
        \caption{$r=0$, Scenaria 1 }
    \end{subfigure}
    \begin{subfigure}[b]{0.32\textwidth}
        \includegraphics[width=\textwidth]{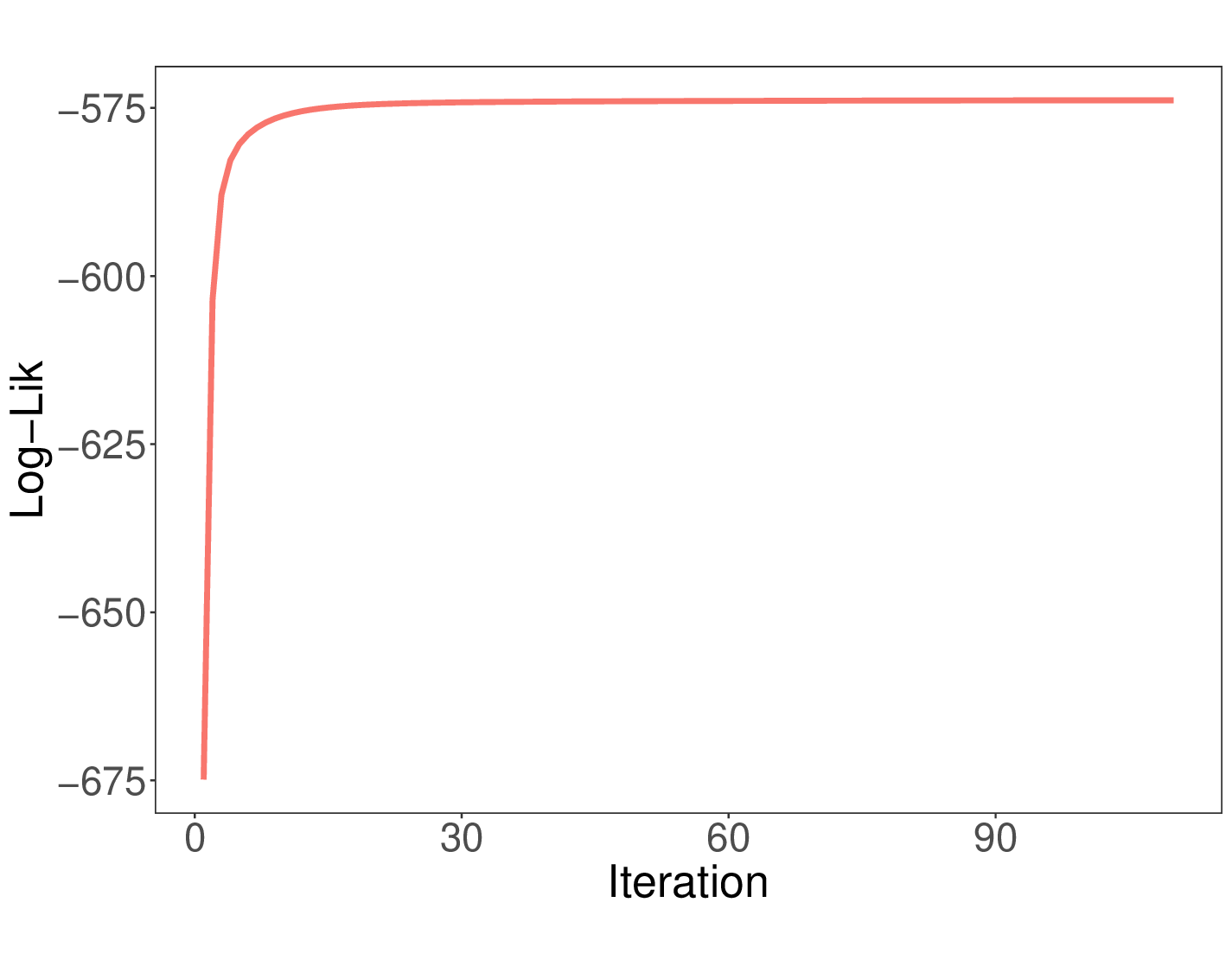}
        \caption{$r=0$, Scenaria 2}
    \end{subfigure}
    \begin{subfigure}[b]{0.32\textwidth}
        \includegraphics[width=\textwidth]{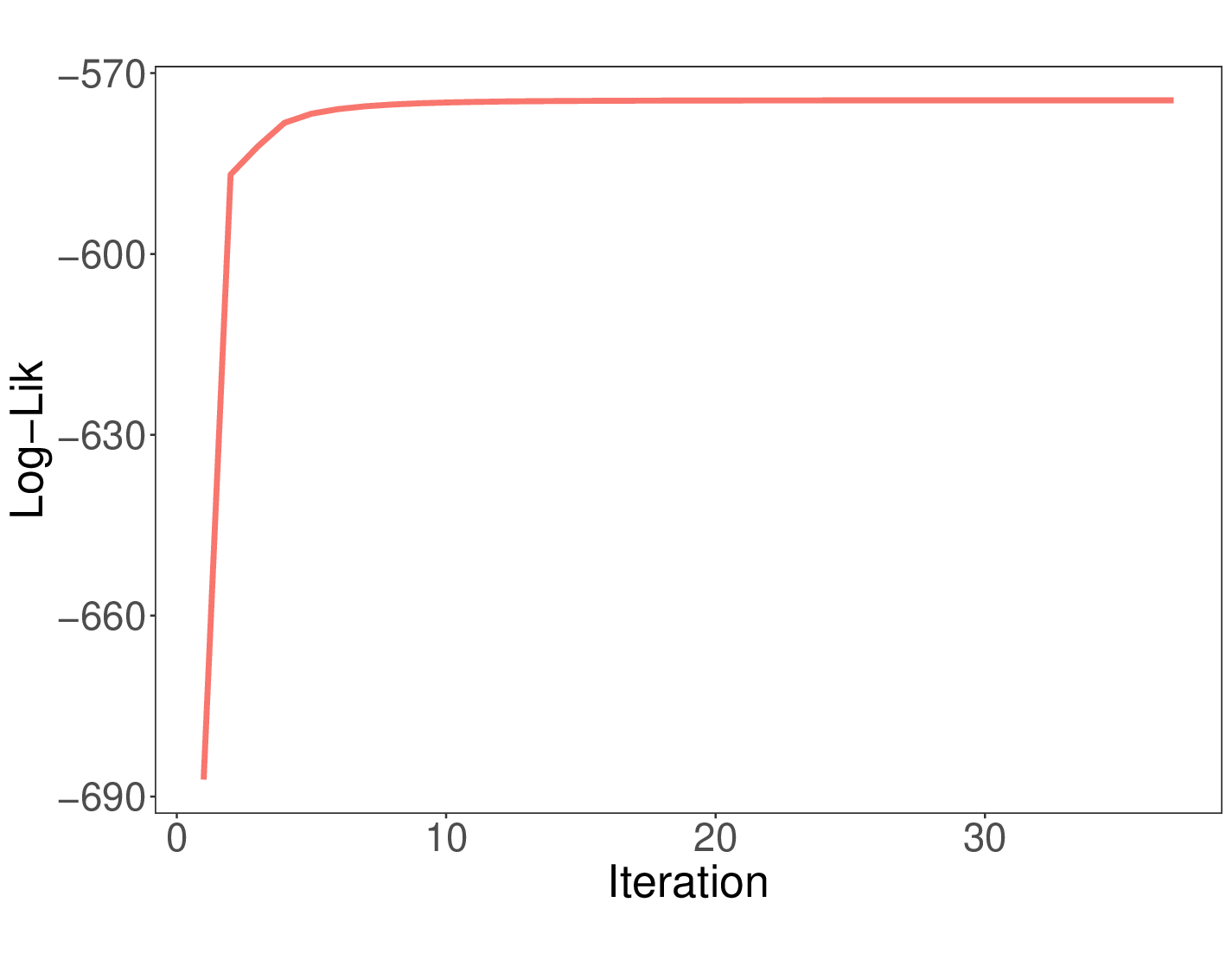}
        \caption{$r=0$, Scenaria 3}
    \end{subfigure}
    \begin{subfigure}[b]{0.32\textwidth}
        \includegraphics[width=\textwidth]{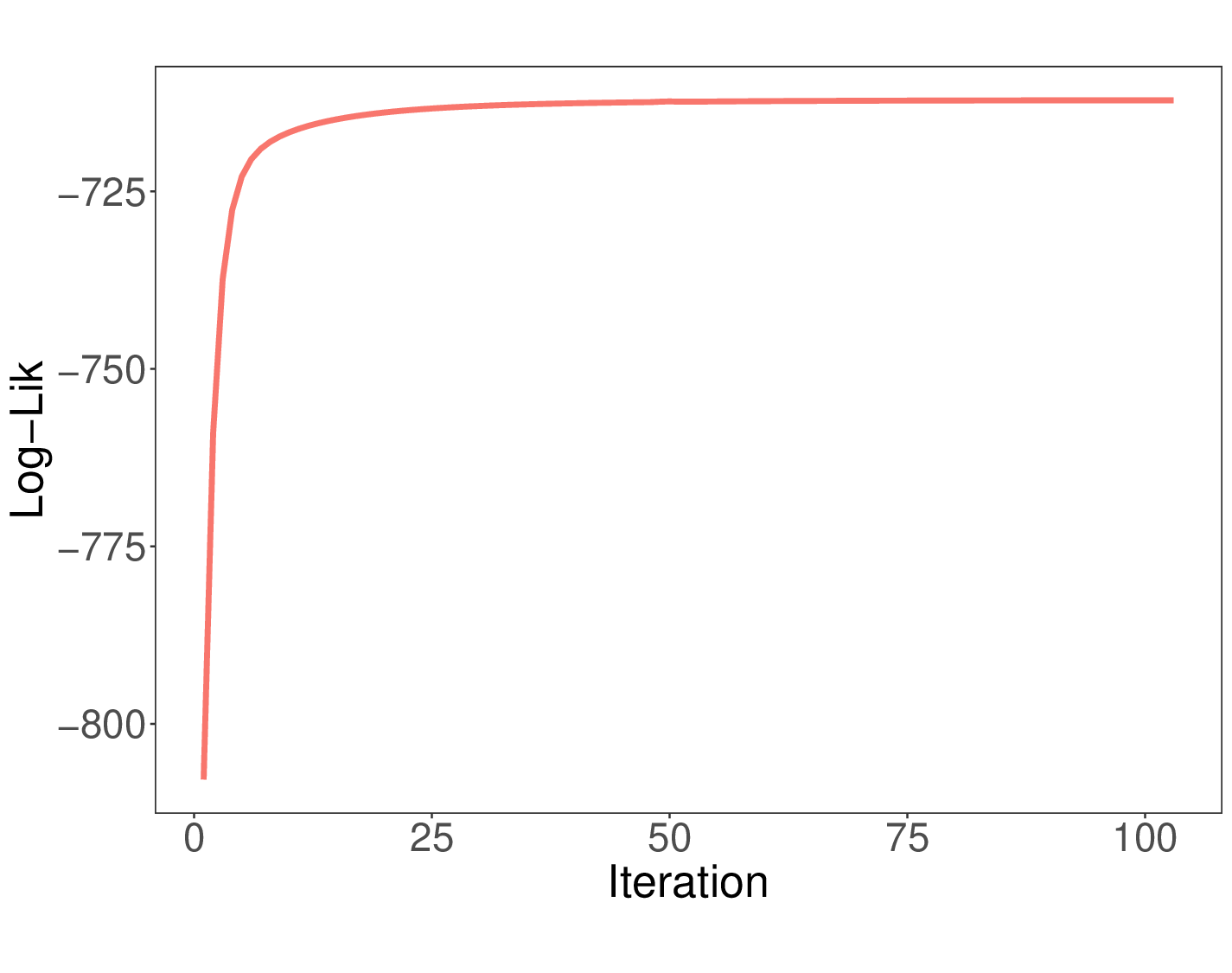}
        \caption{$r=1$, Scenaria 1 }
    \end{subfigure}
    \begin{subfigure}[b]{0.32\textwidth}
        \includegraphics[width=\textwidth]{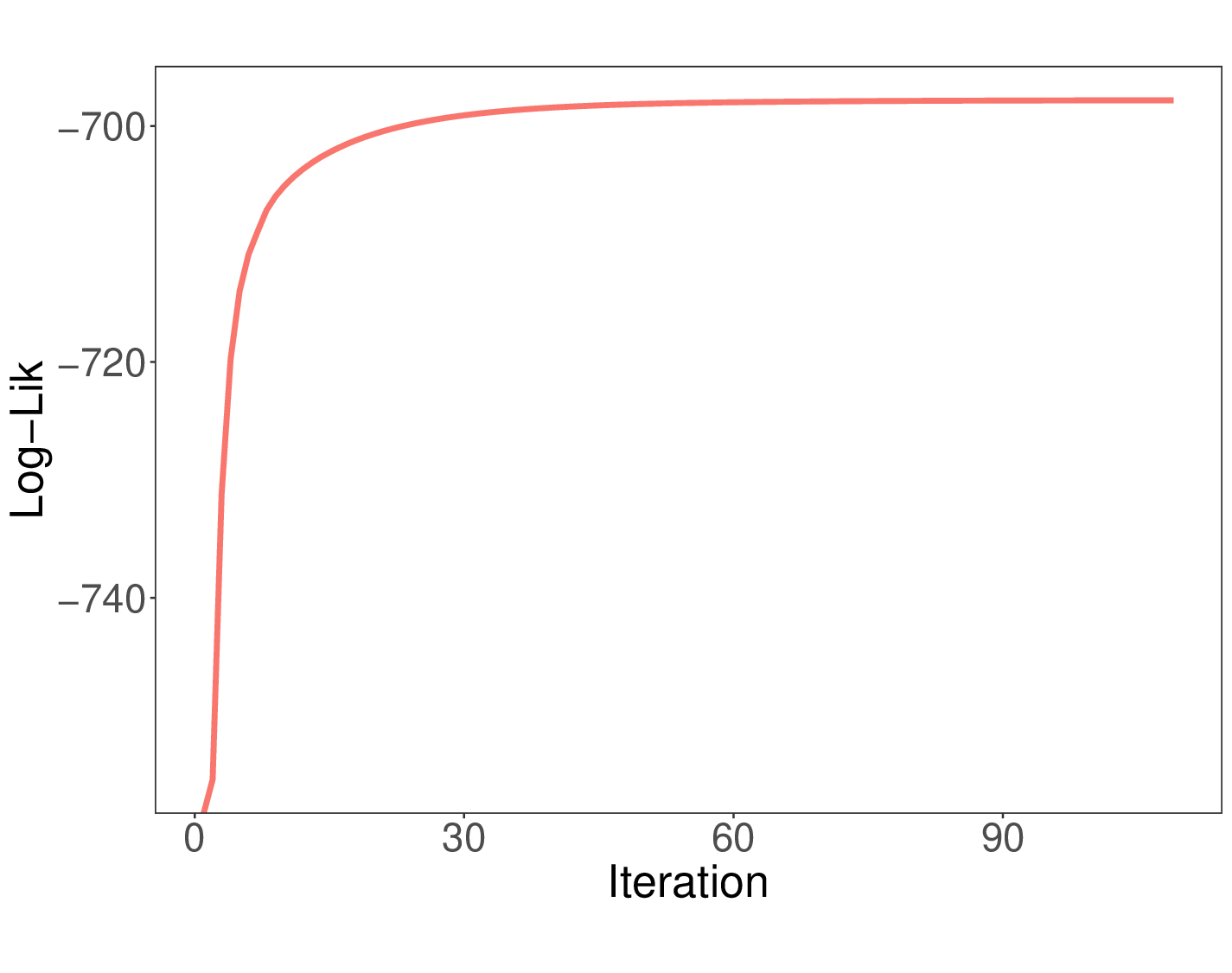}
        \caption{$r=1$, Scenaria 2}
    \end{subfigure}
    \begin{subfigure}[b]{0.32\textwidth}
        \includegraphics[width=\textwidth]{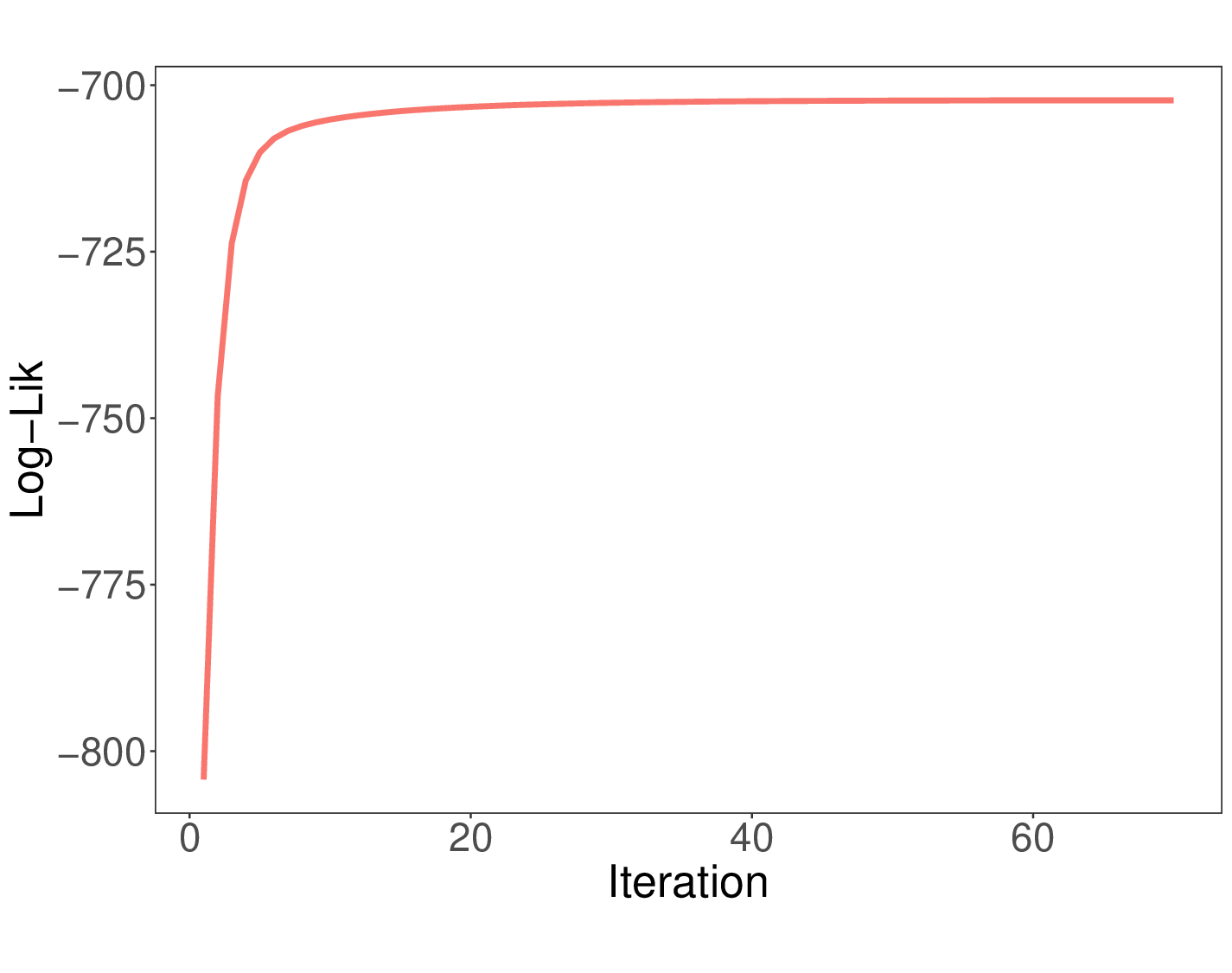}
        \caption{$r=1$, Scenaria 3}
        \end{subfigure}
        \begin{subfigure}[b]{0.32\textwidth}
        \includegraphics[width=\textwidth]{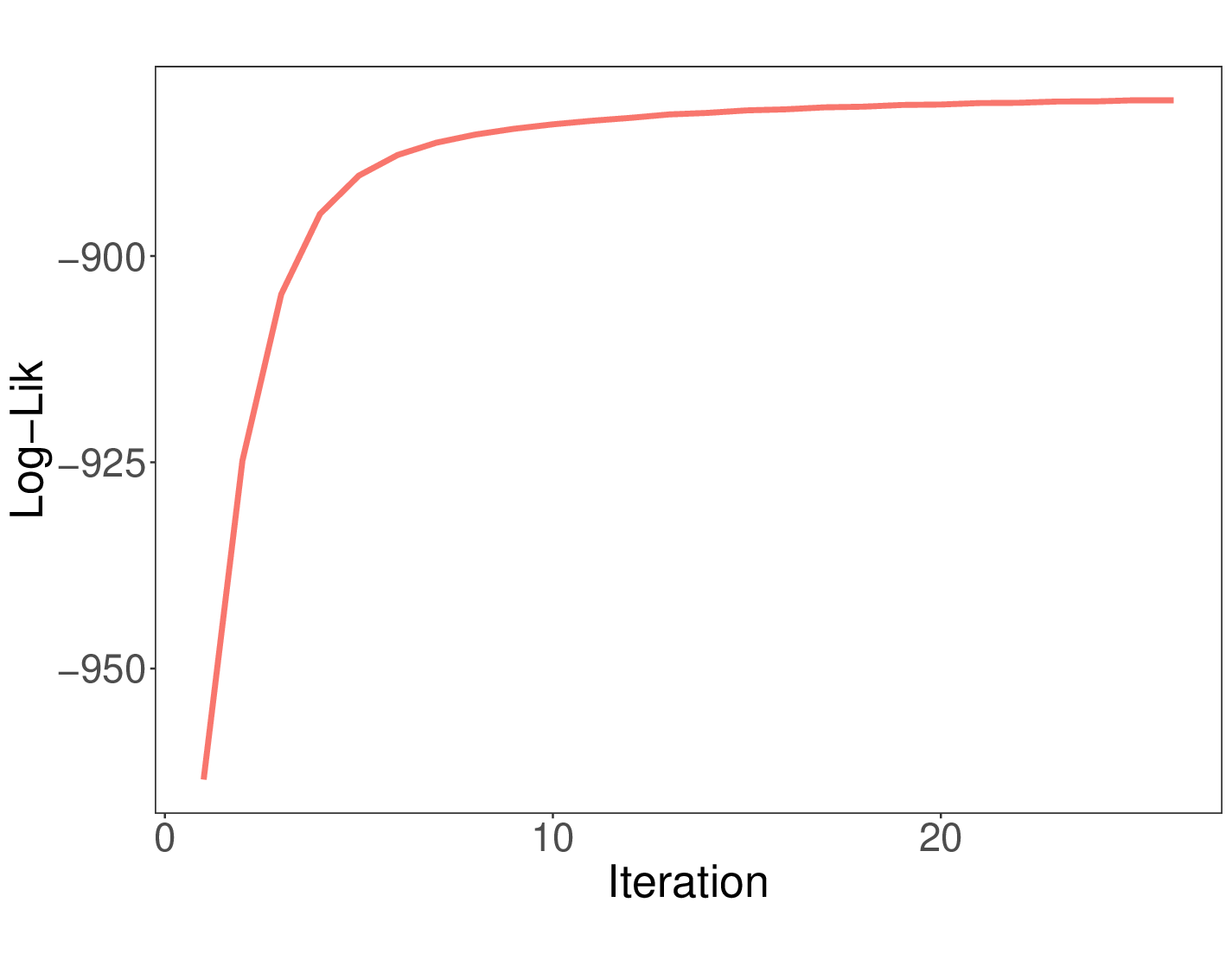}
        \caption{$r=2$, Scenaria 1 }
    \end{subfigure}
    \begin{subfigure}[b]{0.32\textwidth}
        \includegraphics[width=\textwidth]{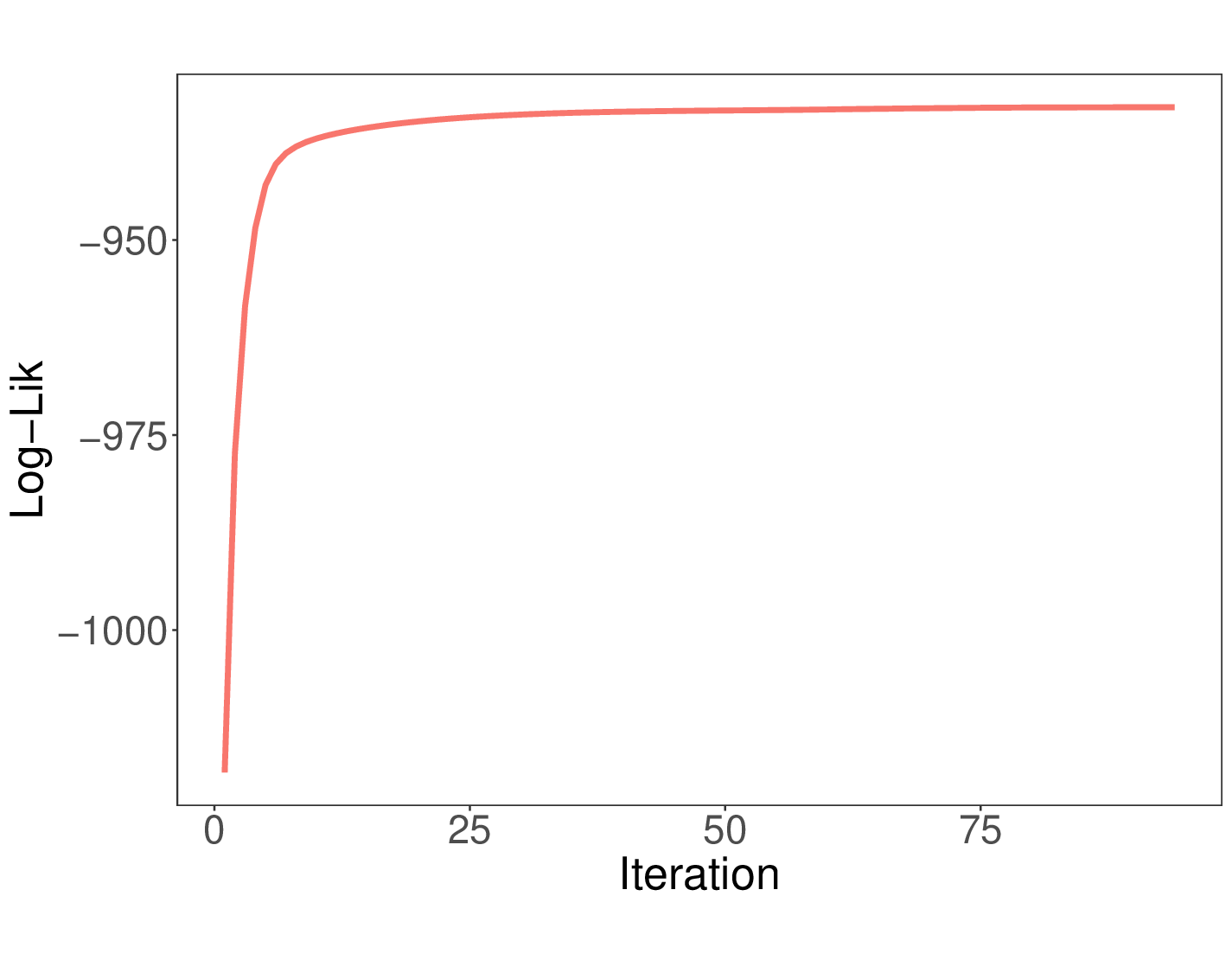}
        \caption{$r=2$, Scenaria 2}
    \end{subfigure}
    \begin{subfigure}[b]{0.32\textwidth}
        \includegraphics[width=\textwidth]{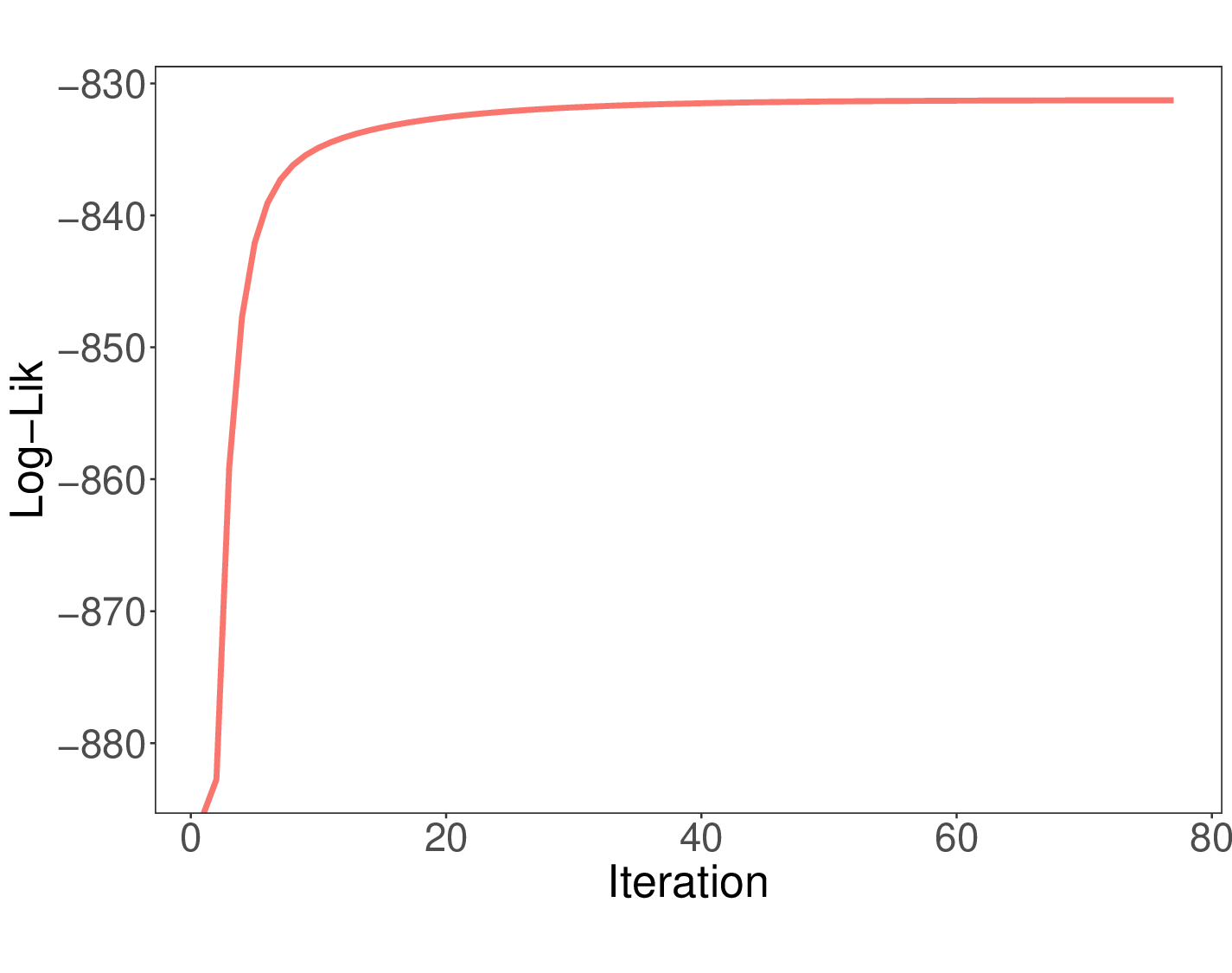}
        \caption{$r=2$, Scenaria 3}
    \end{subfigure}
    \caption{Convergence behavior of the log-likelihood function during SMCI-K algorithm iterations. The first row ($r=0$), second row ($r=1$), and third row ($r=2$) demonstrate how the parameter influences convergence rate and stability in Scenarios 1, 2, and 3 (columns from left to right). }
    \label{log-lik-K}
\end{figure}

\begin{figure}[H]
    \centering
        \begin{subfigure}[b]{0.32\textwidth}
        \includegraphics[width=\textwidth]{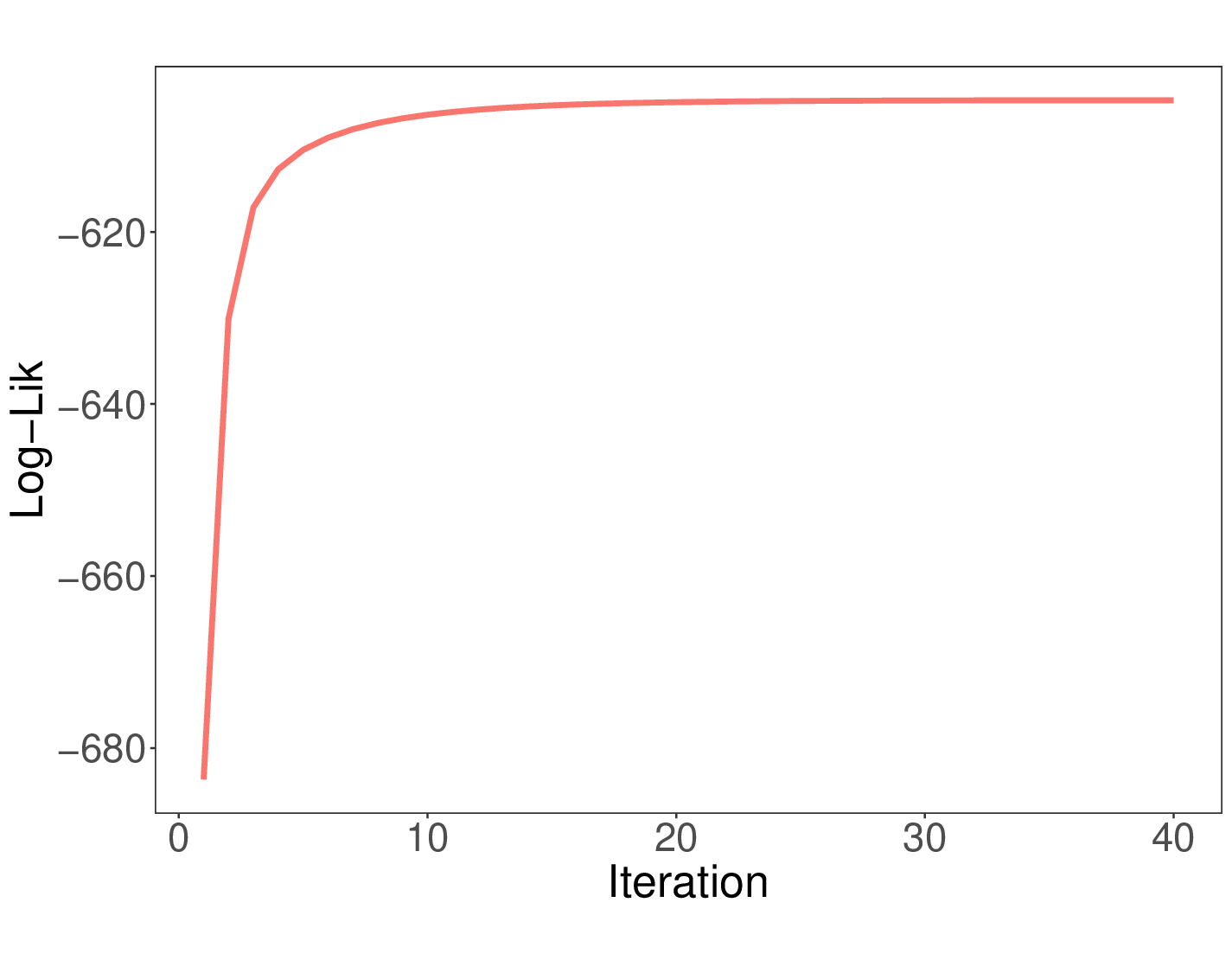}
        \caption{$r=0$, Scenaria 1 }
    \end{subfigure}
    \begin{subfigure}[b]{0.32\textwidth}
        \includegraphics[width=\textwidth]{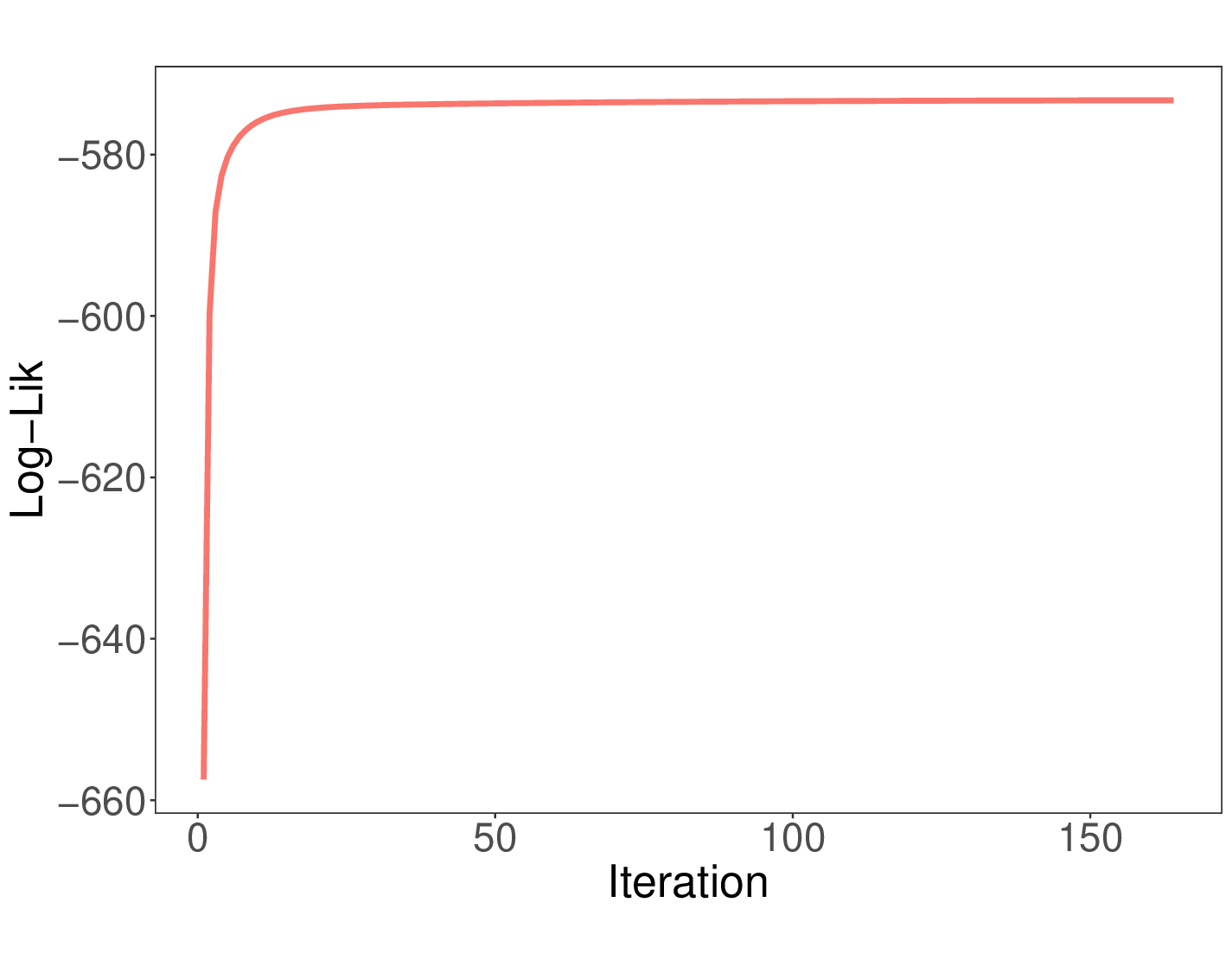}
        \caption{$r=0$, Scenaria 2}
    \end{subfigure}
    \begin{subfigure}[b]{0.32\textwidth}
        \includegraphics[width=\textwidth]{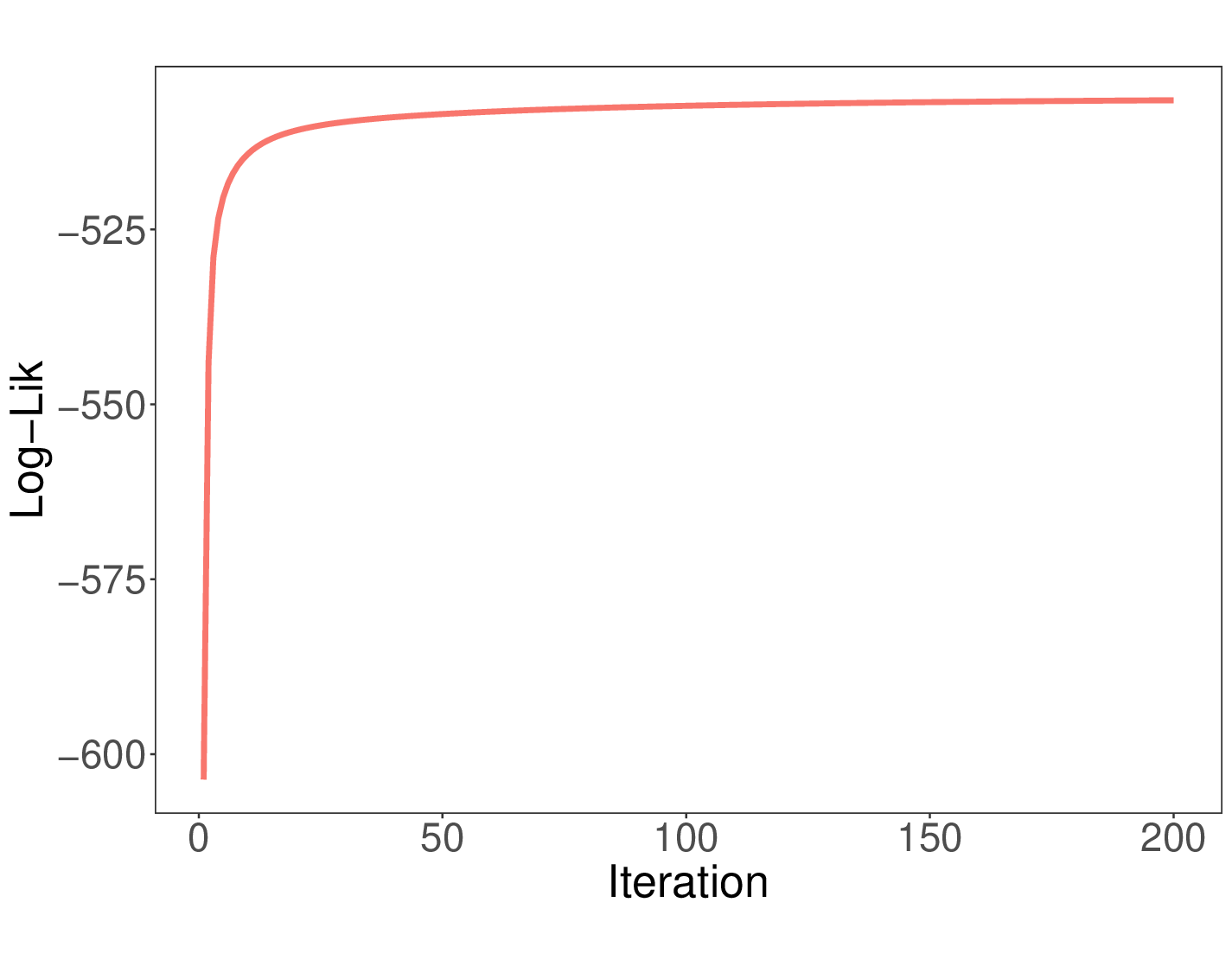}
        \caption{$r=0$, Scenaria 3}
    \end{subfigure}
    \begin{subfigure}[b]{0.32\textwidth}
        \includegraphics[width=\textwidth]{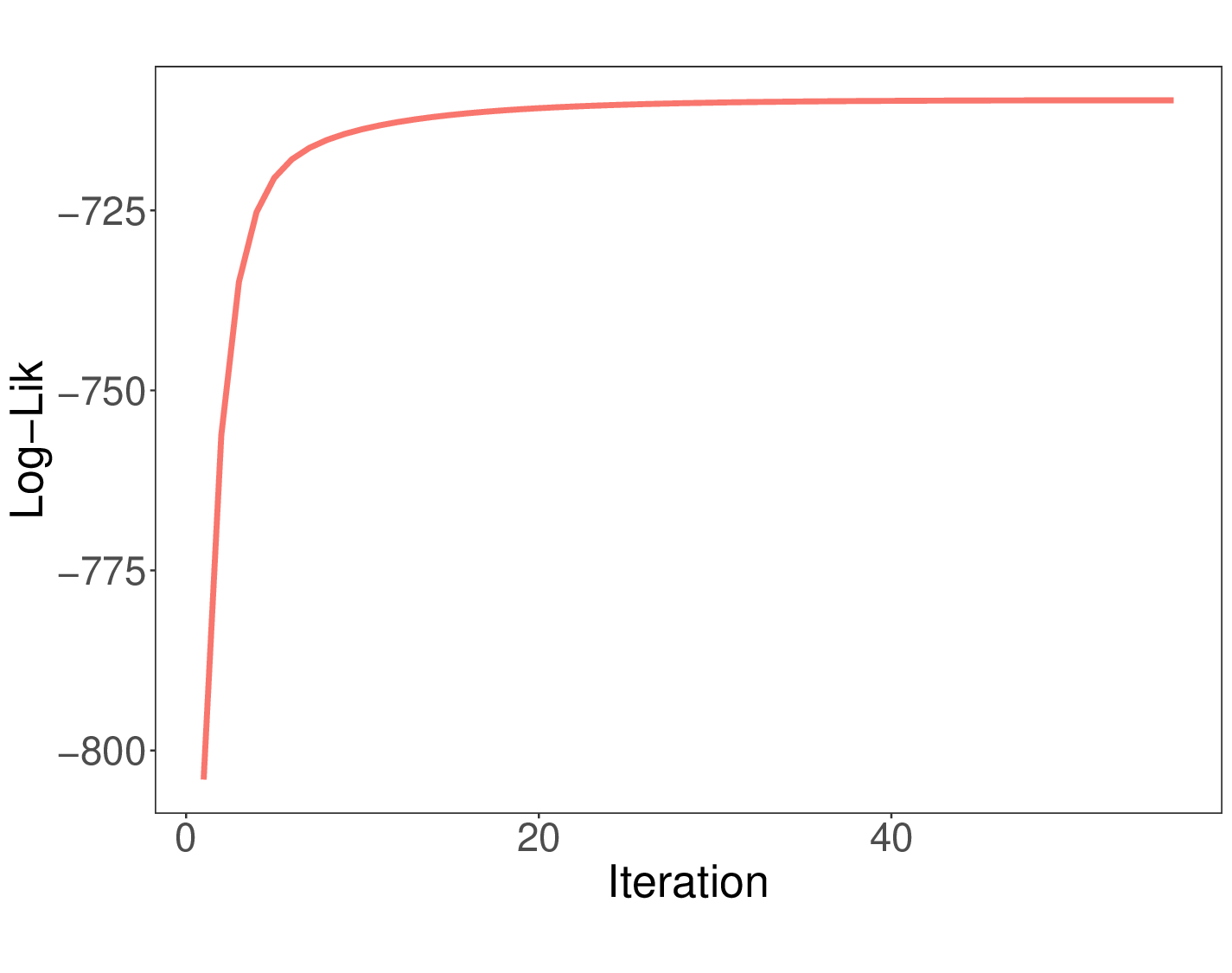}
        \caption{$r=1$, Scenaria 1 }
    \end{subfigure}
    \begin{subfigure}[b]{0.32\textwidth}
        \includegraphics[width=\textwidth]{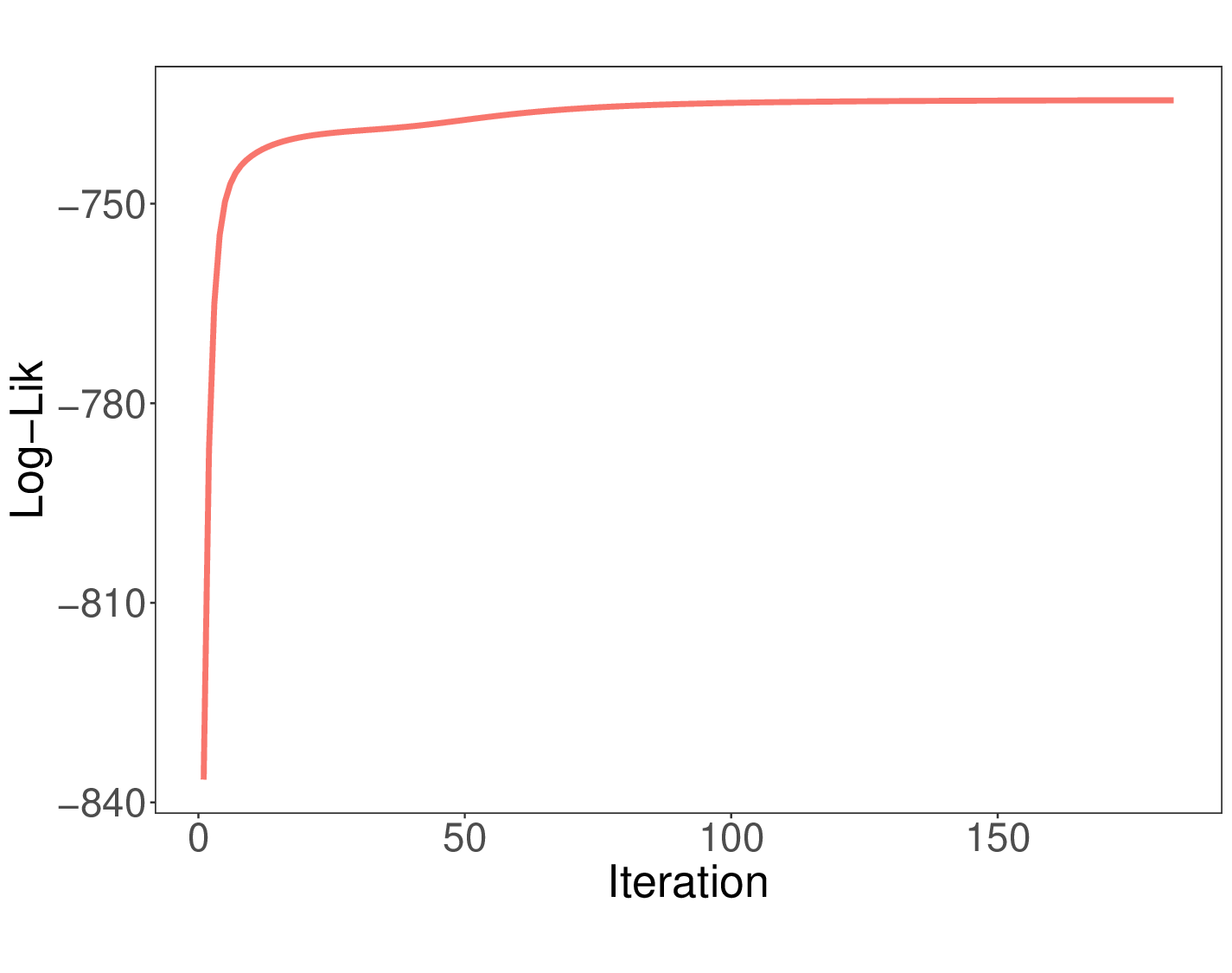}
        \caption{$r=1$, Scenaria 2}
    \end{subfigure}
    \begin{subfigure}[b]{0.32\textwidth}
        \includegraphics[width=\textwidth]{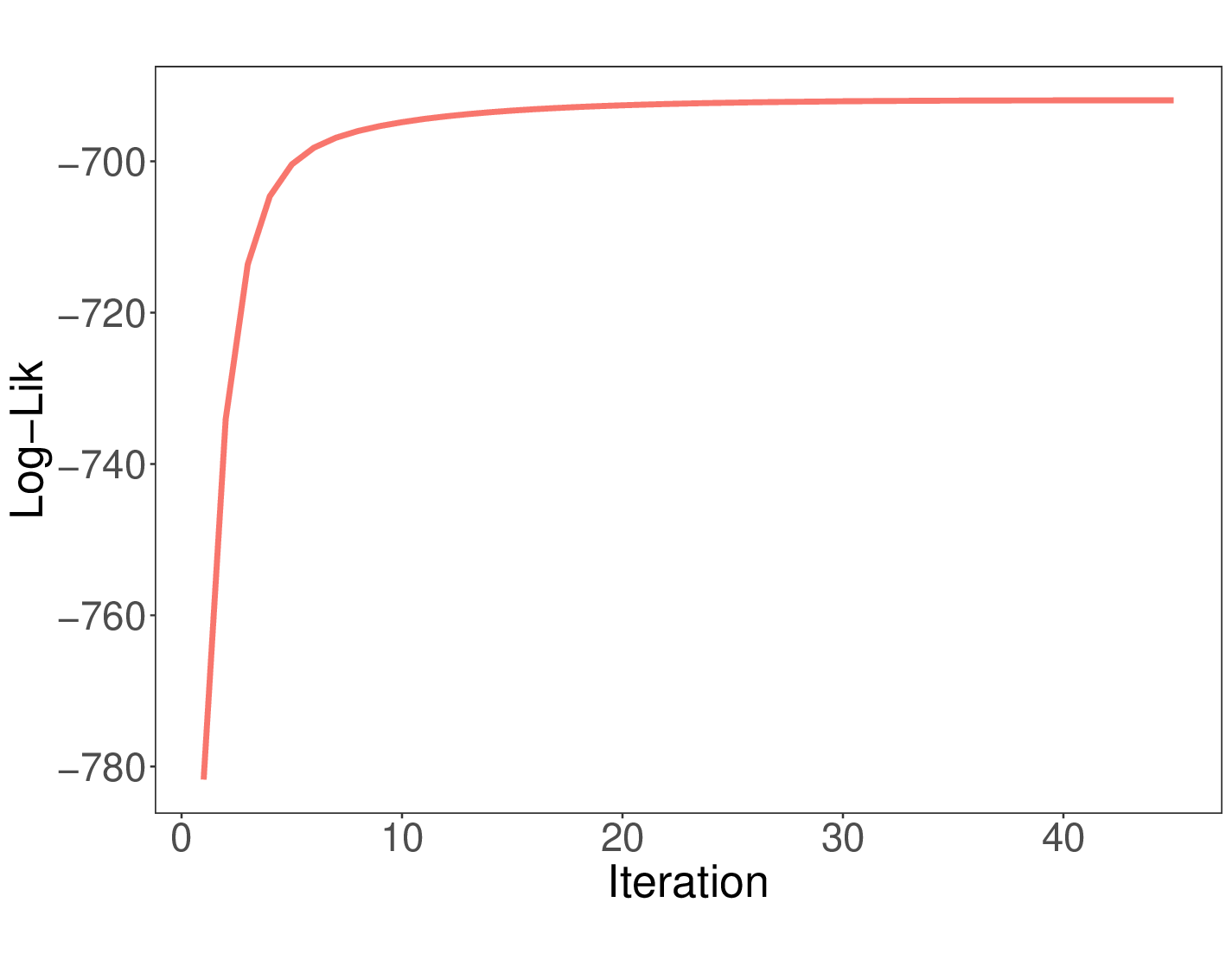}
        \caption{$r=1$, Scenaria 3}
        \end{subfigure}
        \begin{subfigure}[b]{0.32\textwidth}
        \includegraphics[width=\textwidth]{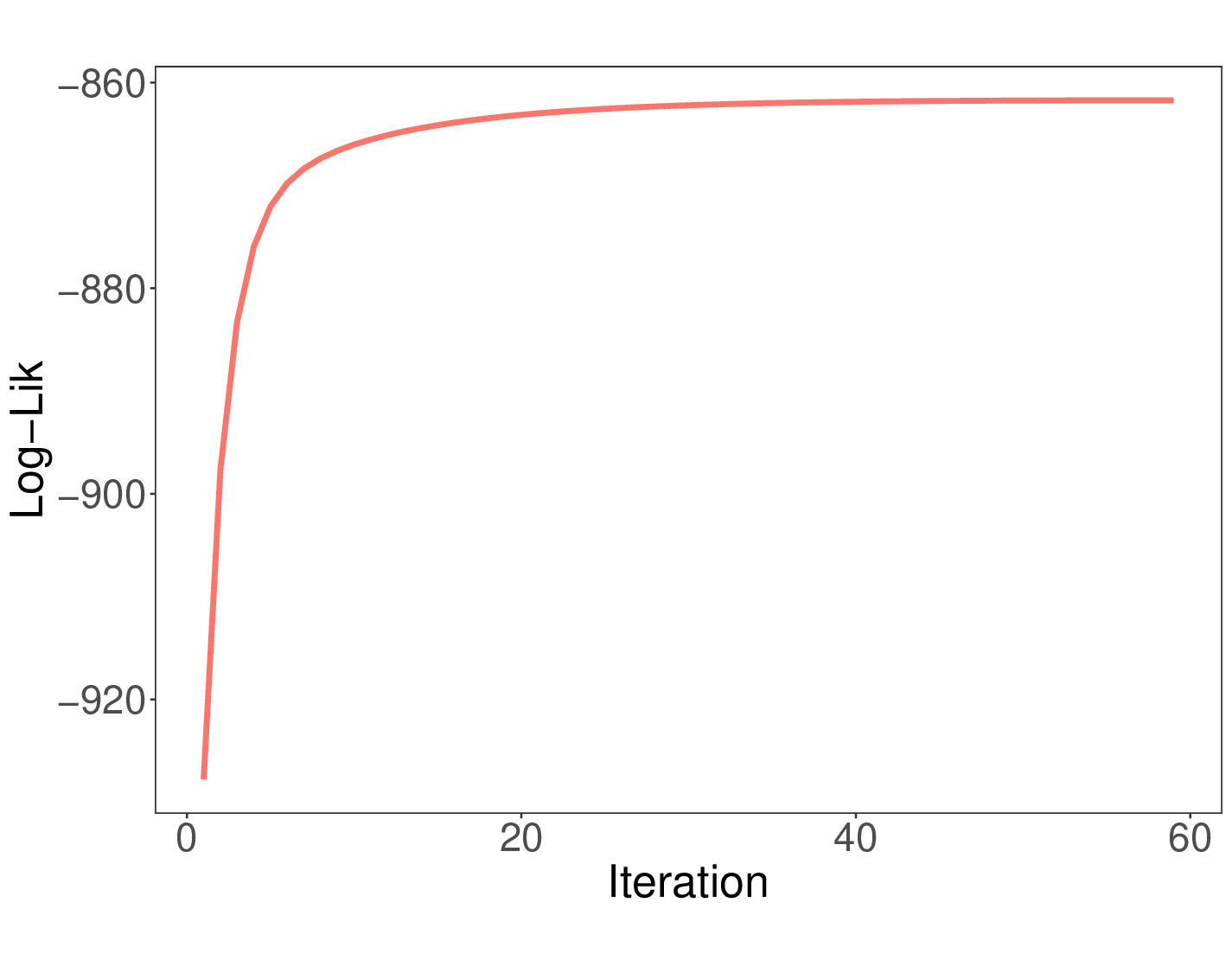}
        \caption{$r=2$, Scenaria 1 }
    \end{subfigure}
    \begin{subfigure}[b]{0.32\textwidth}
        \includegraphics[width=\textwidth]{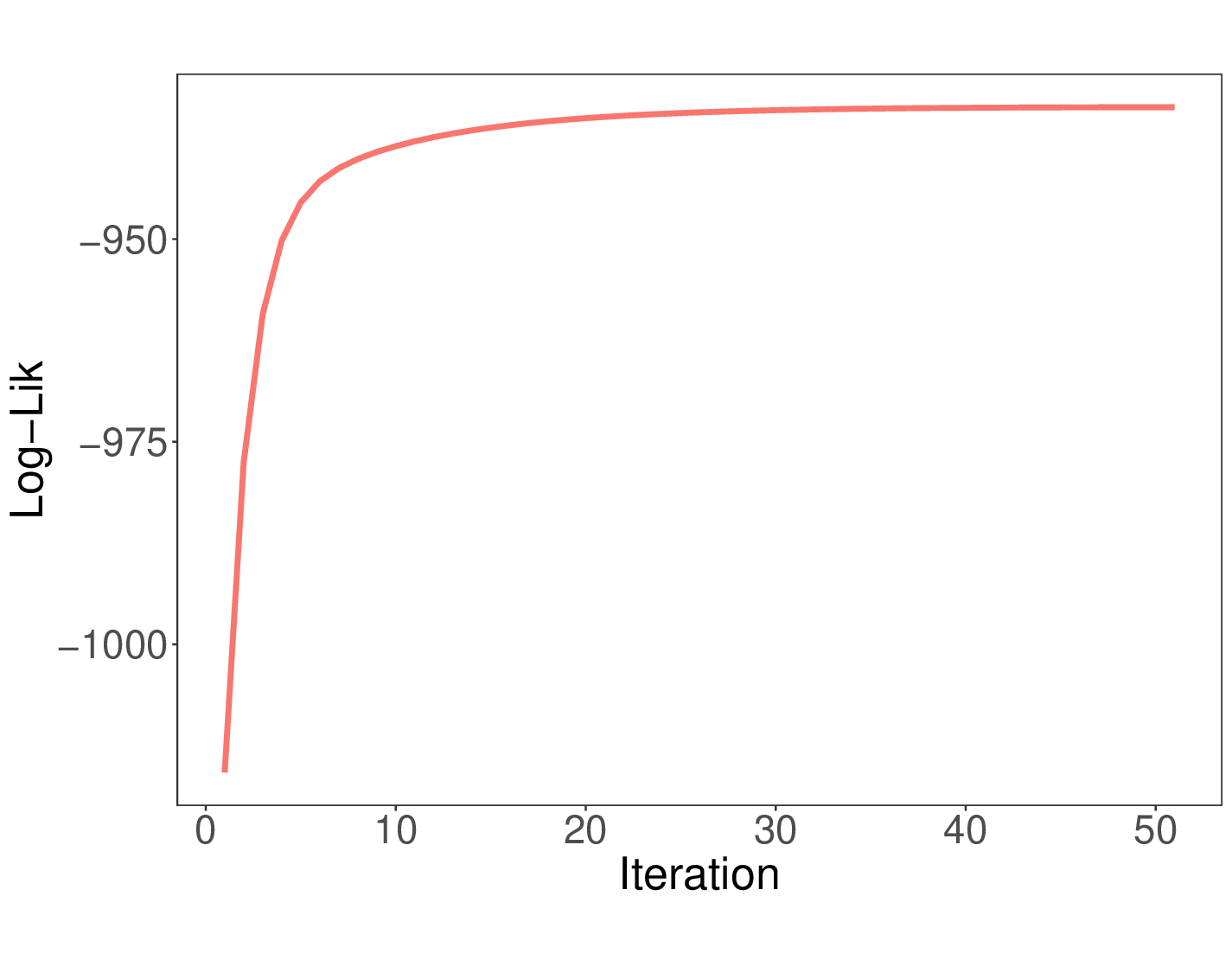}
        \caption{$r=2$, Scenaria 2}
    \end{subfigure}
    \begin{subfigure}[b]{0.32\textwidth}
        \includegraphics[width=\textwidth]{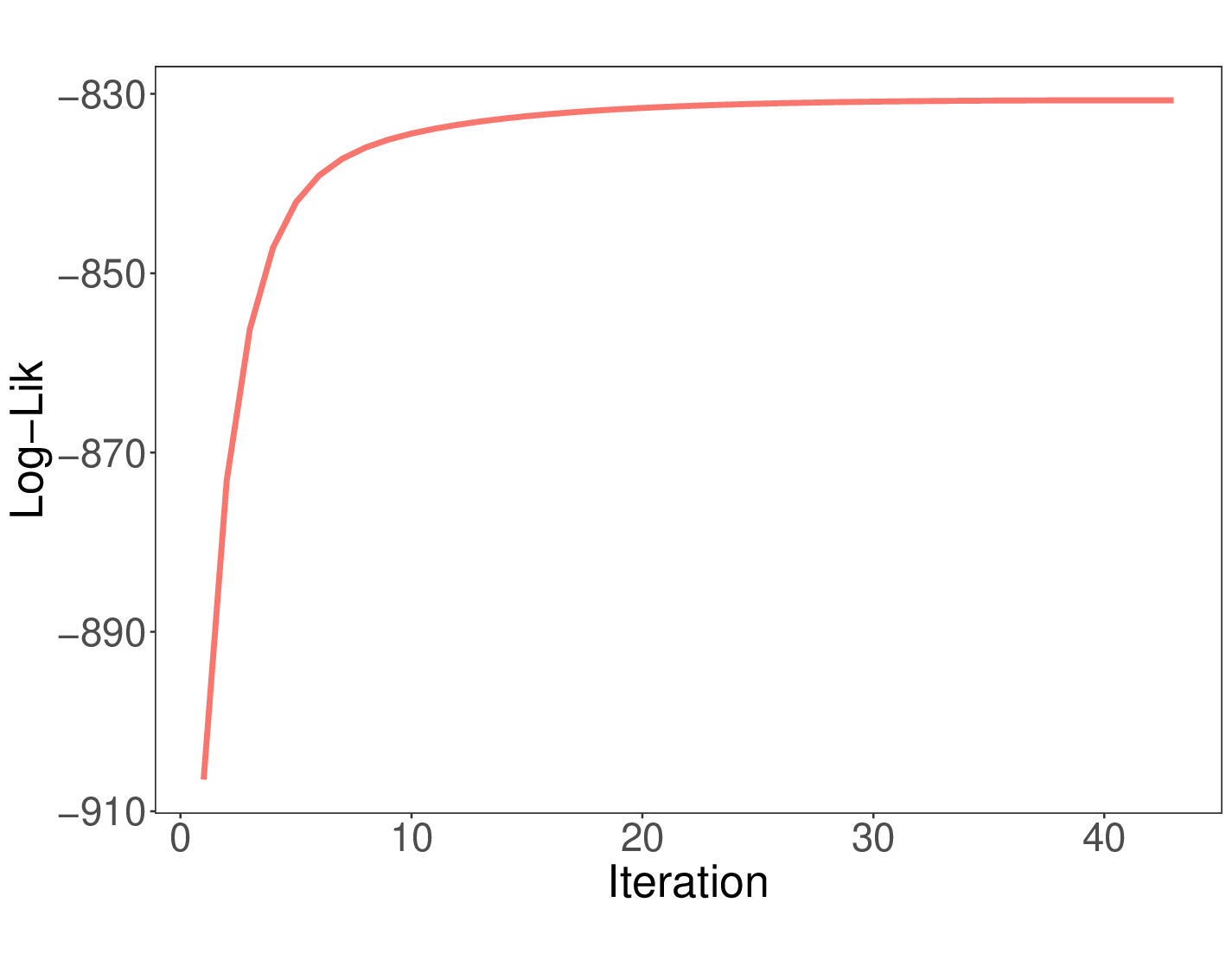}
        \caption{$r=2$, Scenaria 3}
    \end{subfigure}
    \caption{Convergence behavior of the log-likelihood function during SMCI-S algorithm iterations. The first row ($r=0$), second row ($r=1$), and third row ($r=2$) demonstrate how the parameter influences convergence rate and stability in Scenarios 1, 2, and 3 (columns from left to right). }
    \label{log-lik-S}
\end{figure}

{
Furthermore, to assess sensitivity to initial values, 
we performed additional simulation studies with three distinct sets of initial values for $\boldsymbol{\beta}$ and $\boldsymbol{\eta}$: 
\begin{itemize}
    \item Set A : $\boldsymbol{\beta}^{(0)}=(-0.5,-0.5,-0.5)$, $\boldsymbol{\eta}^{(0)}=(0.5,0.5,...,0.5)$
    \item Set B: $\boldsymbol{\beta}^{(0)}=(0.5,0.5,0.5)$, $\boldsymbol{\eta}^{(0)}=(1.5,1.5,\dots,1.5.)$ 
    \item Set C: $\boldsymbol{\beta}^{(0)}=(-0.5,-0.5,-0.5)$ $\boldsymbol{\eta}^{(0)}=(1.5,1.5,\dots,1.5)$
\end{itemize}
Note $\boldsymbol{\gamma}^{(0)}$ is obtained
by fitting the generalized additive model using pseudo values $1-\delta_{R_i}$, thus we don't consider different initial values for $\boldsymbol{\gamma}^{(0)}$. %The results show that all runs converged to nearly identical parameter estimates (maximum relative difference < 0.01\%), confirming the robustness of our estimation procedure.
As summarized in Table \ref{tab:intial values} and Figures \ref{fig1: initial values}-\ref{fig2: initial values}, the resulting parameter estimates are highly consistent and exhibit negligible bias and proper coverage probabilities across all initializations. This consistency strongly suggests that our algorithm is insensitive to initial values.
}

\begin{table}[htpb!]
\centering
\renewcommand{\arraystretch}{1.3}
%\resizebox{0.65\linewidth}{!}{
\begin{tabular}{cccccccc}
  \hline
  $n=500$ & True & Bias & SSD & SE & CP& \\ 
    \hline
\multicolumn{7}{l}{Set A}\\
$\beta_1 $&1 &0.03 &0.18 &0.20 &0.96\\
$\beta_2$  &-1 &-0.02 &0.12 &0.13 &0.96\\
$\beta_3$  &1 &0.02 &0.24 &0.23 &0.96\\
\multicolumn{7}{l}{Set B}\\
$\beta_1 $&1 &0.01 &0.19 &0.20 &0.97\\
$\beta_2$  &-1 &-0.01 &0.13 &0.13 &0.92\\
$\beta_3$  &1 &0.05 &0.22 &0.23 &0.96\\
\multicolumn{7}{l}{Set C}\\
$\beta_1 $&1 &-0.02 &0.20 &0.20 &0.94\\
$\beta_2$  &-1 &-0.02 &0.11 &0.13 &0.96\\
$\beta_3$  &1 &0.01 &0.22 &0.23 &0.96\\
\hline
\end{tabular}
%}
\caption{
Simulation results for Scenario 3  with 
$n=500$ and $r=1$ using three sets of initial values. % with 5 quantile-based knots.
Bias: the estimated bias; ESD:
empirical standard deviation; ESE: empirical standard error estimate; CP: the 95\% empirical
coverage probabilities.}
\label{tab:intial values}
\end{table}

\begin{figure}[H]
    \centering
        \includegraphics[width=0.4\textwidth, angle=-90]{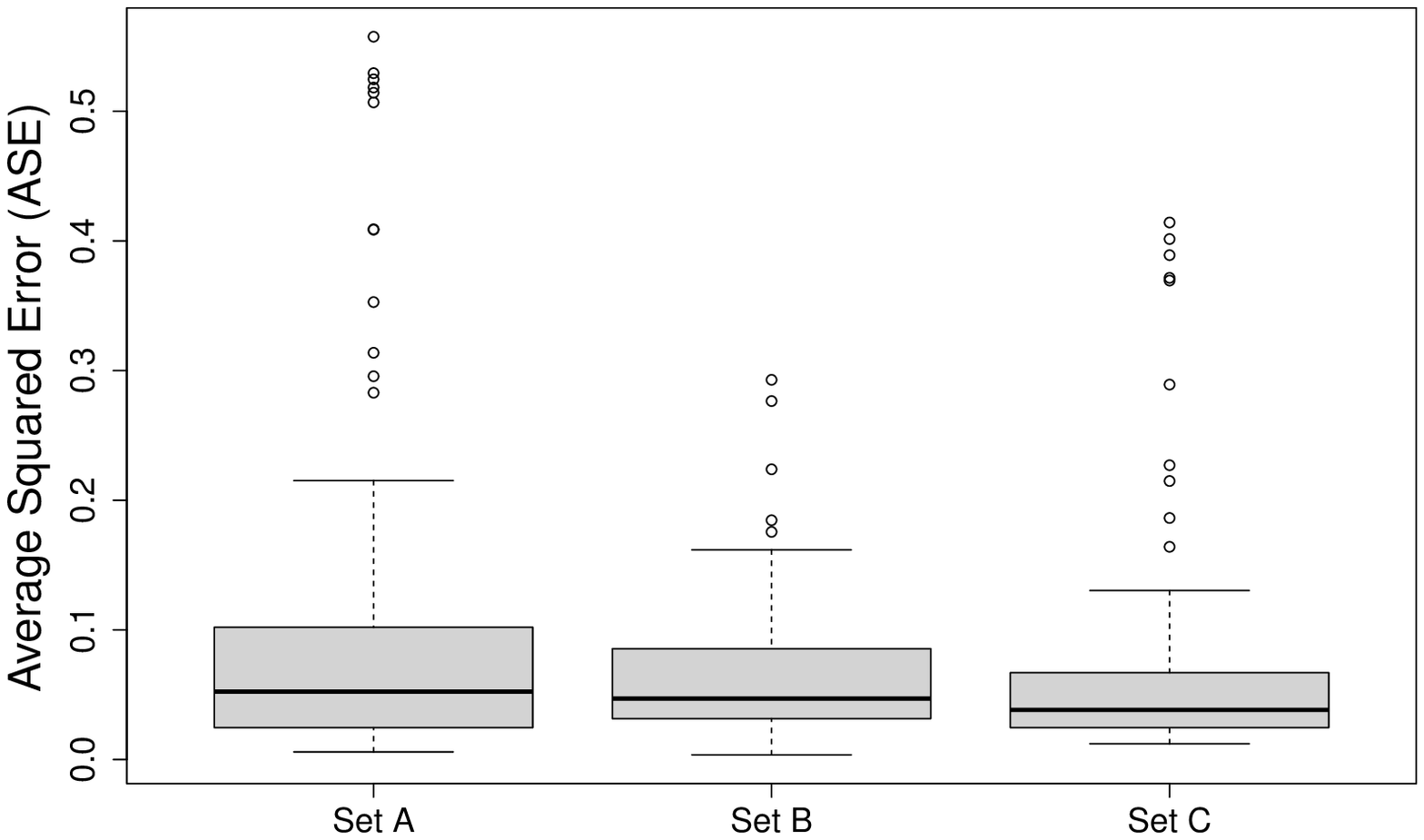}
    \caption{Boxplots of %the average squared error (ASE) 
    ASE for the incidence link function in Scenario 3 with $n=500$ and $r=1$, using three sets of initial values.% (a) ASE distribution across spline orders 2, 3, and 4. (b) ASE distribution across knot numbers 6, 7, 8, and 9. The results are based on 200 replications under the setting $n=200$, Scenario 3, $r=1$.
    }
    \label{fig1: initial values}
\end{figure}

\begin{figure}[H]
    \centering
        \begin{subfigure}[b]{0.32\textwidth}
        \includegraphics[width=\textwidth]{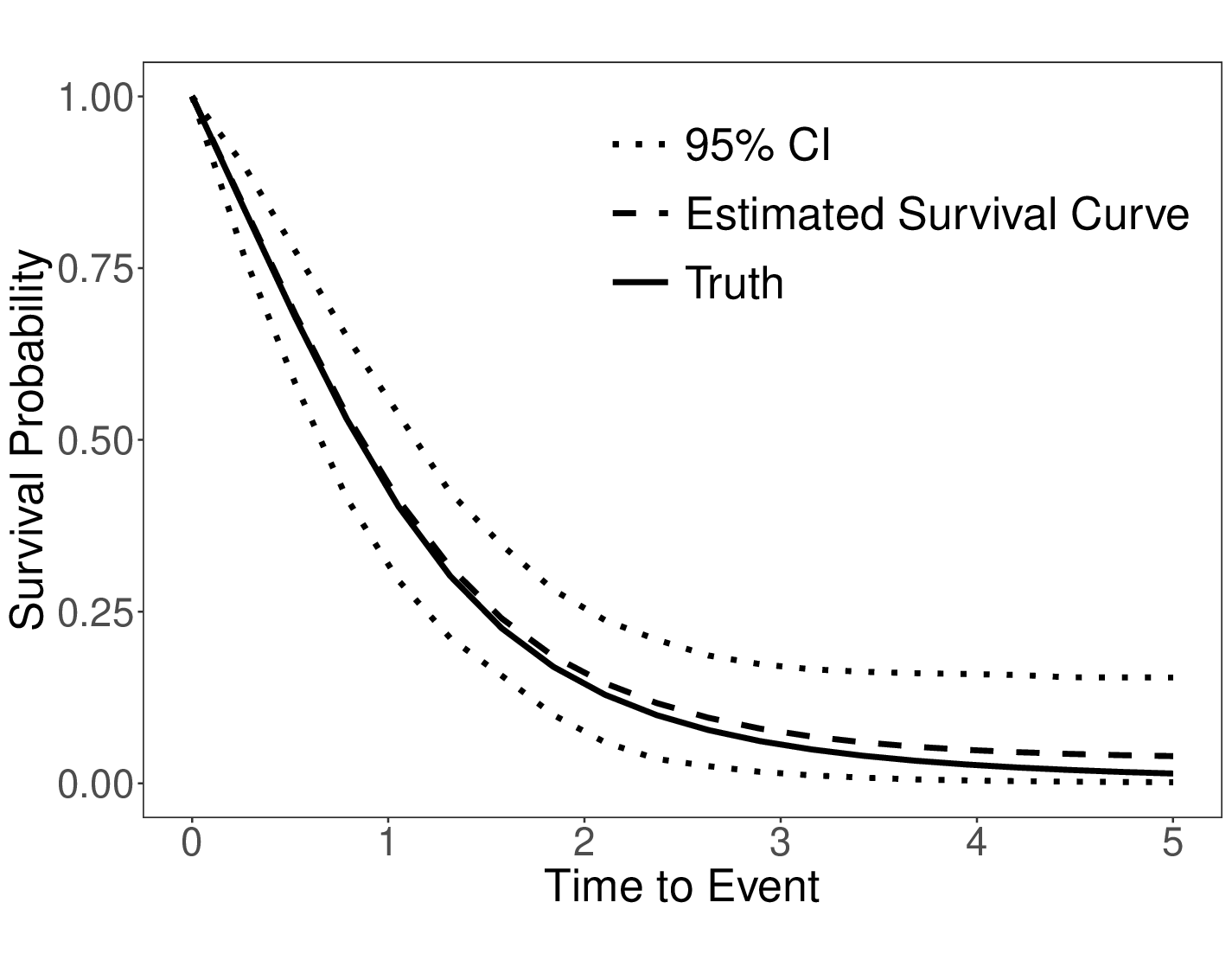}
        \caption{Set A}
    \end{subfigure}
    \begin{subfigure}[b]{0.32\textwidth}
        \includegraphics[width=\textwidth]{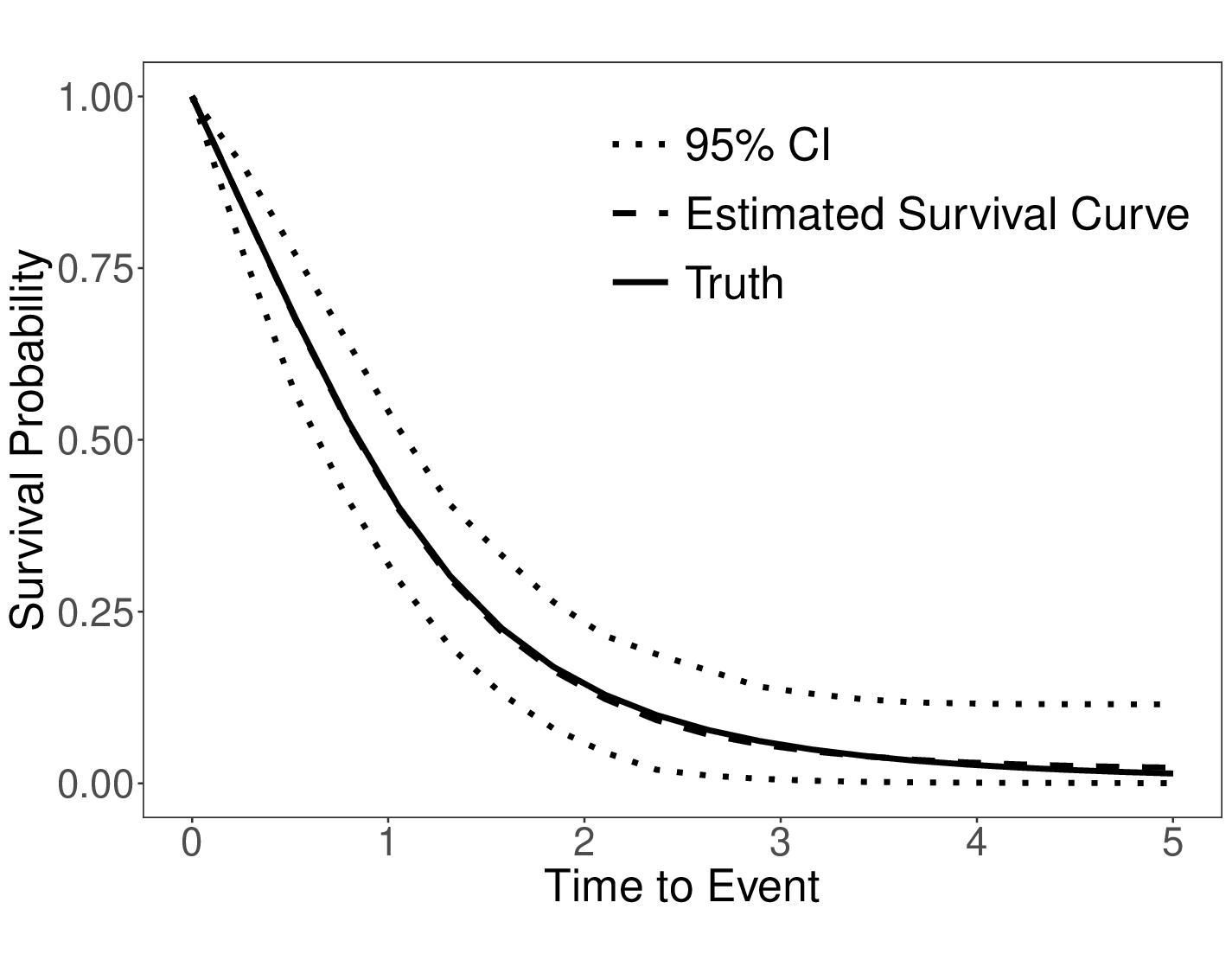}
        \caption{Set B}
    \end{subfigure}
    \begin{subfigure}[b]{0.32\textwidth}
        \includegraphics[width=\textwidth]{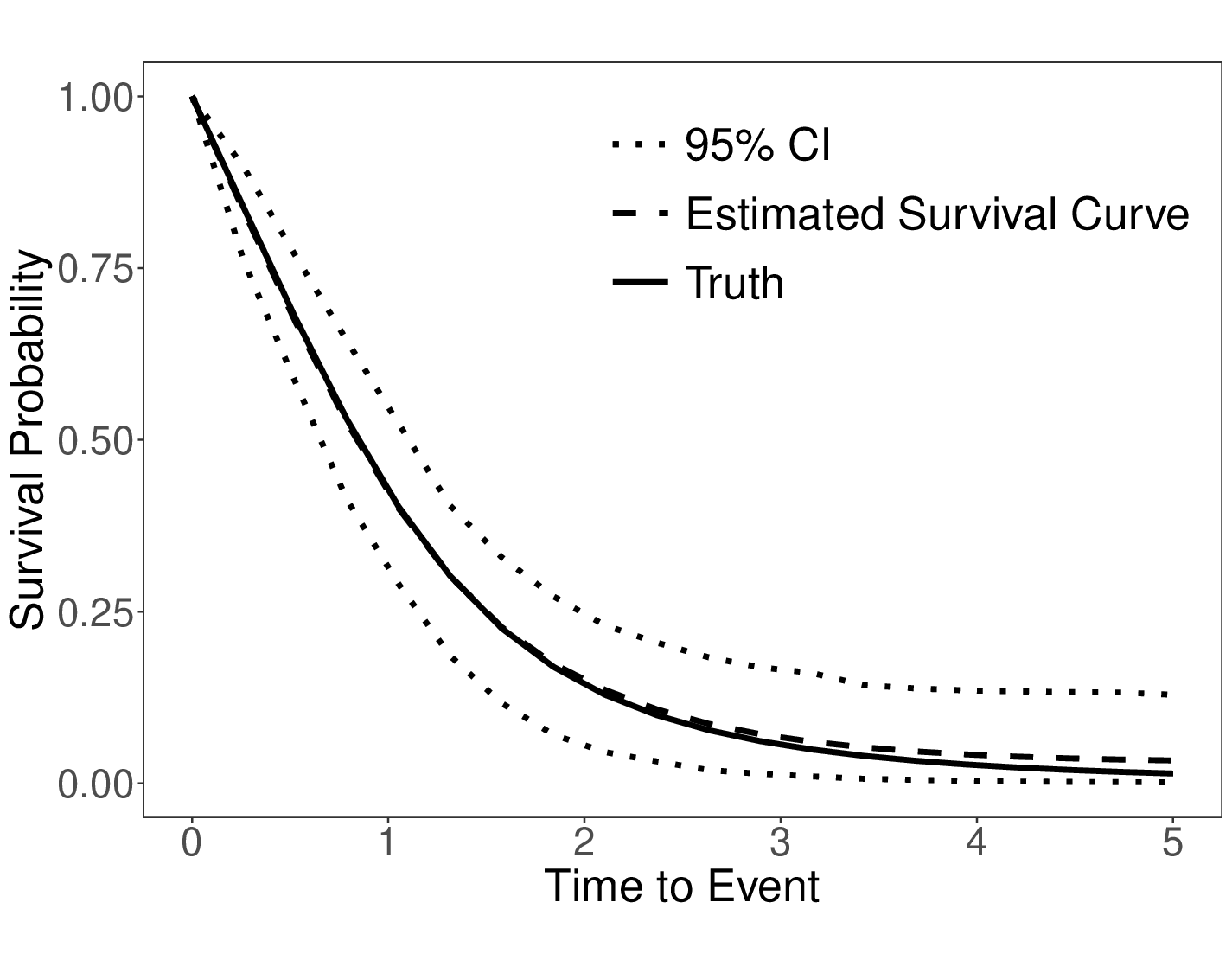}
        \caption{Set C}
    \end{subfigure}
    \caption{Estimated baseline survival functions of Scenario 3 with $n=500$ and $r=1$, using three sets of initial values. % (a) ASE distribution across spline orders 2, 3, and 4. (b) ASE distribution across knot numbers 6, 7, 8, and 9. The results are based on 200 replications under the setting $n=200$, Scenario 3, $r=1$.
    }
    \label{fig2: initial values}
\end{figure}

\end{appendices}

%\section{Reference}
\bibliographystyle{apalike}
\bibliography{smci.bib} 

\end{document}